%% file: forbidden.tex
\documentclass{aa}  
\newcommand{\macc}{$ \dot{M}_{acc}$}

\newcommand{\sii}{[\ion{S}{ii}]}
\newcommand{\sfour}{[\ion{S}{ii}]\,406 nm}
\newcommand{\ssix}{[\ion{S}{ii}]\,673 nm}

\newcommand{\oi}{[\ion{O}{i}]}
\newcommand{\osix}{[\ion{O}{i}]\,630 nm}
\newcommand{\ofive}{[\ion{O}{i}]\,557 nm}

\newcommand{\nsix}{[\ion{N}{ii}]\,658 nm}

\newcommand{\kms}{km\,s$^{-1}$}
\newcommand{\um}{$\mu$m}

\newcommand{\lsun}{L$_{\odot}$}
\newcommand{\msun}{M$_{\odot}$}

\newcommand{\msunyr}{M$_{\odot}$\,yr$^{-1}$}
\newcommand{\rstar}{$R_{\mathrm{*}}$}
\newcommand{\lstar}{$L_{\mathrm{*}}$}
\newcommand{\mstar}{$M_{\mathrm{*}}$}
\newcommand{\teff}{$T_\mathrm{eff}$}
\newcommand{\lacc}{$L_{\mathrm{acc}}$}

\newcommand{\mloss}{$ \dot{M}_{\mathrm{loss}}$}
\newcommand{\mwind}{$ \dot{M}_{\mathrm{wind}}$}
\newcommand{\mjet}{$ \dot{M}_{\mathrm{jet}}$}

\newcommand{\cmt}{cm$^{-3}$}

\usepackage{graphicx}
\usepackage{longtable,lscape,amssymb}
\usepackage{multicol}
\usepackage{color}
\usepackage{supertabular}

\usepackage{lipsum}
\usepackage{placeins}
\usepackage{txfonts}
\begin{document}

   \title{GIARPS High-resolution Observations of T Tauri stars (GHOsT)\thanks{Based on observations made with the Italian {\it Telescopio Nazionale
Galileo} (TNG) operated by the {\it Fundaci\'on Galileo Galilei} (FGG) of the {\it Istituto Nazionale di Astrofisica} (INAF) at the {\it  Observatorio del Roque de los Muchachos} (La Palma, Canary Islands, Spain).}}

   \subtitle{V. New insights into disk winds from 3 \kms\, resolution spectroscopy}

   \author{Brunella Nisini\inst{1}
          \and Manuele Gangi \inst{1,2}
          \and Teresa Giannini \inst{1}
          \and Simone Antoniucci \inst{1}
          \and Katia Biazzo    \inst{1}
          \and Antonio Frasca \inst{3}
        \and \\ Juan M. Alcal\'a  \inst{4}
        \and Carlo F. Manara   \inst{5}   
        \and Michael L. Weber   \inst{6,7}
        }

   \institute{INAF - Osservatorio Astronomico di Roma, Via Frascati 33, I-00078 Monte Porzio Catone, Italy \\ \email{brunella.nisini@inaf.it}
    \and ASI, Italian Space Agency, Via del Politecnico snc, 00133 Rome, Italy 
    \and INAF - INAF-Osservatorio Astrofisico di Catania, via S. Sofia 78, 95123, Catania, Italy
        \and  INAF - Osservatorio Astronomico di Capodimonte - Salita Moiariello 16, 80131 Napoli, Italy 
        \and  European Southern Observatory, Karl-Schwarzschild-Strasse 2, 85748 Garching bei M\"unchen, Germany 
        \and University Observatory, Faculty of Physics, Ludwig-Maximilians-Universit\"at M\"unchen, Scheinerstr. 1, 81679 Munich, Germany
        \and Excellence Cluster 'Origins', Boltzmannstr. 2, 85748 Garching, Germany\\
             }

   \date{Received XX XX, XXXX; accepted XX XX, XXXX}

 
  \abstract
   {}
   {This paper aims to revisit the kinematical and physical properties of the warm ($T \sim$ 5000-10\ 000 K) atomic gas in the inner disk ($<5$ au) region of classical T Tauri stars (CTTs) and relate them to the properties of the outer dusty disk resolved with ALMA. We also want to define constraints for the mass-loss in the inner atomic winds and jets to assess their role in the evolution and dispersal of planet-forming disks.}
   {We used the high resolution (R=115,000, $\sim$ 2.6 \kms) spectra of 36 CTTs observed as part of the GIARPS High-resolution Observations of T Tauri stars (GHOsT) project and analysed the profile and luminosity of the brightest optical forbidden lines, namely \osix , \ofive , \sfour , \ssix , and \nsix . }
   {We decomposed the line profiles into different velocity components, and concentrated our analysis mostly on the so-called narrow low-velocity component (NLVC). 
    We find that about 40 $\%$ of sources display a NLVC peak velocity (V$_p$) compatible with the stellar velocity. 
    These include the transitional disks (TD) and typically show a single low velocity component (LVC), lower mass accretion rates, and the absence of a jet. They therefore might represent later evolutionary stages where the emission from the disk is dominant with respect to the wind contribution. No difference in kinematical properties was instead found between sources with full disks and disks with substructures as resolved by ALMA. 
    The \osix\ profiles peaking at the stellar velocity are well fitted by a simple Keplerian disk model, where the emission line region extends from $\sim$ 0.01 au up to several tens of au in some cases. 
    The \oi\ emission is detected inside the sub-millimetre dust cavities of all the TDs. No correlation is found between $R_{kep}$, derived  from the line half width at half maximum (HWHM), and the size of the dust cavity. 
    We see an anti-correlation between the \oi\ 557/630 nm ratio and $R_{kep}$, which suggests that the \oi\ emitting region expands as the gas dominating the emission cools and becomes less dense. 
    We confirmed previous findings that the line ratios observed in the LVC, if compared with a thermal single temperature and density model, imply $n_e \sim 10^6- 10^8$ \cmt\, and $T_e \sim$ 5000 - 10\,000 K, and additionally constrained the ionisation fraction in the NLVC to be $x_e < 0.1$. We however discuss the limits of applying this diagnostic to winds that are not spatially resolved.} 
   {The emission from the disk should be considered as an important contribution to the forbidden line emission in CTTs. Also, the clearing of warm atomic gas from the upper disk layers does not seem to follow the dispersal of the bulk of molecular gas and dust during late disk evolution. 
   For the outflow component, we estimated the mass-loss for both the disk winds and jets. We conclude that without better knowledge of the wind geometry and spatial extent, and given the limitation of the diagnostics, the mass-loss rates in the wind traced by the blue shifted LVC cannot be constrained better than a factor of 100, with a \mwind /\macc\ spanning between $\sim$ 0.01 and more than 1. 
   When compared with synthetic \osix\, images of X-ray photoevaporation models, the estimated \mwind\, represents a lower limit to the total mass-loss rate of the model, indicating that \osix\, is likely not the best tracer to probe mass-loss in low-velocity winds.}   

\maketitle

\section{Introduction}

Mass outflows represent a fundamental ingredient for  understanding the accretion process in young stellar objects (YSOs) and the evolution of proto-planetary disks (PPD). Jets and winds ejected from the disk at different spatial scales remove both mass and angular momentum, and thus they have a fundamental role in driving accretion and setting the timescale of disk dispersal \citep{pascucci2022}. 

Ejections of matter from the inner part of the disks in the form of extended collimated jets have been studied for decades. Their role in extracting angular momentum and thus driving accretion from the disk to the central star is now well recognised \citep[][and references therein]{frank2014,ferreira06}. 
Recent attention has been given to the large influence that disk winds, launched from disk radial distances $>$ 0.5 au, might have on the evolution and dispersal of PPD. 
In particular, the classical paradigm of viscous accretion has been recently questioned due to the fact that both theoretical simulations and observations show that the level of disk turbulence is not high enough to support viscous disk evolution \citep[e.g.][]{turner2014, flaherty2020}.  
This has led to the recognition that radially extended magnetised winds could be the dominant mechanism driving accretion in planet-forming disks by extracting angular momentum from the disk. Consequently, several theoretical works in recent years have developed wind-driven accretion and disk evolution models \citep{bai2016, suzuki2016, lesur2021, tabone2022}.

In addition to magnetised winds, the importance of photo-evaporative (PE) winds, i.e. winds thermally driven by stellar energetic photons, has also been recognised in late disk evolution. In particular, PE winds are so far the main  
mechanism able to explain the very rapid disk dispersal observed at later stages, when the wind mass-loss rates might become comparable to the disk accretion rates \citep[e.g.][]{alexander14,ercolanopascucci2017}.   
Both type of winds, that is,  magneto hydro dynamic (MHD) and PE winds, might compete with each other in driving the evolution and shaping the disk surface density, and thus also influencing the formation and migration of planets \citep[e.g.][]{suzuki2016, kunitomo2020}. It is therefore clear that determining the properties of disk winds is a critical step towards understanding disk evolution and planetary systems' formation. 

From an observational point of view, atomic forbidden lines in a wide spectral range, from UV to IR wavelengths, represent the brightest and easiest way to access tracers of both collimated jets and radially extended disk winds. These lines have been widely used as diagnostic tools to determine the main physical parameters of collimated jets and infer the mechanism for their launching, as well as their role in mass and angular momentum extraction \citep[e.g.][]{frank2014,pascucci2022}. 
From these previous studies it appears that magneto-centrifugal disk wind models are more suitable to explain the kinematical and dynamical properties of such jets \citep[e.g.][]{ferreira06}, including the $\dot{M}_{\mathrm{jet}}$/$\dot{M}_{\mathrm{acc}}$ ratio between $\sim$ 0.01-0.1 so far derived in large samples of YSOs \citep[e.g.][]{nisini18}. 

Low-velocity disk winds are more elusive, as they are not spatially resolved by the current instrumentation. Their presence and properties have thus far mainly investigated through high resolution spectroscopy, able to kinematically isolate, in the forbidden line emission profiles, the contribution from the slow winds from that of the high-velocity jets. Indeed, since the original spectroscopic works by \cite{edwards13} and \cite{hartigan95}, it has been common practice to decompose the profile of bright lines, such as the \oi\, at 630nm, in a high-velocity component (HVC), which is usually blueshifted with peak velocities V$_p$ larger than 30 \kms , and a low-velocity component (LVC), which presents V$_p$ not exceeding a few \kms . 
As demonstrated by spatially resolved observations, the HVC is associated with the extended collimated jets, while the LVC appears spatially confined on scales of a few tens of au \citep{whelan2021}. Spectroscopic observations at resolution $\sim$ 7 \kms\, have further shown that the LVC itself presents a composite profile that can be assimilated to a superposition of two components, called, following \cite{rigliaco2013}, broad LVC (BLVC) and narrow LVC (NLVC) because of their different widths. 
Further studies conducted on larger samples of classical T Tauri stars (CTTs) suggest that the BLVC and NLVC originate from distinct regions in the disk and that the presence of both components is preferentially found in highly accreting sources, while more evolved disks (for example transitional disks, TD), usually present a single narrow component \citep{simon2016, banzatti2019,mcginnis2018}. Given that the BLVC is mainly observed in conjunction with the HVC, it has been suggested that it is probably associated with inner MHD winds that feed the high-velocity jets \citep{banzatti2019}.  Less clear is the origin of the NLVC, for which both photoevaporative and MHD winds have been proposed to explain its properties \citep{fang18}. Based on the above evidences, \cite{banzatti2019} proposed an evolutionary trend in the line profiles, where the \osix\, emission region recedes to larger radii as the inner disk region becomes depleted. 

The excitation and physical properties of the LVC have been studied in some details by \cite{natta14} and \cite{fang18} through an analysis of the \osix, \ofive\, and \sfour\, line ratios. They derive that the observed ratios are best explained by a thermally excited and neutral gas with electron density $\sim$ 10$^7$-10$^8$ \cmt\, and temperatures $\sim$ 5000-10\,000 K. \cite{fang18} further show that the BLVC and NLVC broadly share similar conditions, although it is very difficult to tightly constrain the physical parameters. 
Even more difficult is to set constraints on the mass-loss rate, which requires the knowledge not only of the physical conditions in the wind but also of its geometry. \cite{natta14}, adopting very crude approximations on the wind geometry, estimates \mwind\, $\approx$ 0.1- 1 \macc . \cite{fang18}, thanks to the higher resolution of their spectra, provide constraints on the \mwind\, in both the NLV and BLV components. They conclude that the mass-loss rate in the BLVC dominates over that of the NLVC, and might be of the same order as the mass accretion rates, hence significantly higher than the HVC and thus jet rates.

In this paper, we aim at further investigating the kinematical and physical properties of the forbidden lines in PPD, and the LVC in particular, through observations at unprecedented spectral resolution ($\sim$ 2.6 \kms). 
For that, we use the database of high resolution optical spectra acquired with the GIARPS instrument at the Telescopio Nazionale Galileo in the framework of the GHOsT (GIARPS High-resolution Observations of T Tauri stars) project \citep[e.g.][Paper III and IV]{alcala2021,Gangi22}, a flux-limited survey of T Tauri Stars in the Taurus-Auriga star forming region. We have already used sub-samples of data from the GHOsT project to address two specific topics related to the mass-loss in young stars. In a first paper \citep[][Paper I]{giannini2019}, a pilot study was performed on five sources known to drive energetic jets in order to constrain the physical parameters of mass ejection as a function of velocity. A second study \citep[][Paper II]{gangi20} was devoted to the comparison of the kinematical properties between the atomic NLVC traced by the \osix\, line, and the molecular winds traced by the H$_2$ 2.12$\mu$m transitions.
The present paper analyses a set of optical forbidden lines from oxygen, sulphur and nitrogen species in the total GHOsT sample of 36 sources. Moreover, this study differs from previous high resolution surveys performed on similar sources \cite[e.g.][]{simon2016,banzatti2019,mcginnis2018,fang18} in the following major ways: i) it is based on data with a spectral resolution higher by at least a factor of two with respect to previous studies, and thus allowing us to better disentangle the different kinematical components; ii) stellar and accretion parameters for the sample have been simultaneously and self-consistently determined from the same dataset; iii) for most of the considered sources their disks have been resolved by ALMA, providing the basic information for seeking possible links between the inner gaseous disk, traced by the forbidden lines, and the properties of the outer disks. 

\section{Sample}\label{sect::sample}

In this work, we analyse the forbidden line emission in the sample of YSOs of the GHOsT project. The sample has been observed with the GIARPS 
instrument that combines the HARPS-N \citep{cosentino2012} and GIANO-B \citep{origlia2014} high-resolution (115\,000 and 50\,000, respectively) spectrographs, simultaneously covering a wide spectral range of 390–690 nm for HARPS-N, and 940–2420 nm for GIANO-B. The GIARPS observations have been complemented with contemporaneous low-resolution spectroscopy and photometry, that have been used to flux calibrate the GIARPS spectra. All the details about the observing strategy and data reduction are fully described in Paper IV and will not be repeated here.
The sample consists of 36 CTTs showing IR excess from the list 
of Taurus-Auriga members of \cite{esplin2014}. The source selection was driven by the sensitivity limits of the GIARPS instrument and resulted in a sample covering masses between 0.2 and 2.2 \msun\, and luminosities between 0.05 and 8.9 \lsun\ . The stellar (\lstar, \mstar, \rstar) and accretion (\lacc, \macc) properties of the sources were consistently derived from the GHOsT flux-calibrated data \citep[][Paper III and IV]{alcala2021,Gangi22} and are listed in Table \ref{tab:sources_param}, together with other information taken from the literature. In particular, the disks of most of these sources have been spatially resolved by ALMA or other high resolution millimetre facilities, and Table \ref{tab:sources_param} reports the inferred disk properties that will be used in the analysis. Based on the disk structure as observed in the ALMA images, disks have been divided into sub-structured (showing dust gaps and rings), transitional (TD, with an inner dust cavity) or full (uniform distribution at the resolution of the observations). We point out that this nomenclature refers only to the disk morphology as revealed at mm wavelengths, which is not necessarily associated with an evolutionary stage of the object. For example with this nomenclature we consider as TD also RY Tau, which is a highly accreting object with a jet, and GG Tau A, whose inner cavity is related to the binary nature of the source. 
The disk of a few of our sources have been also imaged with SPHERE in scattered light, \citep[e.g.][]{menard2020,mesa2022,garufiA&A...628A..68G,keppler2020}, showing additional structures at large scale, such as spirals, streamers and flaring that we consider in the discussion of individual objects.  

\input{Tables/table1}  
\input{Tables/table2.tex}
\section{Spectra of forbidden lines}\label{sect::spectra}

The GIARPS spectral range covers several forbidden lines from abundant atoms/ions \citep[see e.g.][]{giannini2019}. In this paper we consider only the five optical lines most relevant for the analysis and observed with the highest signal-to-noise ratio (S/N) in the HARPS-N spectra, which are listed in Table \ref{tab:lines}. In particular, GIANO IR forbidden lines with high enough S/N have been detected only in the five bright targets analysed by \cite{giannini2019} and will not be considered here.
We corrected the forbidden line profiles for the contribution of the absorption features due to the stellar photosphere, using HARPS spectra of main sequence stars of the same SpT as the sources \citep[see Table C.1 of][]{manara21}. In particular, we divided the source spectrum around the line of interest by the photospheric template, corrected for both the vsin(i) broadening and veiling $r_{\lambda}$ estimated in \cite{Gangi22}. 

Finally, to transform the spectra to the stellar velocity reference frame, we have corrected them by the stellar radial velocity (RV). This latter has been derived by cross-correlating each star optical spectrum with the synthetic BTSettl model \citep{allard2012} that best
matched the \teff\ derived in 
\cite{Gangi22}, adopting the ROTFIT tool \citep{frasca2006}. 
The cross-correlation is carried out separately for 14 spectral segments of 200\,\AA\, each from 4000\,\AA, to 6800\,\AA, after masking the emission lines and the regions strongly affected by telluric absorption lines. The model spectrum is degraded to the spectral resolution of HARPS-N and resampled to the spectral points of the target spectrum. To measure the centroid of the cross-correlation peak, we fitted it with a Gaussian  by the procedure \textsc{curvefit} \citep{Bevington}, taking the cross-correlation function (CCF) noise, $\sigma_{\rm CCF}$, into account.  The latter was evaluated as the standard deviation of the CCF values in two windows at the two sides of the peak.
The RV error per each segment, $\sigma_{\rm RV}$, was estimated as the error of the centre of the Gaussian fitted to the CCF \citep[see also][]{Frasca2017}.   
Table \ref{tab:sources_param} provides the stellar RV and associated uncertainty computed as the weighted mean and weighted standard deviation of the RVs obtained from individual segments, adopting $w=1/\sigma_{\rm RV}^2$ as the weight. The RV uncertainties typically range between 0.2 and up to 2 \kms\ for sources whose spectra are highly affected by emission lines and the depth of the photospheric absorption lines strongly reduced by the veiling. The method could not be applied to RW Aur as very few photospheric lines have been detected. For this source, we shifted the spectrum using the weighted $\lambda_{air}$ = 670.7876\,nm of the \ion{Li}{i} photospheric line as reference (Campbell-White et al. 2023).
The resulting corrected profiles are shown in Figure A.1-A.6 of Appendix A. In Table \ref{tab:lines} we indicate the rate of detection of each line . 

\section{Gaussian profile decomposition}

We adopt here the commonly used procedure to fit each spectral profile as the sum of different Gaussian functions.
Models of emission lines in winds and disks show that the resulting profiles can be rarely described as a single Gaussian, being in most cases asymmetric or with double-peaked profiles \citep[e.g.][]{weber2020}. Nevertheless, Gaussian fitting helps in the identification of different kinematical components and allows us to compare our results with those of the recent literature. The limitations of the applicability of the Gaussian fitting will be discussed in the following sections. 

The Gaussian fitting was already applied by \cite{gangi20} on the \osix\, profile on a sub-sample of our sources. We have re-done the profile decomposition in all the sources adopting a slightly different fit procedure. 

\subsection{Decomposition procedure}
The procedure consists of an IDL code based on the $\chi^2$ minimisation with the MPFIT method and allows to perform multi-Gaussian fit of complex line morphology's. The procedure provides, for each Gaussian component, width, peak velocities and peak intensity values from which flux and luminosity are computed. For a given profile, the total number of Gaussians is determined as the minimum number of Gaussians yielding a $\chi^2$ stable within 20\% of its minimum value, a criterion already adopted in previous similar studies \citep[e.g.][]{banzatti2019}. Errors on fit parameters were estimated by following a Gaussian randomisation approach. In short, we simulate for each line profile 10$^4$ datasets having a random Gaussian distribution with central value the observed spectral points and standard deviation the local S/N. Each simulated profile is then decomposed and the errors are determined as the sigma of the fit parameter distributions. The error on the fitted peak velocity has been then added in quadrature to the uncertainty of the stellar RV, and reported in Table \ref{tab:sources_param}.
Our procedure was already successfully applied on both absorption \citep{Gangi2021b} and emission line profiles (Gangi et al. 2023), from the UV to the NIR spectral range. 

\begin{figure*}
\begin{center}
\includegraphics[trim=0 0 0 0, width=2.\columnwidth, angle=0]{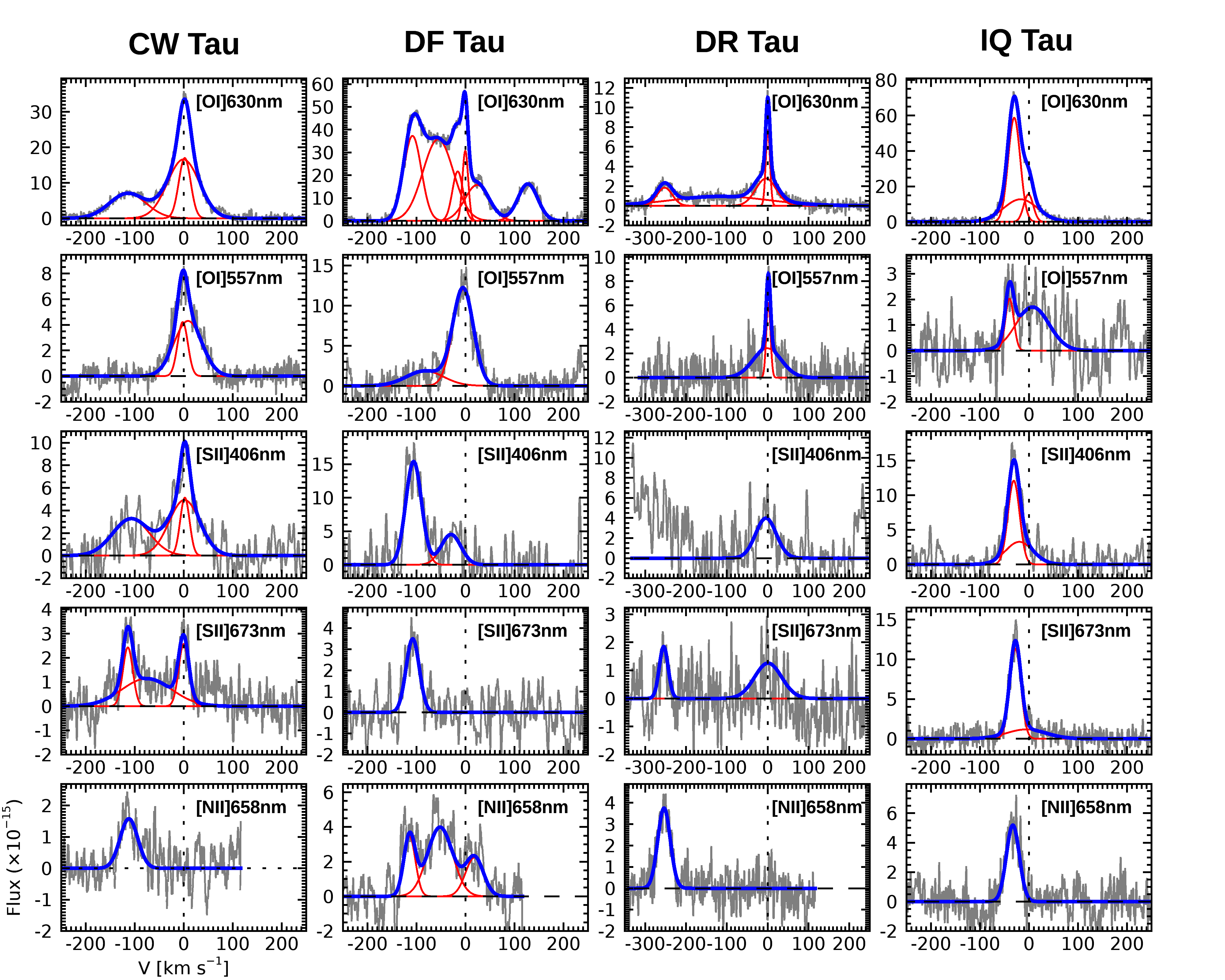}
\end{center}
\caption{Representative continuum-subtracted spectra of the forbidden lines observed in our survey. Flux units are erg\,s$^{-1}$\,cm$^{-2}$\,$\AA^{-1}$. Red lines shows the Gaussian profiles of the individual fitted components, while the blue line shows the combined fitted Gaussians. }
  \label{fig:decompose}
\end{figure*}

In this work, we firstly apply the procedure to the \osix\ line, which is the one observed at highest S/N. Next, we assume the solution of the \osix\ decomposition as the initial input for the fit of the other lines. Fig. \ref{fig:decompose} shows the result of the procedure for some of the sources in which all the considered forbidden lines have been detected. The decomposition for all the other sources is given in Appendix A.
Tables \ref{tab:line_fluxes1} and \ref{tab:line_fluxes2} give the result of the decomposition for the different line, reporting the kinematical information ($V_p$ and FWHM, this latter deconvolved for the instrumental resolution) and the luminosity of each component. This latter has been determined from the extinction corrected integrated flux, using the distance and the A$_V$ value given in Table \ref{tab:sources_param} and applying the extinction law by \cite{cardelli89}.

\subsection{Velocity components}

For the different Gaussian components we adopt the name applied in previous similar studies, i.e.  Narrow Low velocity Component (NLVC) if $\lvert{V_p}\rvert <$ 30 \kms\, and FWHM $\la$ 50 \kms\ , Broad Low velocity Component (BLVC) if $\lvert{V_p}\rvert <$ 30 \kms\, and FWHM $\ga$ 50 \kms\ and High Velocity Component if $\lvert{V_p}\rvert >$ 30 \kms. 

The separation between the HVC and LVC has a physical justification based on the evidence that the HVC is associated with spatially resolved high-velocity jets \citep[e.g.][]{hirth97} and it has excitation conditions different from the LVC, pointing to a physical distinct region. The distinction between the NLVC and BLVC is more subtle and the adopted width value that separates a narrow component from a broad component is somewhat arbitrary. This is even more evident when different lines of the same sources are considered, since a component which is classified as narrow in the \osix\, line can have a larger width in the \sfour\ or \ofive\ lines. 
In Table \ref{tab:line_fluxes1}, we adopt 55 \kms\ as a separation value between a broad and a narrow LVC, which roughly corresponds to the maximum width displayed by \osix\, profiles when a single LVC component is detected. The only exceptions are DS Tau and IP Tau, where we detect a single line having a width of 90 and 160 \kms , respectively. Observations with MUSE in these two sources show indeed that the \osix\ emission is extended and can be associated with a large scale jet \citep{floresrivera2023}. 

\subsection{Profile types of the \osix\, line}
\begin{figure}
\includegraphics[trim=0 0 0 0, width=1\columnwidth, angle=0]{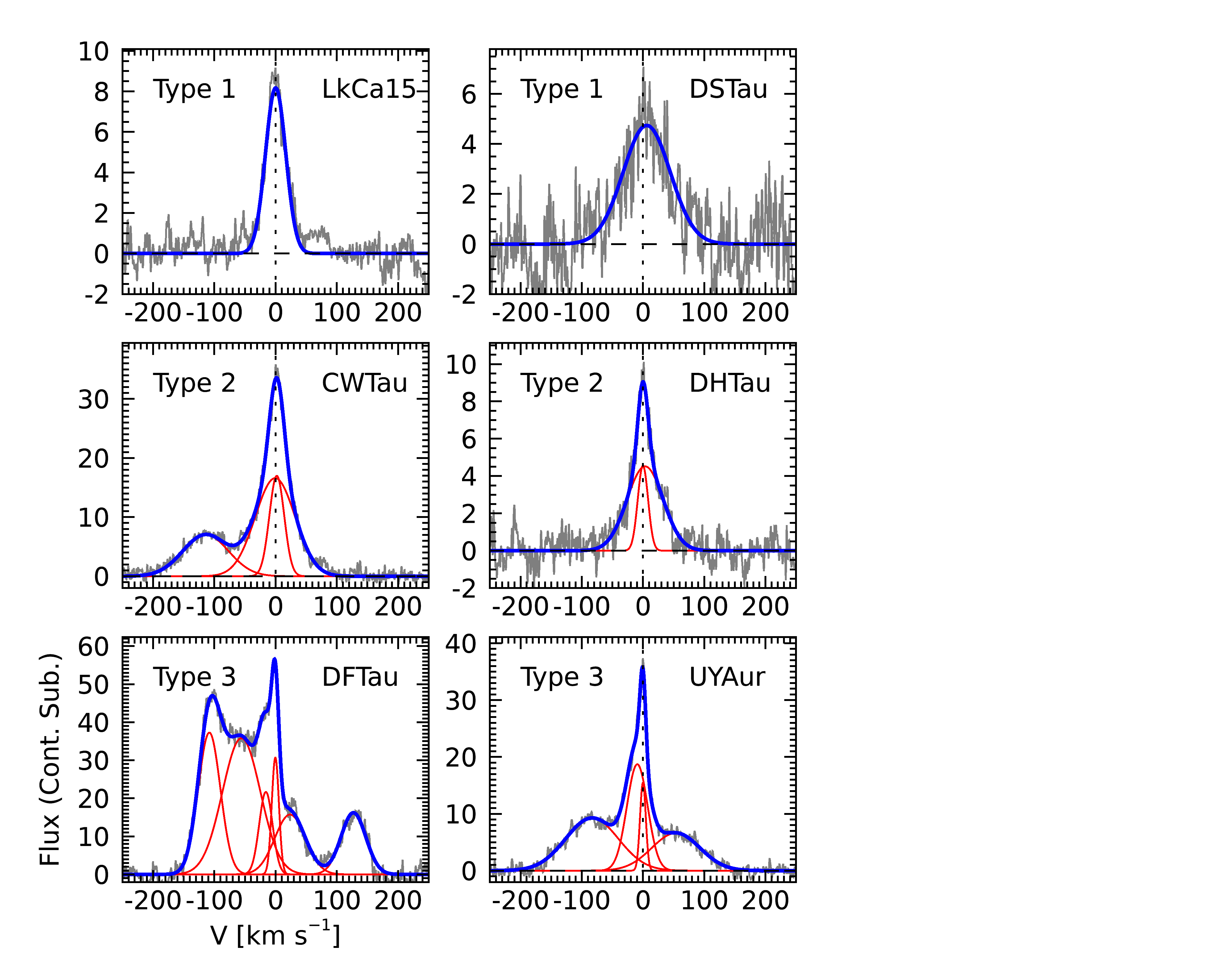}

\caption{Examples of \osix\, profile types for the LVC. Type 1 is a single low velocity profile. Type 2 is a double low velocity profile, that could be fitted either with a double component (one narrow and one broad) or with a Lorentzian profile. In Type 3 profiles, the LVC is composed by more than one component, not necessarily one narrow and one broad, peaking at different velocities.  }
  \label{fig:types}
\end{figure}

The above definitions, while having the advantage of giving a simplified interpretation of the profiles, might not reflect the complex dynamical processes of the different outflow manifestations.
In more general terms, the observed \osix\ line profiles can be classified into three different types: Type 1 - a single component profile peaking at low velocity, such as those shown in Fig.~\ref{fig:types}, upper panel. This is observed in 10 sources (CY Tau, DN Tau, DS Tau, GG Tau, GM Aur, IP Tau, LkCa15, MWC480, V409 Tau and V836 Tau). 
Type 2 - triangular-like profiles whose LVC is decomposed with two Gaussians, one narrow and one broad, peaking at similar, close to zero, velocity. They could be alternatively fitted with a single component having a Lorentzian shape.
They can have or not also a HVC as in the case of CW Tau shown in Fig. ~\ref{fig:types}, middle panel. We remark that emission from a disk extending at small ($<1$ au) radii could in principle take into account this kind of profiles without invoking an additional component, as we show later.  
Type 3 - profiles where the LVC is composed by more than one component, not necessarily one narrow and one broad, peaking at different velocities. 
These represent the majority of the observed profiles.
Here the distinction between a BLVC and a HVC is less compelling, as the separation between the LVC and HVC set at 30 \kms\, is somewhat arbitrary and depends also on the velocity correction for the disk inclination. In addition, at the higher resolution of our observations, some of the profile components that were previously fitted with a single Gaussian appear even more structured, with more than one NLVC at increasing velocity shifts, as in the case of DF Tau, DG Tau, and HN Tau.

\subsection{Comparison between different lines}

The separation between BLVC and NLVC is often more difficult to be recovered in lines different from the \osix .  In particular, sources having both low velocity components in the \osix\, line often have a single profile in the other lines. This is partially due to the lower S/N ratio that prevents to distinguish the two components. In general the LVC is detected less frequently in the [\ion{S}{ii}] lines, and in half of these cases only the BLVC is observed (see Table \ref{tab:lines}). 

\begin{figure*}
\begin{center}
\includegraphics[trim=0 0 0 0,width=1\columnwidth, angle=0]{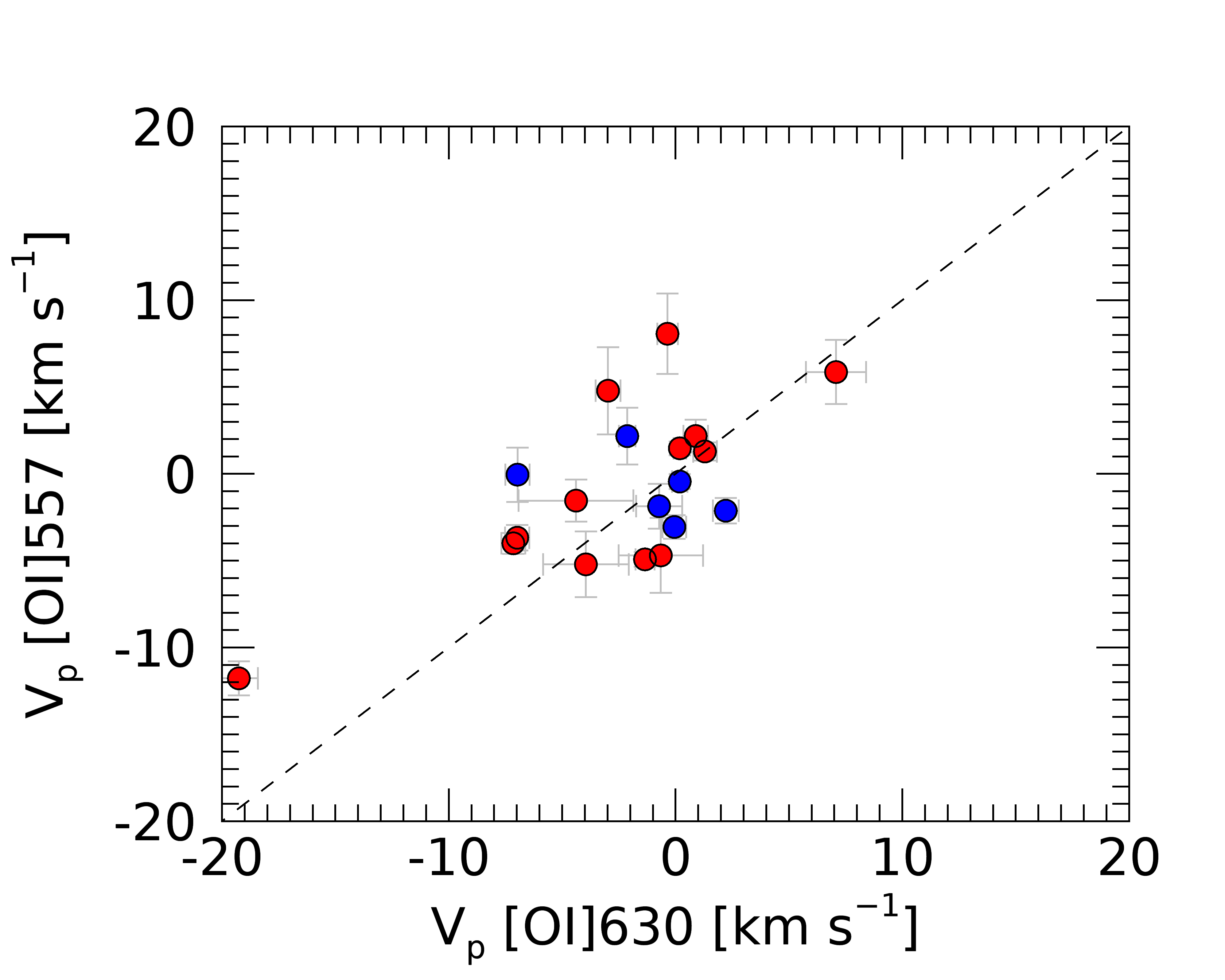}
\includegraphics[trim=0 0 0 0,width=1\columnwidth, angle=0]{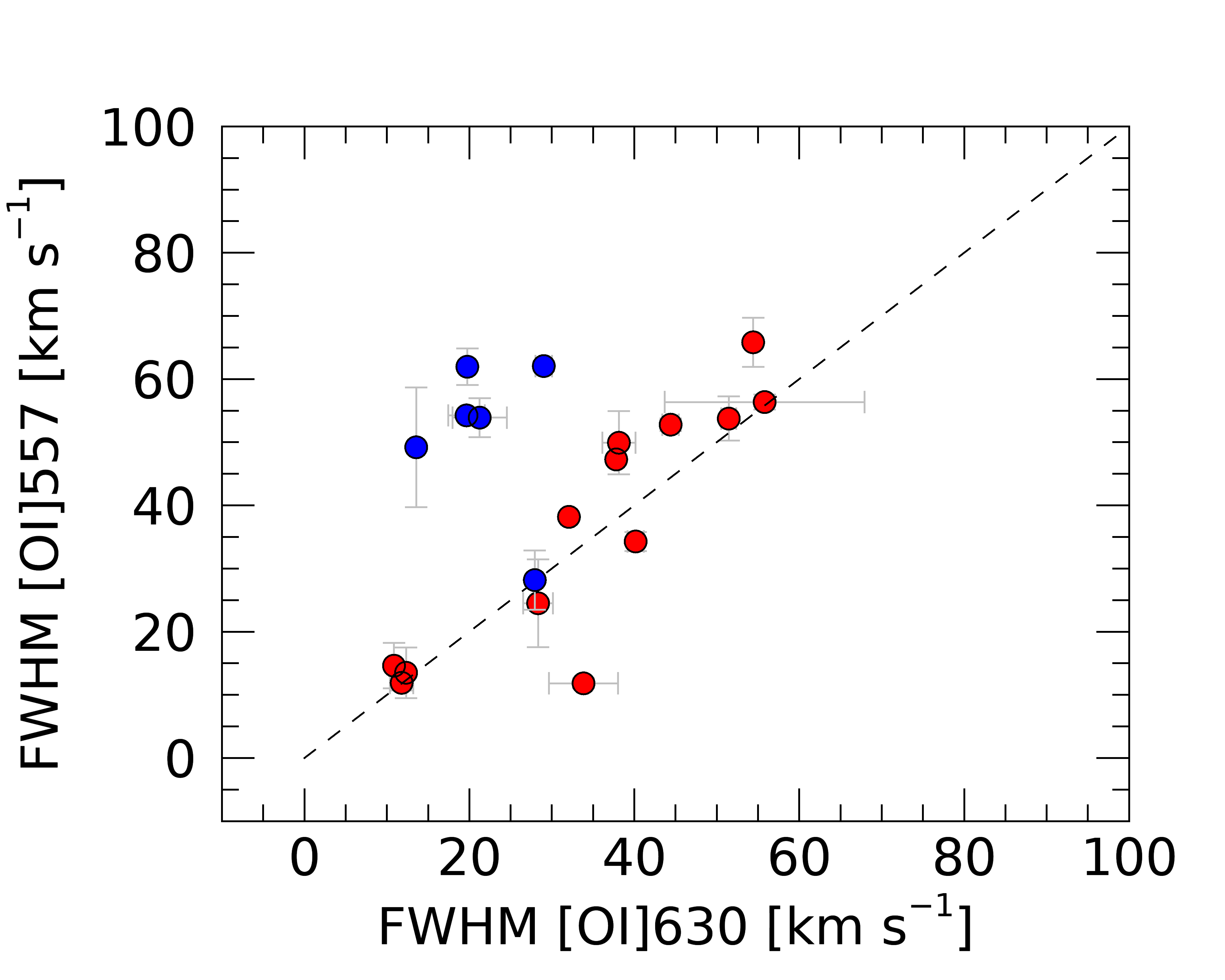}
\\
\includegraphics[trim=0 0 0 0,width=1\columnwidth, angle=0]{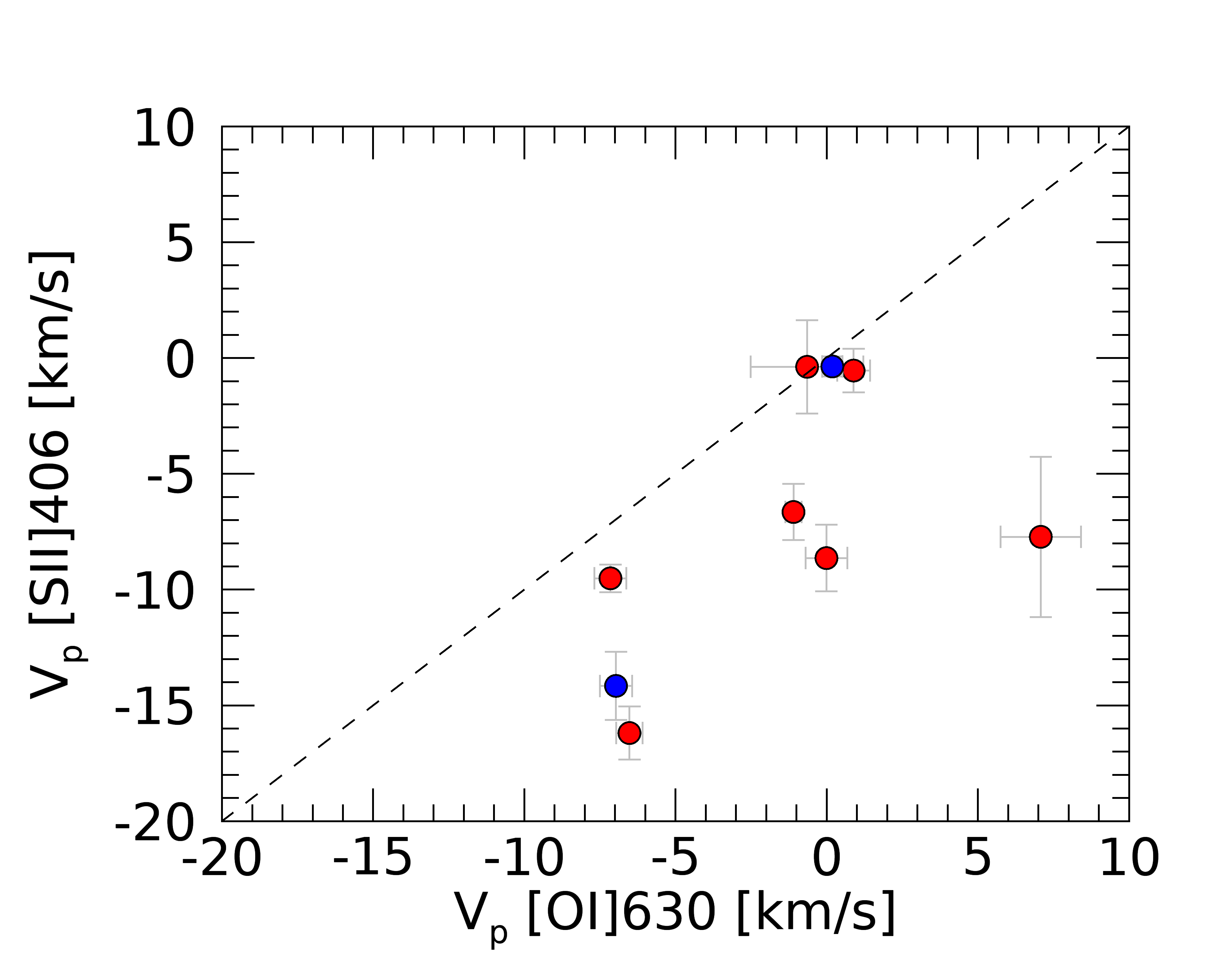}
\includegraphics[trim=0 0 0 0,width=1\columnwidth, angle=0]{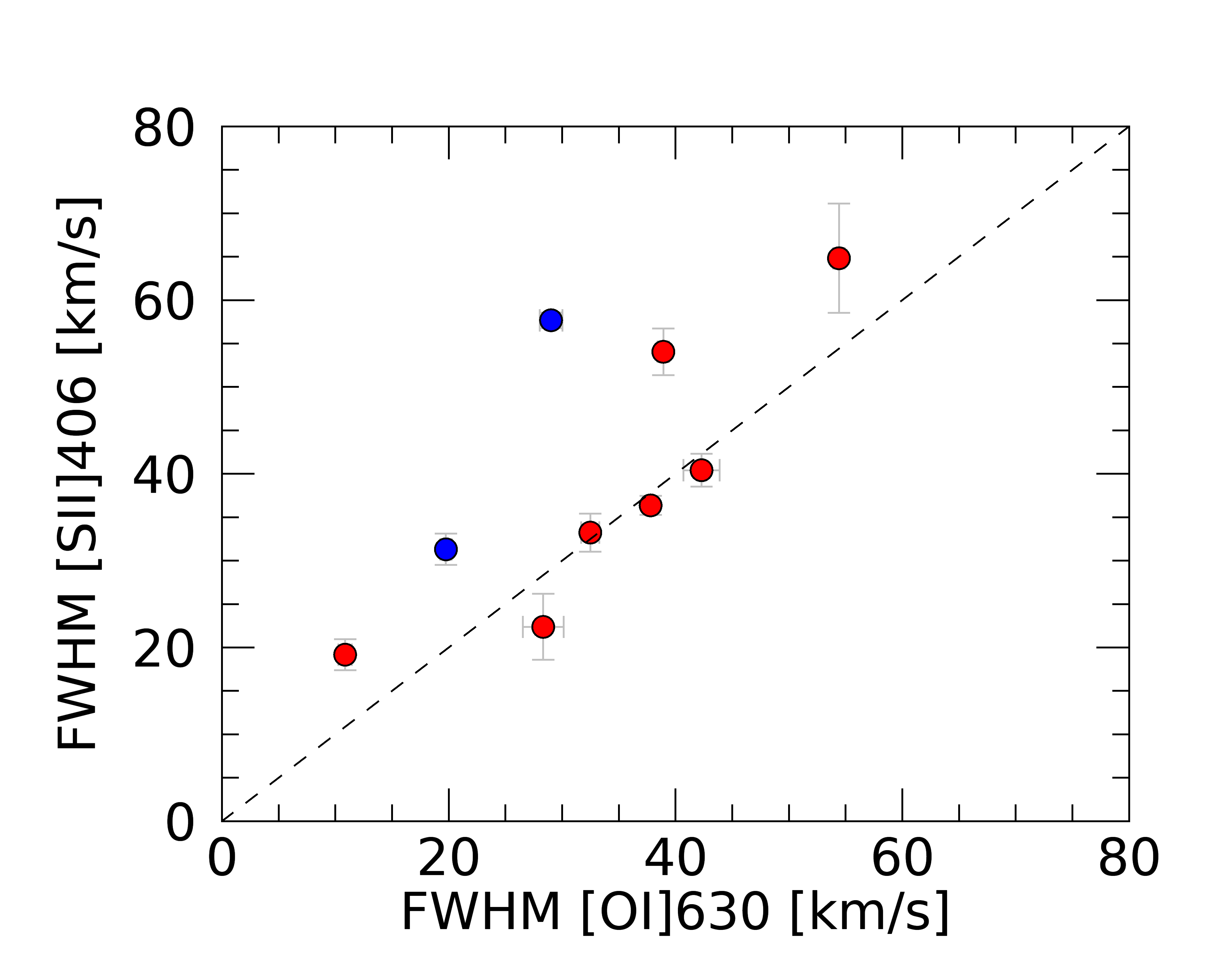}

\end{center}
\caption{Comparison of V$_p$ and deconvolved FWHM between the \osix\, line and the \ofive\,(upper panels) and \sfour\, (bottom panes) lines for the NLVC. Red dots indicate sources where the NLVC has been clearly identified in both lines, while blue dots refer to sources where the LVC in the \osix\, line has been decomposed into a NLVC and a BLVC, while a single component has been identified in the other lines.}
  \label{fig:comp}
\end{figure*}

To better analyse whether the adopted decomposition retrieves the same NLVC components for the three brightest lines, i.e. \osix\ , \ofive\ and \sfour ,  Fig. \ref{fig:comp} shows a comparison of the peak velocities and deconvolved FWHM of the \sfour\, and \ofive\, lines, vs those of the \osix\, line. In these plots, we mark in blue the sources where the \osix\, line has been deconvolved in a NLV and a BLV component while the other lines were fitted with a single component. 
As indicated before, for 9 sources only a LVC has been identified in \sfour . 
Looking at Fig. \ref{fig:VPvsFWHM} we note that their centroid velocities tend to be more blue shifted than the \osix\, line, while the FWHM is wider than that of the \osix\, line in half of the sources. 
For the \ofive\, line, the peak velocity is mostly consistent with that of \osix , with few exceptions where the \ofive\, line is less blue shifted than \osix . 
This latter effect has been already noted \citep[e.g.][]{hartigan95, giannini2019} and interpreted considering the different critical densities of the two lines (see Tab.~\ref{tab:lines}): when the lines originate from an accelerating wind, \ofive , which has a critical density an order of magnitude larger than that of \osix , is less blue shifted as it comes from gas at lower velocity in the inner and denser region of the wind. This interpretation cannot be however applied to the \sfour\, line, which has a critical density similar to that of \osix . 
Fig. \ref{fig:VPvsFWHM} also shows that FWHM(\ofive) is compatible with FWHM(\osix) except for the sources where it was not possible to separate the broad and narrow components from the \ofive\, LVC. 

Regarding the other two considered lines, \nsix\, is detected only in the HVC, as a consequence of the higher ionisation fraction attained in the jets with respect to the neutral gas constituting the LVC \citep[e.g.][]{giannini2019}.
Finally, The \ssix\, line has a critical density an order of magnitude lower than the other lines, and thus intercepts the low density gas in the extended jet better than the LVC high density gas. Hence, its detection in the LVC is also limited (i.e. 14 sources in total).

\section{Kinematical properties of the NLVC}

We concentrate here on the kinematic properties of the NLVC, seeking any link between them and the properties of corresponding sources and their disk. We use the \osix\, for this analysis, as it has the largest number of detections at the highest S/N.

\begin{figure*}
\begin{center}
\includegraphics[trim=0 0 0 0,width=1\columnwidth, angle=0]{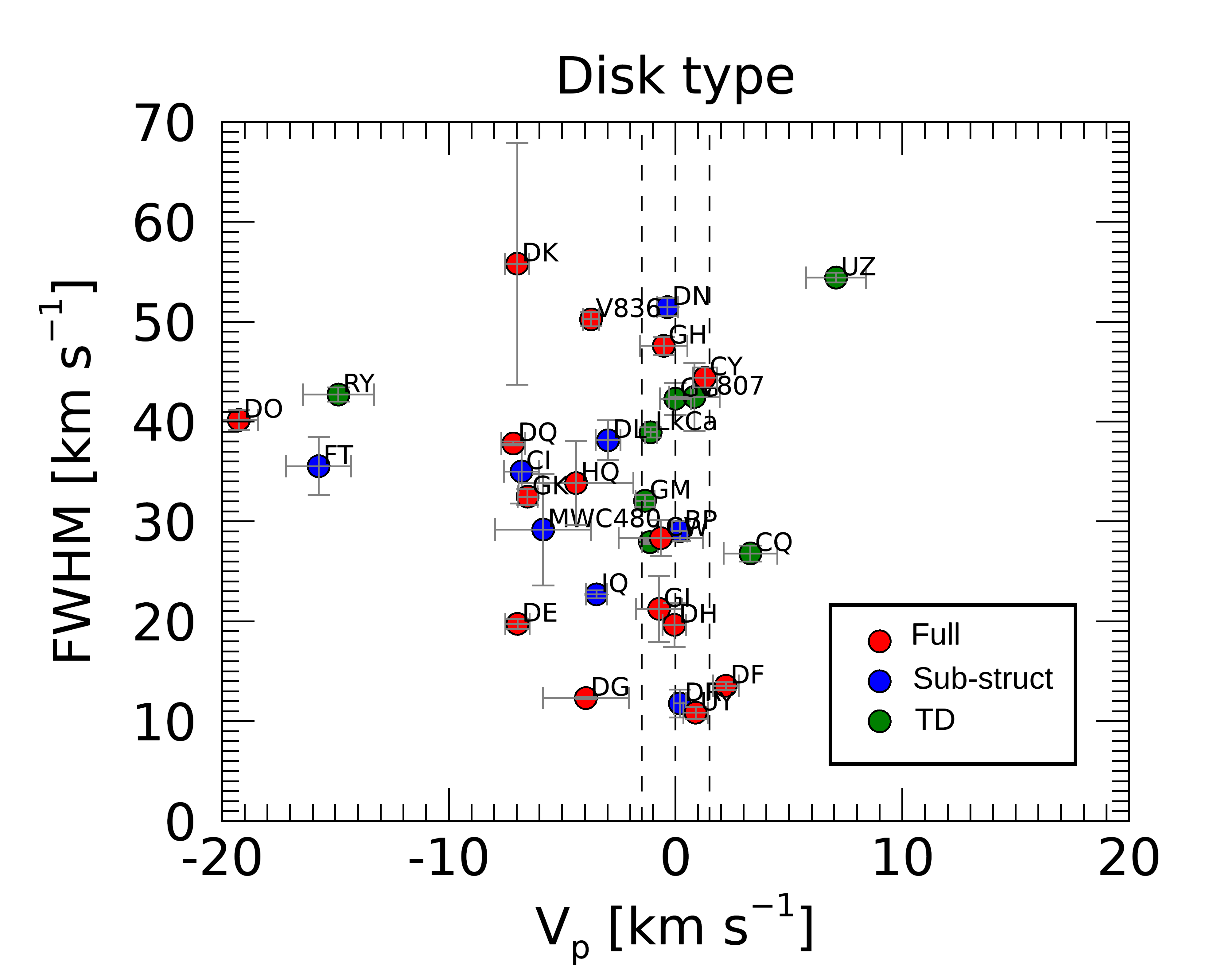}
\includegraphics[trim=0 0 0 0,width=1\columnwidth, angle=0]{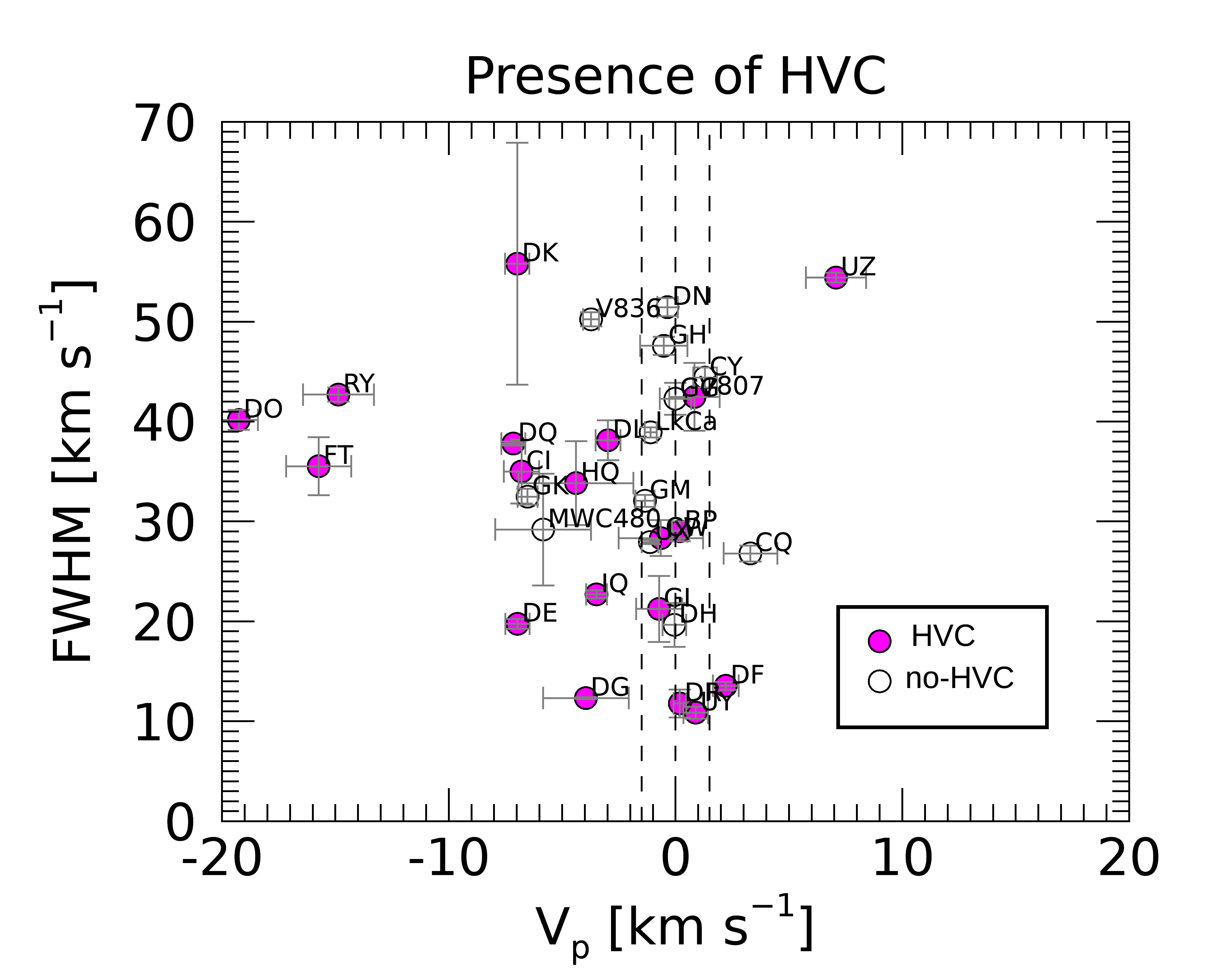}
\\
\includegraphics[trim=0 0 0 0,width=1\columnwidth, angle=0]{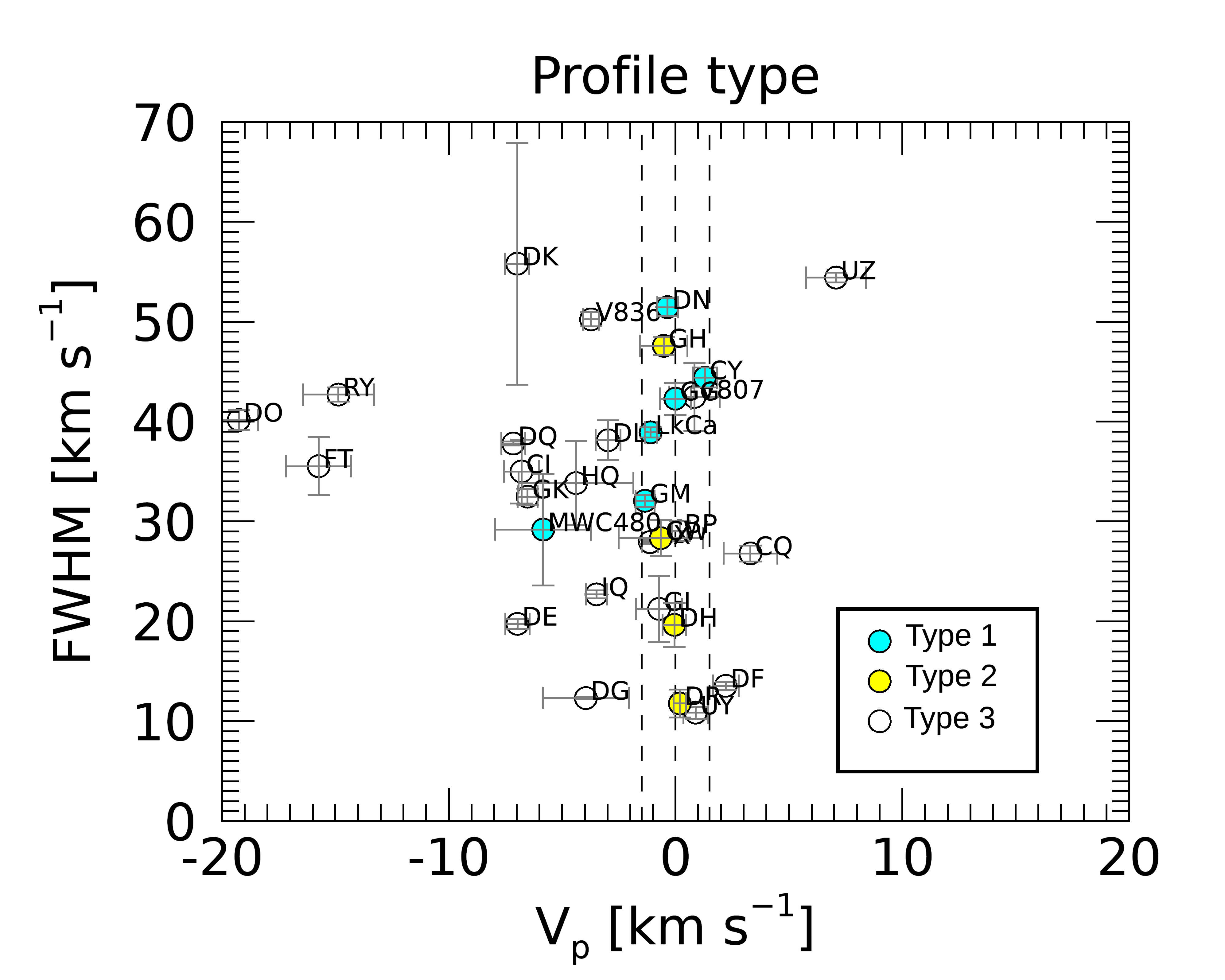}
\includegraphics[trim=0 0 0 0,width=1\columnwidth, angle=0]{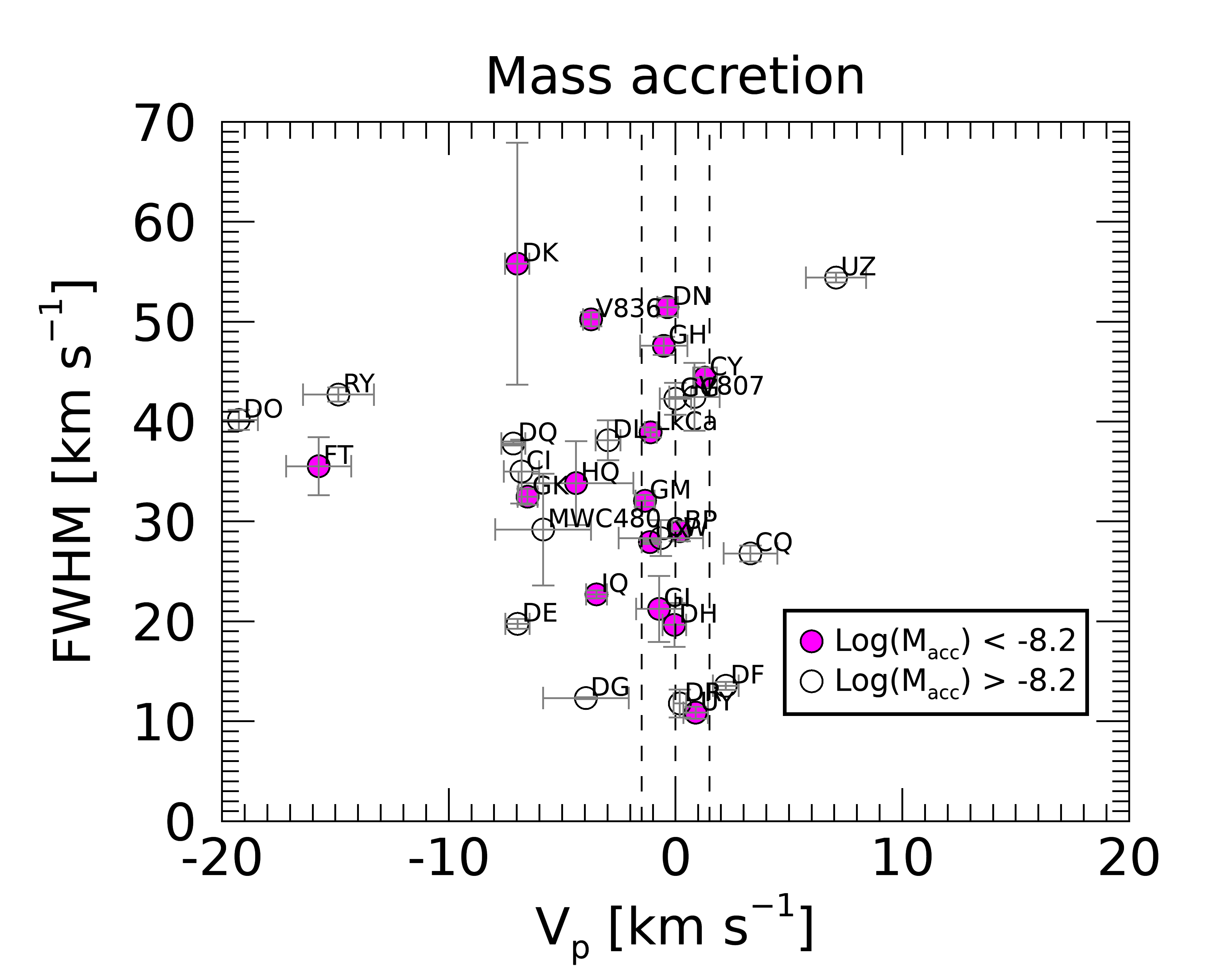}

\end{center}
\caption{Deconvolved FWHM vs line peak velocity $V_p$ for the NLVC. The four panels highlight the distribution of sources with different properties. Upper left: disk type, upper right: presence of a HVC, bottom left: type 1,2 and 3 profiles as described in the caption of Fig. 2, bottom right: value of \macc\, below (filled circles) or above (empty circles) the median value for our sample (Log \macc = -8.2)}
  \label{fig:VPvsFWHM}
\end{figure*}

Fig. \ref{fig:VPvsFWHM} presents the deconvolved FWHM of the line versus its peak velocity. Each of the four panels highlights the distribution of sources with different properties. 
The two vertical dashed lines delimit the maximum uncertainty in velocity centroid attained by our observations. The figure shows that there is no correlation between the line width and line velocity, as already reported in other studies \citep[e.g.][]{banzatti2019,simon2016}. It also shows that in about 15 of the sources, the NLVC $V_p$ is compatible with zero velocity, i.e. for 40\% of the sources of our sample there is no direct evidence that the NLVC originates in a wind. 
Inclination effects have little influence in this trend, as only four of these sources have disk inclined by more than 50 degrees, that would shift their peak velocity by a factor $>$ 1.5. 

Panel (a) shows that full disks and disks with sub-structures are equally distributed in this plot, but noticeably all the known TD sources, with the exception of RY Tau and UZ Tau do not show any appreciable velocity shift. We point out that both RY Tau and UZ Tau are rather atypical TD source as, ad variance with the other TDs, they drive strong jets \citep{agra-amboage2009}.
In addition, panel (b) shows that all the sources with a blue shifted NLVC, but MWC480 and GH Tau, are those with a high-velocity outflow. On the other side, sources with not appreciable velocity shift also comprise several objects with a prominent HVC, such as CW Tau.  From panel (c) we also see that all the type 1 and 2 profiles, i.e. single LVC profiles, or profiles having a triangular-like shape, fall in the category of lines with no-velocity shifts.  
Finally, in panel (d) we distinguish high and low accretors (i.e. sources with a mass accretion rate above or below the median value of the \macc\ = log (\macc) $< 8.2$ \msunyr). It can be noted that on average sources with low $V_{peak}$ have a lower mass accretion rate with respect to the rest of the sample. 

In conclusion, we derive that the NLVC with no-velocity shifts are also those having single peaked or Lorentzian profiles and lower mass accretion rates, and can be preferentially found among TD sources or sources with no high-velocity outflows. Full disks or disks with sub-structures do not show any clear segregation in the plots.  

This result supports the view, proposed by \cite{banzatti2019} and recently confirmed by \cite{fang2023} on a survey of YSOs of the 5-10 Myr old Upper Sco population, that as disks evolve and accretion starts to settle down, the \osix\, profile tends to have 
a single narrow component while the peak velocities move from being mainly blue shifted to being closer to the stellocentric velocity.  

The evidence that in more evolved disks the NLVC has peak centroids compatible with zero velocity implies that in these systems the emission mostly originates from gas bounded in the disk or from a very low velocity disk wind, poorly accelerated above the disk surface. 
Velocity shifts less than $\sim$ 3 \kms\, are accounted for by the X-ray PE wind models of \cite{ercolano10} only for low inclination angles. More symmetric profiles with respect to zero velocity are predicted for disks with dust inner cavities such as in TD. In these case the wind is seen in both sides within the cavity as its red shifted emission is not blocked by the dusty disk.
According to the \cite{ercolano10} simulations however, this effect is appreciable only for sufficiently large inner holes (i.e. larger than about 8 au) and disk inclinations below about 50 degrees, which would not explain the symmetric profiles observed in TD with small inner dust holes or in full disks imaged at high resolution by ALMA \citep[e.g. DR Tau or DN Tau,][]{long2019}. 

\cite{ballabio2020} compared their analytical PE model with kinematical properties of the \osix\, emission  concluding that observed low V centroids can be reproduced only assuming sound speeds around 3-5 \kms . These are however inconsistent with the observed FWHM which are better reproduced with larger sound speeds of the order of 10 \kms .

Finally, also in MHD winds the velocity centroid is significantly more blue shifted except for favourable wind inclinations \citep{weber2020}. 

Therefore wind models are not able to account for the large number of sources with peak velocities close to zero and different disk inclination angles. 
This might suggest that as the inner disk gas evolves, the contribution of the emission from bound gas becomes dominant with respect to wind emission. In such a case, the emission region should extend to large disk radii, to explain the lack of double peaks in our spectra. This will be further discussed in Section 7.2.

\section{Comparison with previous studies}

\citet{banzatti2019} analysed the kinematical properties of the \osix\, line in 22 out of the 36 sources of our sample, while \citet{fang18} conducted an analysis of the different components of 8 of our sources based on the \osix , \ofive\, and \sfour\, lines. 
With respect to these previous studies, our observations are conducted at higher resolution (i.e. 2.5 \kms\, against 6 \kms) allowing one to better isolate the different velocity components and identify winds at lower velocities. In addition, through a comparison with the previous observations, we can analyse any kinematical variability on a temporal scale of a decade or more.

\begin{figure*}[!t]
\includegraphics[trim=0 0 0 0,width=1.5\columnwidth, angle=0]{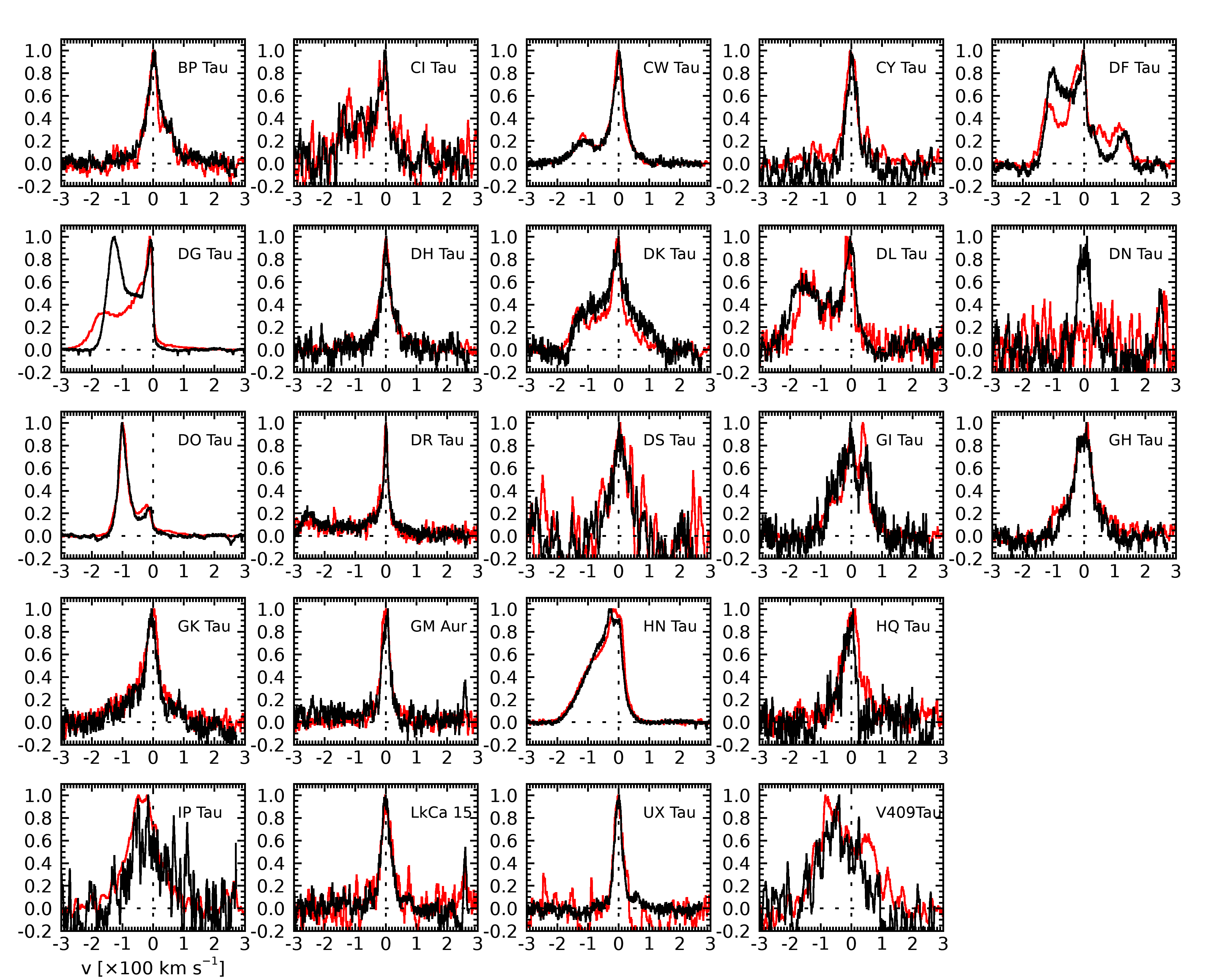}
\begin{center}\caption{\label{fig:B19_prof_comp} Comparison of continuum-subtracted and peak normalised \oi\ line profiles observed in this work (black) and in \cite{banzatti2019} (red). }\end{center}
\end{figure*}

Fig. \ref{fig:B19_prof_comp} compares the profiles obtained in our observations with respect to those of \cite{banzatti2019} for the 22 sources in common. The figure shows similarities in most profiles, where small differences are mainly due to different resolution/signal to noise ratio, but also remarkable differences attributable to variability. As also shown by \citet{simon2016}, who compared their observed line profiles with those presented by \citet{hartigan95} almost 20 years before, the most important changes are observed in the HVC, while the profile of the LVC generally remains fairly constant. Sources with complex HVC profiles extending to very large velocities, such as DF Tau, DG Tau and DK Tau, are those with larger variability, both in the relative intensity of the components and in the terminal velocity. These can be partially due to different regions of the extended jets intercepted by the slit widths or fiber size of the instruments, but also to real variability of the ejection mechanism. 

Indeed collimated jets consists of moving knots ejected from the star with different velocities and at irregular time intervals, that can account for the different observed peak intensity of the HVC in different epochs \citep[e.g.][]{takami2022}.
DG Tau, in particular, is known to have episodic ejection phenomena on temporal scales of about 1.5 yrs \citep[e.g.][]{white14a} that can be explained by the different observed peak intensity of the HVC. In addition, the terminal velocity of the DG Tau jet has considerably decreased in time. Indeed, the line profile shown by \citet{hartigan95}, as well as the velocity resolved spectral images obtained more than 20 yrs ago \citep{maurri14,lavalley1997} have detected maximum velocities in excess of 400 \kms, compared with the less than 200 \kms\, measured now.

Noticeable is also the case of DN Tau, where the \osix\, line remained undetected in the \cite{banzatti2019} observations. The line was however detected by \citet{hartigan95} with a similar profile as the one we observe with GIARPS. 

\begin{figure}
\begin{center}
\includegraphics[trim=0 0 0 0,width=1\columnwidth, angle=0]{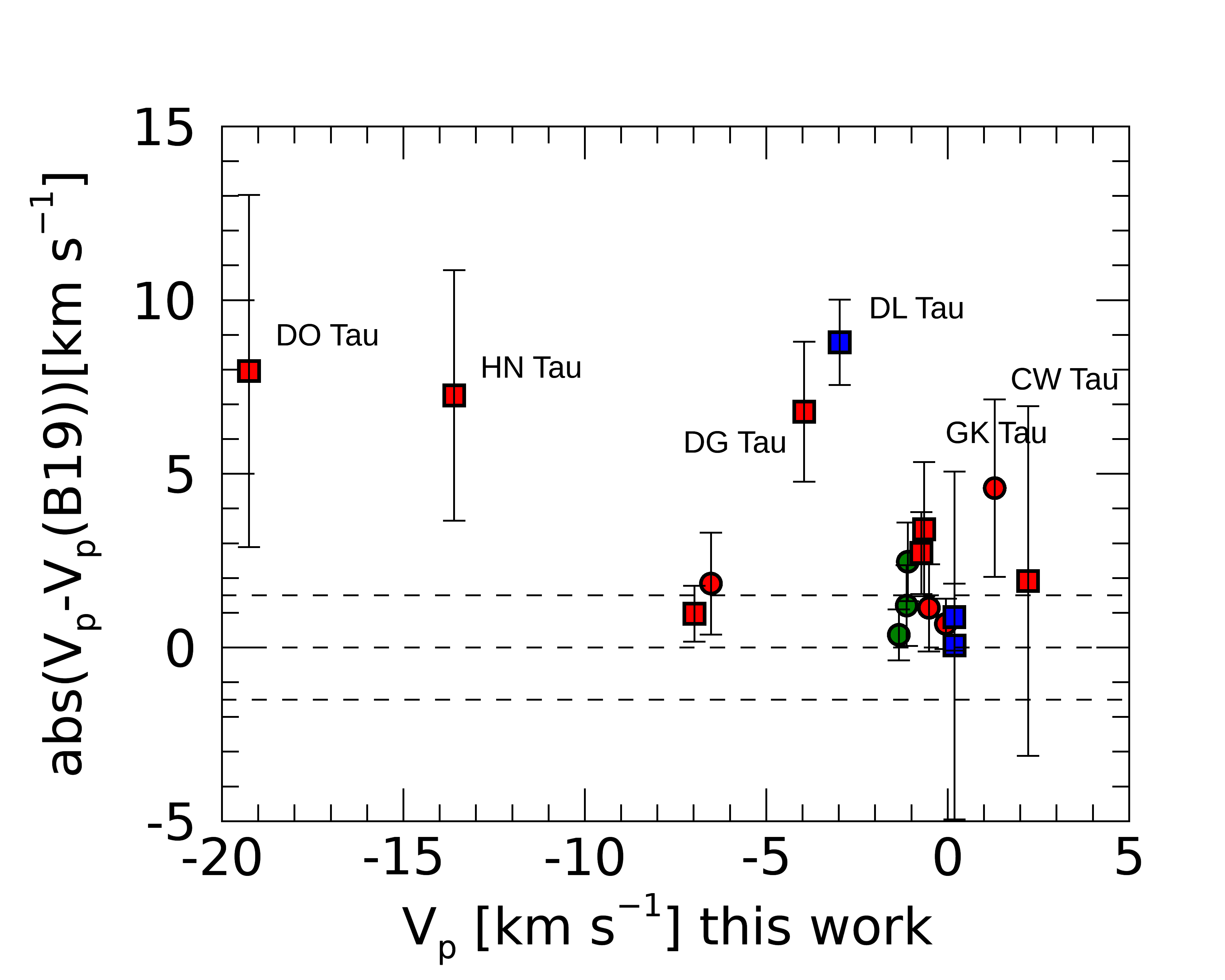}
\end{center}
\caption{Difference between the peak velocity measured in this work (V$_p$) for the NLVC and that measured by \cite{banzatti2019} (V$_p$[B19]), plotted against V$_p$. The V$_p$ of \cite{banzatti2019} have been corrected by the slightly different adopted RV with respect to our work. The two dashed lines at $\pm$ 1.5 \kms\, indicate the velocity uncertainty of our measurements.  Symbols colours refer to the disk type (red: full disks, blue: disk with sub-structures, green: TD) while squares indicate sources with a HVC.}
  \label{fig:VP_banz}
\end{figure}

Fig. \ref{fig:VP_banz} shows the difference in peak velocity of the NLVC between \cite{banzatti2019} and this work, as a function of $V_p$. 
For this comparison, the $V_p$ values reported in \cite{banzatti2019} have been corrected by the slightly different adopted RV with respect to our work. Different colour symbols indicate both the disk type and the presence of a HVC. Differences in $V_p$ are never larger than a few \kms\, and could be due either to errors in wavelength calibrations or intrinsic variability. We note in particular that the larger differences arise in sources where we measure an appreciable blue or red shifted emission, while sources with a centroid close to zero are more consistent with the \cite{banzatti2019} values. We also note that peak differences are more common in sources with a HVC. On this bases, we believe that this variability is real and indicates that the kinematics of low-velocity winds is also affected by variations in the jet.  

\section{Correlations with disk structure and inclination}

We discuss here whether there is any relationship among the properties of the forbidden line emission, and in particular the \osix\, line, and those of the gas and dust distribution in the circumstellar disks, which, for the majority of cases, have been resolved through ALMA observations. Table \ref{tab:sources_param} reports the disk inclination and type,  while Table \ref{tab:disk_param} includes additional information relative to the TD. For these latter, it has been shown that inner and outer disks might be misaligned \citep[e.g.][]{bohn2022,francis2020} therefore this table reports the properties of both inner and outer dusty disks, when available, as well as the size of the  dust cavity. Measurements of the inner disk inclination, when available, have much larger uncertainties (up to 20-30 degrees) with respect to the outer disk inclinations estimated from ALMA resolved images. Therefore for the analysis we consider $i_{disk,in}$ only when it differs from $i_{disk,out}$ outside the errors, which basically is the case only for GG Tau and UX Tau.

In general, considering also the kinematical analysis described in Section 5, we did not find any difference in the properties of the forbidden lines between full disk and disks with structures (gaps/rings). Both types of disks share the same kinematical properties in terms of LVC widths and velocity shifts. In addition, jets traced by the HVC are observed in sources with both type of disk structure. Therefore, as for  the presence of outflows concerns, there is no evidence of any evolutionary trend between full and structured disks. 
This also confirms the result shown in \cite{Gangi22} that there are no differences in accretion properties connected with the structure of the outer disk.  

Hereafter, we focus on possible relationship between the \osix\, NLVC emission line region and the properties of the outer dusty disk. 

\input{Tables/table3.tex}  

\subsection{{\rm \osix}\, Keplerian radius}

If the NLVC comes from a slow wind or from gas bound in the disk, then it must be subject to broadening due to Keplerian rotation.
In such an hypothesis the line width, corrected for the disk inclination, has been often used to have a rough estimate of the characteristic radius of the line emission \citep[e.g.][]{simon2016,banzatti2019,mcginnis2018}.

\begin{figure}
\begin{center}
\includegraphics[trim=0 0 0 0,width=1\columnwidth, angle=0]{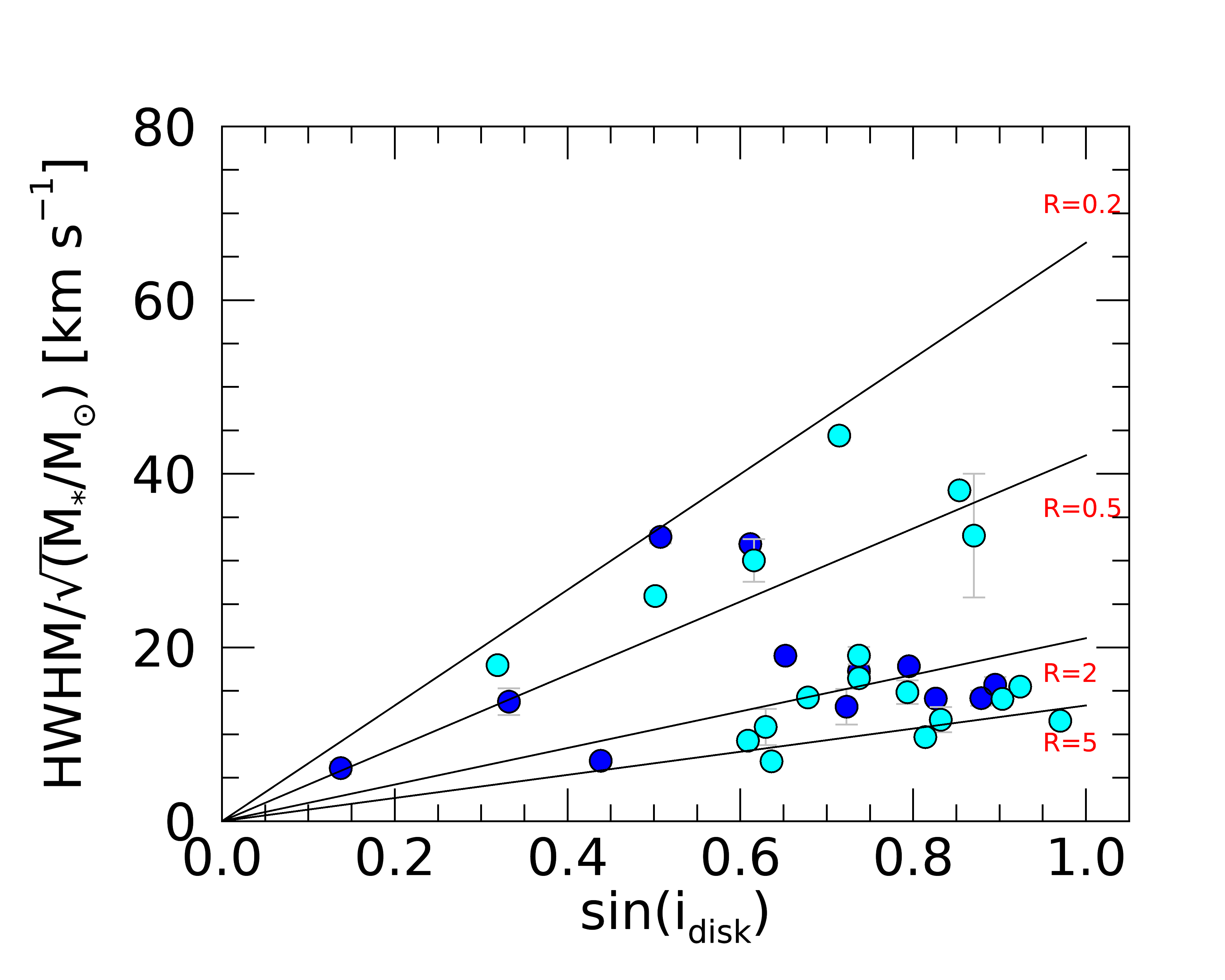}
\includegraphics[trim=0 0 0 0,width=1\columnwidth, angle=0]{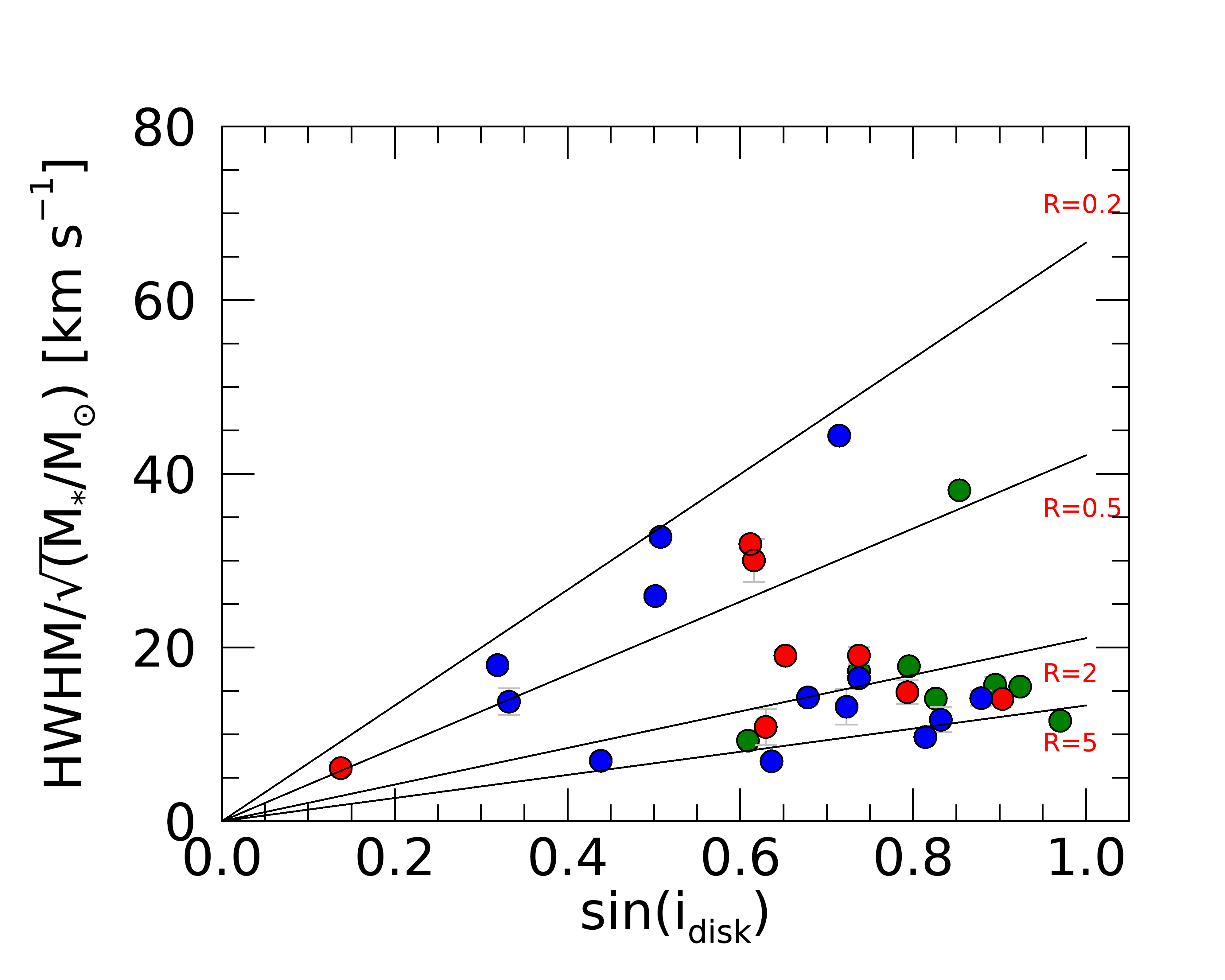}
\end{center}
\caption{Relation between the deconvolved HWHM normalised by the stellar mass and sin($i_{disk}$) for the NLVC. Black lines indicate line widths as a function of inclination at disk radii of 0.2, 0.5, 2 and 5 au, in the assumption of Keplerian motion. In the upper panel, blue and cyan symbols distinguish sources where the V$_p$ of the line is consistent with the origin in the disk or consistent with a wind, respectively. In the bottom panel, the distribution with respect to the disk type is shown, with colours code as in Fig. \ref{fig:VPvsFWHM}}
  \label{fig:fwhm_incl}
\end{figure}

To compare our results with those of these previous studies, we plot in Fig. \ref{fig:fwhm_incl} the HWHM divided by the square root of the stellar mass against the sine of the disk inclination, marking lines of constant
Keplerian radii (i.e. R$_{kep}$ = (sin$(i)$/HWHM)$^2$\,G\,M$_\star$) at
0.2, 0.5, 2 and 5 au. In the figure, blue symbols indicate sources having a peak velocity compatible with gas bounded to the disk within 2$\sigma$, while cyan symbols are sources with blue shifted velocity centroids, and thus having a clear wind component. The figure shows that sources with zero centroid velocity have Keplerian  radii between 0.5 and 5 au, in line with previous studies \citep[e.g.][]{simon2016,mcginnis2018}. Sources with blue shifted peaks are more scattered extending further in the inner 0.2 au. We remind that our definition of the NLVC is broader than that of previous works, including sources with a single LVC with FWHM $<$ 55 \kms\, while in previous studies a limit of 40 \kms\, was adopted. 
In the lower panel of the Figure, colour symbols identify disks with different structure. 
In TD disks, the \osix\, emitting extends towards larger radii, with R$_{kep}$ between 1 and 5 au, while there is no distinction for the other disks.
Errors shown in Fig. \ref{fig:fwhm_incl} take into considerations only the errors on the fitting procedure. The final uncertainty on the R$_{kep}$ values should also consider the uncertainty on the disk inclinations and source mass. The latter is of the order of 0.15 in log \mstar \citep{Gangi22}, while the error on $\rm i_{disk}$ is typically 5 degrees, if measured with ALMA. This results in a final relative uncertainty varying between $\sim$ 30 to 60\% . 

\cite{weber2020} pointed out that in line profiles resulting from wind models, the decomposition in a BLVC and a NLVC might not reflect the presence of distinct physical components originating in separate regions of the disk. This is because a single Gaussian component will contain emission from many different parts of the wind, and this effect is more prominent at large inclinations. However, Type 1 and 2 profiles should be less affected by this problem, as basically a single component is detected and thus contamination by emission at different velocities should be minimised.

Even in the case of emission from the disk, however, the emitting region will extend from an inner to an outer radius, and this latter should be rather large since we do not detect the double peaked profile typical of disk emission. It is therefore important, for a correct interpretation of the R$_{kep}$ parameter, to evaluate to which extent it can be used as a proxy for the \osix\ emission region. We analyse this issue in details in the next section.

\subsection{Disk radial distribution of the \osix\ NLVC and significance of the R$_{kep}$ parameter.}

In this section we analyse whether a simple model of emission from bound gas in Keplerian rotation in the disk can account for the line profiles of LVCs having $V_p$ consistent with zero velocity. 
This analysis will also allow us to understand to which extent the R$_{kep}$ parameter can be considered an indicator of the size of the line emission region, as often assumed. 
To this aim, we modelled the observed line profiles   
assuming that the line intensity follows a power-law distribution as a function of the radial distance from the star $R$ of the form $I(R) = I(R_i)\times (R/R_i)^{-\alpha}$, where $I(R_i)$ is the intensity at the inner radius \citep{fedele2011}.

\begin{figure*}[h]

\includegraphics[width=0.3\textwidth]{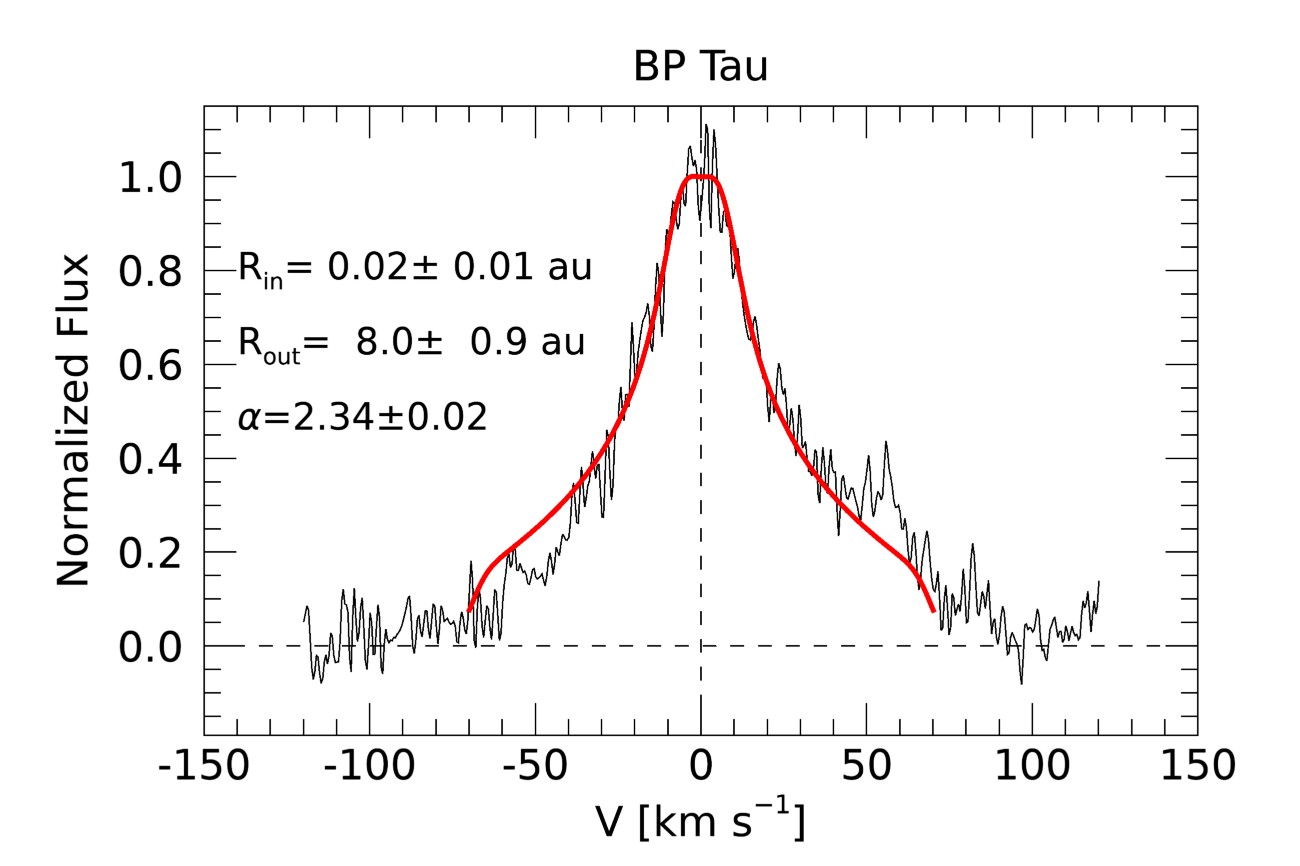}
\includegraphics[width=0.3\textwidth]{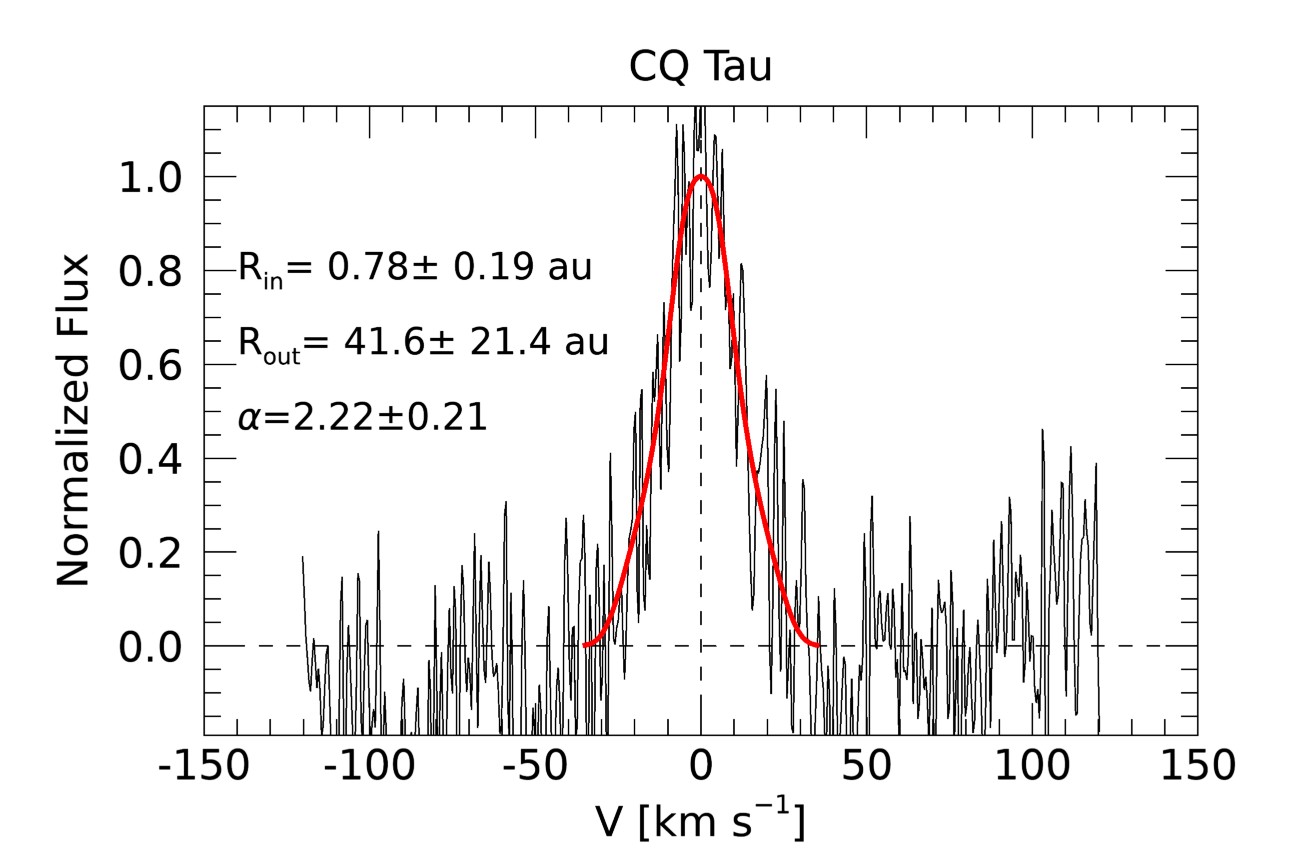}
\includegraphics[width=0.3\textwidth]{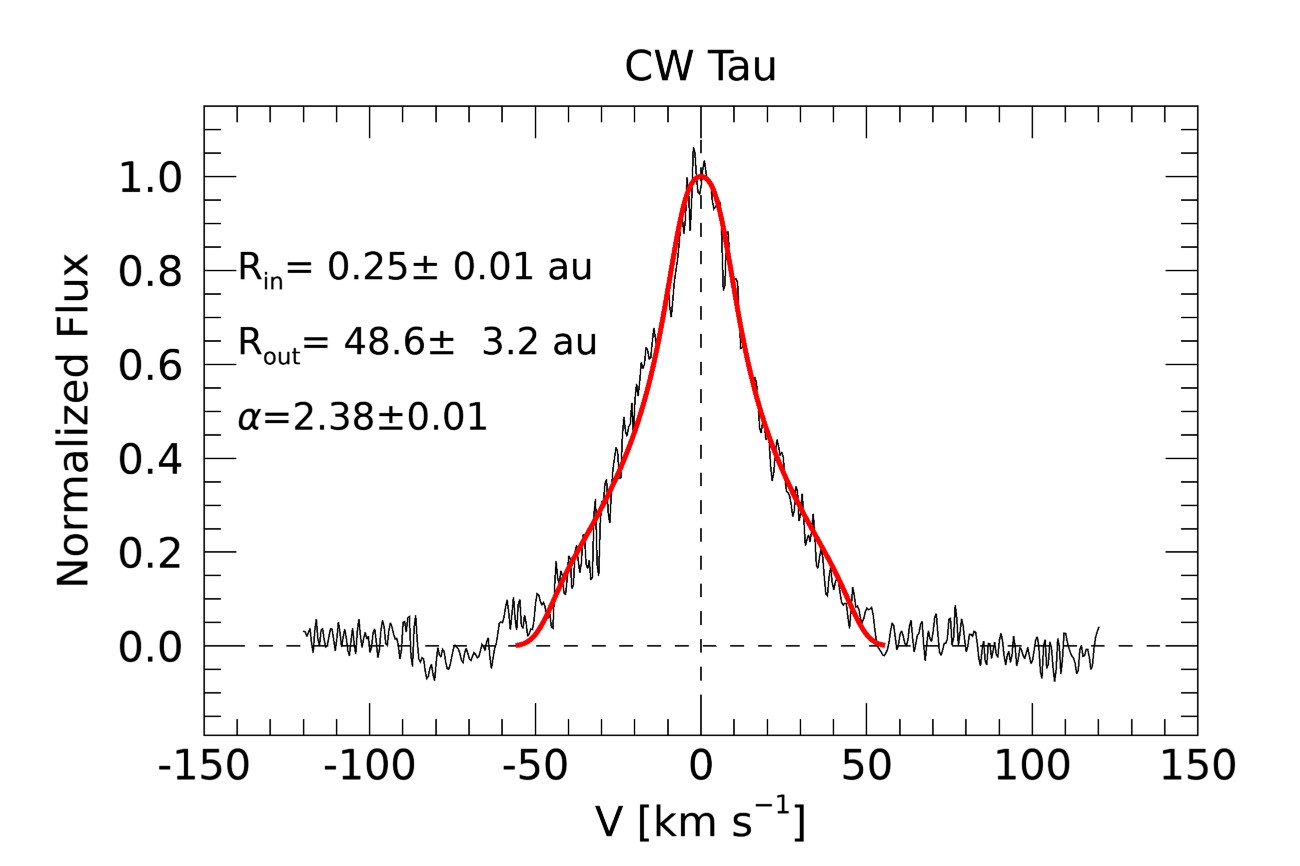}

\includegraphics[width=0.3\textwidth]{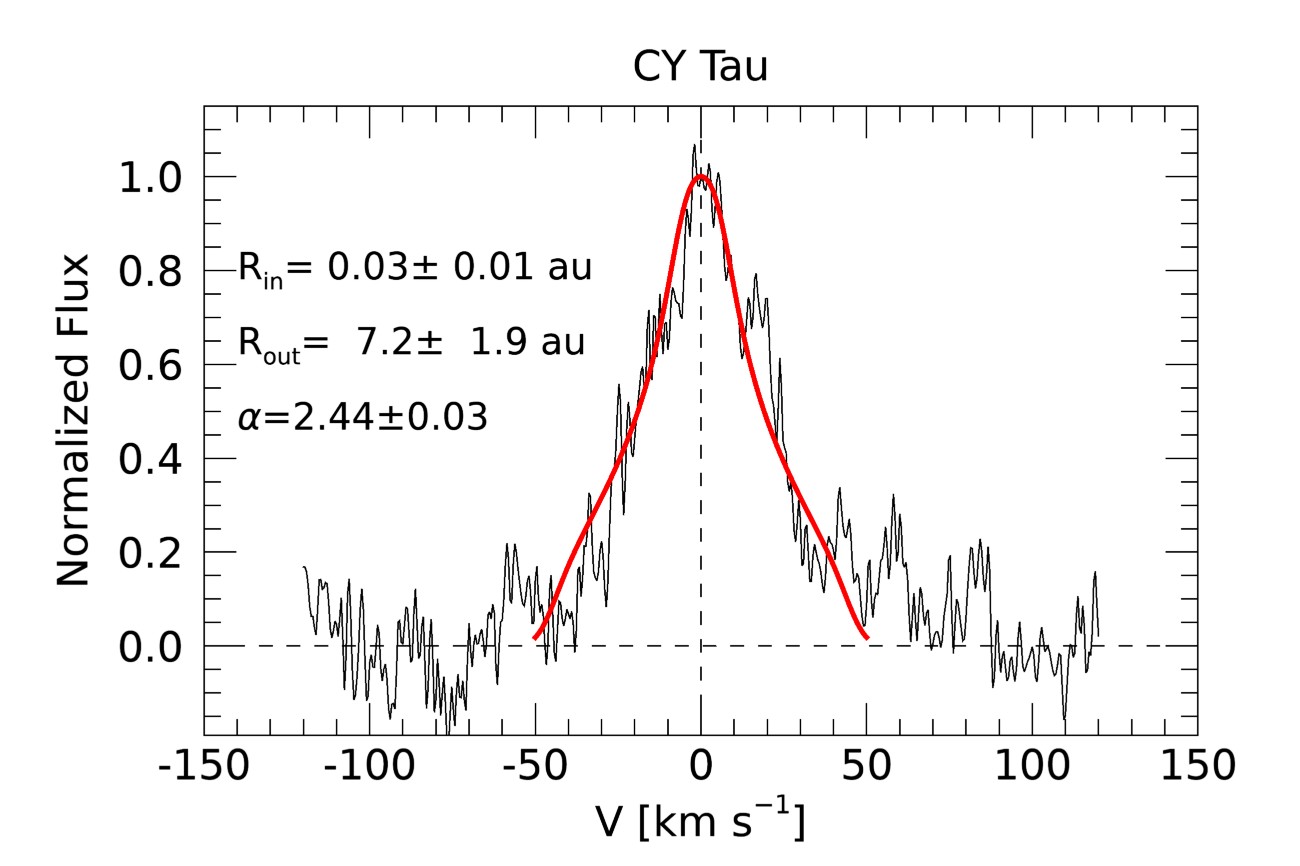}
\includegraphics[width=0.3\textwidth]{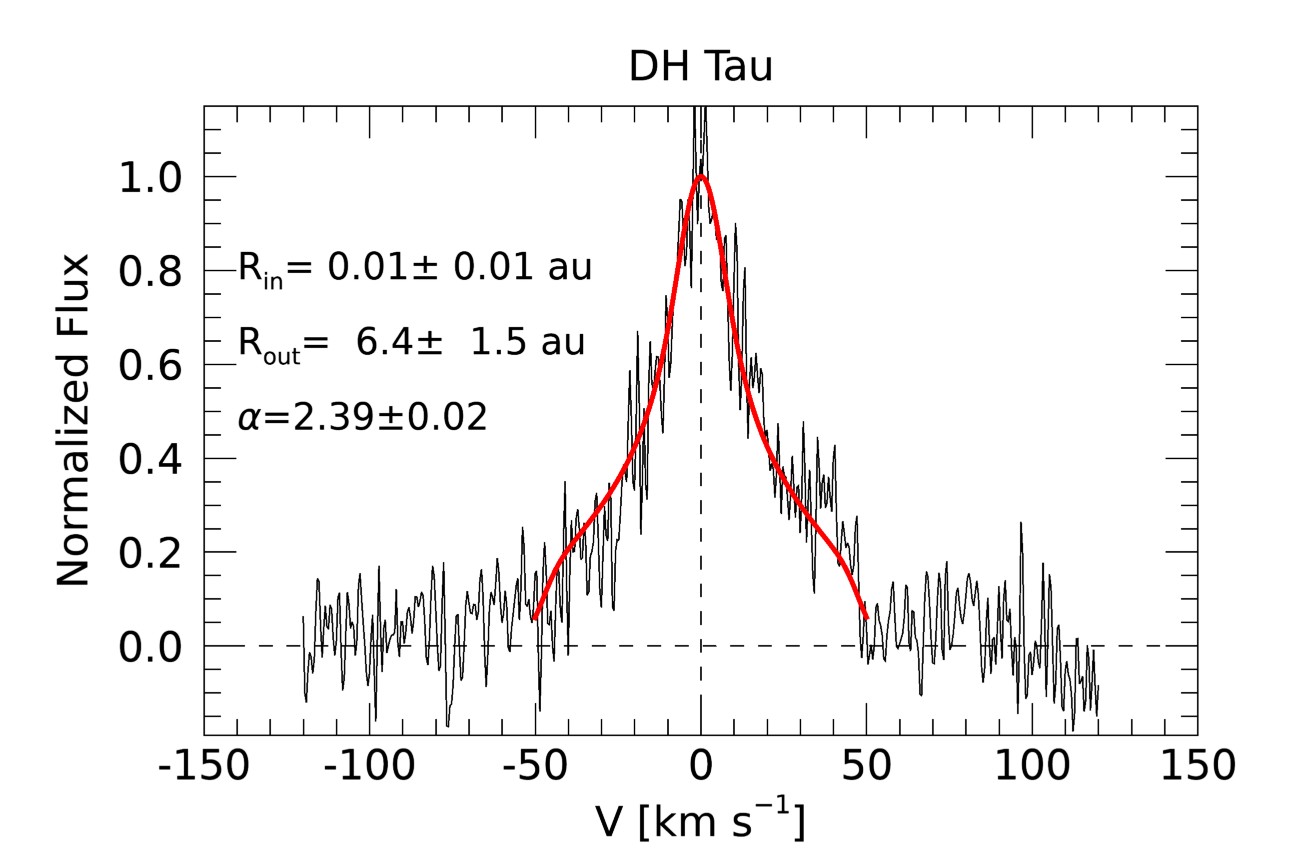}
\includegraphics[width=0.3\textwidth]
{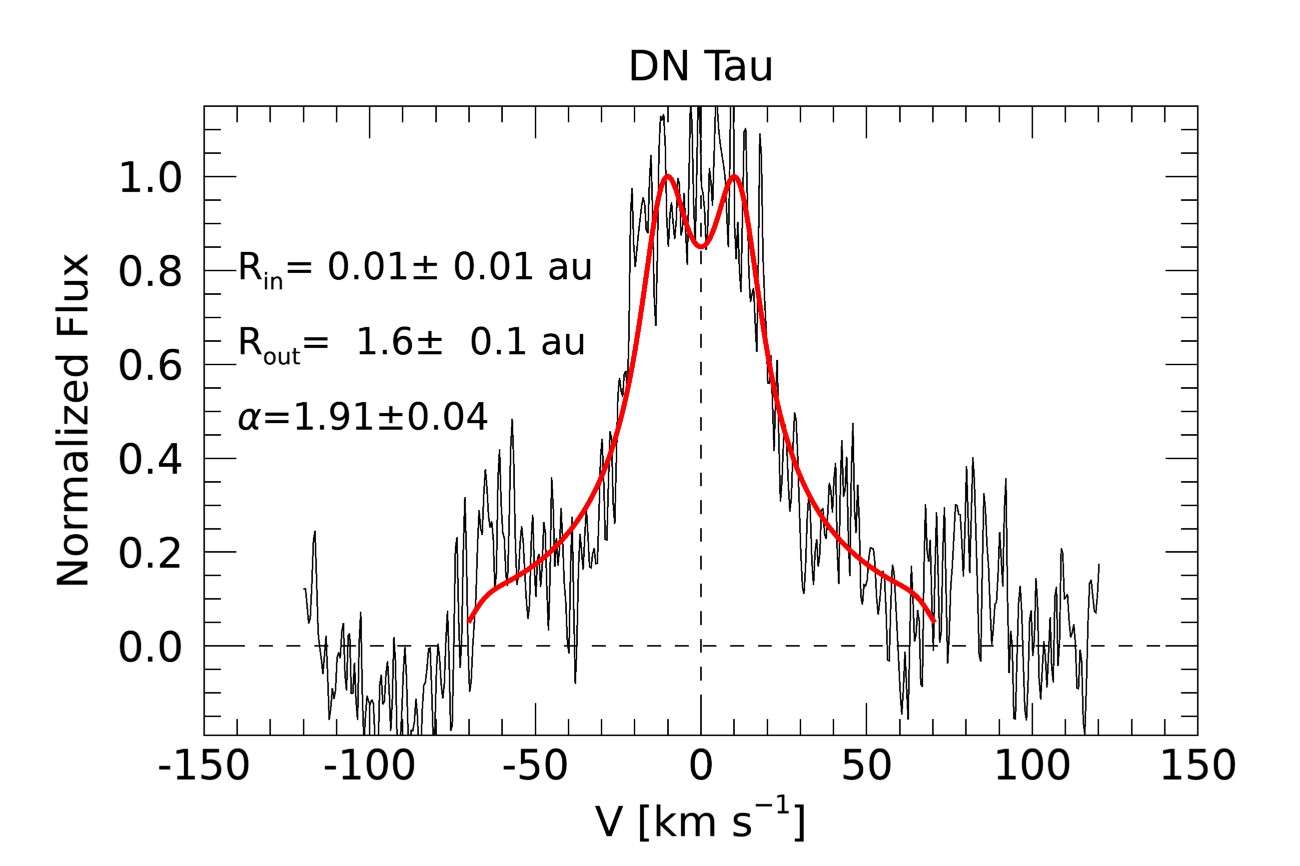}

\includegraphics[width=0.3\textwidth]{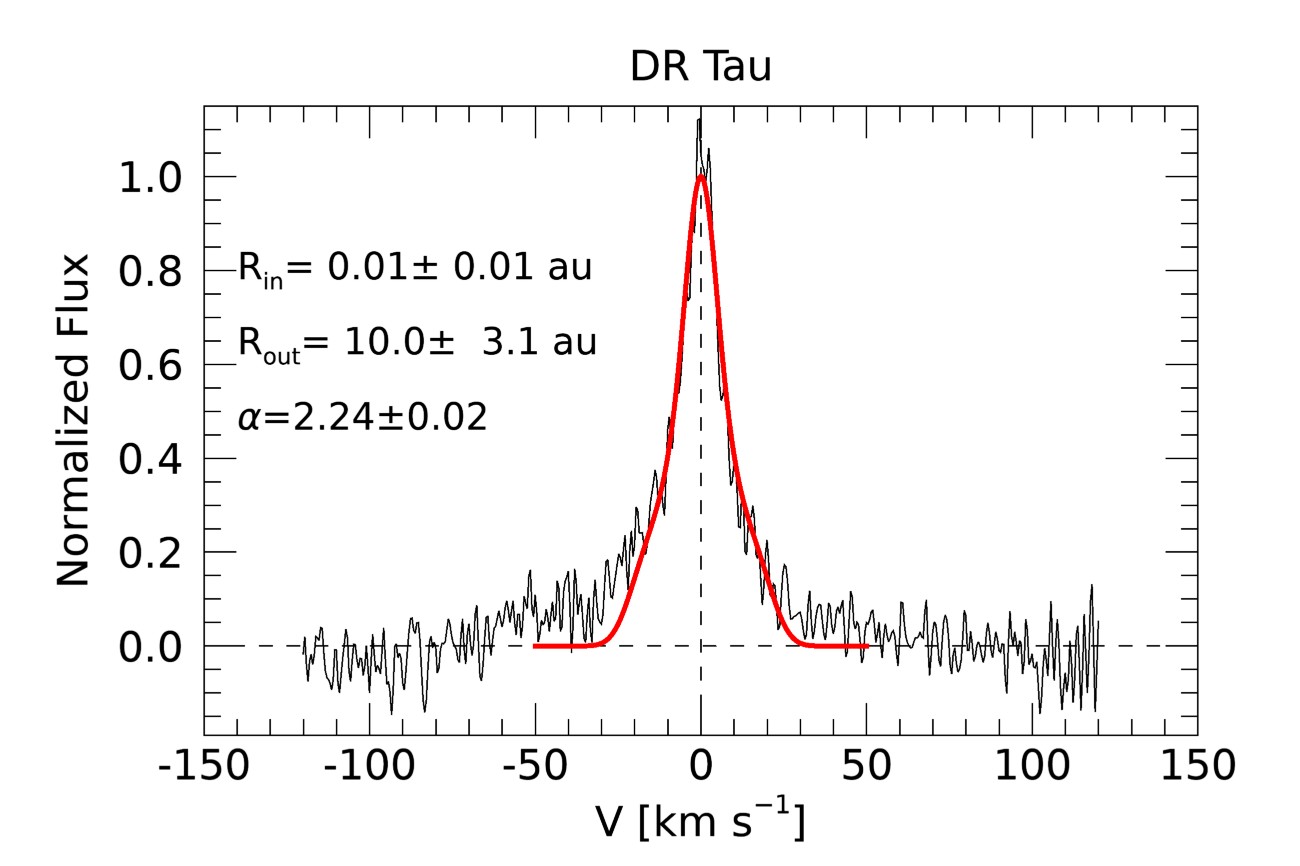}
\includegraphics[width=0.3\textwidth]
{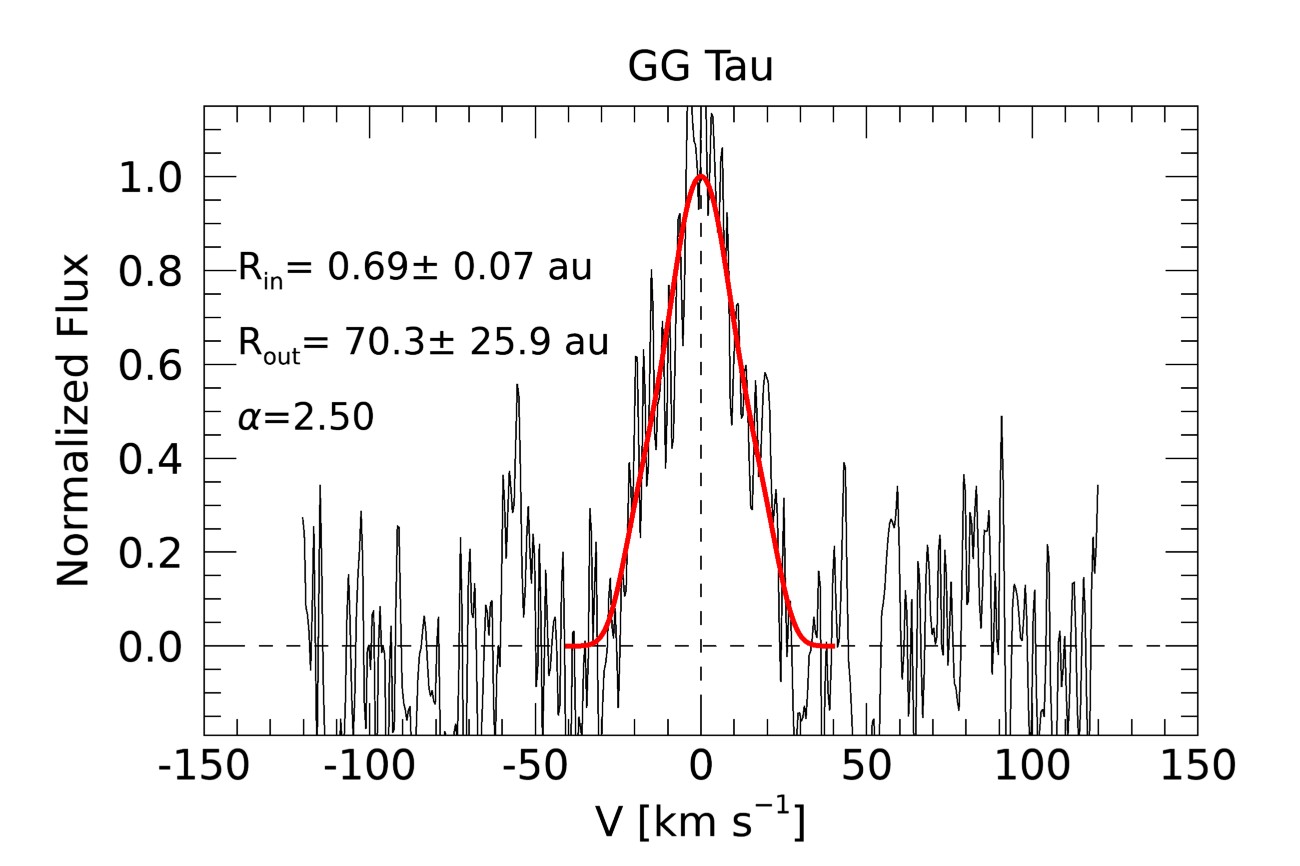}
\includegraphics[width=0.3\textwidth]{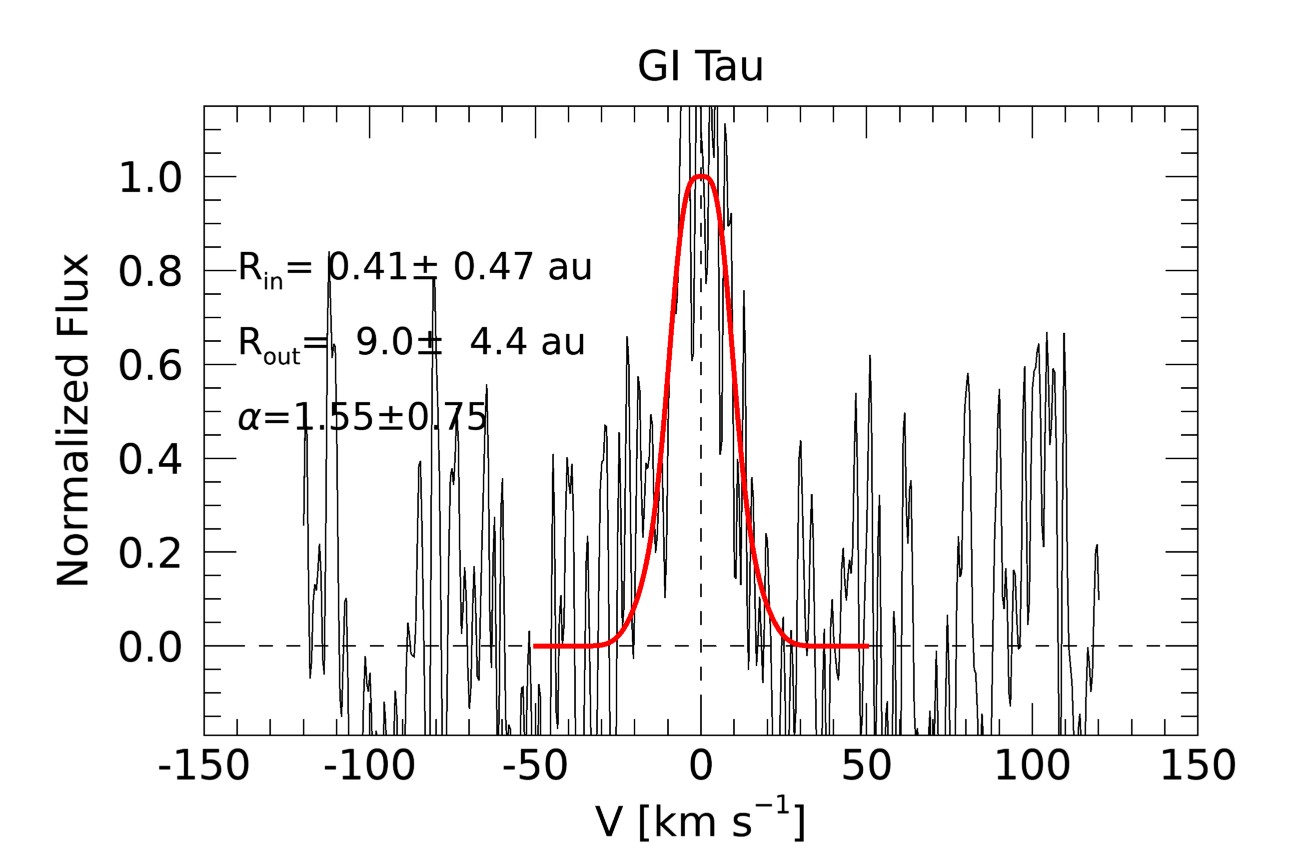}

\includegraphics[width=0.3\textwidth]{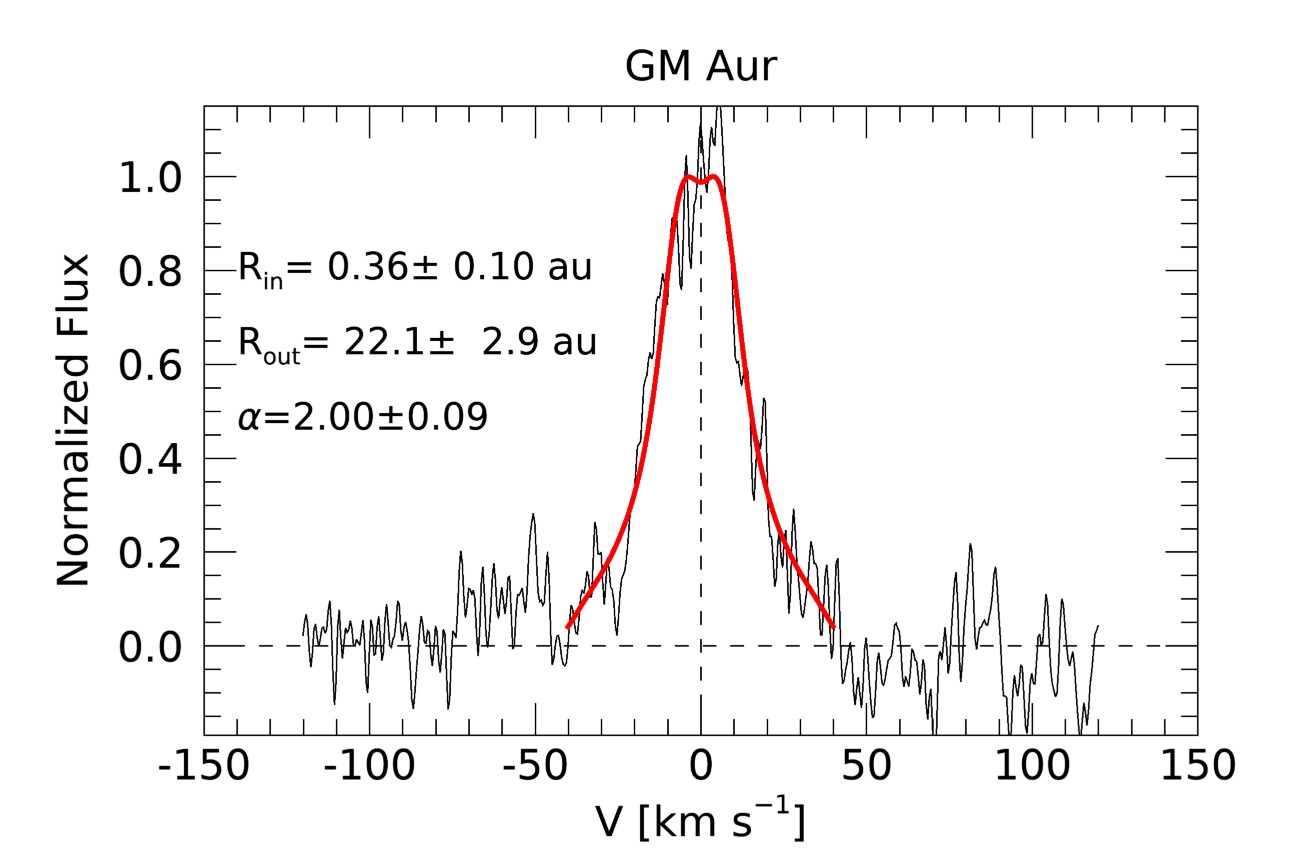}
\includegraphics[width=0.3\textwidth]{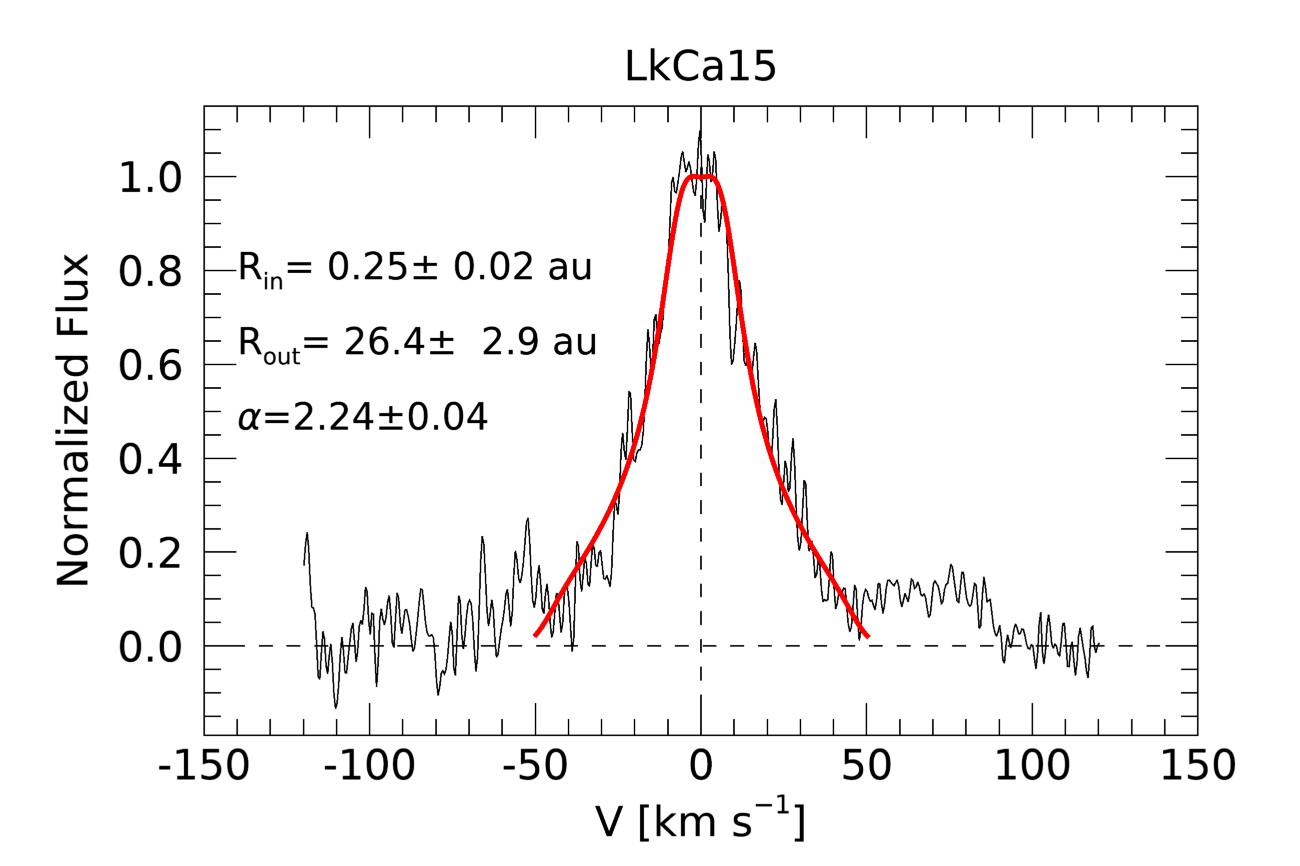}
\includegraphics[width=0.3\textwidth]{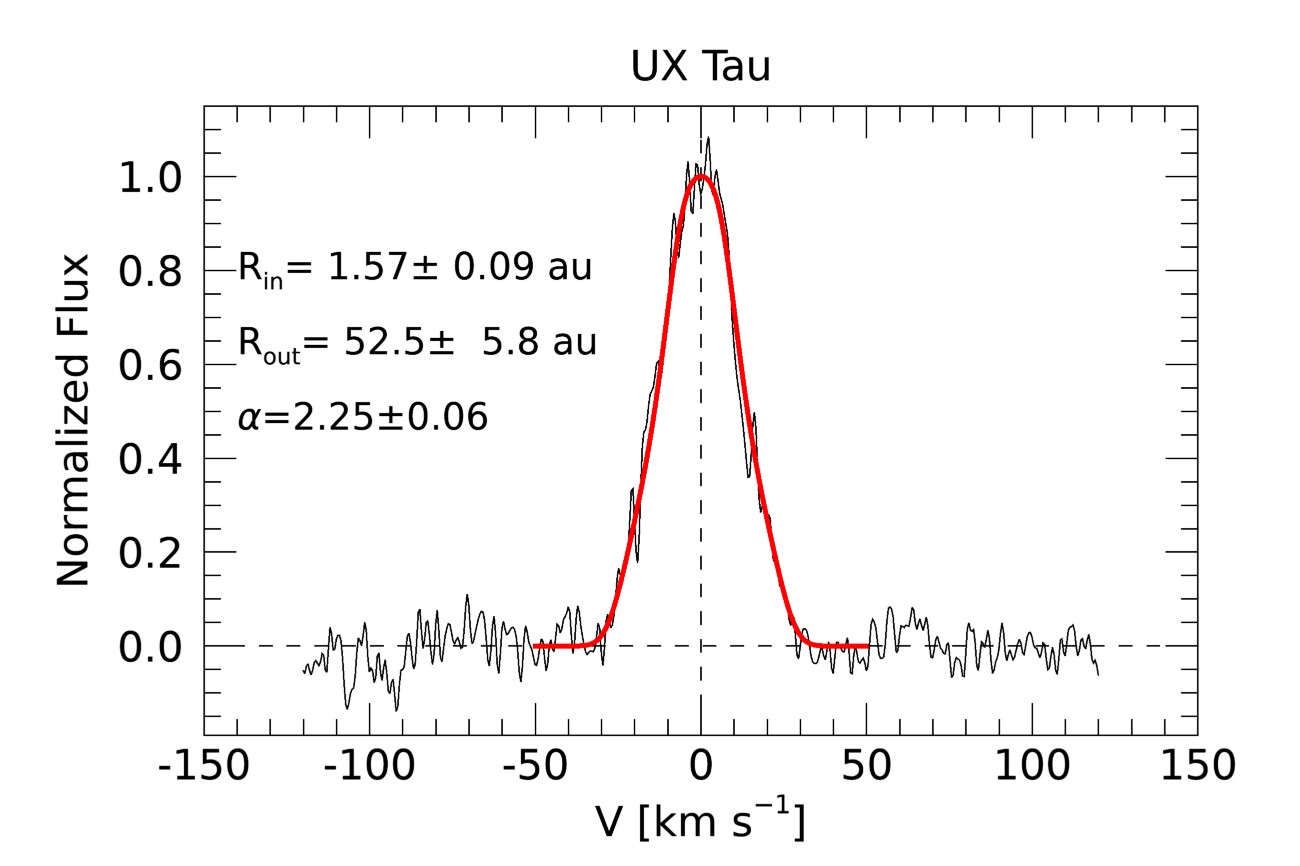}

   \caption{Comparison between the \osix\ line profiles having V$_p$ close to zero, and a model of emission from a Keplerian disk. In the observed profiles the NLVC (or the NLVC+BLVC in the cases of Type 2 profiles) has been isolated by subtracting the Gaussian fit derived for the other components at different velocities. The fitted parameters of the model are indicated in each panel.}
   \label{fig:fit_profiles}
  
\end{figure*}

The radial profile is converted into a velocity profile assuming that the gas is in Keplerian rotation. Finally, the resulting line profile is convolved with a velocity
width $V = \sqrt{V_{in}^2 + V_{th}^2}$, where $V_{in}$ is the instrumental
broadening and $V_{th}$ is the thermal broadening, computed assuming a gas temperature of 8000 K. For the Keplerian rotation and line convolution we
used the IDL codes \textit{keprot} and \textit{convolve} described in \cite{acke2005}. We then used the IDL routines \textit{mpfitfun} to
find the best fit to the observed line profiles, adopting an
uncertainty equal to the rms on the continuum around the line. 
The parameters that are varied in the
fitting procedure are the inner ($R_{i}$) and outer ($R_{o}$) radii of the emitting gas and the power-law index $\alpha$. The inner radius is constrained by the maximum velocity reached in the line wings. On the other hand, the outer radius regulates 
the extent of the double-peak profile characteristic of the disk emission: the larger is the extent of $R_o$, the smaller is the separation between the two peaks of the line. We do not detect any unambiguous double peak in our line profiles, which implies that $R_o$ should be large or, alternatively, that an additional wind component at very low velocity is present.
We applied the model to a sub-set of sources, observed at a sufficiently high S/N, where the NLVC is peaking at $V_p \sim 0$ and can be easily identified with respect to the other velocity components, and for which we know the inclination angle, needed to model the Keplerian rotation.
 With these criteria we selected 12 sources. 
For targets with velocity components in addition to the LVC, we subtracted the corresponding Gaussian fits to isolate the NLVC only. 
 For the sources with Type 2 profiles, we apply the model to the entire LVC (thus to the NLV+BLV components), to analyse if a simple disk model can account for the triangular-like profile of the LVC. To perform the comparison with the model, a small velocity shift (up to a maximum of 3 \kms) has been applied to sources with $V_p$ not strictly equal to zero in order to have a profile symmetric around zero velocity.
The list of analysed sources is given in Table \ref{table:disk_fit} together with the fitted parameters. Fig. \ref{fig:fit_profiles} shows the fitted profiles superimposed on the observed ones. The simple adopted model reproduces quite well the profiles. Looking at the fitted parameters, we see that the radial extent of the emission goes from $r < 1$ au to very large outer radii, up to more than 50 au, to reproduce the lack of double peak in the profiles. 
However, for some of the extreme cases of very large $R_0$, such as CQ Tau, GG Tau and UY Aur, the outer radius is poorly constrained as testified by the large errors provided by the fit. This reflects the fact that above a certain value of $R_o$ the $\chi^2$ of the fit remains rather constant. On the other side, DN Tau has a rather noisy flat peak that can be fitted with a double peak within the large errors, consistent with a $R_0 <$ 2 au. On the other side, the smallest $R_{i}$ values are found in sources with the triangular-like profile and thus having a large full width at zero intensity (FWZI).  The power-law index is always between 2 and 2.5, with the exception of a couple of cases with $\alpha < 2$. 

We point out that the fit provides unrealistically high $R_0$ values for some of the sources, most notably CW Tau and UX Tau, as they are larger than the size of the corresponding disk as resolved by ALMA \citep[e.g.][]{ueda2022,bacciotti18}. Although the actual gaseous disk can be larger than the ALMA dusty disk, it remains to understand how the \osix\ line can be excited at such large radii. The presence of a flared disk that intercept the stellar photons at large distances could be considered, however high-contrast imaging SPHERE observations of some of these disks \citep[e.g.][]{menard2020, keppler2020} do not provide evidences for them. 
Alternatively, we should consider that the absence of a double-peak in the \osix\ profile, at least in some of the sources, is not due to the emission extending at larger radii but rather to an additional wind component at very low radial velocity. For CW Tau and UX Tau, this hypothesis is supported by the rather high inclination angles of their inner disks, which are 59 and 74 degrees respectively.

Using the fitted disk surface brightness we calculate what is the percentage of flux arising within our empirically estimated $R_{kep}$ radius. The total line flux within $R_{kep}$ is given by 
$F(R_{kep}) = {\int_{R_i}^{R_{kep}}}{I(R)\times \Omega (R)}$ = 
2$\pi$\,cos($i$)/D$^2\times {\int{I(R)\ R\ dR}}$, where $\Omega (R)$
is the solid angle subtended by a disk ring at the radius $R$, and $D$ is 
the distance of the source. This value is then compared with the total flux 
$F_{tot}$ calculated between $R_i$ and $R_o$. The corresponding $F(R_{kep})/F_{tot}$ ratio is reported in Table \ref{table:disk_fit}. 

We see that, according to the model, a fraction between 40\% and 80\% of the line emission is confined within $R_{kep}$. 
Taking into consideration the uncertainties on the parameters, 
we can say that $R_{kep}$ roughly indicates the size of the region within which, on average, about 50\% of the emission arises.

\cite{simon2016} applied a similar disk model to the \osix\, profiles of nine sources observed at relatively lower resolution with respect to our data. We find fairly similar results for the sources in common, the main differences being a 
larger $R_o$ driven by the fact that we still do not see any double peak even at the high resolution of our spectra, and a smaller $R_{i}$ in those cases, such as BP Tau and DR Tau, where we fit the NLVC and the BLVC as a single line profile. 

\input{Tables/table4.tex}

\subsection{Correlation of R$_{kep}$ with disk properties}

\subsubsection{Properties of \osix\ in TD disks}

In eight sources of our sample, high resolution sub-mm observations have resolved the structure of an inner dust cavity and in some cases also studied the gas distribution inside it. 
Table \ref{tab:disk_param} summarises the main parameters derived for these disks. As we have pointed out in Section 2, we name all of them as TD, even if the presence of the cavity might not be necessarily related to an evolutionary effect.

\begin{figure}
\begin{center}
\includegraphics[trim=0 0 0 0,width=1\columnwidth, angle=0]{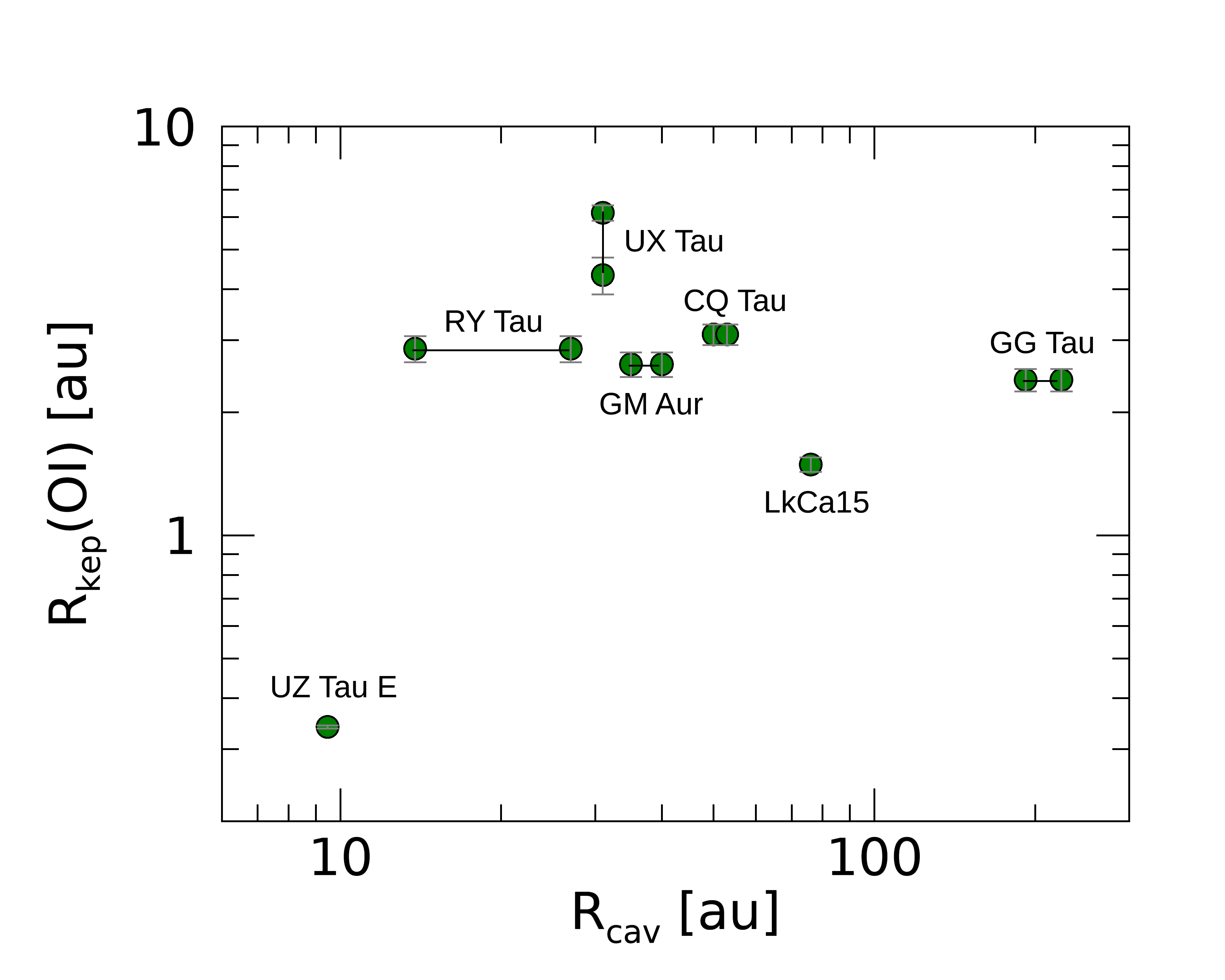}
\end{center}
\caption{Keplerian radius of the \osix\, NLVC emission region for the TD of our sample is plotted against the size of the dusty disk inner cavity. For sources where multiple determinations of the inner cavity are given in literature, they are both displayed and connected with a line. } 
  \label{fig:Rkin_Rhole}
\end{figure}

We have already shown that the \osix\, emission in TD generally has a negligible velocity shift and a relatively large Keplerian radius R$_{kep}$, between 1 and 5 au.  Fig. \ref{fig:Rkin_Rhole} compares R$_{kep}$ with the size of the dusty cavity R$_{cav}$. When two determinations for this latter are available in the literature, both of them are plotted in the figure. For all sources R$_{kep}$ is smaller than R$_{cav}$. Considering that only about 50\% of the emission arises from within R$_{kep}$, this result indicate that most of the cavity is not fully depleted of gas yet.
The figure does not show any direct correlation between $R_{kep}$ and the size of the dust cavity resolved by sub-mm observations.
\cite{banzatti2019}, found a correlation between R$_{kep}$, and the spectral energy distribution spectral index between 13 and 30\um\, for sources with a single component profile, and interpreted it as evidence that gas and dust in the inner disk are both being depleted in an inside-out fashion. A similar correlation has been also shown for other samples \citep{mcginnis2018,fang2023}. 
While we confirm the general trend of a larger median $R_{kep}$ radius for TD with respect to sources with full disks, we show that the \osix\, launching radius is maintained fairly constant among the TD sample and thus it is not directly related to the size of the resolved sub-mmm dust cavity. 
In this respect, it is noticeable that the only peculiar source in this plot is UZ Tau, which has R$_{kep} < 1$ au. We included this source among the TD as a small inner cavity of 9.5 au has been resolved by \cite{long2019}. This cavity however has only a factor of two of dust depletion, therefore this source is a TD less evolved than the others of our sample.
This implies that the general trend delineated by the correlation of R$_{kep}$ with the spectral index, which is a parameter related to long scale time evolution of the disk structure, is not preserved once the a well defined dust cavity is formed.

It should be also noted that the inner dusty cavity of TD disks is, at least partially, depleted of cold ($<$ 100 K) gas, as shown by ALMA observations, and sometimes also depleted of warm molecular gas \citep{carmona2017,vanderplas2009}. For example, CO mm observations of CQ Tau indicates a gas depletion factor between 10$^{-3}$ and 10$^{-1}$ inside a cavity of $\sim 20$ au (i.e. more than a factor of two lower than the dust cavity) \citep{uberia2019}. Also GM Aur shows an heavily depleted gas inner cavity of $\sim 20$ au \citep{bosman2021}. 
In spite of that, for the TD of our sample the \osix\, luminosity scales with respect to the accretion luminosity in the same way as for disks without inner cavity \citep{Gangi22}, indicating no significant differences in the temperature and density of the warm atomic gas on the surface layers of disk. 

In fact, as seen in Section 8.2, the ratio \ofive/\osix\, in TD sources is similar to those of full disks, suggesting similar physical conditions in the upper layers of the disk from where the lines originate. These evidences indicate that the innermost regions of the disk (i.e. $< 5$ au) are still not severely depleted by atomic gas, as also testified by the still relatively high \macc\, measured in these and other TD sources \citep[e.g.][]{Gangi22,manaraPPVII2022}.  

\subsubsection{Correlation between R$_{kep}$ and inner dusty disk}

\begin{figure}
\begin{center}
\includegraphics[trim=0 0 0 0,width=1\columnwidth, angle=0]{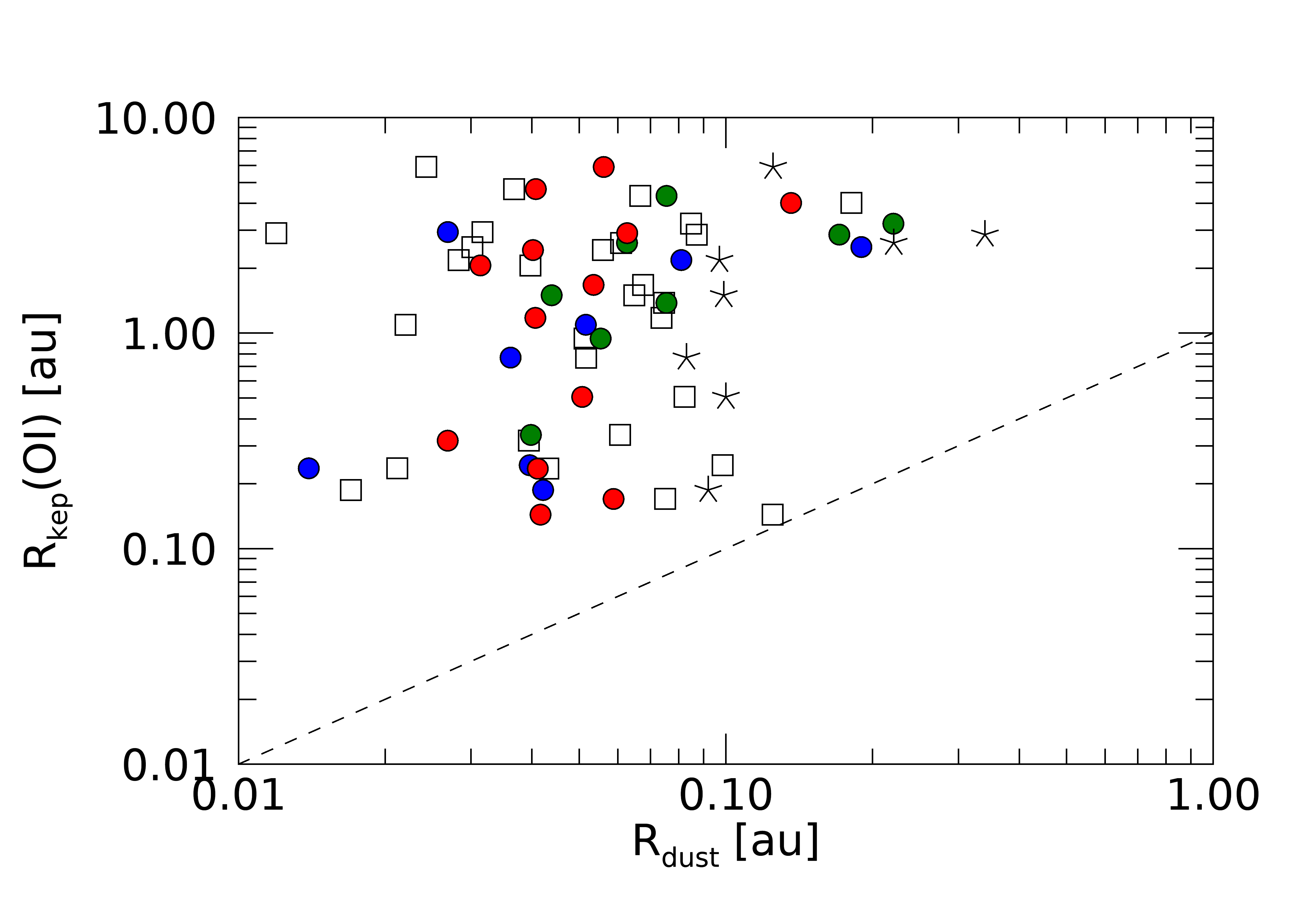}
\end{center}
\caption{Keplerian radius of the \osix\, NLVC emission region is plotted against the estimated and/or measured radius of the dusty disk inner rim, following different approaches. Color circles refer to the dust sublimation radius (colour code as in left upper panel of Fig. 4), 
squares refer to the magnetospheric radius,
while asterisks are the dust inner rim radius measured through near-IR interferometric observations (see text for more details).} 
  \label{fig:Rkin_Rsub}
\end{figure}

We here discuss how the \osix\, emitting region is located with respect to the inner dusty disk, by showing, in Fig. \ref{fig:Rkin_Rsub} the \osix\, Keplerian radius versus the dust inner radius. This latter is estimated through different assumptions. In particular, the colour filled symbols in the figure consider the dust sublimation radius, derived  from the relationship $R_{subl} = \sqrt{\frac{L_{\star}+L_{acc}}{4\pi \sigma T^{4}_{subl}}}$ \citep{dullemond2001}, where $T_{subl}$ is the sublimation temperature that ranges between $\sim$ 1500 and 2300 K, which we take here equal to 1700 K \citep[e.g.][]{alcala2021}. All other parameters are taken from \cite{Gangi22} and listed in Table \ref{tab:sources_param}. Empty squares refer instead to the magnetospheric truncation radius, i.e. the radius where the disk is truncated by the magnetic field lines and gas material is directly funnelled from here through the accretion columns \citep[e.g.][]{koenigl1991}. This is estimated taking the stellar parameters from Table \ref{tab:sources_param}, and assuming a stellar magnetic field of 1 kG \citep[e.g.][]{johns-krull2007}.  

Finally, with asterisks we consider the direct measurements of the inner dust radius provided by infrared interferometric facilities, taken from \cite{akeson2005, perraut2021, eisner2007}.
Interferometric measurements in T Tauri stars measure inner disk radii which are a few times larger than the sublimation radii predicted by disk models \citep[e.g.][]{eisner2007,alcala2021}. \cite{eisner2007} have shown that 
in sources with relatively low mass accretion rates the derived large interferometric radii can be accounted for by magnetospheric truncation. In any case, the \osix\, Keplerian radii are always significantly larger than the other estimates, which give values typically below  than 0.2 au. 

We however remark that 
in Section 7.2 we estimated that the \osix\ emission can extend below the $R_{kep}$, down to inner radii ranging between 0.01 and 0.6 au, and thus might extend within the dust free inner disk region. Inner radii $<$ 0.1 au are found for sources where the triangular profile has been fitted with a single model that needs to take into account the high-velocity wings with gas at a relatively high Keplerian velocity. Similarly, most of the broad components of Type 3 profiles, if interpreted as Keplerian broadening, imply emission within 0.02 and 0.5 au \citep[][e.g.]{banzatti2019,mcginnis2018} and thus might similarly extend within the inner dusty disk. These considerations indicate that the \osix\ emission covers a disk region starting from within the gaseous disk and extending well beyond the dust sublimation radius. 

Except for sources resolved with interferometry, this plot clearly does not consider cases where dust has been already heavily depleted in the inner region, such as the TD sources, for which we have already shown in Fig. \ref{fig:Rkin_Rhole} that the \osix\, emitting region is always inside the dusty hole. 

\section{Line diagnostics in the LVC}

 Ratios of forbidden lines have been used in different studies to infer the physical conditions of the emitting gas in the LVC. In particular, \cite{natta14} considered the excitation conditions of the LVC concluding that the gas should be very dense ($>10^8$ \cmt), warm ($T = 5000-10\,000$ K) and mostly neutral. The limited resolution of their observations prevented them to separately investigate the NLVC and the BLVC. \cite{fang18} were able to separately analyse the NLVC and the BLVC, finding similar conditions for both components in agreement with \cite{natta14}. Both studies conclude that the observed line emission is consistent with thermal excitation, in contrast to a possible origin for the \osix\, emission from photodissociation of OH, as previously suggested for some of the sources \citep{gorti2011,rigliaco2013}. 
\cite{giannini2019} further analysed the variations of physical conditions as a function of velocity in well known jet-driving sources, finding that gas at the lowest velocity is colder and denser than the high-velocity gas, and almost neutral. 

Here we aim at addressing again the issue of the diagnostic potential of forbidden lines ratios, taking in particular advantage of the higher resolution of our observations that allows us to better isolate the NLVC with respect to the other components. In addition, we also consider in the analysis the observation of the \nsix\ line, whose ratio with the \oi\ can provide upper limits on the ionisation fraction \citep[e.g.][]{giannini2019}. 

To this aim, our dereddened line luminosity ratios have been compared with a statistical equilibrium model for electron collisions of a gas at a single temperature and density \citep[see also][]{natta14,giannini2019}. 
The model assumes that the forbidden emission originates from the wind or from the uppermost layers of the disk where molecules are dissociated, and therefore considers that all oxygen is in atomic form and all sulphur singly ionised. These assumptions are considered valid for the LVC also on the bases of {\it i)} the non-detection of near-IR lines of [\ion{S}{i}] in some of the sources \citep{giannini2019} and  {\it ii)} the low ionisation fraction pertaining to the NLVC implying that the gas is mostly neutral \citep[see][and next section]{giannini2019}. Solar abundance values from \cite{asplund05} have been assumed. Only collisions with electrons are assumed, taking into consideration that electronic collisional rates are more than three order of magnitudes larger than those for H atoms. \cite{gorti09} have for example shown that electrons always dominate collisions if the fractional ionisation $x_e > 4\times 10^{-4}$.

We separately consider and discuss the three line ratios.

\subsection{The \sfour/\osix\, ratio}

\begin{figure}
\begin{center}
\includegraphics[trim=0 0 0 0,width=0.8\columnwidth, angle=0]{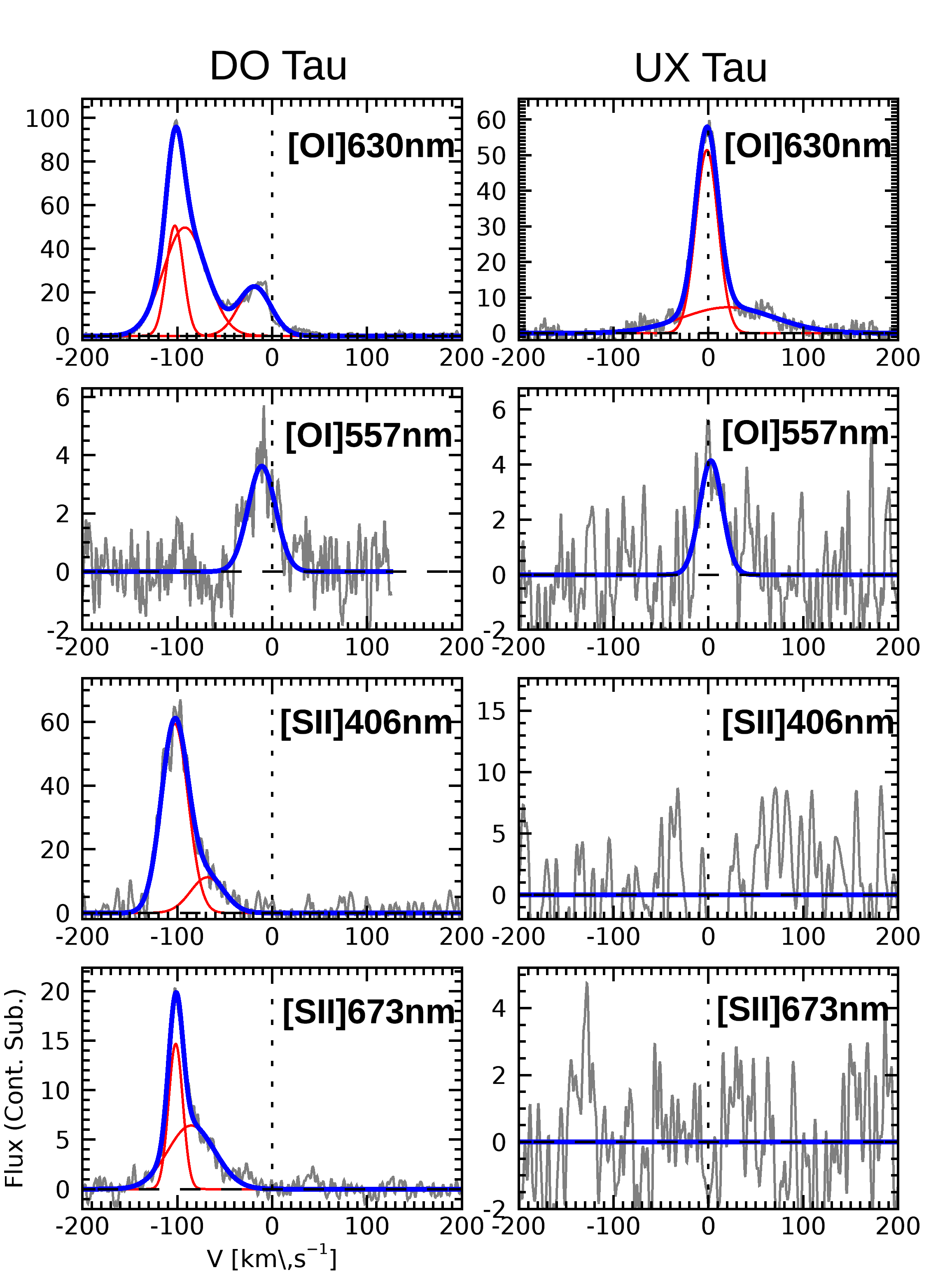}
\end{center}
\caption{Continuum-subtracted spectra of the \oi\, and [\ion{S}{ii}] lines in DO Tau and UX Tau. The NLVC is well detected in the \oi\, lines but not in the [\ion{S}{ii}] lines, with a \osix/\sfour\, ratio $>$ 10. }
  \label{fig:nosulphur}
\end{figure}

\begin{figure}
\begin{center}
\includegraphics[trim=0 0 0 0,width=1\columnwidth, angle=0]{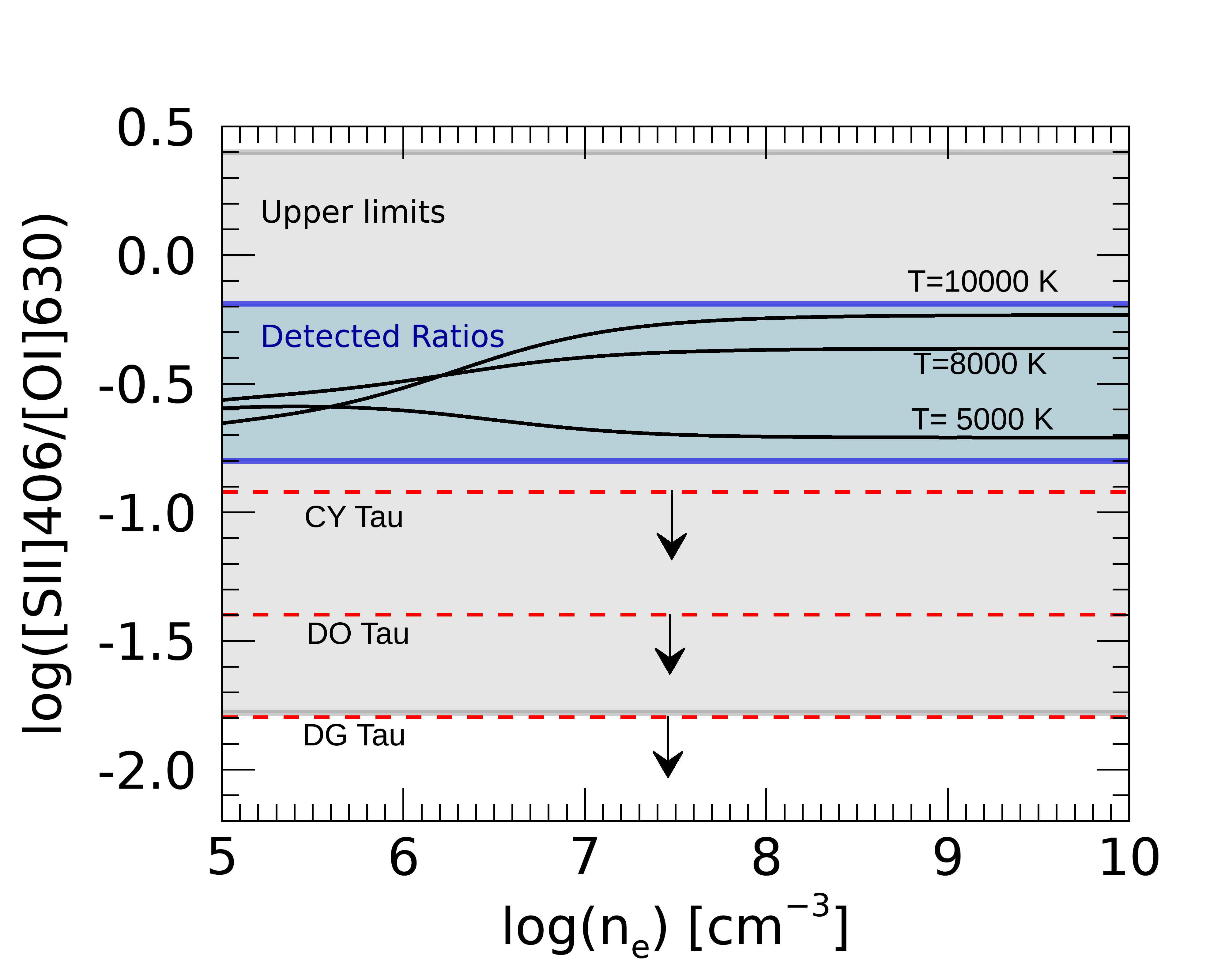}
\end{center}
\caption{Slab model prediction for the \sfour/\osix\, ratio for collisions with electrons. Solid lines refer to gas temperatures of 6000, 8000, and 10000\,K. The cyan area indicates the location of the range of values observed in the LVC (both NL and BL) for the sub-sample of sources where the \sfour\, line has been detected. The grey area shows the location of the upper limits measured in all the other sources.
Horizontal lines indicate the more stringent upper limits.}
  \label{fig:406_dens}
\end{figure}

The \sfour\ LVC is detected in 19 sources (Tab. 2 and 6), and in only nine this detection can be attributed to a NLVC.
In Fig. \ref{fig:nosulphur} we show the example of DO Tau and UX Tau, where bright \oi\ lines are detected in the NLVC but not the sulphur lines. 
Fig. \ref{fig:406_dens} plots the expected \sfour/\osix\ ratio as a function of the electron density $n_e$. Since \osix\ and \sfour\ have a similar critical density and differ for  $\sim$ 10\,000 K in excitation temperature, their ratio varies very little with density, and only a factor of three for temperatures in the range 5000-10\,000 K. In fact, Figure \ref{fig:406_dens} shows that the ratio is between 0.2 and 0.6 for $T_e \approx 5000-10\,000$ K and $n_e$ between 10$^5$ and 10$^{10}$ \cmt. We see from the figure that these values are consistent with the detected ratios and a large fraction of upper limits. Some of the upper limits are however below the values predicted by the thermal excitation model.  We also point out that if an oxygen abundance lower than the Solar value is assumed, as in \cite{fang18}, the discrepancy between observed ratios and the slab model would even be enhanced.

If we consider more physical models, such as the PE and MHD winds model by \cite{weber2020}, they predict a \sfour/\osix\, and a \ssix/\osix\, ratio larger than observed by at least a factor of 10 and also larger by a factor greater than five than those predicted by a simple statistical equilibrium model. Therefore we have in general evidence that sulphur lines are sub-luminous in the NLVC of some of the sources or, alternatively, that the \osix\, emission is enhanced  with respect to both wind models and statistical equilibrium codes for collisional excitation. 

Also in the \cite{fang18} study there are a few sources whose \sfour/\osix\, is overpredicted by models. 
One of these sources is TW Hydrae, for which \cite{gorti09} suggested a non-thermal excitation for the \oi\, lines due to OH dissociation. The detected \sfour/\osix\, ratio in TW Hydrae is 0.06, which is in line with some of our observed upper limits and much below the ratio expected from thermally excited gas. 
Therefore we cannot ruled out that OH photo-dissociation partially contributes to the \osix\, emission, at least for some of the targets such as CY Tau, CW Tau and DG Tau (see Fig. \ref{fig:406_dens}). 

Also, \cite{nemer2020} investigated the possibility that the \osix\, emission is affected by a contribution due to fluorescence excitation by the stellar UV photons, which has been recently suggested by \cite{gangi2023} by comparing the \osix\, and H$_2$ UV line emissions in young stars of Orion OB1 and s-Ori association, and $\sigma$ Ori clouds. 

Another possibility is that part of sulphur is depleted in the disk with respect to the solar value. Significant sulphur depletion in dense clouds with respect to the ISM is a well known observational result \citep[e.g.][]{jenkins2009}. In addition,  Spitzer observations suggest that in embedded protostars only 5 to 10\% of total sulphur is in the gas phase \citep{anderson2013}. Sources for which we find the largest discrepancy between the observed \sfour/\osix\, ratio and the thermal excitation model are full disks (e.g. DO Tau and DG Tau), where the NLVC originates from the dusty disk region where depletion might be significant. A similar effect is observed with iron, whose abundance drops to less than 10\% with respect to the solar value in the low velocity wind components \citep[e.g.][]{agra-amboage11,giannini2019,nisini2016}.

\subsection{The \ofive/\osix\, ratio}

\begin{figure}[t]
\begin{center}
\includegraphics[trim=0 0 0 0,width=1\columnwidth, angle=0]{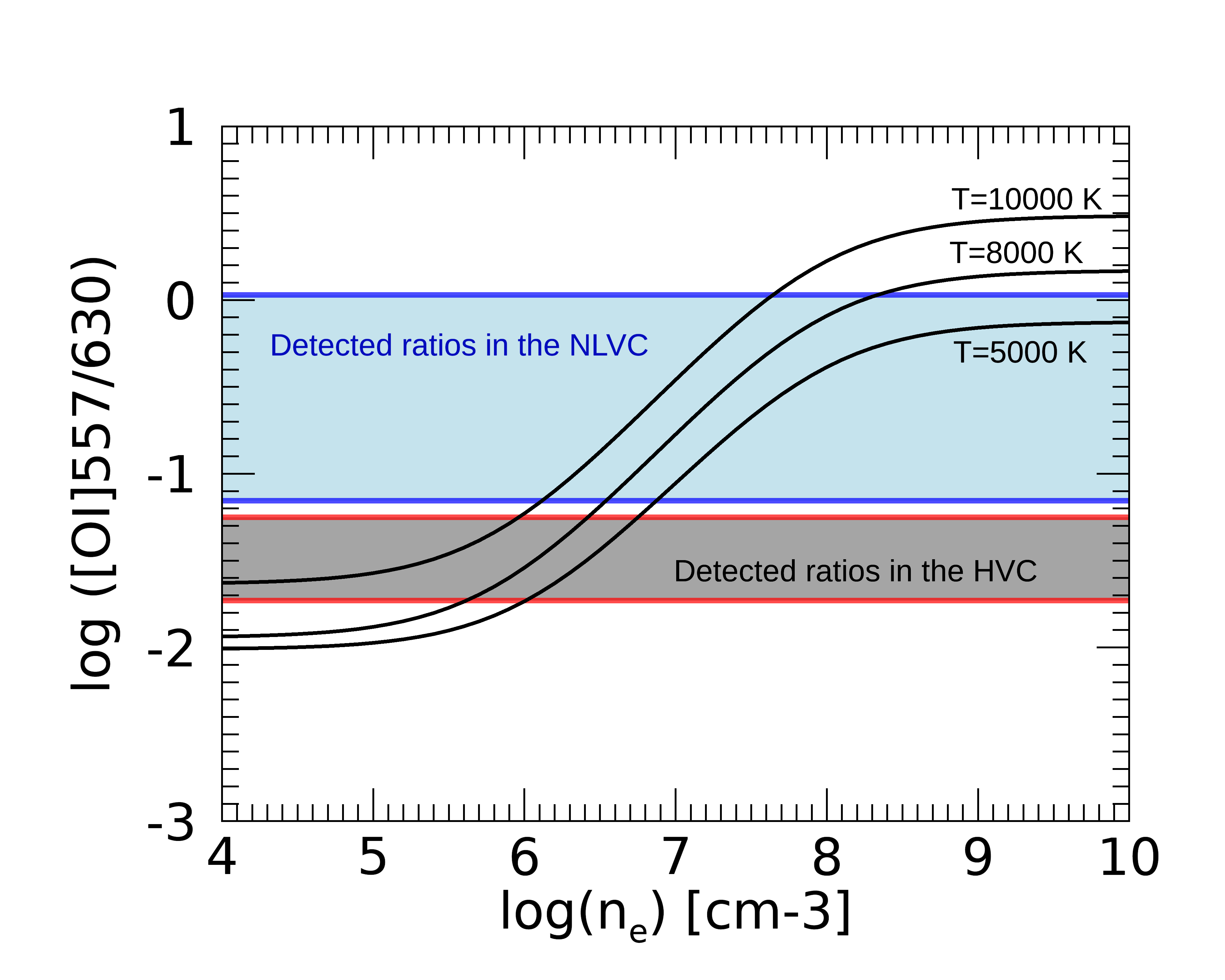}\\
\end{center}
\caption{Slab model prediction for the \ofive/\osix\, ratio for collisions with electrons. Solid lines refer to gas temperatures of 5, 8 and 10$\times 10^3$ K. The cyan area indicates the location of the range of values observed in the NLVC for the sub-sample of sources where the \ofive\, line has been detected in this component. The grey area indicates instead the location of the detected ratios in the HVC.}
  \label{fig:557_dens}
\end{figure}

The profiles of the \ofive\, and \osix\, lines in the LVC are very similar and their luminosity ratio span between 0.1 and 1. 
Figure \ref{fig:557_dens} compares the observed  \ofive/\osix\, ratios against $n_e$ in the statistical equilibrium model. The observed range implies a gas at density $> 10^6$ \cmt\, for a temperature in the range 5000-10\,000 K, in line with previous studies. 

In this plot, we also show the range of values for the \ofive/\osix\, ratio detected in the HVC. This ratio is a few times lower than in the LVC, and implies electron densities $\la$ 10$^6$ \cmt\, for electron temperatures $>$ 8000 K.  

\begin{figure}[t]
\begin{center}
\includegraphics[trim=0 0 0 0,width=1\columnwidth, angle=0]{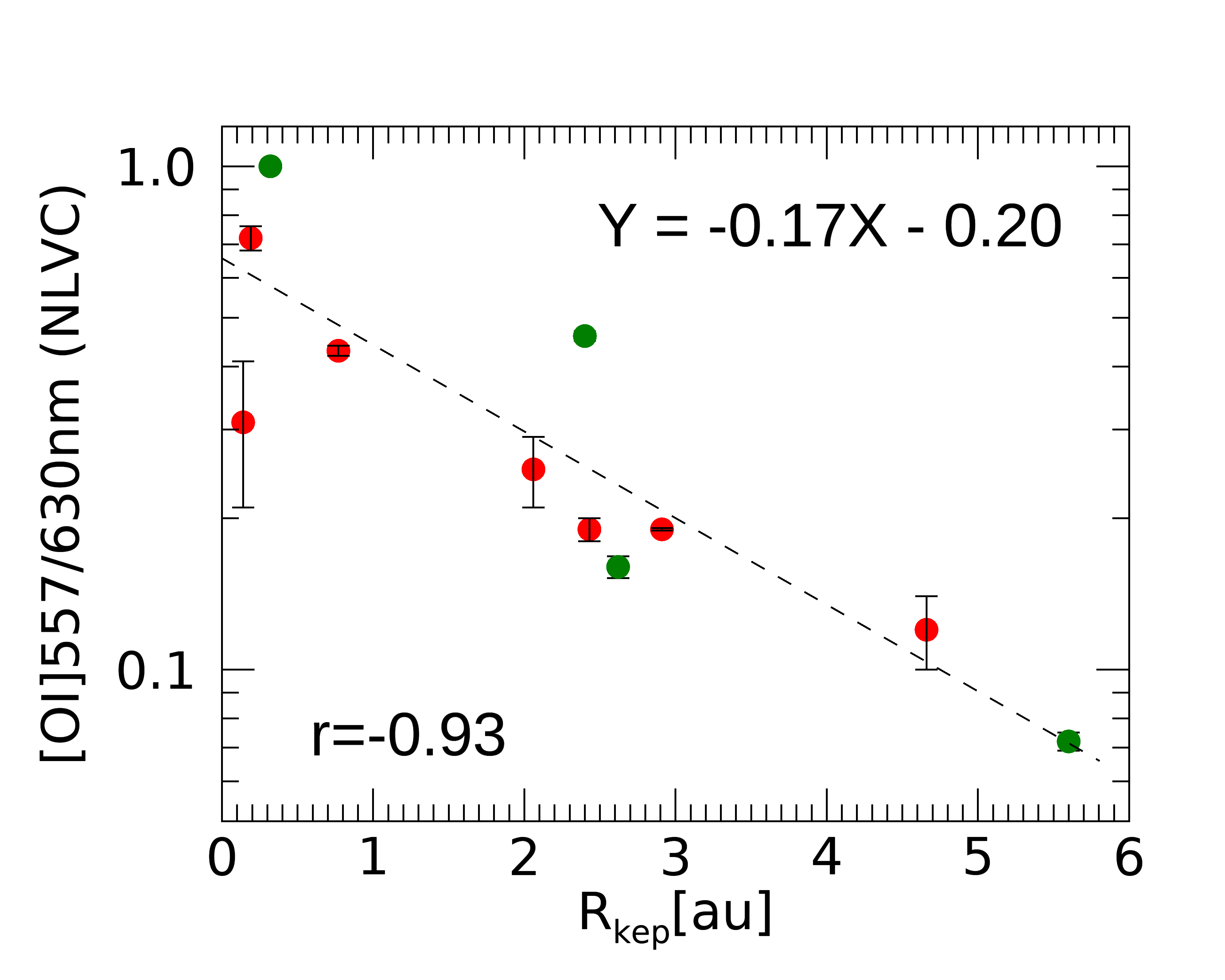}
\end{center}
\caption{
The \ofive/\osix\, ratio observed in the NLVC with no velocity-shifted peak is plotted against the estimated \osix\, emission radius assuming Keplerian rotation. Green symbols indicate TD sources. The dashed line represents the linear fit through the points, whose analytical expression is indicated in the figure. The Pearson correlation coefficient is also displayed in the bottom right of the figure.}
  \label{fig:557_corr}
\end{figure}

Given that there is an order of magnitude spread in the ratio for the NLVC, we investigated whether this spread is due to some intrinsic property of the source. 
We did not find any correlation between the observed \ofive/\osix\, ratio and  L$_{acc}$ or \macc . 
We see however an inverse correlation between the observed ratio and $R_{kep}$(\oi) for those sources having a $V_p$ close to zero, i.e. compatible with emission from the disk (Fig. \ref{fig:557_corr}, Pearson correlation coefficient -0.93). This trend was also found by \cite{banzatti2019} and can be interpreted in terms of a decrease of both disk density and temperature as a function of radius. As already remarked in Section 7.3.1, we see that TD sources follow the same relationship as the other sources, indicating that the physical conditions of warm gas within the dusty-free cavity are similar to those of sources where dust is still present.
A linear fit through the points in Fig. \ref{fig:557_corr} suggests a relationship of the form: \ofive/\osix = $(-0.17\pm 0.05) R_{kep} - (0.20\pm 0.12)$. 

\subsection{The \nsix/\ofive\, ratio}

The \nsix/\ofive\, ratio can be used to constrain the degree of ionisation in the wind. This is a very important parameters both to derive the total gas density ($n_H=n_e/x_e$), and to constrain models that predict a different degree of ionisation in the wind. 
The \nsix\, line is detected in our spectra only in the HVC, in agreement with the finding that the LVC is mostly neutral \citep[e.g.][]{hartigan95,natta14} and that the ionisation increases with the jet velocity \citep[e.g.][]{giannini2019,bacciotti00}. We can therefore use 
the upper limits on the \nsix\, in the NLVC to derive limits on the ionisation degree in this component. 

To this aim, we have used a ionisation model that consider as processes collisional ionisation, radiative and dielectronic recombination, and direct and inverse charge-exchange with hydrogen. Details are given in \cite{giannini15}. 

\begin{figure}[t]
\begin{center}
\includegraphics[trim=0 0 0 0,width=1\columnwidth, angle=0]{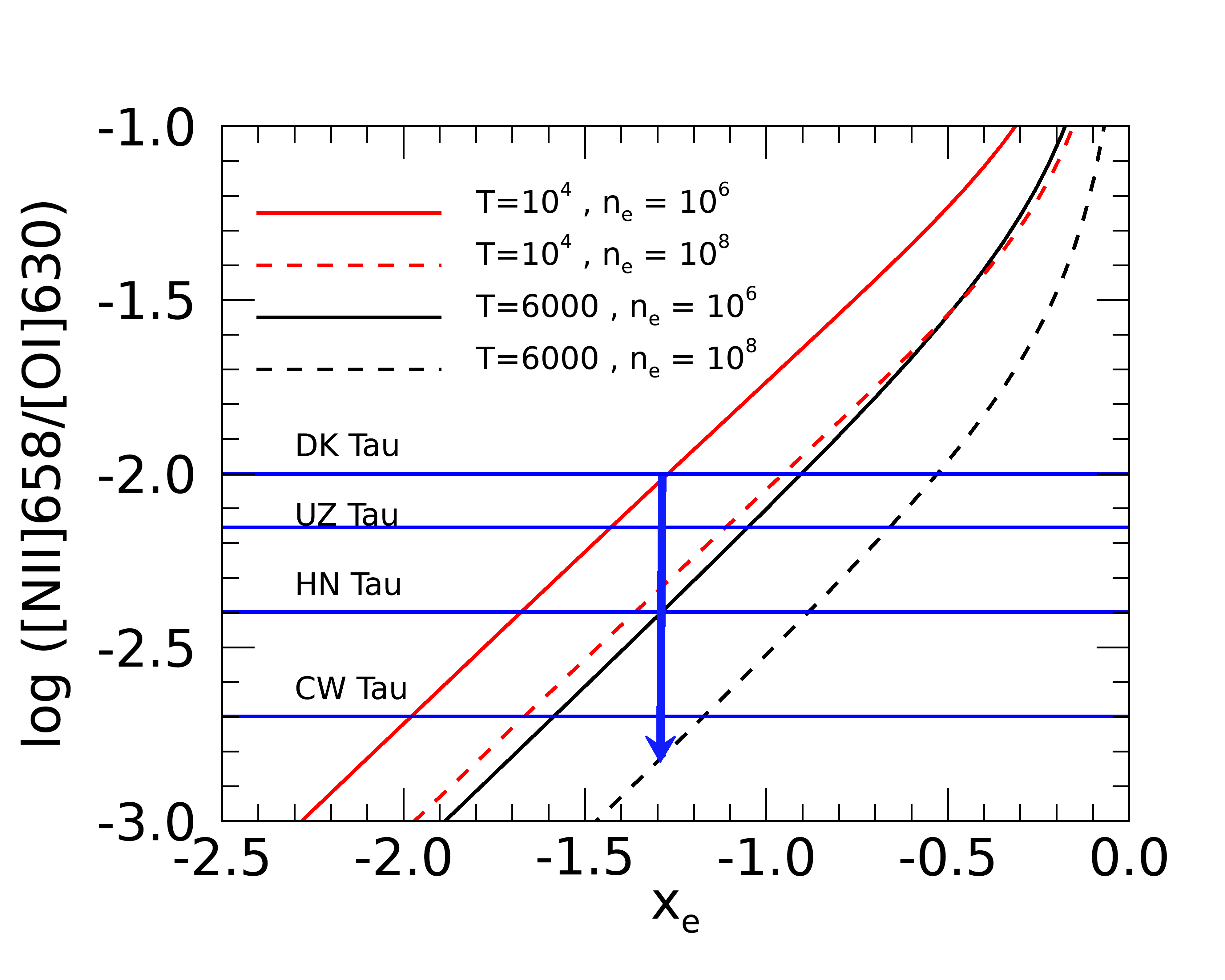}\\
\includegraphics[trim=0 0 0 0,width=1\columnwidth, angle=0]{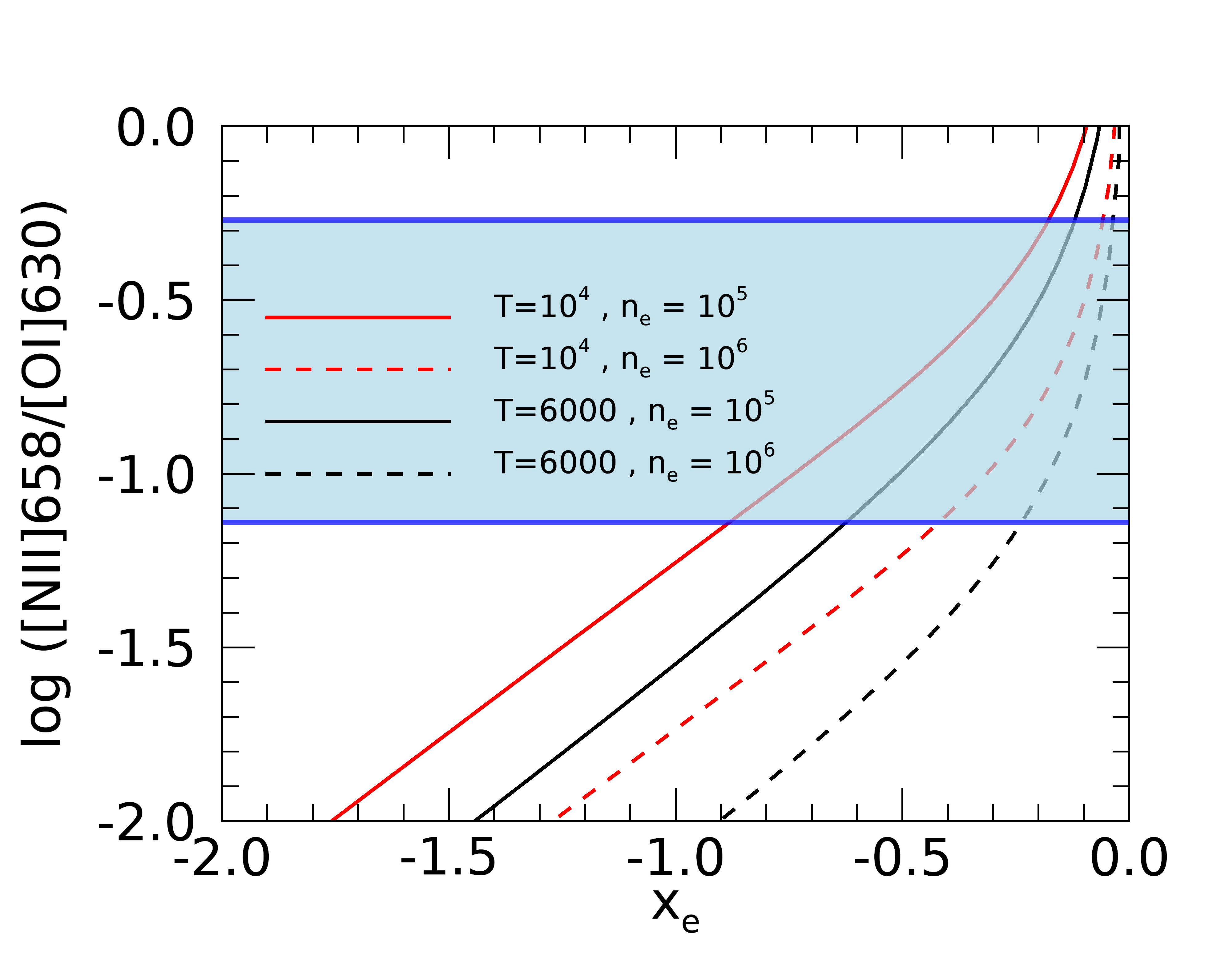}\\

\end{center}
\caption{Ionization equilibrium model predictions for the \nsix/\osix\, ratio for collisions with electrons, plotted as a function of the $x_e = n_e/n_{H}$ ratio. Different lines refer to different values of electron temperature and density as indicated in the legend. 
In the upper panel, horizontal lines indicate some of the more stringent upper limits of the  \nsix/\osix\, ratio in the NLVC, while in the lower panel the coloured region indicates the range of the \nsix/\osix\, ratio observed in the HVC.}
  \label{fig:658_xe}
\end{figure}

Fig. \ref{fig:658_xe} plots the ionisation equilibrium model predictions for the \nsix/\osix\, ratio as a function of  $x_e$, for different values of gas temperature and density. In this plot, horizontal lines with arrows indicate some of the upper limits derived from our data. They show that for the most stringent limits (e.g. in CW Tau), $x_e$ is constrained to be $<$ 0.1 for a density of 10$^8$ \cmt\, and T=6000 K or $<$ 0.01 for n=10$^6$ and T=10\,000  K.  These upper limits are consistent with the values of $x_e$ between 0.05 and 0.1 derived in the LVC of DG Tau, HN Tau and DO Tau from the analysis of multiple optical/IR forbidden lines \citep{giannini2019}. 
These values also imply total densities of the order of 10$^8$-10$^9$ \cmt . For the objects where the gas emission originates in the disk, this density corresponds to that predicted by thermo-chemical models in the inner disk region ($R <$ a few au) \citep[e.g.][]{woitke2019}.

The bottom panel of Fig. \ref{fig:658_xe} shows that in the HVC, where the \nsix\, line is often detected, the \nsix/\osix\, ratio implies $x_e$ between $\sim$ 0.1 and 0.5, for a density 10$^5$-10$^6$ \cmt\, and $T$ = 6000-10\,000 K,  in line with the values estimated from optical observations of resolved jets, which are in the range 0.3-0.5 \citep{frank2014}. 

\subsection{The \ssix\, line}

The \ssix\, line has a critical density two order of magnitudes lower than the \oi\, and \sii\, lines, and therefore it is able to intercept low density wind regions located further out with respect to the other lines. This line has been detected more frequently in the HVC (36\% of sources) while only five sources show the NLVC. In these cases, the \ssix\, profile is similar to that of \sfour\ both in peak velocity and FWHM. However, the observed \sfour/\ssix\ ratio in the range 1-10  is consistent with a density  $\sim$ 10$^5$-10$^6$ \cmt\, i.e. lower by at least 1-2 order of magnitudes with respect to what implied by the \ofive/\osix\, ratio. Presumably \ssix\ intercepts larger regions of the wind at a too low density to significantly excite \osix\ . This has been also demonstrated by spectro-astrometric observations of \osix\, and \ssix\ performed on RU Lup \citep{whelan2021}, showing that the \ssix\  in the NLVC is extended in a wind region of 40 au, up to five times larger than \osix\ . 
Both the photoevaporative and MHD models by \cite{weber2020} predict a \ssix/\osix\ ratio close to 1, while in the five sources where \ssix\ is detected the observed ratio is in the range 0.1-0.2. As for the case of the \sfour\ line, the general lower than expected ratio of sulphur lines with \osix\ can be explained either assuming that sulphur is depleted by dust in the upper layer of the dusty disk atmosphere or by a non negligible contribution of UV pumping to the \osix\, excitation.

\section{The mass-loss rate in winds and jets}\label{sect::mloss}

The mass-loss rate is one of the fundamental parameters governing the evolution of proto-planetary disks as it sets the timescales for disk dispersal. Deriving the mass-loss rate for all the components gives us a measure of how much material is removed from the disk upper layers as a function of the radial location on the disk.
The derivation of the mass-loss rate of the atomic outflow components from single on-source spectra is severely hampered by the lack of spatial information needed to resolve the jets/winds. Nevertheless, flux-calibrated high resolution spectroscopic data are very powerful to investigate the mass-loss from the different kinematic components of the wind in large sample of sources \citep[e.g.][]{hartigan95, nisini18,fang18}. 

Here we discuss the methods used to infer the mass-loss rate in both the low and high-velocity components from single source high resolution spectra, discussing their caveats and limitations and how they might affect the results.

\subsection{\mwind\, in the LVC}\label{sect::mloss}

The determination of the mass-loss rate in the wind traced by the LVC is subject to large uncertainties due both to the poorly determined physical conditions but also to the unknown physical extent and geometry of the wind.

Rough estimates of \mwind\, in the LVC have been provided by \cite{natta14}, who computed the volume of the \osix\, line emission and considered the extreme cases of spherical and cylindrical geometry. A similar approach was used by 
\cite{hasegawa22}. \cite{fang18} separately estimated the \mwind\, from the NLVC and the BLVC, assuming that the \osix\, line emission is in LTE and that the wind length is proportional to the Keplerian radius inferred from the line FWHM. They concluded that on average, the \mwind/\macc\, ratio in the NLVC is around 0.01-0.1, if one assumes a wind length comparable with the wind radius at the base (i.e. around 1-2 au). This work also suggests that the \mwind\, in the BLVC is higher than that of the NLVC, basing this latter statement on the assumption that the broad component has the same or smaller wind length than the narrow component.

Here we revisit the \mwind\, estimate in the NLVC using a different approach, which is also subject to large caveats but allows us to analyse the issue from a different perspective.

The mass-loss rate can be written as the mass that pass through a cross-sectional area in the unit of time 

\begin{equation}
    \dot M_{wind} =  \mu\,n_H\,\pi\,R_w^2\,{\rm v_w}
\end{equation}

where $n_H$ is the total density and $R_w$ the radius at the wind base.

\begin{figure}[h]
\begin{center}

\includegraphics[trim=0 0 0 0,width=1\columnwidth, angle=0]{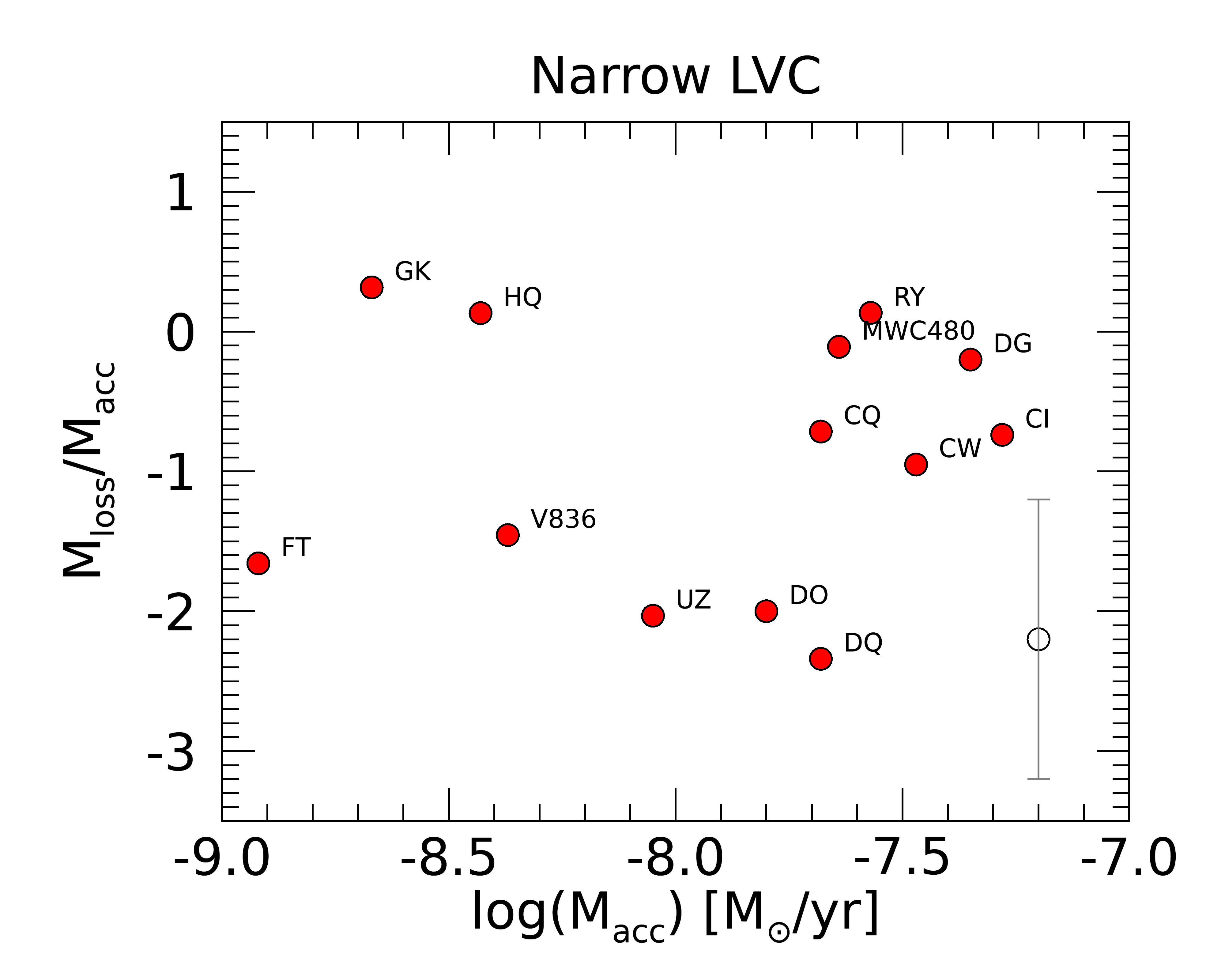}\\
\includegraphics[trim=0 0 0 0,width=1\columnwidth, angle=0]{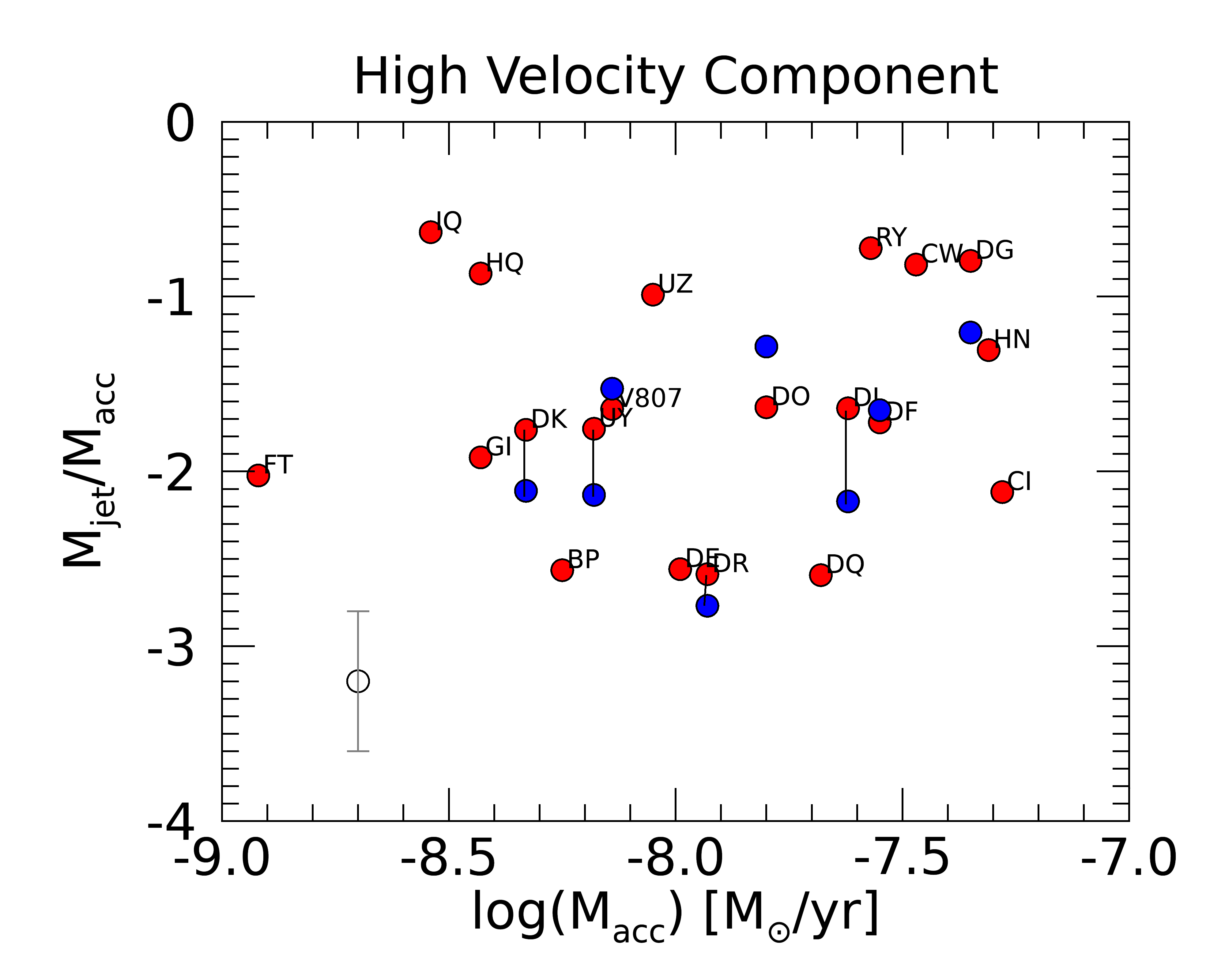}\\
\end{center}
\caption{The ratio $\dot M_{loss}/\dot M_{acc}$, plotted as a function of the source $\dot M_{acc}$. {\bf Upper panel}: $\dot M_{loss}$ for the narrow low velocity component estimated as in Section 7.6 and assuming a fractional ionisation $x_e = 0.1$. {\bf Lower panel}: One sided $\dot M_{loss}$ for the HVC. In cases where the HVC can be decomposed into two additional components at medium and high-velocity, the mass-loss has been separately estimated, with the medium velocity indicated with blue symbols. 
  \label{fig:eta}}
\end{figure}

For our rough estimate we assume that R$_{w}$ is the emitting radius estimated as the Keplerian radius from the \osix\, line width. 
The total density is given by $n_H = n_e/x_e$, where $n_e$ is the density at the wind base, that we derive from the \ofive/\osix\, ratio assuming T=8000 K (see Sect. 7.2).  We consider $x_e = 0.1$, which is the upper limit to the 
ionisation fraction derived in Sec. 7.3,  while the wind velocity v$_w$ is taken from the wind peak velocity corrected for the inclination angle.
We measure the mass-loss rate only for those sources having a blue shifted velocity peak, outside the wavelength calibration error.
In Fig. \ref{fig:eta} we plot the so derived mass-loss efficiency, i.e. the \mwind/\macc\, ratio, as a function of the source mass accretion rate. The mass-loss rate in most cases is between 0.1 and 1 times \macc\, although a few sources show lower values of the order of 0.01, without a clear dependence on \macc .

The uncertainties associated to the above derivation are however large. First of all the total density is poorly constrained since we have only an upper limit on the fractional ionisation.  \cite{giannini2019} estimated, from an analysis that takes into consideration a larger set of lines in DG Tau, HN Tau and DO Tau, that $x_e$ in the LVC can be as low as 0.03, which would increase the final \mwind\, value by up to a factor of 3-4. On the other hand, the electron density giving rise to the \osix\, emission could be overestimated if the \ofive\, line, with a critical density an order of magnitude larger than \osix\, , traces a much more compact region in wind with a large density gradient (see Discussion in Sect. \ref{sect::PE_model}).
Also, the estimated mass-loss rate would be lower if the emitting area is confined to a smaller section of the disk around the wind launching region. For example, if we assume that the region involved in the launching is $\Delta R/R \sim 0.1$, then \mwind\, would decrease by about an order of magnitude. Overall the above uncertainties bring uncertainties on \mwind\ to more than one order of magnitudes on either sides. 

We also consider that the line luminosity can be used to measure the volume of the oxygen gas giving rise to the observed emission \citep[see e.g.]{natta14}:

\begin{equation}
    V = \frac{L({\rm [\ion{O}{I}]630})}{n_H\,\epsilon (\rm [\ion{O}{I}]630)\,X(O)}
\end{equation}

where $\epsilon ({\rm [\ion{O}{I}]630})$ is the emissivity of the \osix\, line 
in erg/s, and $X(O)$ is the total abundance of neutral oxygen, for which we consider the solar value.
With the approximation of a cylindrical geometry and a wind area at the base equal to $\pi\,R_{w}^2$, we  have from here a rough estimate of the wind length above the disk. The estimated wind length is small, in many cases only a fraction of au, with a typical $l_w/R_w$ ratio between 0.1 - 1. 

Our estimate of \mwind\, for the NLVC is typically larger than the values derived by \cite{fang18}, mainly due to the fact that we infer a $l_w/R_w < 1$ while \cite{fang18} assume that the wind height is larger than the corresponding Keplerian radius.  
We have however pointed out that given the large caveats in the determination, any estimate for \mwind\, in the NLVC can span about two order of magnitudes, from $\sim$0.05 to more than $\sim$1 the \macc, value.

\subsubsection{Empirical determination of \mwind\, from a PE model}\label{sect::PE_model}

In order to check how well our determination of \mwind\, from the \osix\, line can reproduce the actual mass-loss in the wind, we performed a test on the models presented in Weber et al. (2020, W20), who post-processed the output of the X-ray PE wind model by \cite{picogna2019}. 
To this aim, we apply the procedure described in Sect. \ref{sect::mloss} to the synthetic line profiles and luminosities given as the output of the model, and verify a posteriori whether the derived parameters are consistent with the theoretical physical and kinematical structure of the wind as predicted by the model.

\begin{figure}[h]
\begin{center}
\includegraphics[trim=0 0 0 0,width=1\columnwidth, angle=0]{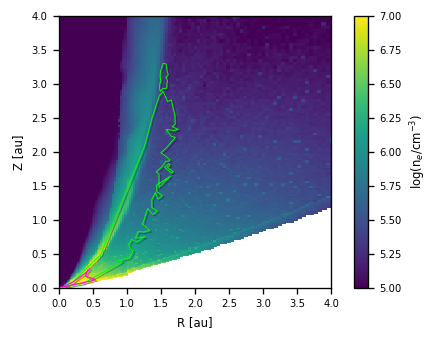}
\end{center}
\caption{Map of the electron density of a photo-evaporative wind model from W20, with overlaid contours of the 80\% emission regions of \osix\, (green) and \ofive\, (magenta)}. 
  \label{fig:model}
\end{figure}

As described in details in W20, the various forbidden lines trace different regions of the wind, depending on their excitation conditions. To illustrate this point, we show in Fig. \ref{fig:model} a density map for the model output that considers irradiation from a source with accretion luminosity \lacc=2.6\,10$^{-2}$\lstar\, (model PE1 in Table 1 of W20). 
Green and magenta lines delineate the wind regions where the \osix\, and \ofive\, lines emit 80\% of their luminosity, respectively. It can be seen that the \ofive\, line is emitted from a region much more compact than \osix\ and thus it traces only the very base of the wind. Table 1 reports the physical parameters obtained by averaging the values within the emitting regions of the two lines, for two models with different \lacc, i.e. \lacc=2.6\,10$^{-2}$\lstar\, and \lacc=\lstar , which also roughly correspond to the extreme values that we estimate for our sample.

The average densities significantly differ for the emitting regions of the two lines, and consequently the $n_e$ value estimated from the \ofive/\osix\, ratio is in between the two values, and thus overestimating the density in the \osix\, emitting region, typically by a factor of ten. 

To test the method described in Sect.\ref{sect::mloss}, 
we have derived $R_{w}$ from the Gaussian fit of the synthetic line profile and the electron density from the \osix/\ofive\, predicted luminosities, and applied the formulae (1) and (2) to derive \mwind\, and the length of the emission region. For this test, the W20 model that assumes a total wind mass-loss rate \mwind = 10$^{-8}$\msunyr\, and a disk inclination angle of 60 degree has been considered. 

Table \ref{table:model} summarises the output of this test for the two accretion luminosity models. For the model with low accretion luminosity, assuming the density from the \ofive/\osix\, ratio and $x_e = 0.1$, the total density is overestimated by a factor about 10 with respect to the average in the model. On the other hand, the Keplerian radius estimated from the line width gives a slightly underestimated size of the emission region. Consequently, the final mass-loss rate derived from equation (1) is similar to the mass-loss in the \osix\, emitting region predicted by the model, i.e. $\sim$ 2\,10$^{-10}$ \msunyr ,  but a factor of 100 lower than the total wind mass-loss rate of the model. This is because the wind is much more extended than the line emission (up to > 100 au) and the \osix\, line traces only the inner part of it (1-2 au). If we then consider our estimate of the emitting volume and length for the \osix\, emitting region, we see that we underestimate both quantities, 
as we adopt a higher total density with respect to the average value in the model. In particular, the \osix\, emitting region extends up to about 3 au above the disk, while the measured length with our approximation is about 10 times smaller. This discrepancy is also due to the fact that the 'empirical' volume is compared to the 80\% emission volume of the model under the assumption of uniform emission, when in reality the emission in the model is higher in the parts close to the star, which contribute only little to the volume, and lower in the outer parts of the emission region, which contribute the most to the volume. 

Considering now the other extreme case of \lacc=\lstar, we see that a higher accretion luminosity increases the extension of the \osix\, emission region (see also discussion in W20). 
In this case, the \mwind\, inferred from observations is about 15 times larger than the mass-loss rate in the \osix\,  emitting region predicted by the model, but, noticeably, very close to the total mass-loss rate in the entire wind. 

\input{Tables/table5.tex}

Therefore our conclusion is that the empirical estimates given in Section \ref{sect::mloss} can be considered as an upper limit on the mass-loss rate in the \osix\, emitting region, but a lower limit to total wind mass-loss rate. We remark that although this test has been performed in the assumption that the X-ray PE wind model by W20 gives a close to real representation of the physical structure of the observed low velocity wind, the results are more general. In fact, the main assumption here is that the physical parameters, and density in particular, have a large gradient in the wind, therefore the information inferred from a single, spatially unresolved line are intrinsically limited. 

\subsubsection{Limitations of \osix\, as a diagnostics for the LVC mass-loss rate }

Given the above discussion, we summarise here the limits and uncertainty of the mass-loss rate determination based on the \osix\, line emission.

In general, the main limitation is that the wind geometry and extent are not known, which prevents to correctly estimate the length scale for the mass transfer. A second important limitation is the fact that large gradients in physical conditions are expected in the wind, and thus the lines whose ratio is used to derive the density intercept portions of the wind with different densities and temperatures. Finally, the choice of the tracer for the mass-loss is also important. In particular, we have shown that the \osix\, line might trace only the very inner wind region, which not necessarily gives the main contribution to the mass-loss. 

Given the above considerations, our estimate for \mwind\, in the NLVC might easily span about two orders of magnitude, from $\sim$0.05 to more than $\sim$1 the \macc, value. If one takes the X-ray PE model described in \cite{weber2020} as a reference, these values should be considered only as lower limits to the global mass-loss in the wind. Therefore we cannot exclude the possibility that the mass-loss in winds ejected at radial distances larger than 0.5 au is of the same order of magnitude or even larger than the mass accretion rate. 

\cite{fang18} suggested that the \mloss/\macc\, in the BLVC is higher than that of the NLVC and thus that most of the wind mass-loss occurs close to the central star. 
 This last conclusion is mainly driven by the higher luminosity and peak velocity of the BLVC with respect to the NLVC. While this is an objective observational difference, we think that the limitations discussed above do not allow one to draw solid conclusions on the difference in the mass-loss between the NLVC and the BLVC. In particular, the \cite{fang18} analysis is based on the assumption that the gas in both the NLVC and the BLVC is in LTE, which implies $n_e > $ 10$^7$ \cmt . However, if the density in the \osix\, emitting region is lower and different in the two components that trace different disk regions, then the \mwind\ estimated from the line luminosity will scale accordingly.  

 Also, the contribution to the mass-loss rate from different gas components, not traced by the optical lines, might be significant in the outer disk regions, for example in regions where the temperature drops below about 5000 K. 
For example, wide angle uncollimated molecular winds have been resolved in the $H_2$ near-IR lines \citep[e.g.][]{agra-amboage11,beck2019} and show similar peak velocities as the \osix\, NLVC lines, but narrower FWHM, indicating they originate at larger disk radii \citep{gangi20}. Also, recent high resolution ALMA observations in HH30 and DG Tau B have shown the presence of cold CO molecular wide winds, whose estimated mass-loss rates are of the same order or even exceed the mass accretion rates for the sources \citep{louvet2018,devalon2020}.   
In conclusion, we think that our knowledge of the mass-loss rate in winds based on observations of optical forbidden lines is too uncertain to allow one to discriminate among different models and to evaluate the impact of the atomic winds on the overall disk dispersal. 

\subsection{\mloss\, in the HVC}

We can use the luminosity and kinematical properties of the HVC to infer the mass-loss rate in the jet. This can be expressed as $\dot M_{jet} = M_{gas}\times V_{jet}/l_{jet}$, where $V_{jet}$ and $l_{jet}$ are the jet velocity and length, respectively. As for the LVC, here the jet velocity $V_{jet}$ is estimated from the radial velocity of the line in exam, corrected for the jet inclination angle. We can however now remove the uncertainty on the geometry if we assume that the jet is more extended than the instrumental FoV, so to use the aperture size as the scale length in which $M_{gas}$ is measured.

Unless estimates of jet width and total density is available for the mass measurement, $M_{gas}$ is usually derived directly from the luminosity of the forbidden lines, assuming uniform temperature and density in the considered aperture. \cite{hartigan95} described in details the method for optical lines, which has been then applied in several subsequent studies.

Unless only a HVC red shifted component is detected, as in the case of BP Tau and GI Tau, we consider here only the blue shifted component and use half of the aperture width, which will give us the mass-loss rate only in the one-sided jet. In the assumption that the jet is symmetric but the red shifted is poorly visible due to a higher extinction or occlusion by the disk, the derived \mjet\, value needs to be multiplied by 2.
For all our sources, the blue shifted HVC can be decomposed in one or two Gaussians. We separately derive \mjet\, for the observed components, that we generically name high- and medium- velocity components (as done in previous studies for resolved jets, as for example Agra-Amboage et al. 2011).

The multi-line analysis performed by the pilot study of \cite{giannini2019}, shows that 
the typical conditions pertaining to the HVC are 
$n_e \sim 4\times10^4-10^5$ \cmt and $T_e \sim 10\,000 - 15\,000$ K. We assume here $n_e = 5\times 10^4$ \cmt\, and $T_e$ = 10\,000 K, on the basis that the line ratios in other objects do not differ within the error from those of this pilot study, and to be also consistent with the \mjet\, values measured in a similar way by \cite{nisini18} in a sample of Lupus and Chamaeleon sources. Considering however the emissivity variations in the range of density and temperature values, the uncertainty on the derived \mjet\, is almost a factor of ten. 

\input{Tables/table6.tex}

\input{Tables/table7.tex}

Table \ref{table:MHVC} reports the derived \mjet\, for the sources showing a HVC, together with the derived values of the total jet velocity. 
In Table \ref{table:mlosscomp} we collect \mjet\, derivations from other studies 
based on spatially resolved observations of the jet, to discuss whether they are comparable with our measurement of the mass-loss done based on individual spectra.
Variations among different determinations for the same jet are typically a factor of a few and are mostly due to different methods, tracers and assumptions. For example the largest differences are observed on RW Aur A and CW Tau, where the \mjet\, in the resolved jet was estimated assuming the electron density from the ratio of the \ssix\ doublet, which provides only a lower limit to the actual gas density \citep[e.g.][]{maurri14}.  
We should however keep also in mind that differences in \mjet\, could be due to time variability of the mass-loss phenomenon \citep[e.g.][]{takami2022}, as seen in Section 5 for some of our sources. For example, the DG Tau \osix\, luminosity in the high-velocity component has more than doubled with respect to the value prior to 2012.
For this specific case, however, the terminal velocity in 2012 was also higher than our measured value ($\sim$ 400 \kms\, vs 200 \kms). Therefore, the linear momentum $M\times V$, and thus the \mjet , remained almost constant in time.

In Fig. \ref{fig:eta}, bottom panel, we plot the ratio \mjet/\macc\, for the HVC with respect to the source mass accretion rate. We separate with red and blue symbols the mass-loss rates values relative to the high and medium velocity components, respectively. As also found by \cite{nisini18}, the ratio spans more than an order of magnitude and it is mostly confined between 0.01 and 0.1, with a no clear dependence on the mass accretion rate. 

\section{Conclusions}
We have analysed the profile and luminosities of optical forbidden lines in 36 CTTs of the Taurus-Auriga cloud whose disks span a range of different characteristics, and compared the properties of the inner atomic gas with the structure and properties of the outer disks. Following previous studies, we have decomposed the line profiles into HVC, NLVC, and BLVC components, focusing our analysis in particular on the NLVC.  We used ratios of different forbidden lines to constrain the physical conditions of the atomic gas (temperature, density and ionisation fraction). For sources whose \osix\ profile peaks at the stellar velocity, we fitted the profile with a simple Keplerian disk model, giving constraints on the size of the emission line region. For those sources with blue shifted velocity centroids, indicative of an origin in slow winds, we constrained the wind mass-loss rate discussing the origin of all uncertainties in its determination. We finally derived the mass flux rate in the HVC and compared it to the values derived in the same sources from spatially resolved observations of the jets. Our main results can be summarised as follow: 
\begin{itemize}
    \item We find that in about 40\% of our sources the peak velocity (V$_p$) of the NLVC is compatible with the stellar velocity. Such sources comprises all the transitional disks (TD) and are typically those having a single LVC, lower mass accretion rates and absence of a HVC. They therefore might represent later evolutionary stages where the emission from the disk is dominant with respect to the wind contribution. Full disks and disks with sub-structures do not show differences in the forbidden line properties.

    \item Assuming, for the lines with no-velocity shift, a Keplerian disk model with a brightness scaling as a power-law index,  we derive that the emission line regions extend from $\la$ 0.01 au, and thus within the dust sublimation radius in some of the cases, up to several tens of au in a few sources. In these models, the characteristic radius $R_{kep}$, estimated from the HWHM of the line, represents the region within which on average 50\% of the emission is confined. 
   
    \item We find that the $R_{kep}$(OI) in TD does not increase with the size of the sub-mm dust cavity, therefore the clearing of inner warm atomic gas probing the upper disk surface layer does not follow the dispersal of cold molecular gas and dust during the late disk evolution. We have found, in addition, an anti-correlation between the \oi 557/630nm line ratio and $R_{kep}$, which suggests that the emitting region of \oi\ is moving outwards in radius as the gas dominating the emission becomes cooler and less dense.
 
    \item We confirm previous findings that the line ratios observed in the LVC, if compared with a thermal single temperature and density model, implies $n_e \sim 10^6- 10^8$ \cmt\, and $T_e \sim 5000 - 10\,000$ K, and we moreover constrain the ionisation fraction in the NLVC to be $x_e < 0.1$. We however discuss the limits of such a diagnostic if applied to the emission in spatially not resolved winds, where physical conditions have large spatial gradients, as the different lines probe different regimes of density and temperature. 
     \item  We constrain the mass-loss rate in the wind traced by the blue shifted LVC finding a \mwind/\macc, ratio that might span between $\sim$ 0.01 and more than 1. When compared with synthetic \osix\, images from X-ray photoevaporation models, the estimated \mwind\, represents a lower limit to the total rate of mass-loss of the model, indicating that the \osix\, line is probably not the best tracer to probe the mass-loss in winds.
     \item For the HVC, we found \mjet/\macc\, ratios in the range 0.01-0.1, in line with previous findings in other star forming regions (for example Lupus and Cha, \cite{nisini18}). The \mjet\, derived from the on-source spectra are consistent, within a factor of a few, with the values inferred in spatially resolved jet observations, providing that similar assumptions are used. 
 \end{itemize}    

Two main conclusions can be drawn from this work. Firstly, we have shown that contribution from disk emission, in addition to disk winds, to the forbidden lines in T Tauri stars can be significant, especially for more evolved sources. 
We also conclude that without a better knowledge of the wind geometry and spatial extent, and given the limitation of the diagnostics, mass-loss rates in the wind traced by the blue shifted LVC cannot be constrained better than a factor of 100. 
    
For a step forward in our understanding of this issue, one would need observations at higher spatial resolution, to significantly constrain the wind emitting region, but also of tracers at lower excitation likely probing winds at larger disk radii. In this respect, future JWST observations will allow us to investigate in detail the range of conditions, between 1000 and 5000 K, presently poorly traced by ground-based observations. 
In addition, the determination of the mass-loss rate should involve modelling of both line profiles and luminosities of the different tracers with appropriate wind models.


\begin{acknowledgements}
We thank Andrea Banzatti for providing the Keck \osix\, spectra and for useful discussions, and Davide Fedele for providing the model of line profiles from a Keplerian disk. We also thank Barbara Ercolano for helpful insights about photo-evaporative wind models. We finally thanks the anonymous referee for the useful comments and suggestions.
This work has been supported by the projects PRIN-INAF 2019 "Spectroscopically Tracing the Disk Dispersal Evolution (STRADE)", PRIN-INAF 2019 "Planetary systems at young ages (PLATEA)" and by the Large Grant INAF 2022 YODA (YSOs Outflows, Disks and Accretion: towards a global framework for the evolution of planet forming systems). CFM is funded by the European Union (ERC, WANDA, 101039452). Views and opinions expressed are however those of the author(s) only and do not necessarily reflect those of the European Union or the European Research Council Executive Agency. Neither the European Union nor the granting authority can be held responsible for them. This work benefited from discussions with the ODYSSEUS team (HST AR-16129), \url{https://sites.bu.edu/odysseus/}.
\end{acknowledgements}

%

   \bibliographystyle{aa} 
   \bibliography{forbidden} 

\begin{appendix}

\section{Profiles decomposition}

Figures \ref{fig:profiles1}-\ref{fig:profiles6} show, for each of the 36 sources of our sample, the continuum subtracted spectra of the five lines analysed in this paper, with overlaid the derived fit to the profiles obtained summing up one or more Gaussians. 

\newpage

\begin{figure*}[h]
\includegraphics[trim=80 0 80 0,width=0.2\textwidth]{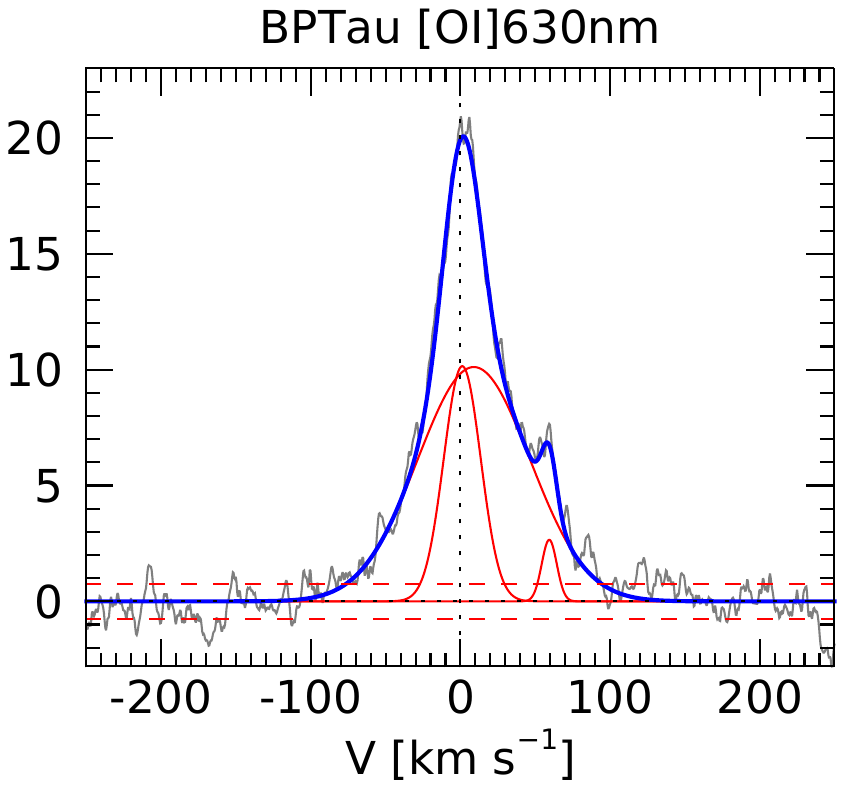}
\includegraphics[trim=80 0 80 0,width=0.2\textwidth]{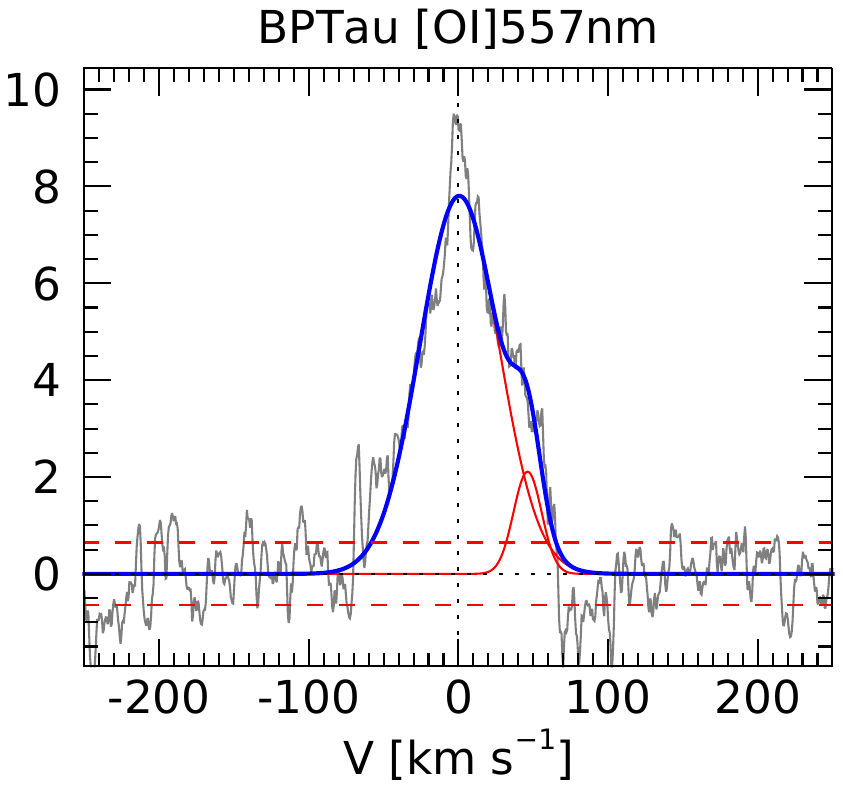}
\includegraphics[trim=80 0 80 0,width=0.2\textwidth]{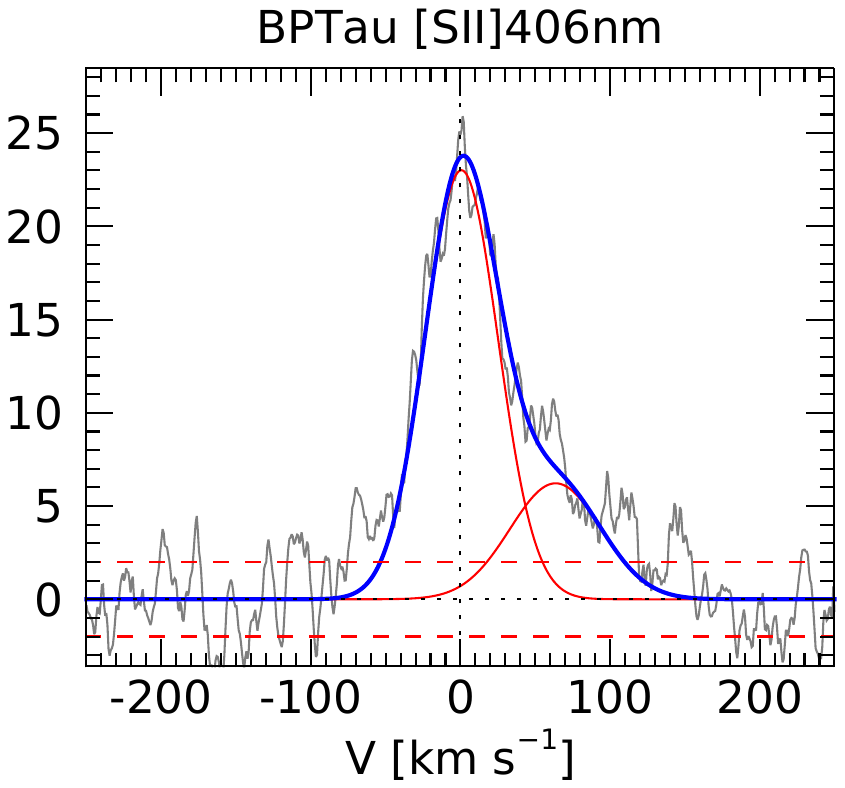}
\includegraphics[trim=80 0 80 0,width=0.2\textwidth]{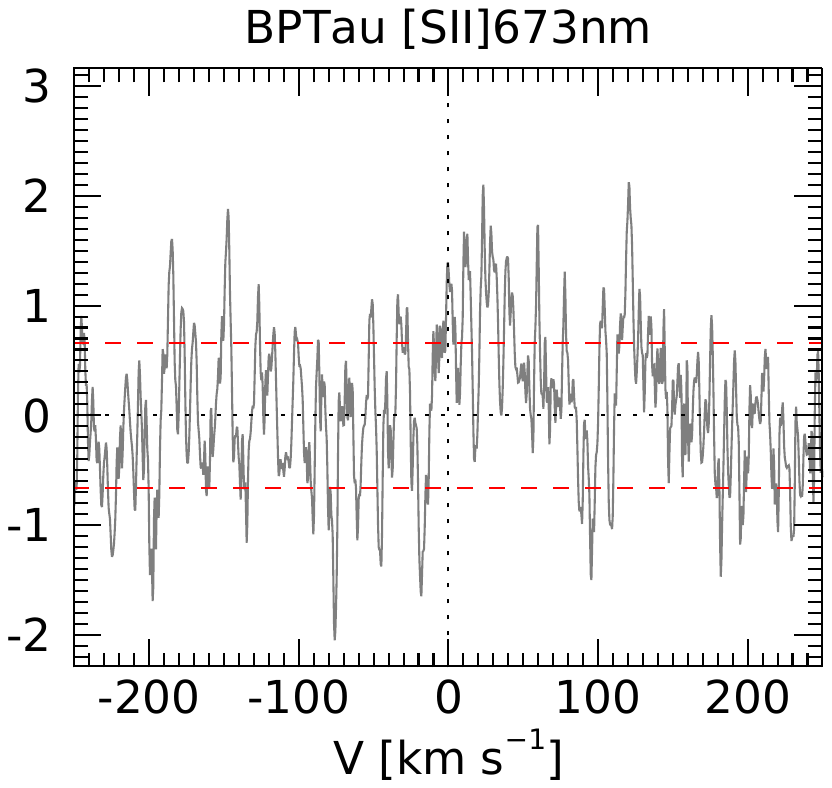}
\includegraphics[trim=80 0 80 0,width=0.2\textwidth]{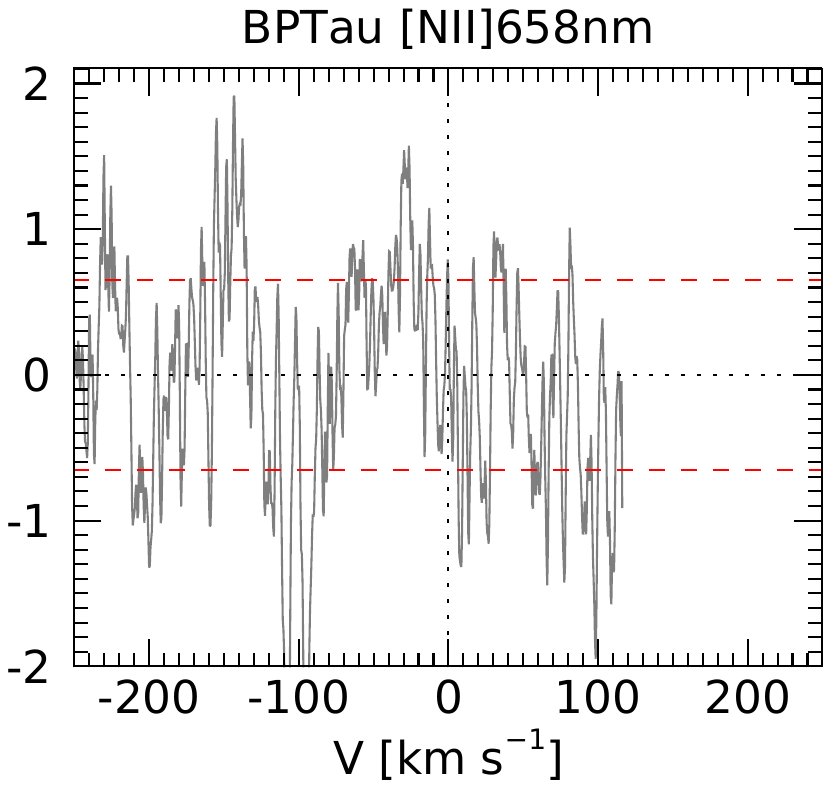}

\includegraphics[trim=80 0 80 400,width=0.2\textwidth]{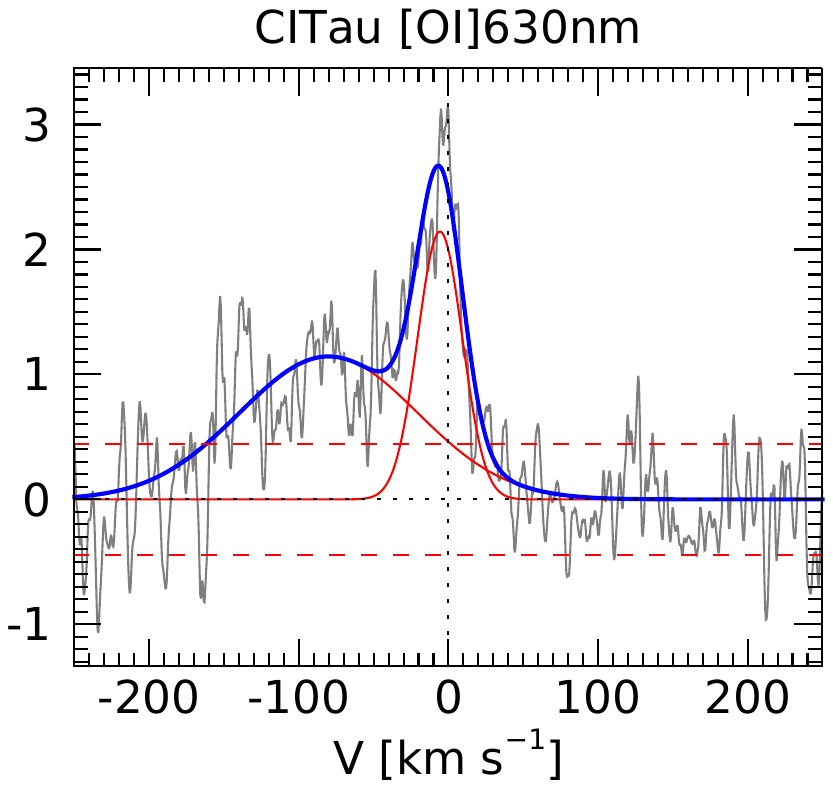}
\includegraphics[trim=80 0 80 400,width=0.2\textwidth]{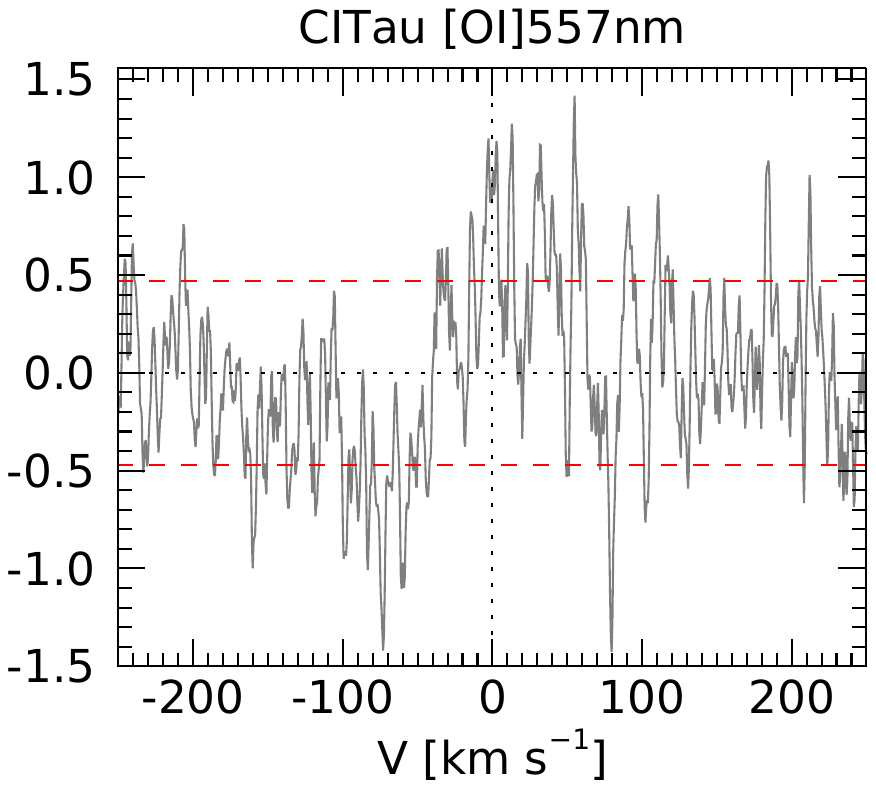}
\includegraphics[trim=80 0 80 400,width=0.2\textwidth]{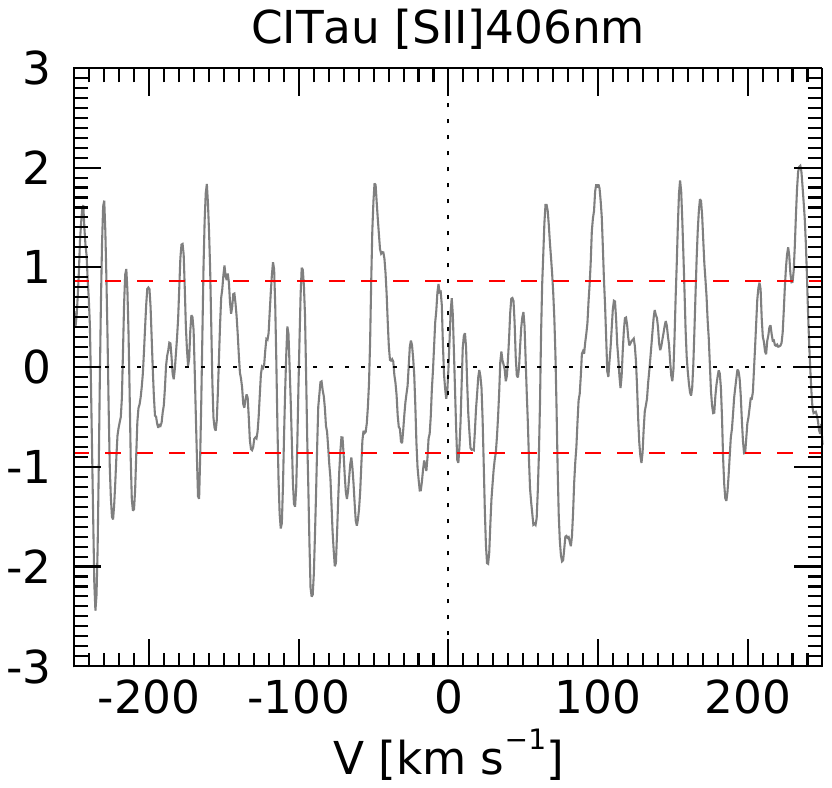}
\includegraphics[trim=80 0 80 400,width=0.2\textwidth]{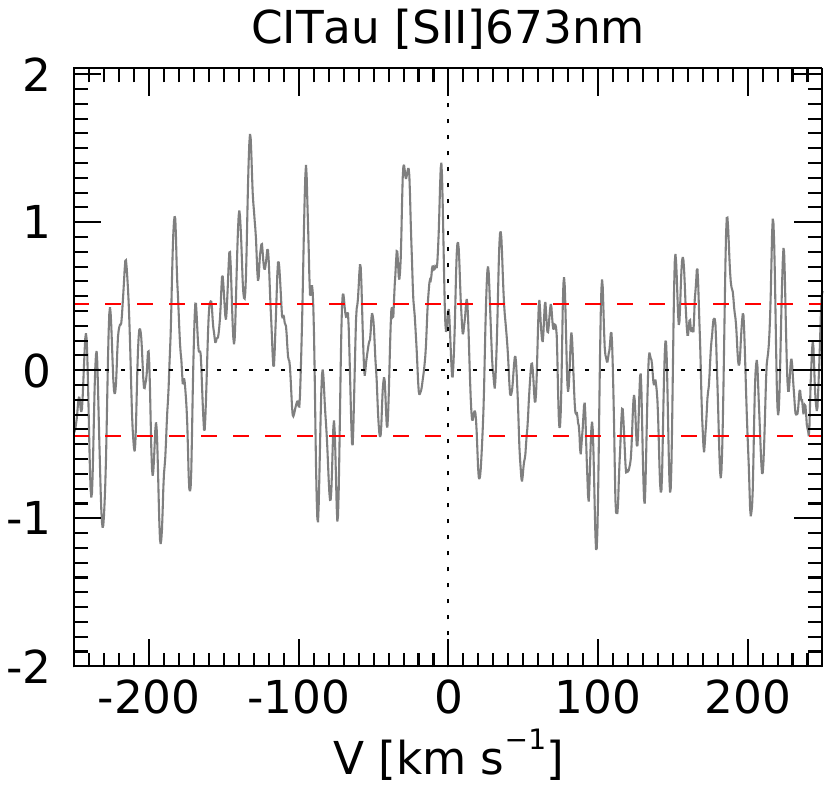}
\includegraphics[trim=80 0 80 400,width=0.2\textwidth]{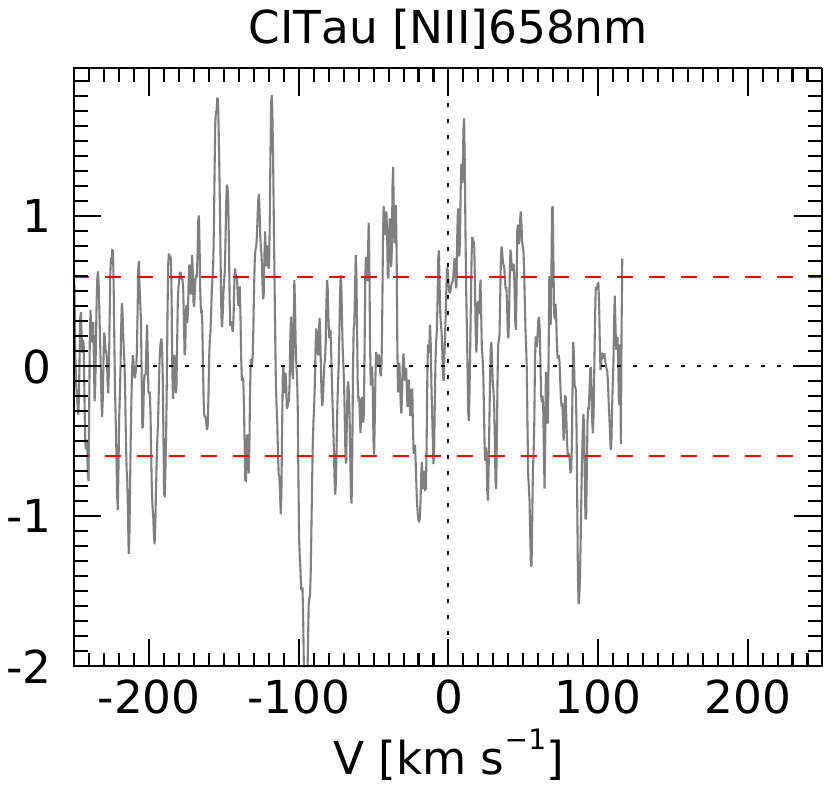}

\includegraphics[trim=80 0 80 400,width=0.2\textwidth]{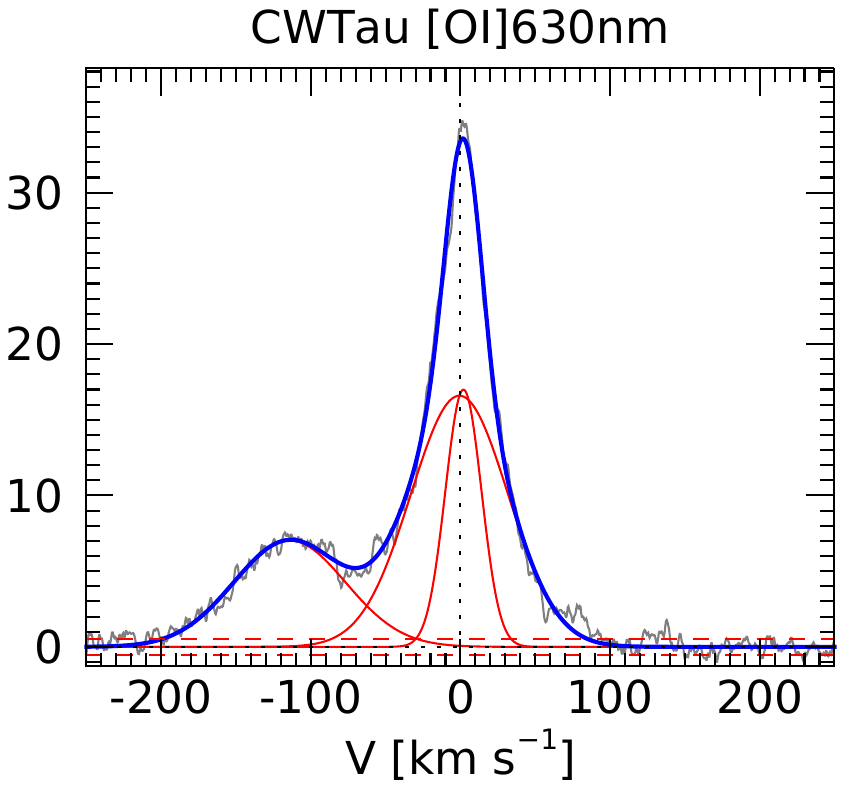}
\includegraphics[trim=80 0 80 400,width=0.2\textwidth]{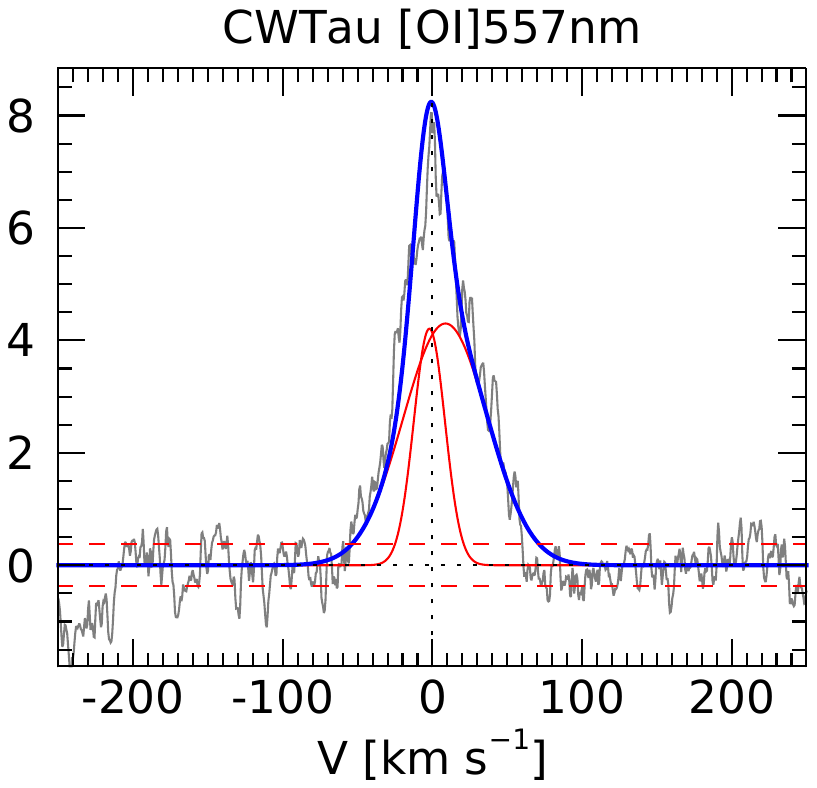}
\includegraphics[trim=80 0 80 400,width=0.2\textwidth]{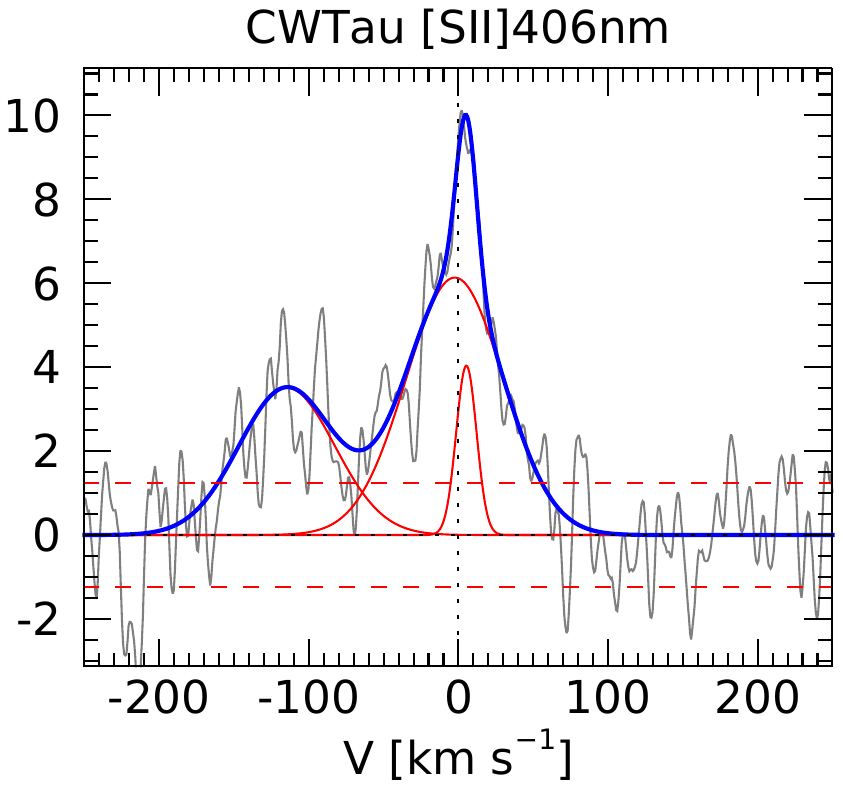}
\includegraphics[trim=80 0 80 400,width=0.2\textwidth]{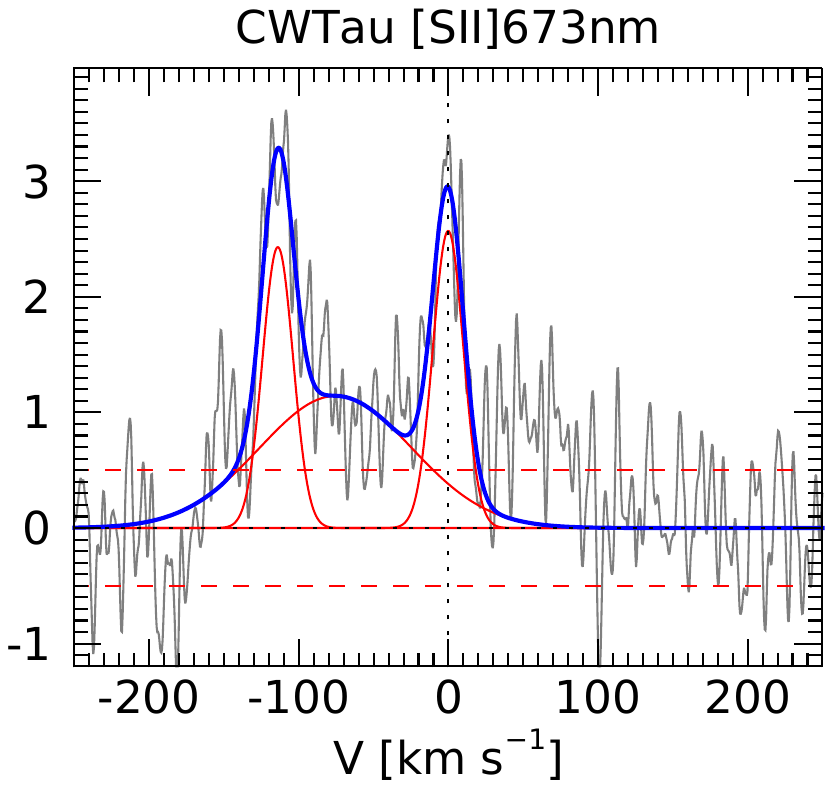}
\includegraphics[trim=80 0 80 400,width=0.2\textwidth]{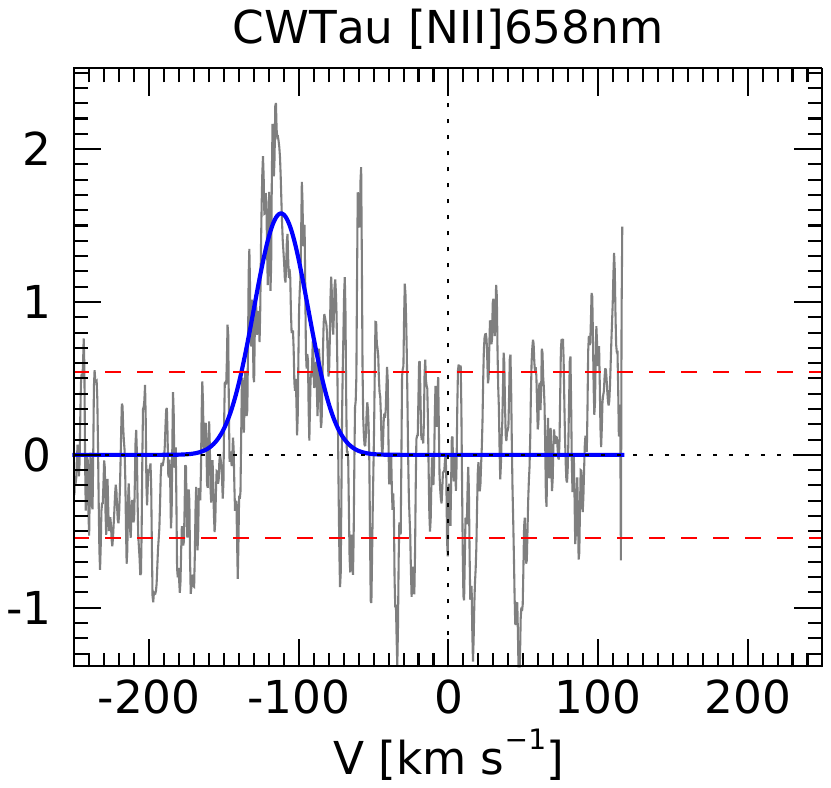}

\includegraphics[trim=80 0 80 400,width=0.2\textwidth]{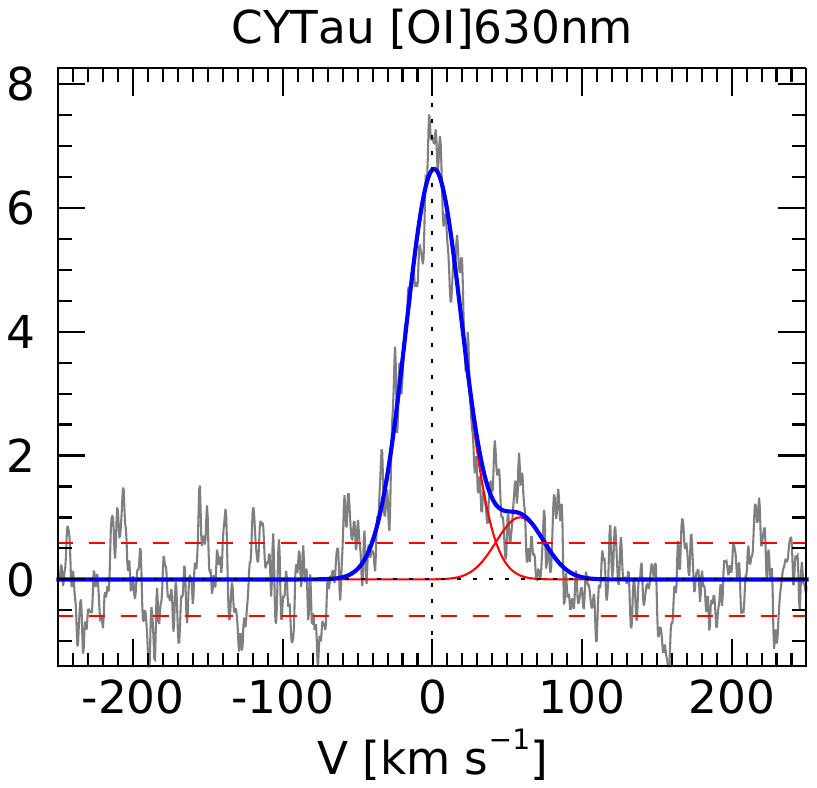}
\includegraphics[trim=80 0 80 400,width=0.2\textwidth]{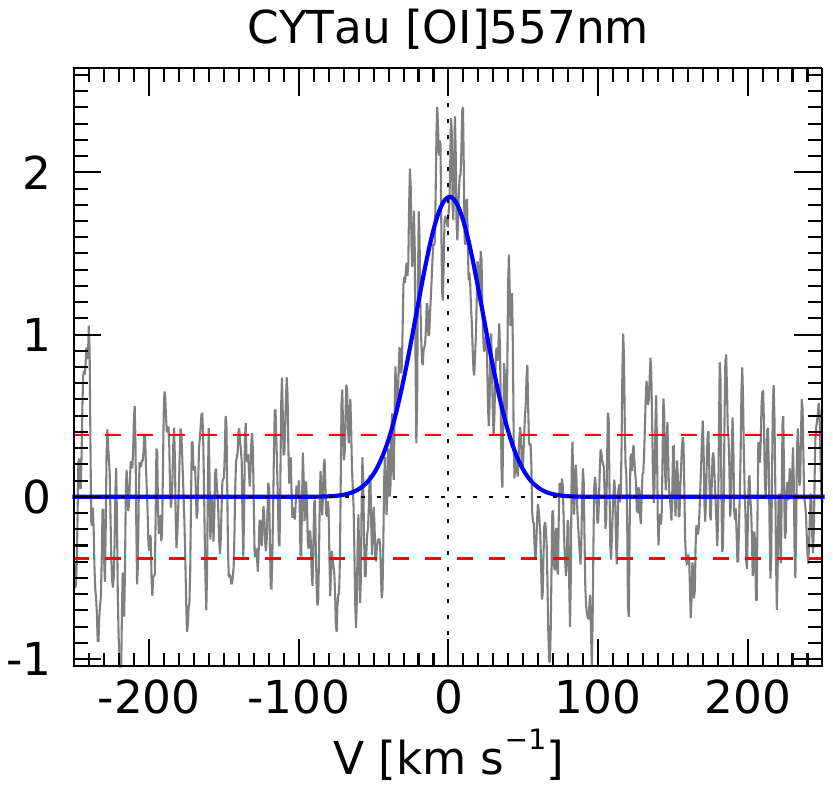}
\includegraphics[trim=80 0 80 400,width=0.2\textwidth]{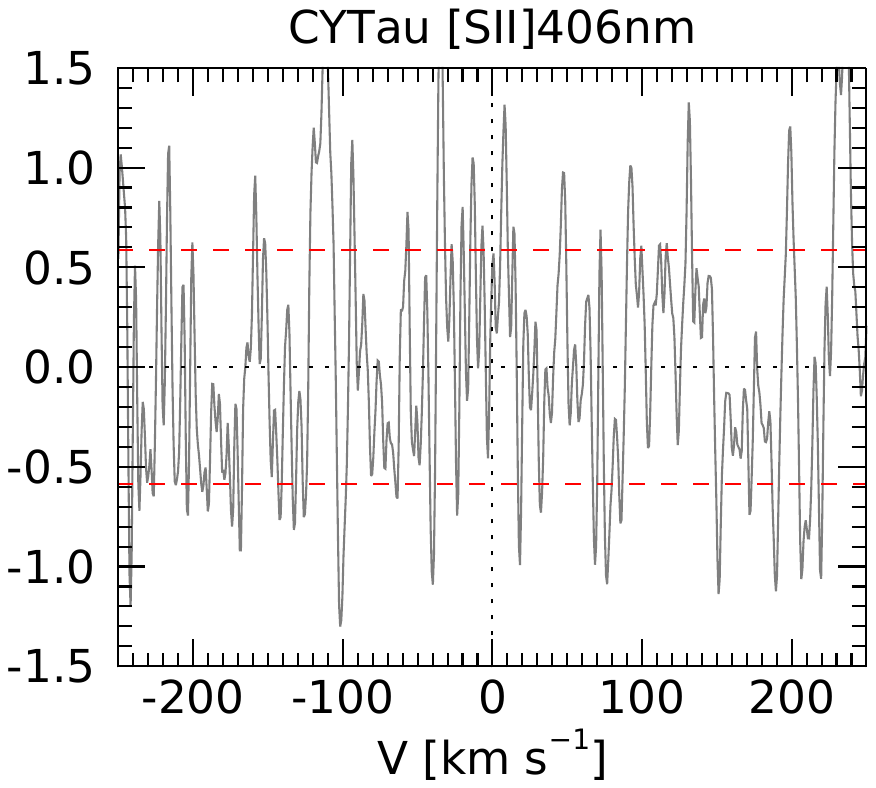}
\includegraphics[trim=80 0 80 400,width=0.2\textwidth]{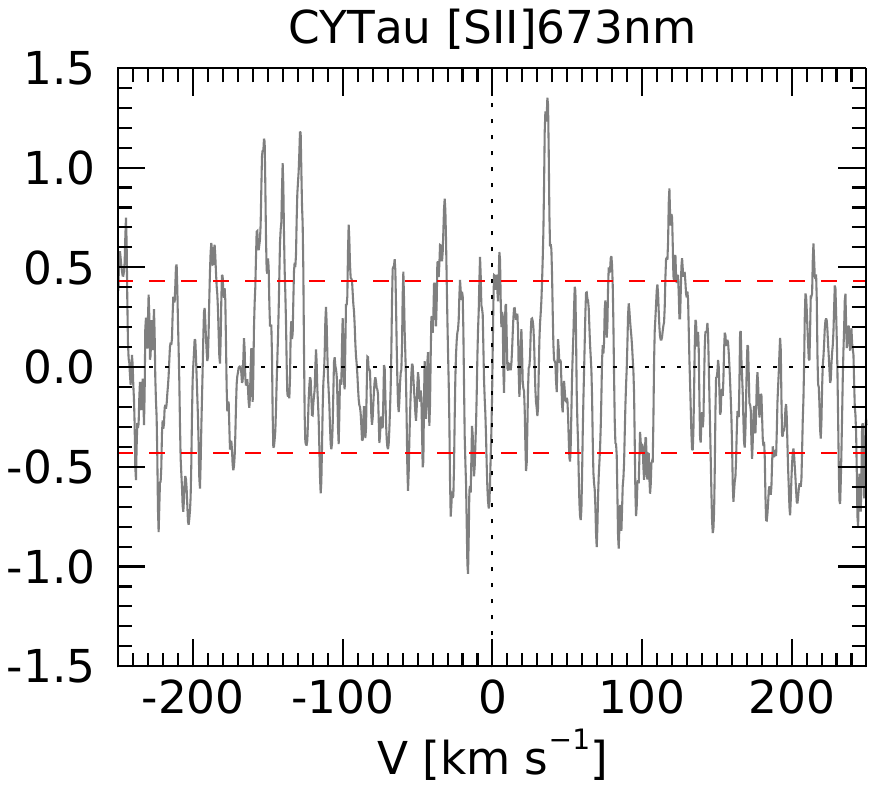}
\includegraphics[trim=80 0 80 400,width=0.2\textwidth]{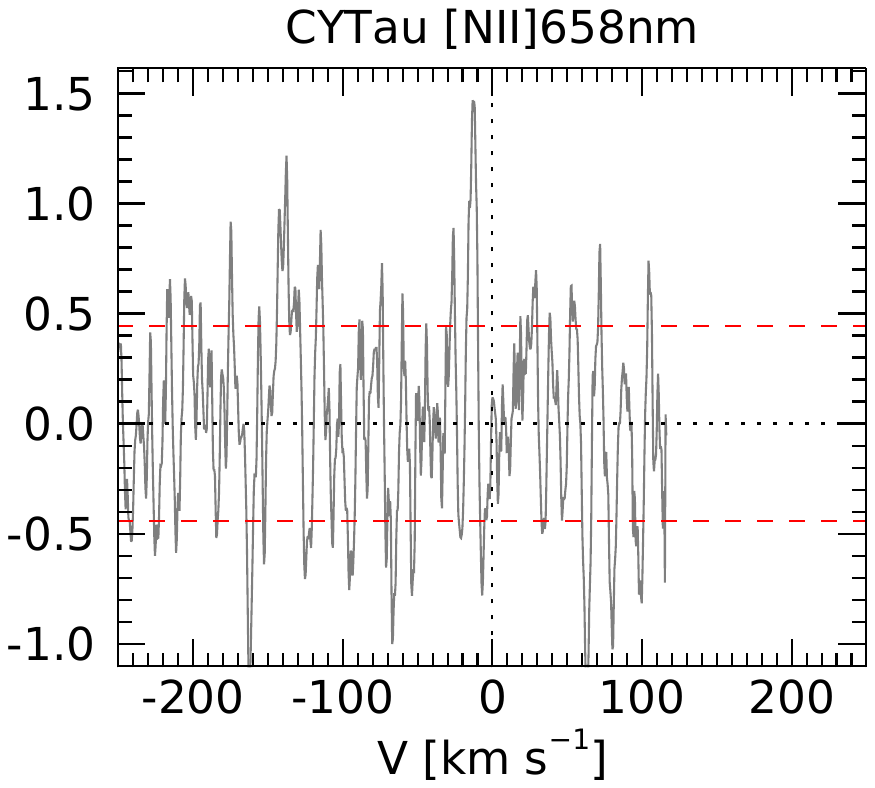}

\includegraphics[trim=80 0 80 400,width=0.2\textwidth]{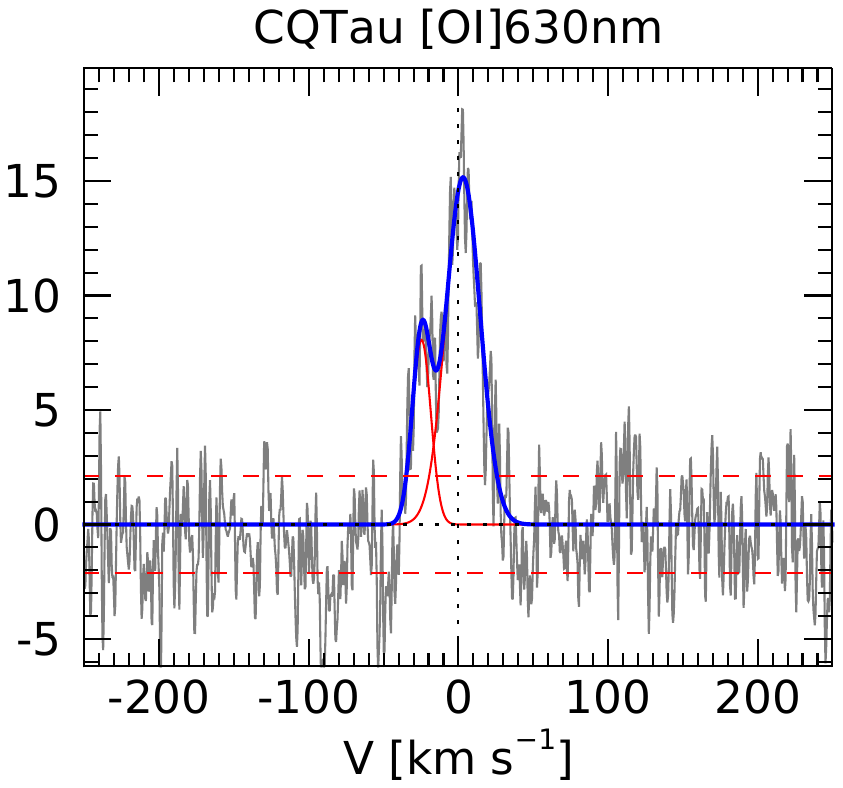}
\includegraphics[trim=80 0 80 400,width=0.2\textwidth]{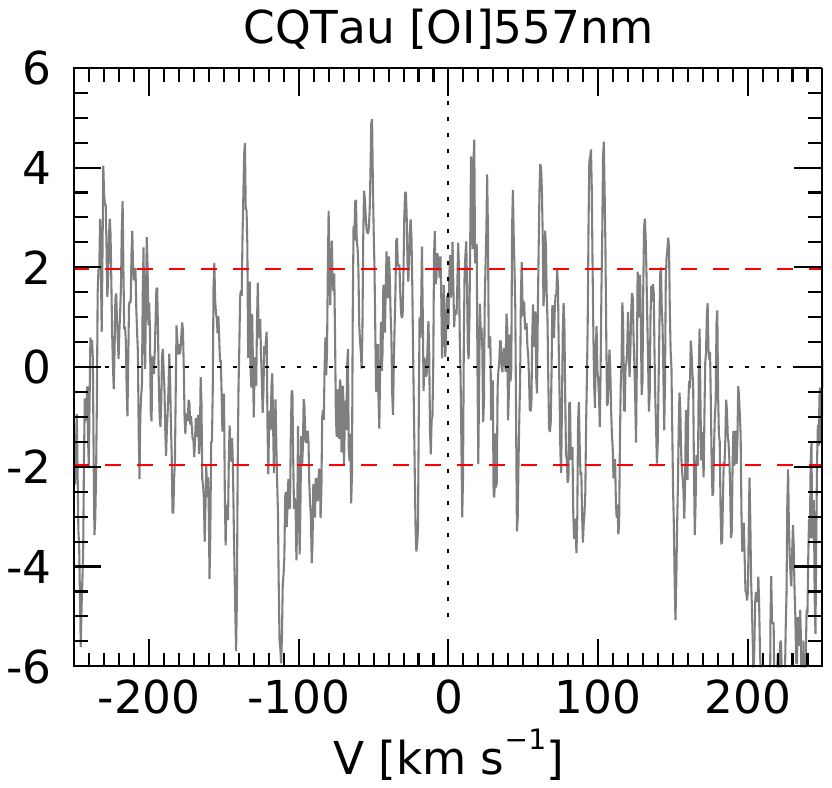}
\includegraphics[trim=80 0 80 400,width=0.2\textwidth]{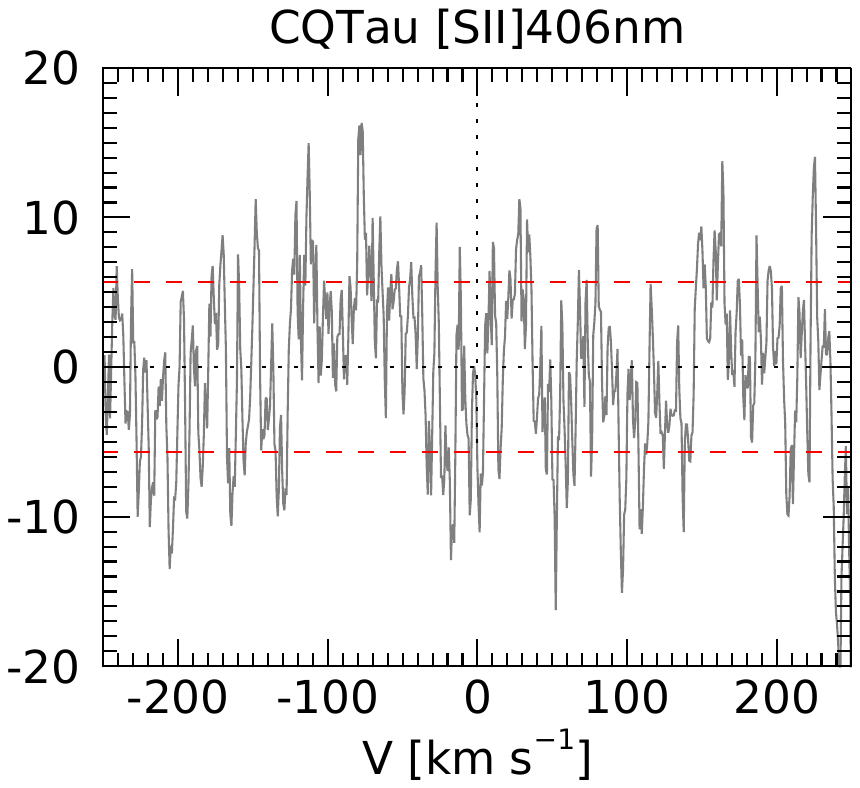}
\includegraphics[trim=80 0 80 400,width=0.2\textwidth]{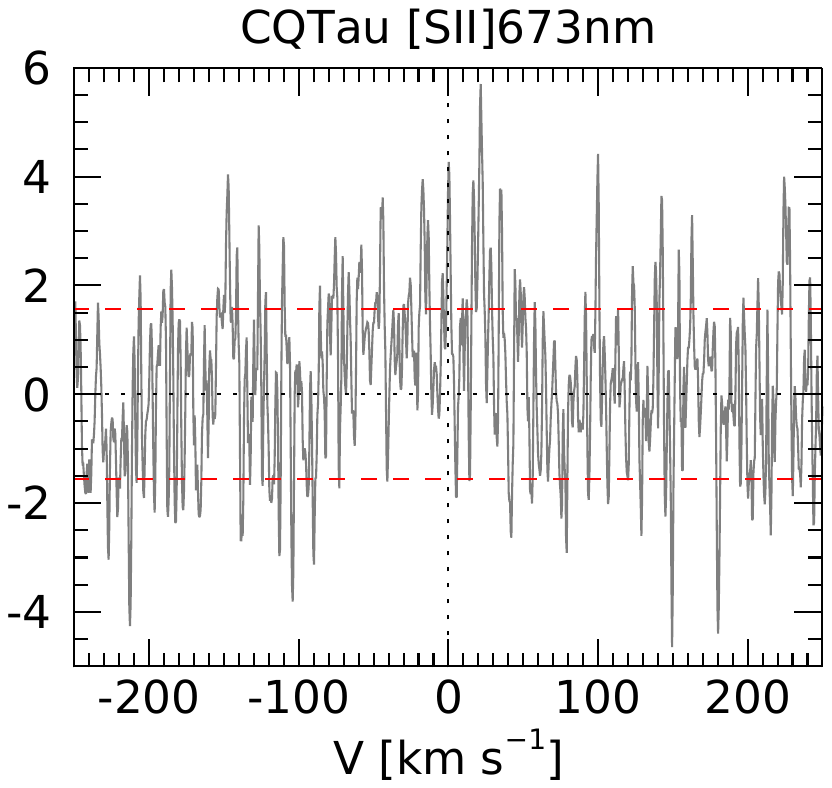}
\includegraphics[trim=80 0 80 400,width=0.2\textwidth]{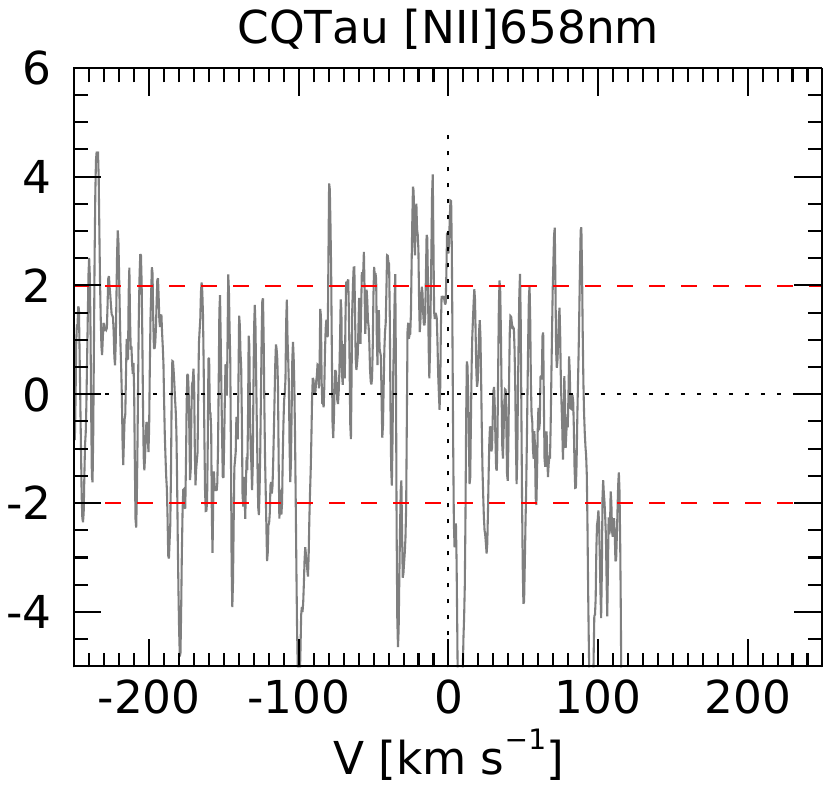}

\includegraphics[trim=80 0 80 400,width=0.2\textwidth]{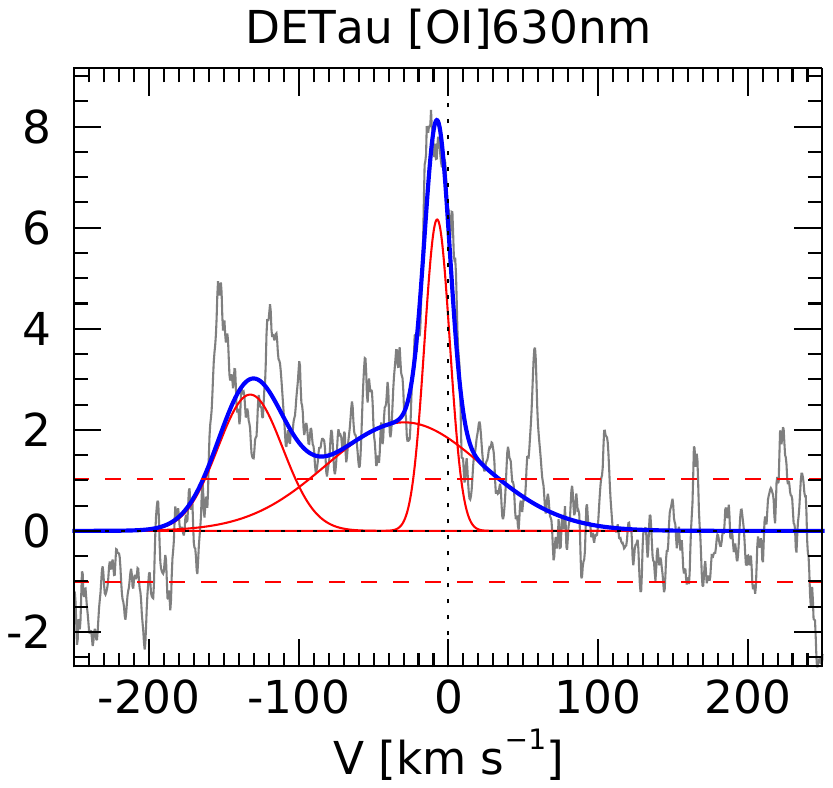}
\includegraphics[trim=80 0 80 400,width=0.2\textwidth]{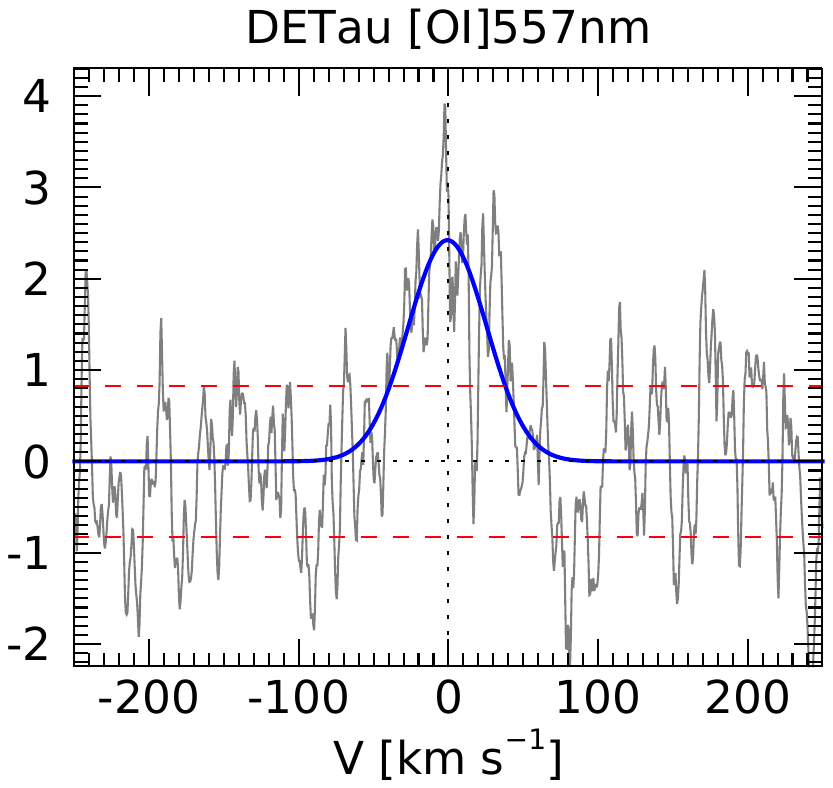}
\includegraphics[trim=80 0 80 400,width=0.2\textwidth]{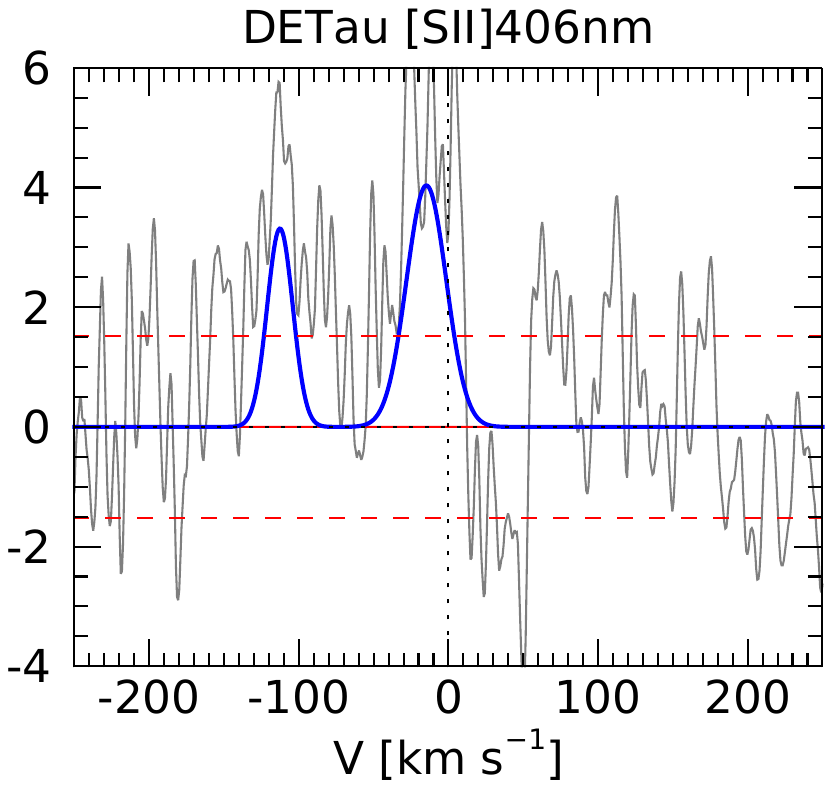}
\includegraphics[trim=80 0 80 400,width=0.2\textwidth]{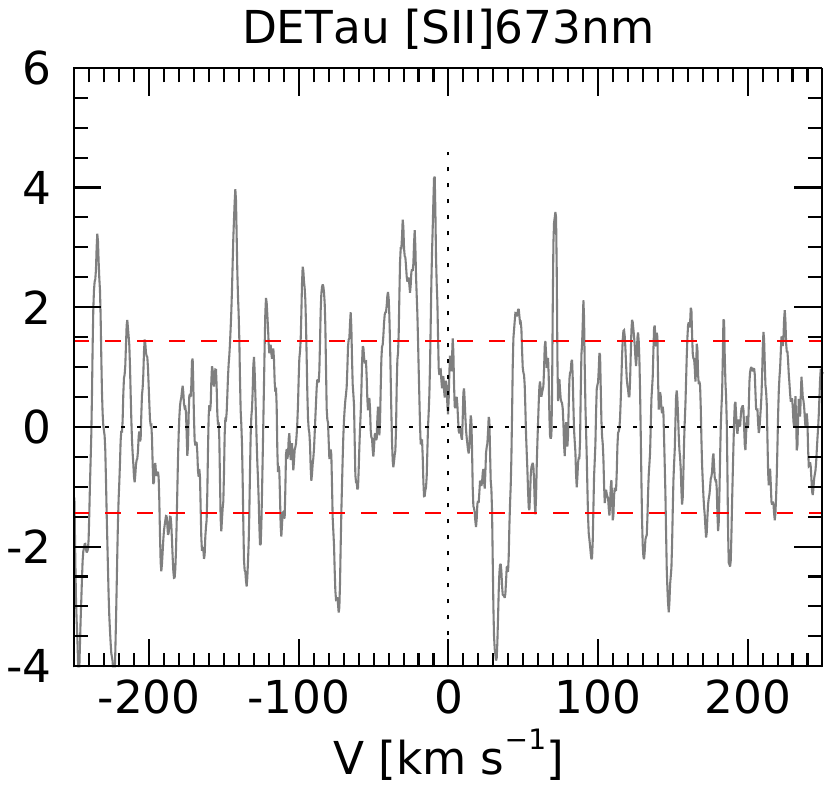}
\includegraphics[trim=80 0 80 400,width=0.2\textwidth]{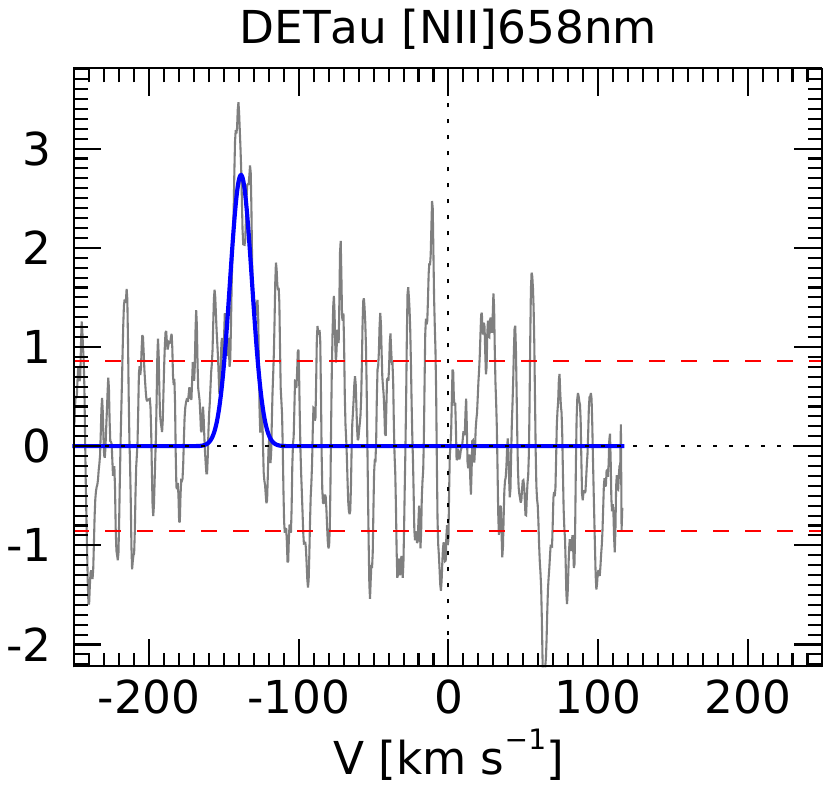}
   \caption{Continuum-subtracted line profiles. Fluxes are in units of 10$^{-15}$ erg\,s$^{-1}$\,cm$^{-2}$\, \AA$^{-1}$. The fit to the profile is shown in blue and it is obtained by adding the single or multiple Gaussians indicated in red. The two horizontal red dashed lines indicate $\pm 1\sigma$.}
   \label{fig:profiles1}
\end{figure*}

\newpage

\begin{figure*}[h]

\includegraphics[trim=80 0 80 0,width=0.2\textwidth]{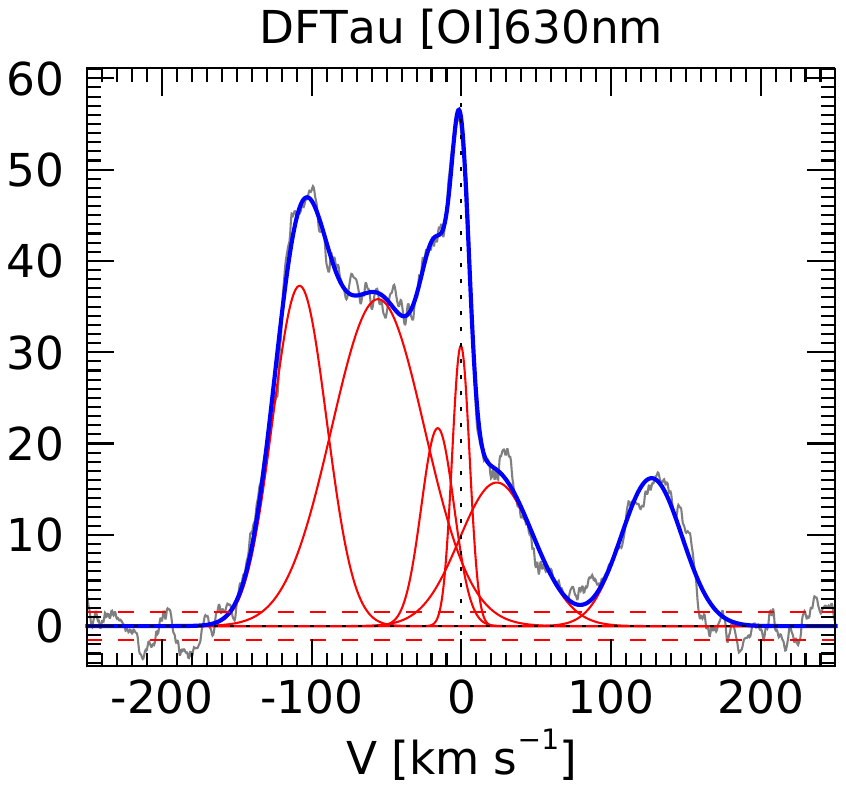}
\includegraphics[trim=80 0 80 0,width=0.2\textwidth]{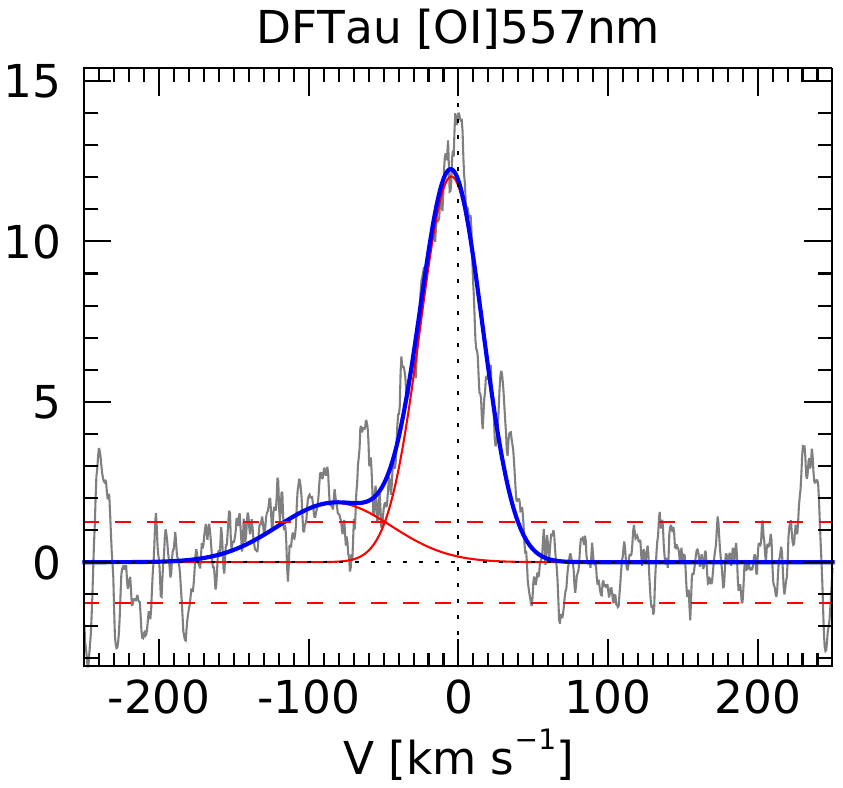}
\includegraphics[trim=80 0 80 0,width=0.2\textwidth]{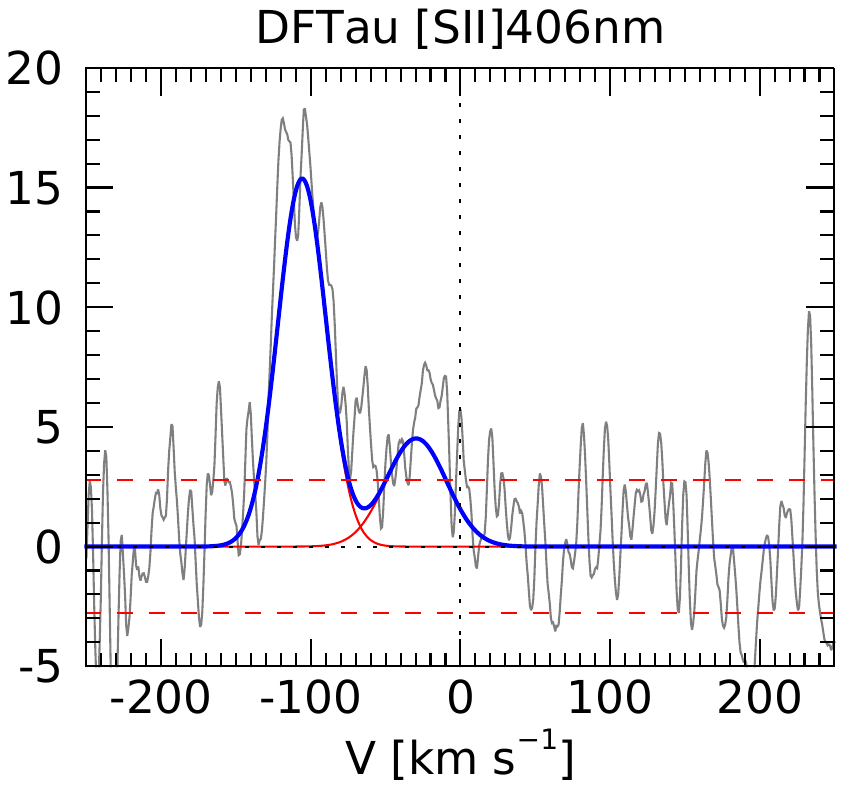}
\includegraphics[trim=80 0 80 0,width=0.2\textwidth]{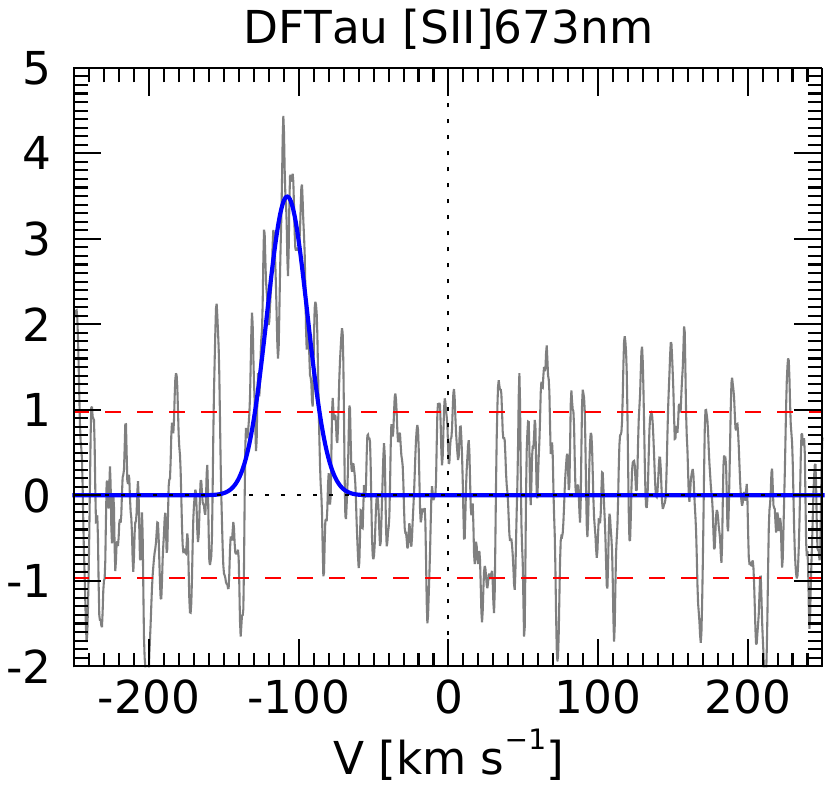}
\includegraphics[trim=80 0 80 0,width=0.2\textwidth]{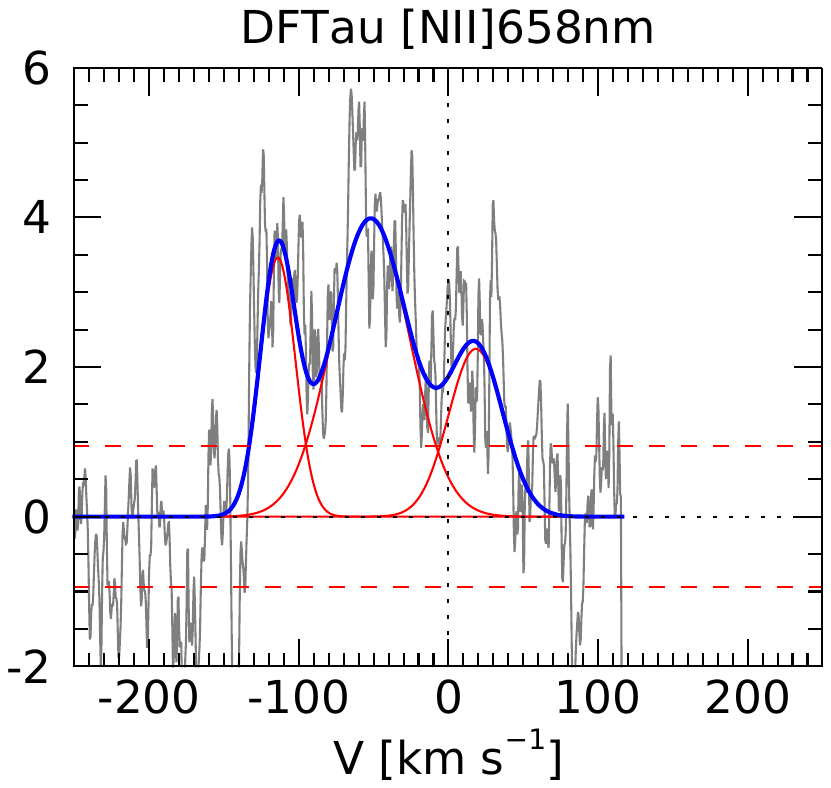}

\includegraphics[trim=80 0 80 400,width=0.2\textwidth]{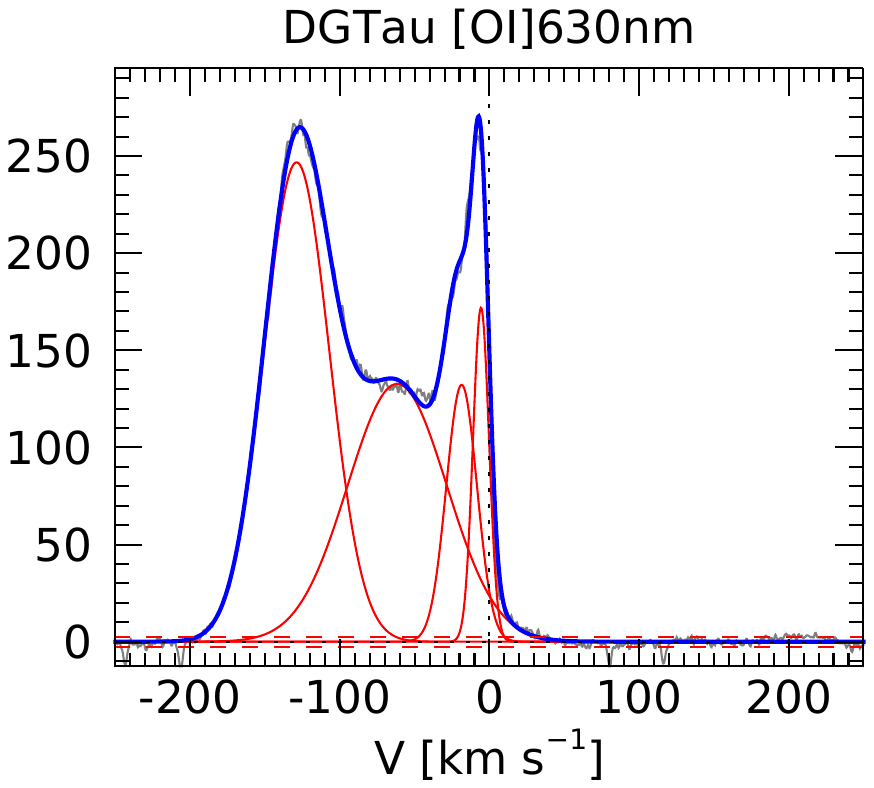}
\includegraphics[trim=80 0 80 400,width=0.2\textwidth]{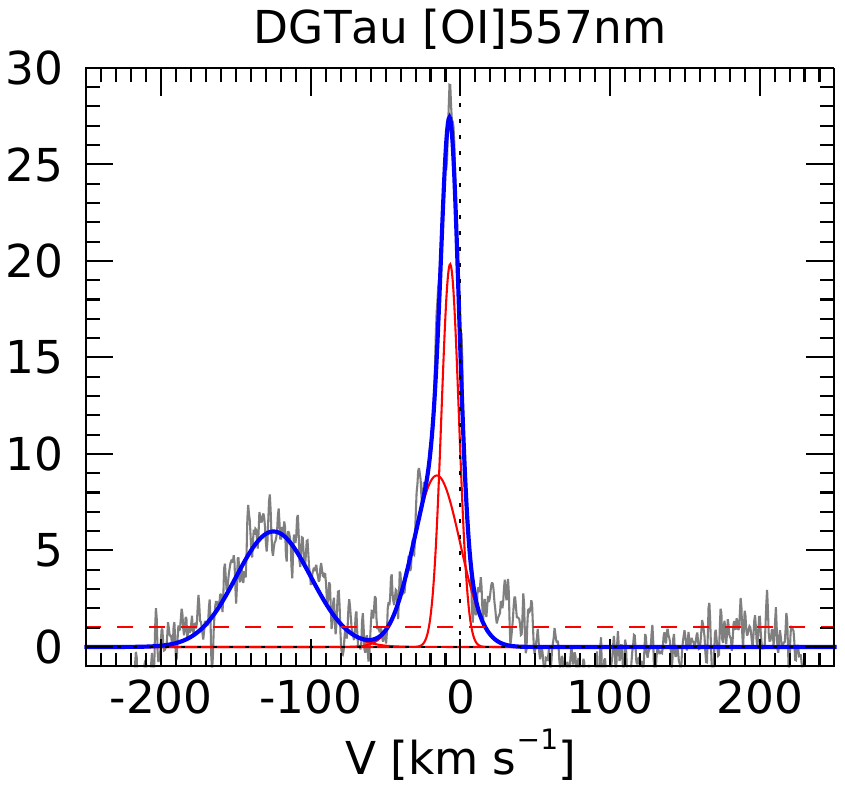}
\includegraphics[trim=80 0 80 400,width=0.2\textwidth]{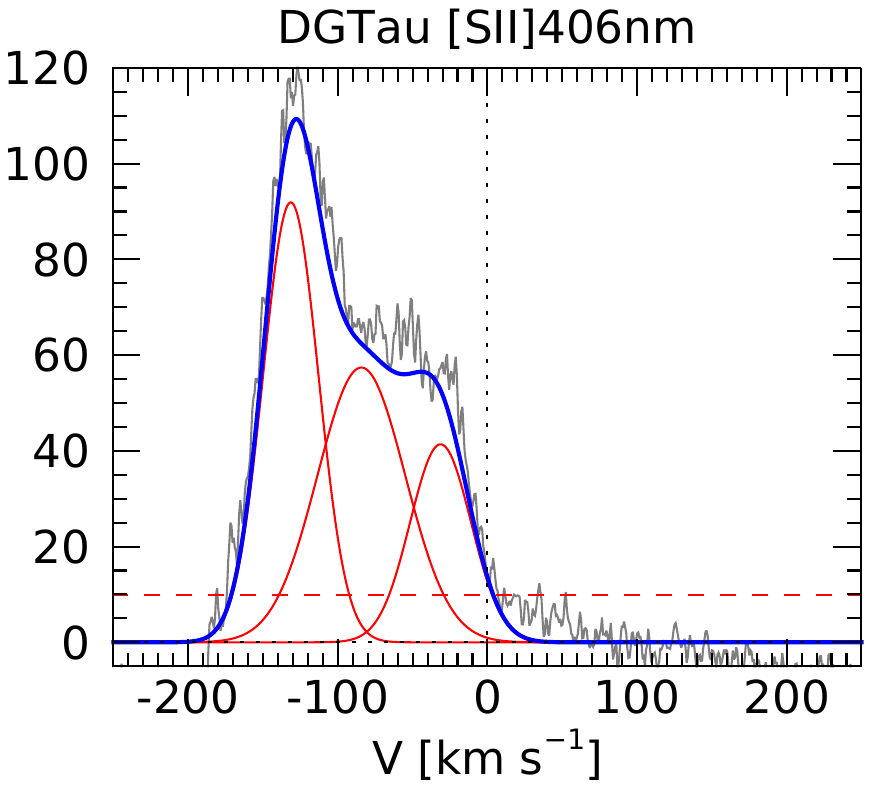}
\includegraphics[trim=80 0 80 400,width=0.2\textwidth]{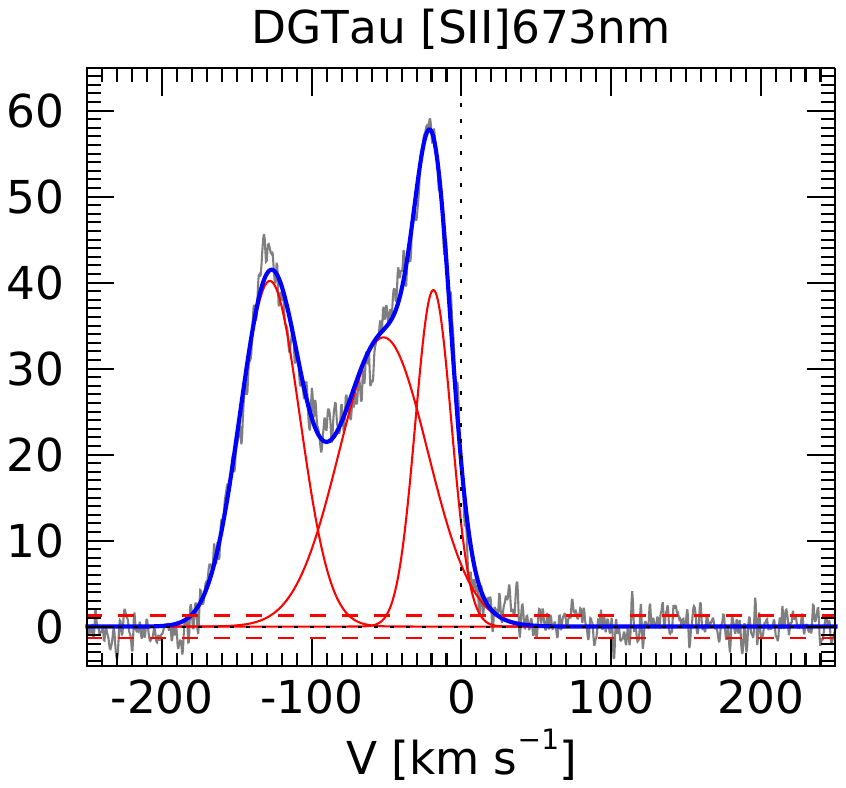}
\includegraphics[trim=80 0 80 400,width=0.2\textwidth]{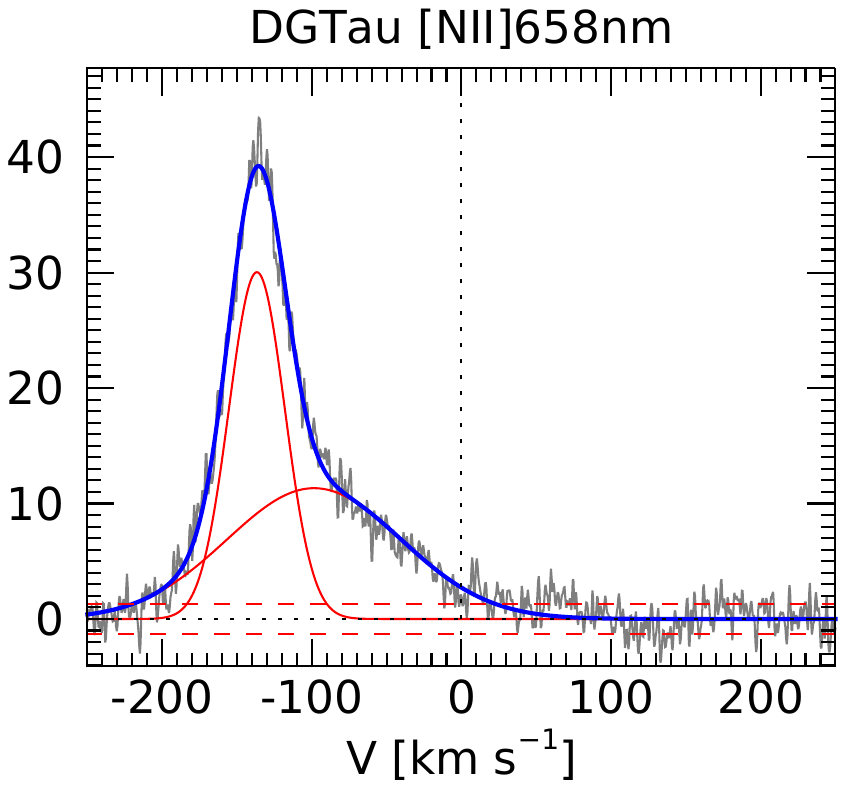}

\includegraphics[trim=80 0 80 400,width=0.2\textwidth]{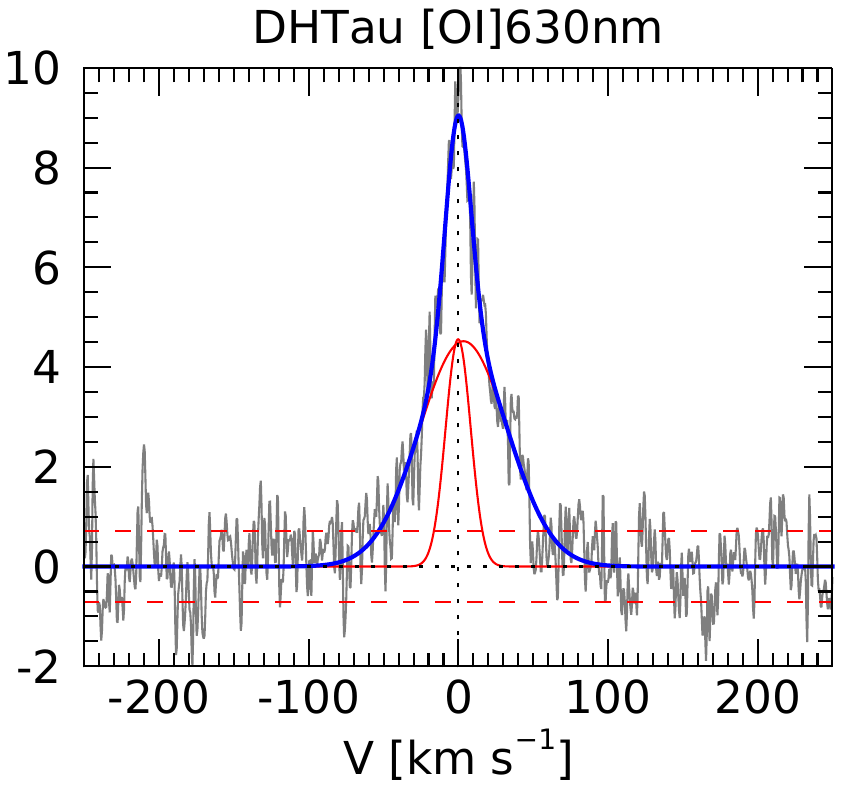}
\includegraphics[trim=80 0 80 400,width=0.2\textwidth]{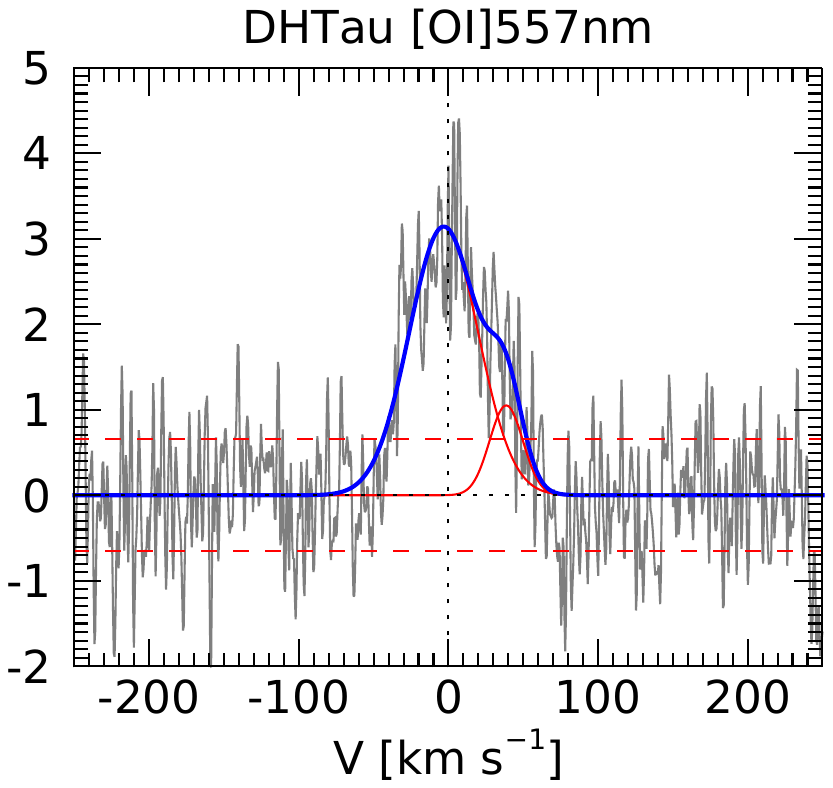}
\includegraphics[trim=80 0 80 400,width=0.2\textwidth]{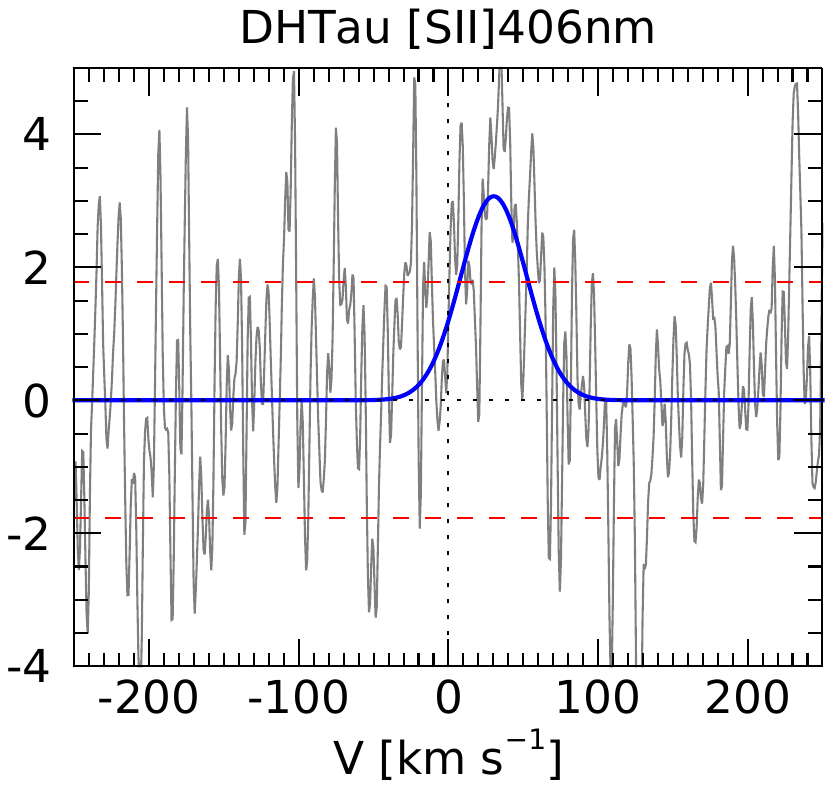}
\includegraphics[trim=80 0 80 400,width=0.2\textwidth]{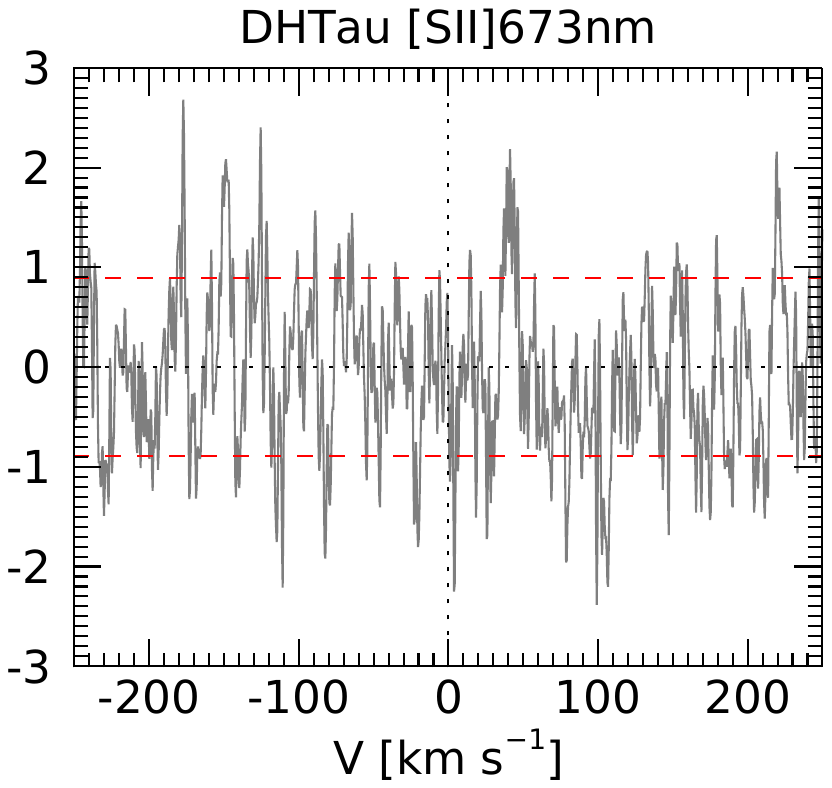}
\includegraphics[trim=80 0 80 400,width=0.2\textwidth]{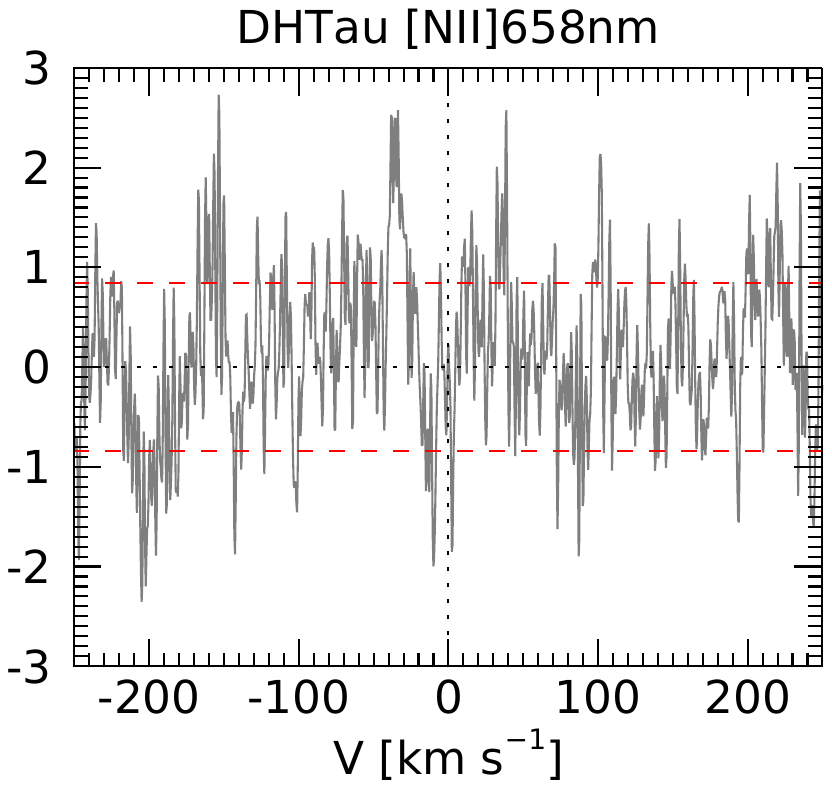}

\includegraphics[trim=80 0 80 400,width=0.2\textwidth]{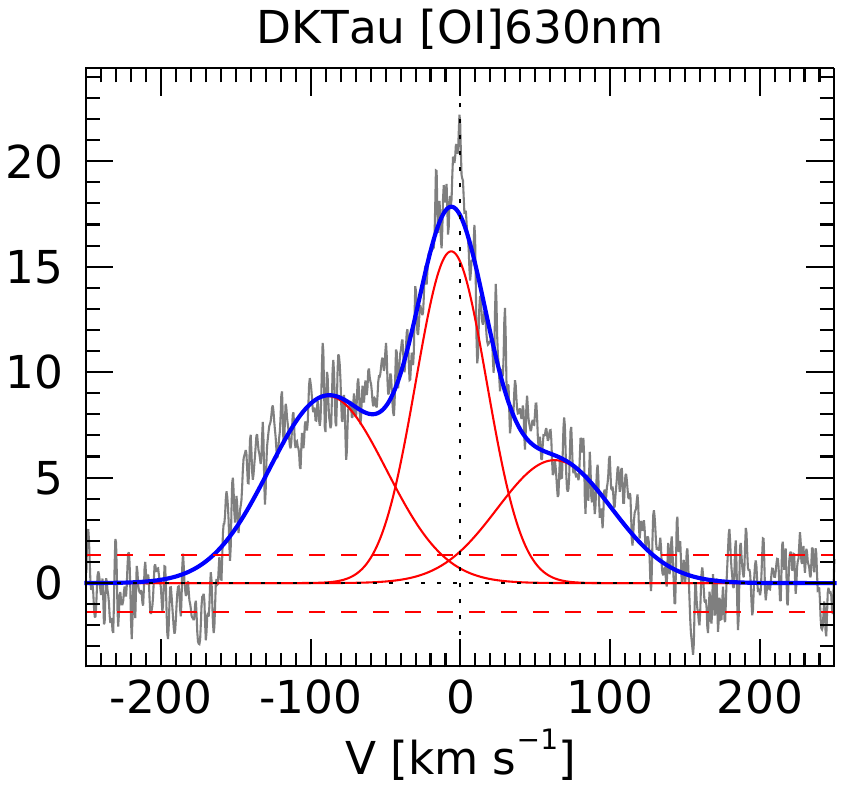}
\includegraphics[trim=80 0 80 400,width=0.2\textwidth]{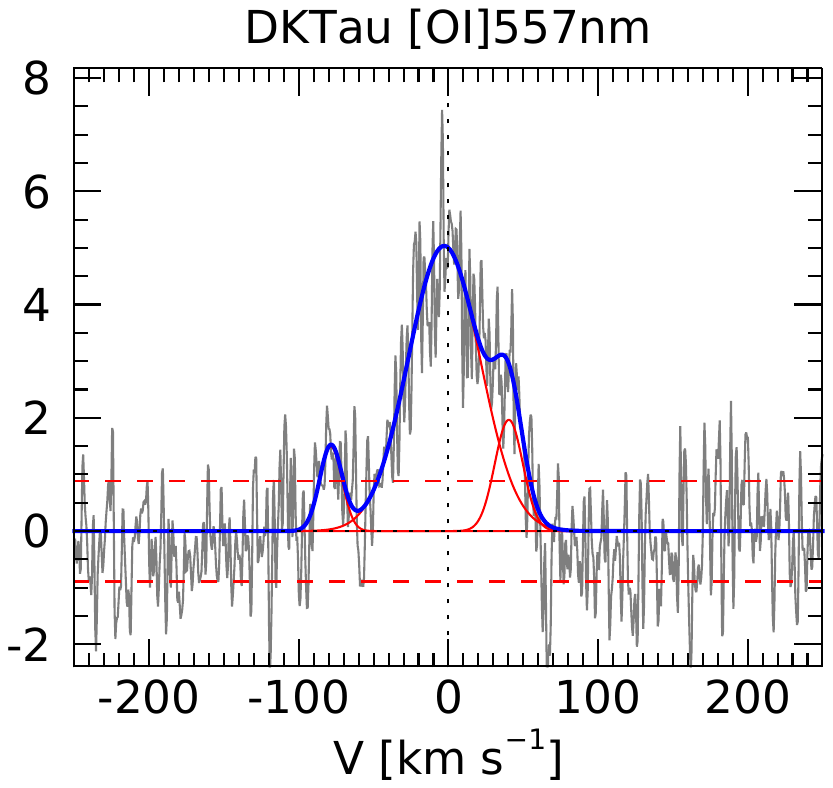}
\includegraphics[trim=80 0 80 400,width=0.2\textwidth]{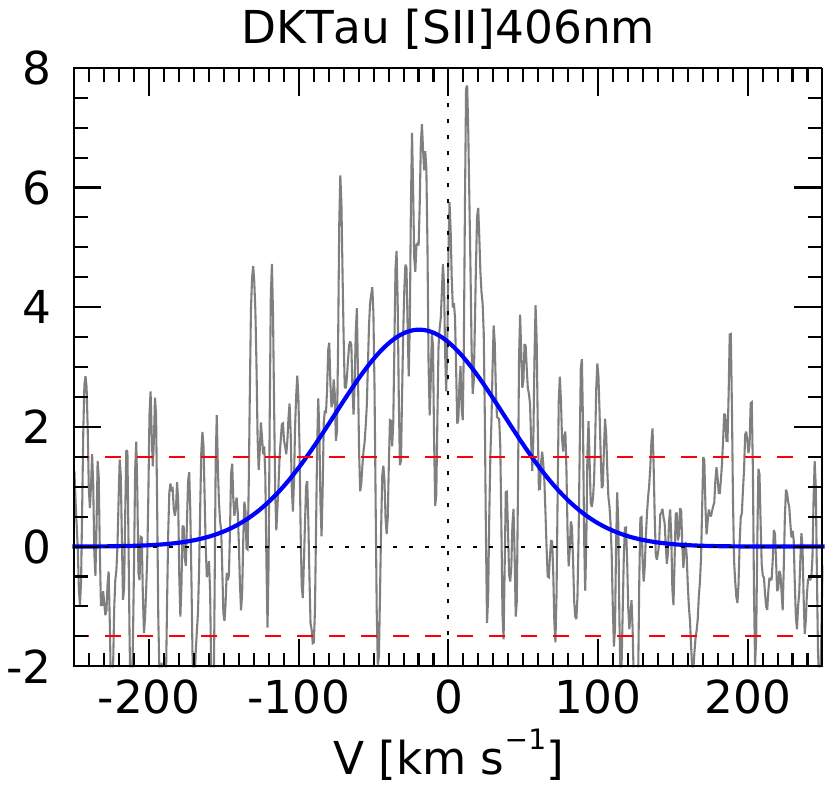}
\includegraphics[trim=80 0 80 400,width=0.2\textwidth]{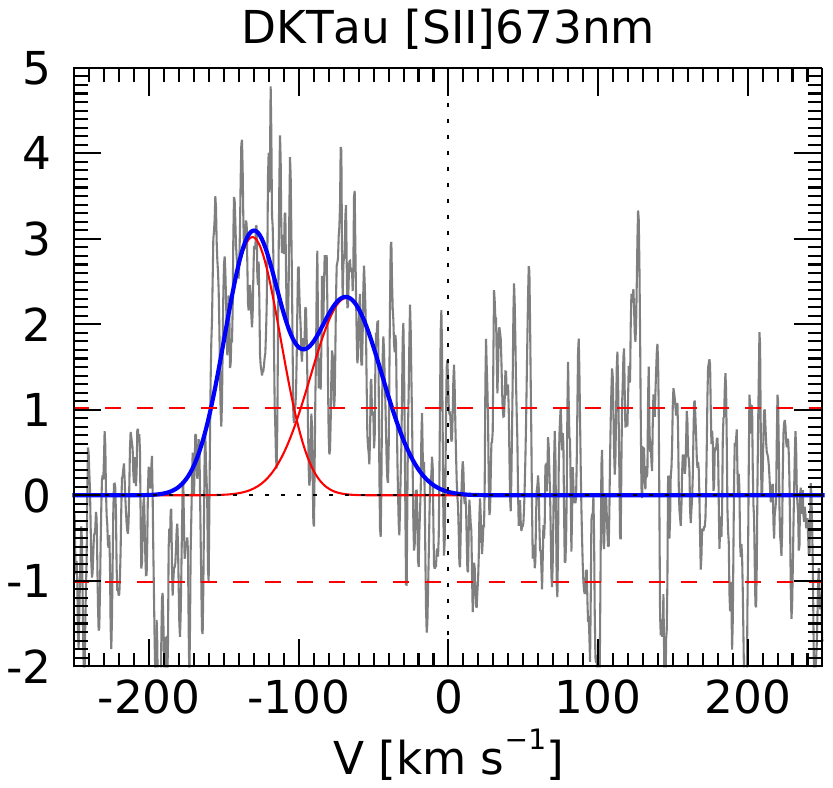}
\includegraphics[trim=80 0 80 400,width=0.2\textwidth]{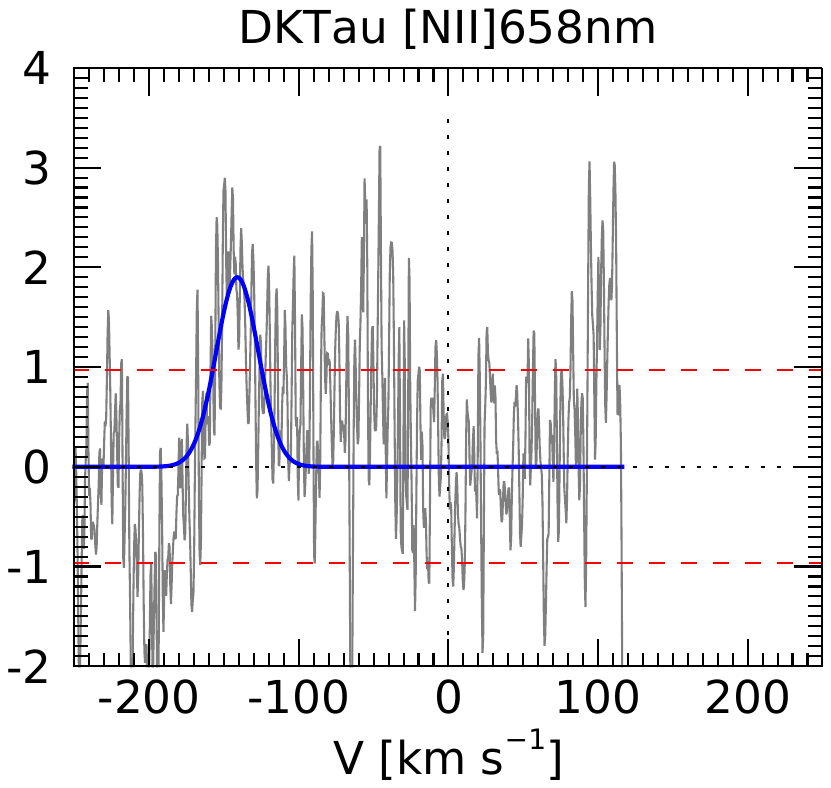}

\includegraphics[trim=80 0 80 400,width=0.2\textwidth]{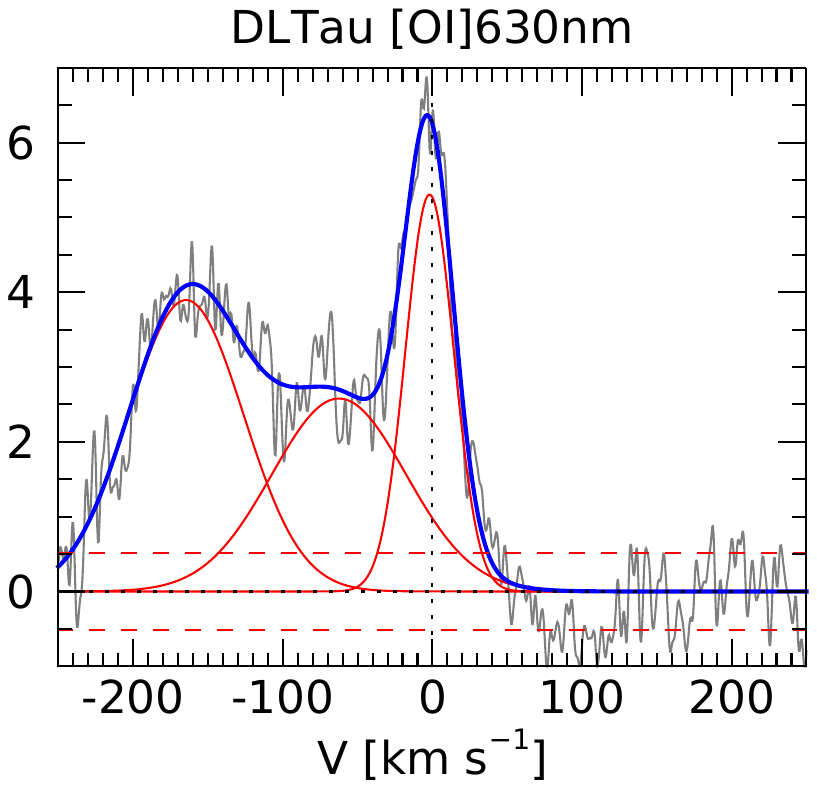}
\includegraphics[trim=80 0 80 400,width=0.2\textwidth]{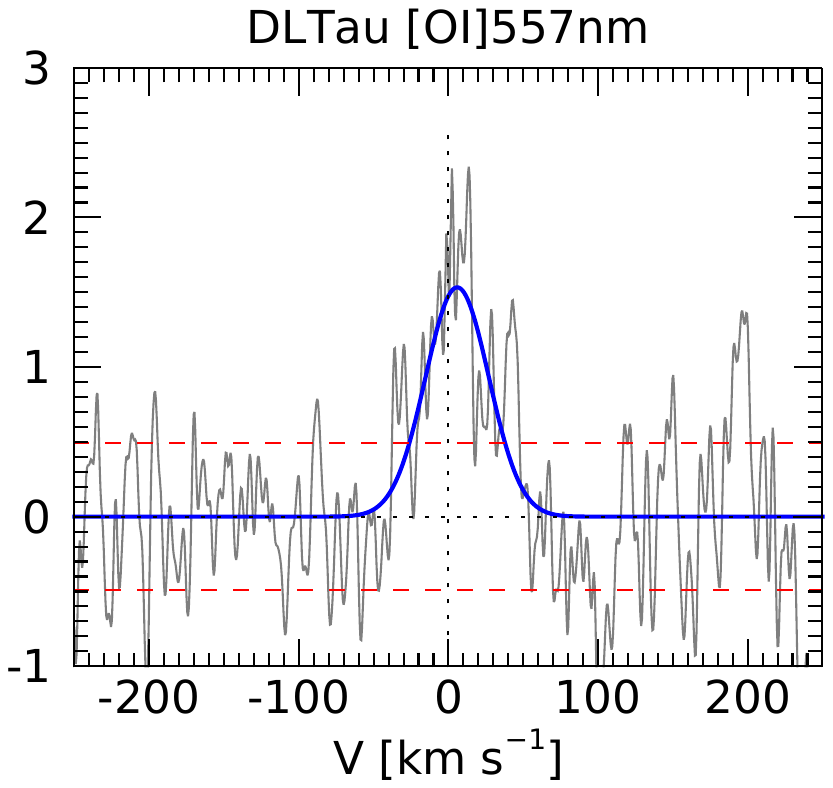}
\includegraphics[trim=80 0 80 400,width=0.2\textwidth]{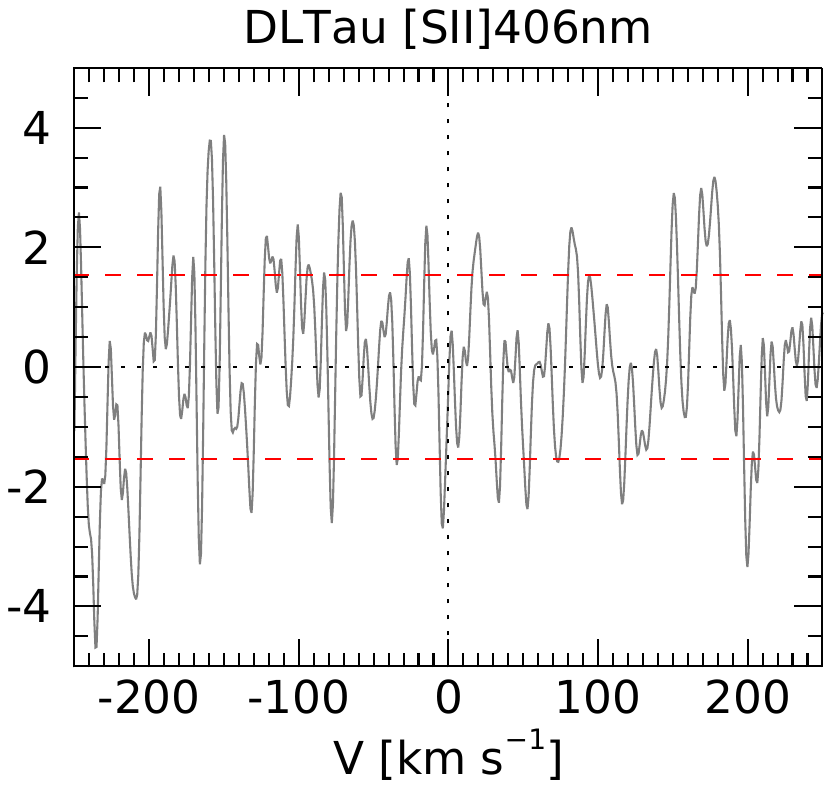}
\includegraphics[trim=80 0 80 400,width=0.2\textwidth]{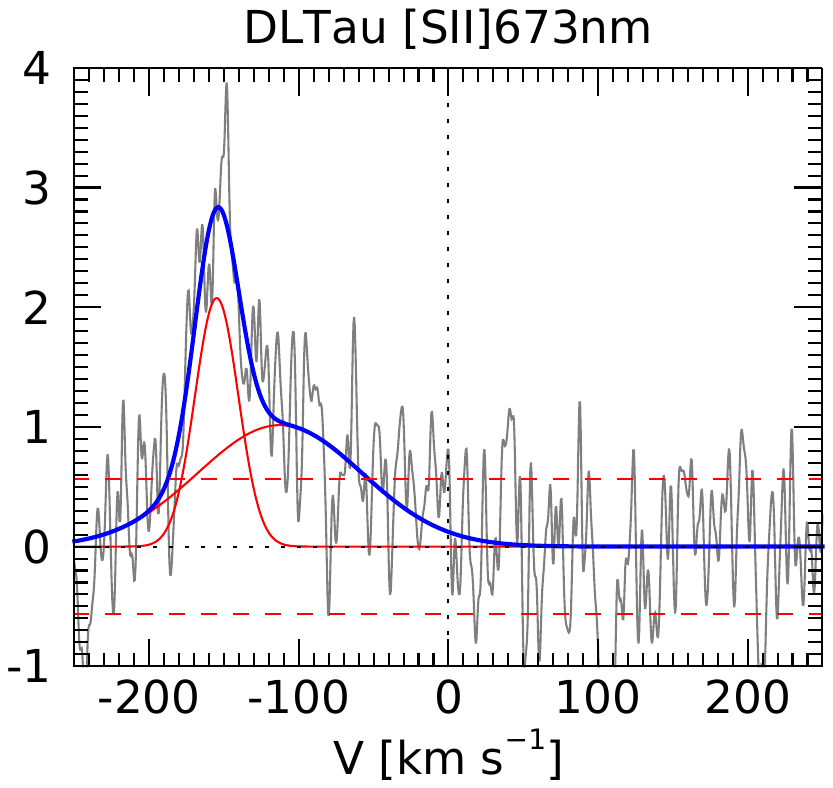}
\includegraphics[trim=80 0 80 400,width=0.2\textwidth]{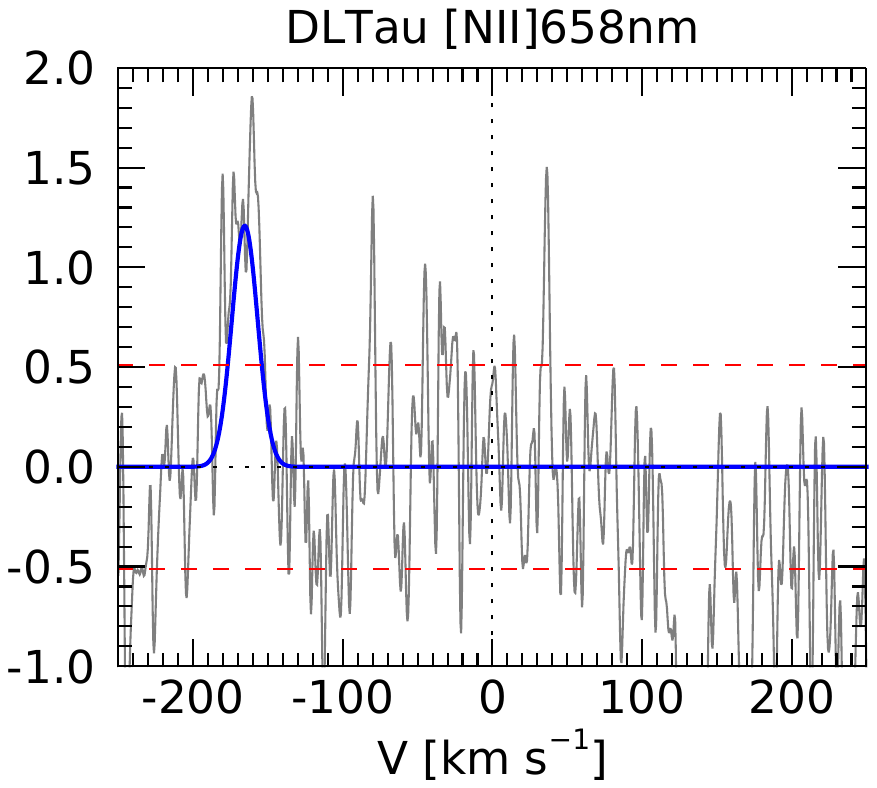}

\includegraphics[trim=80 0 80 400,width=0.2\textwidth]{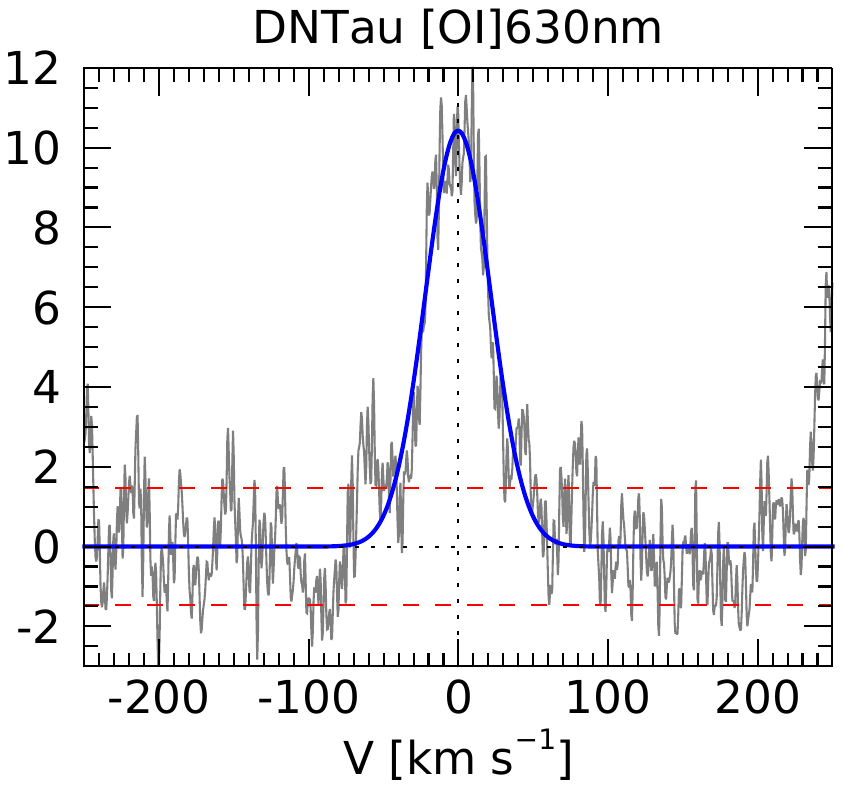}
\includegraphics[trim=80 0 80 400,width=0.2\textwidth]{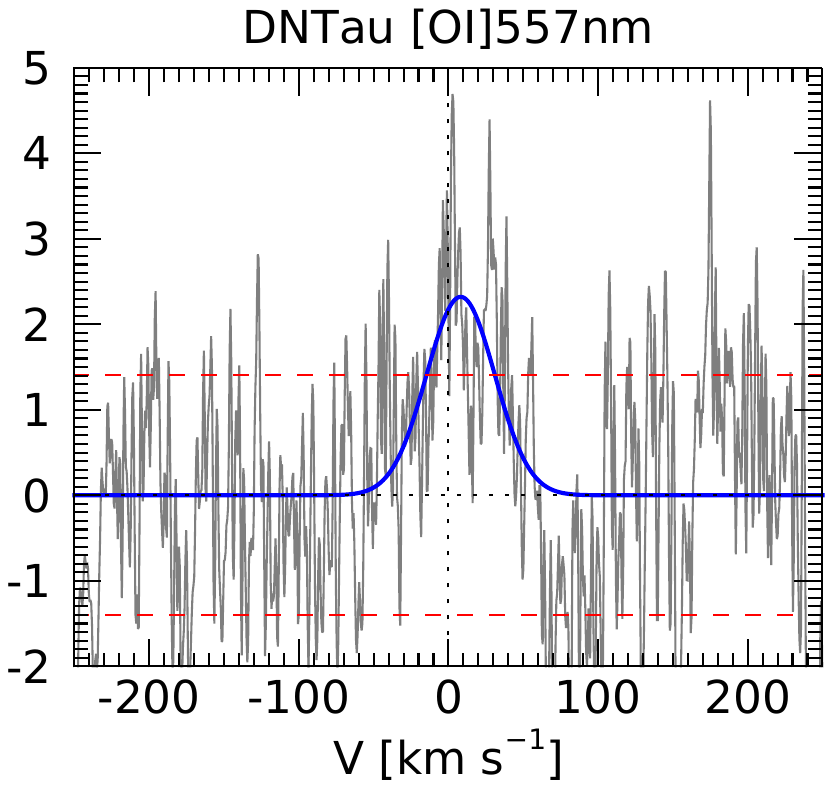}
\includegraphics[trim=80 0 80 400,width=0.2\textwidth]{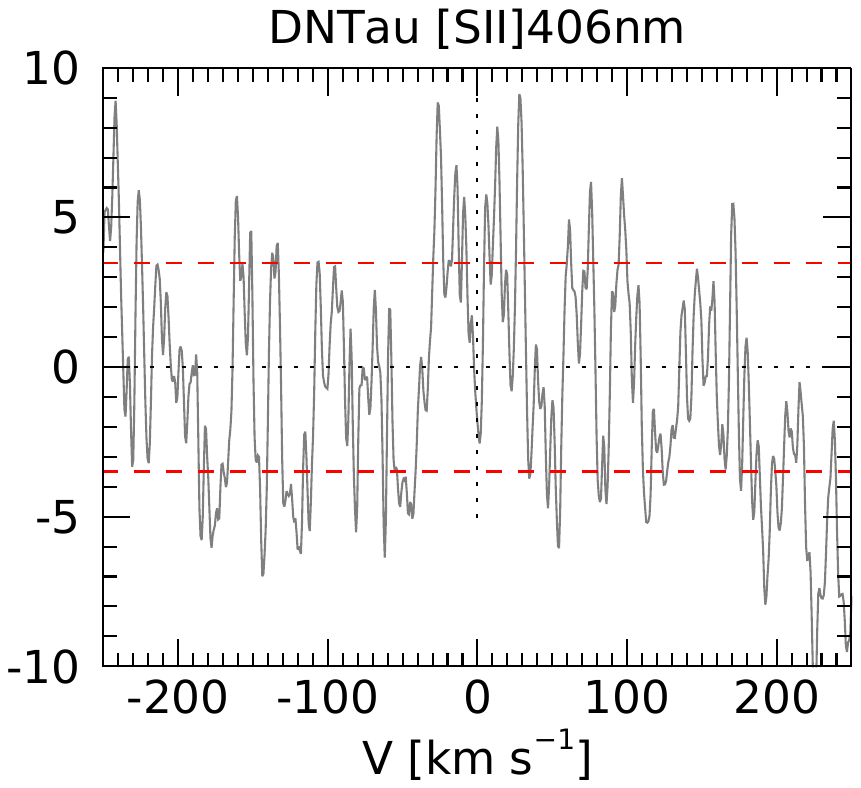}
\includegraphics[trim=80 0 80 400,width=0.2\textwidth]{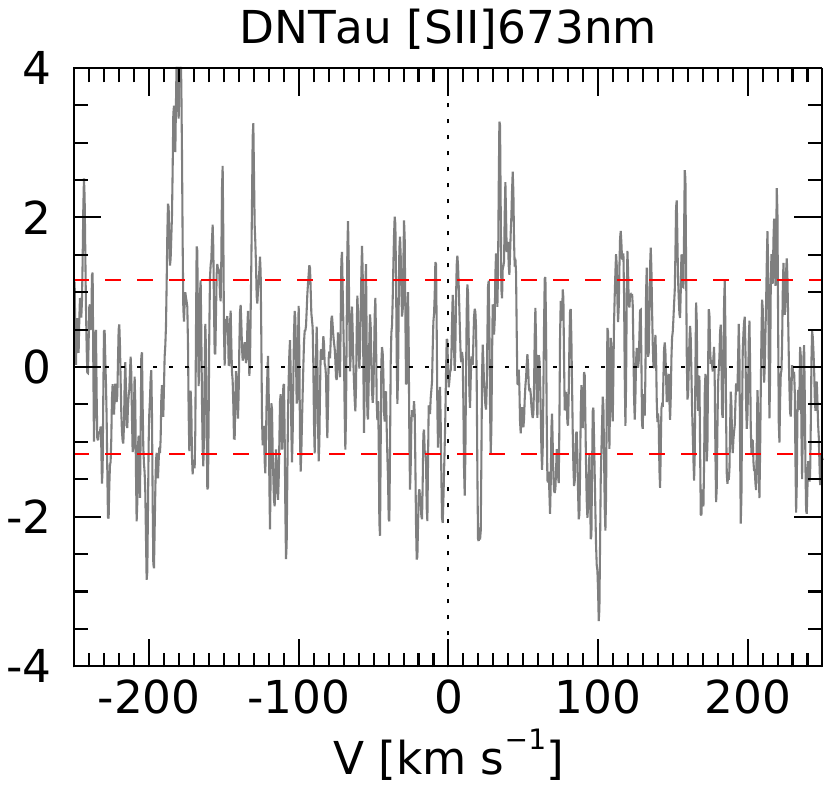}
\includegraphics[trim=80 0 80 400,width=0.2\textwidth]{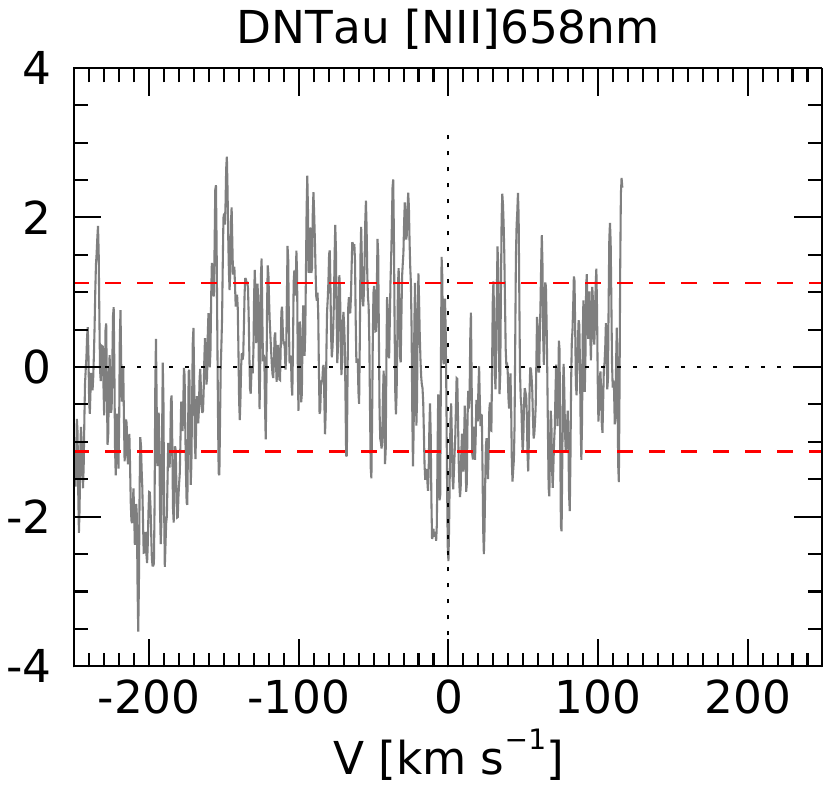}
   \caption{Continued}
   \label{fig:profiles2}
\end{figure*}

\newpage

\begin{figure*}[h]

\includegraphics[trim=80 0 80 0,width=0.2\textwidth]{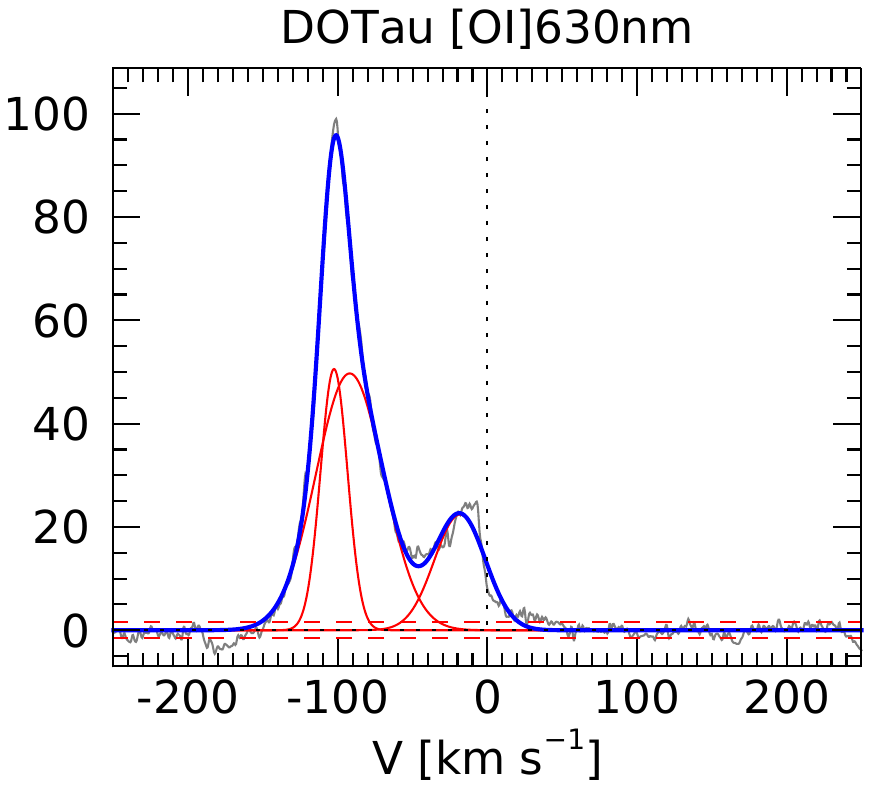}
\includegraphics[trim=80 0 80 0,width=0.2\textwidth]{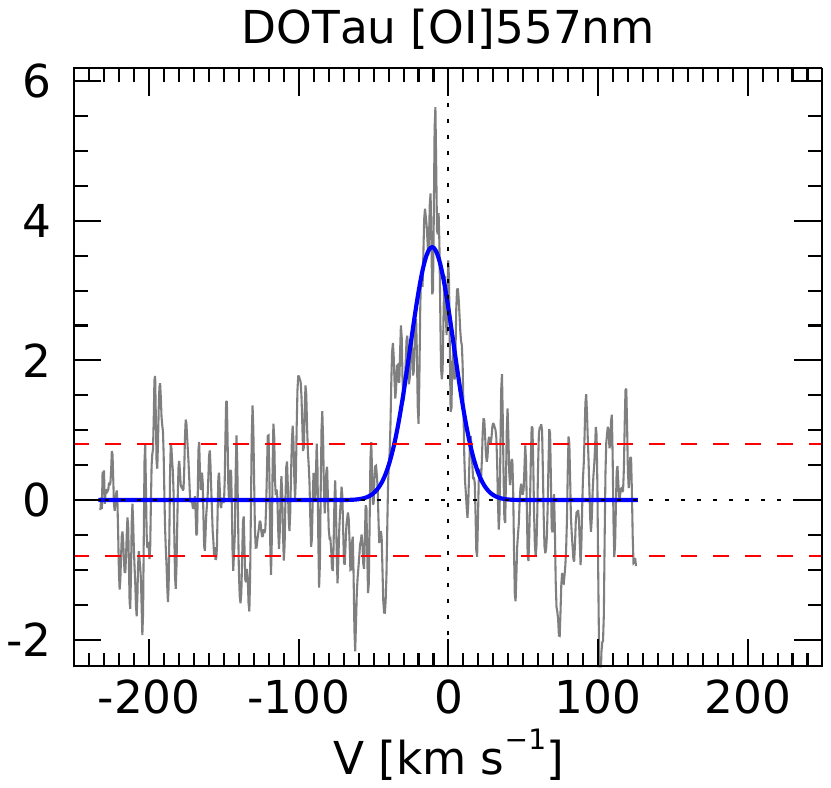}
\includegraphics[trim=80 0 80 0,width=0.2\textwidth]{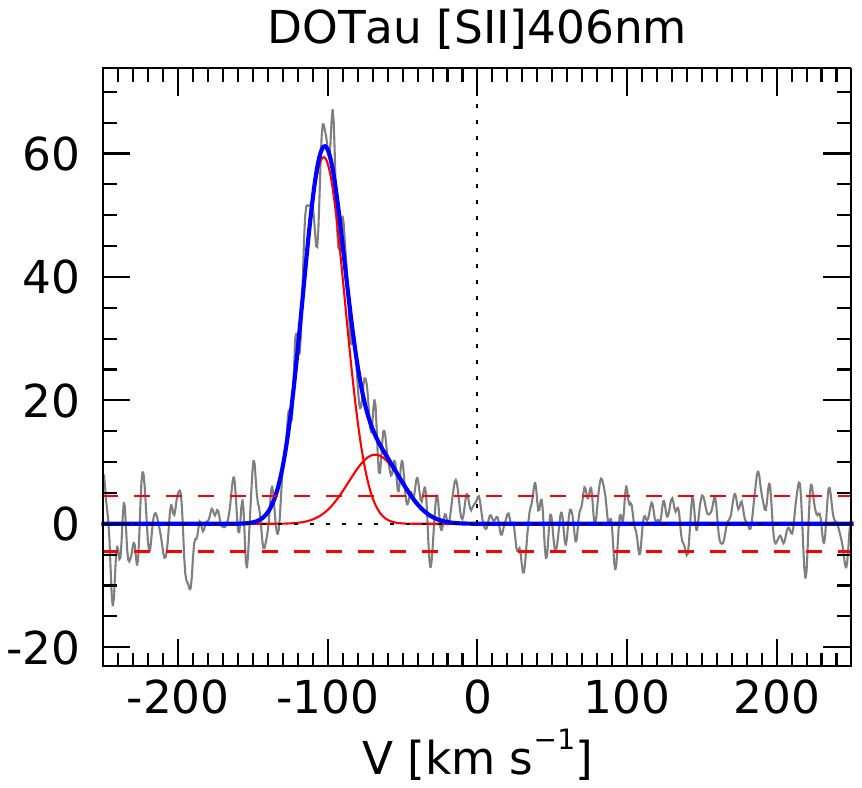}
\includegraphics[trim=80 0 80 0,width=0.2\textwidth]{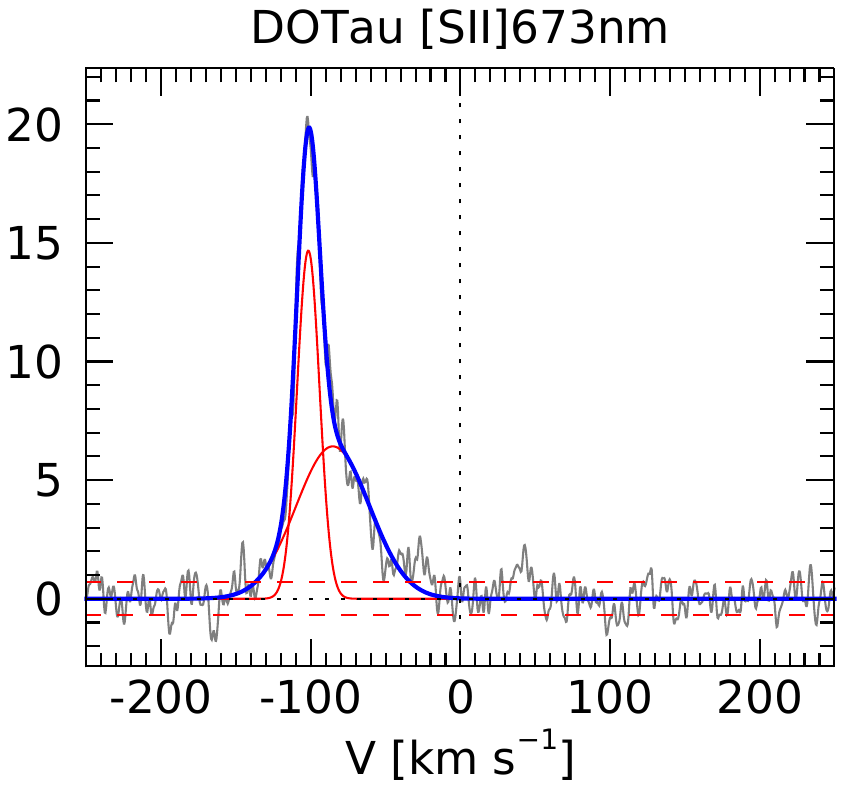}
\includegraphics[trim=80 0 80 0,width=0.2\textwidth]{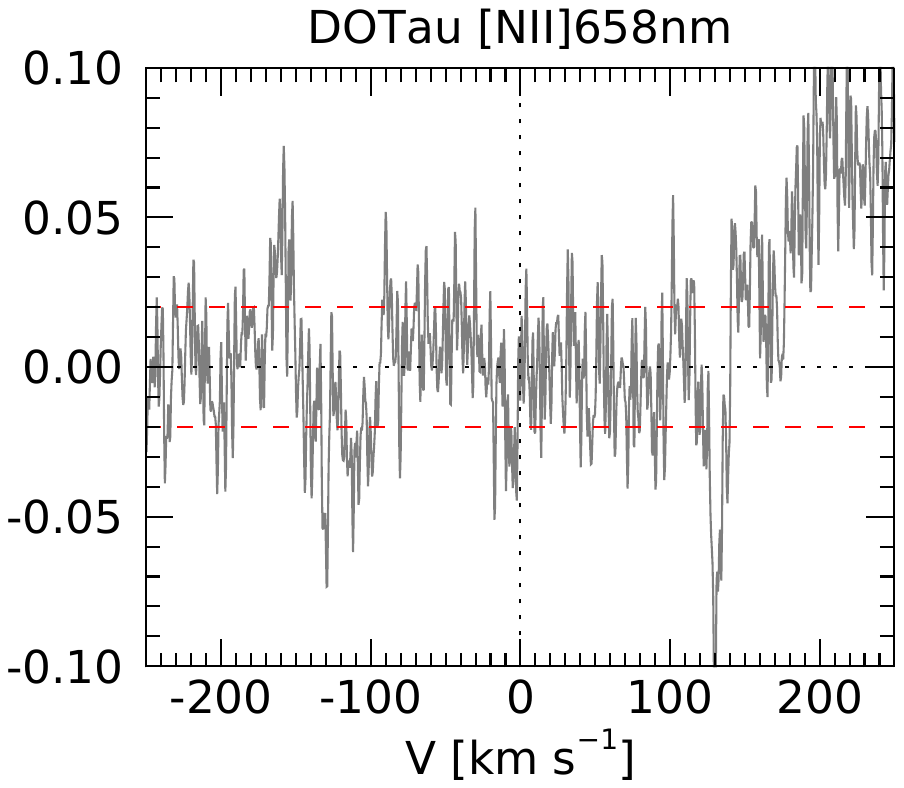}

\includegraphics[trim=80 0 80 400,width=0.2\textwidth]{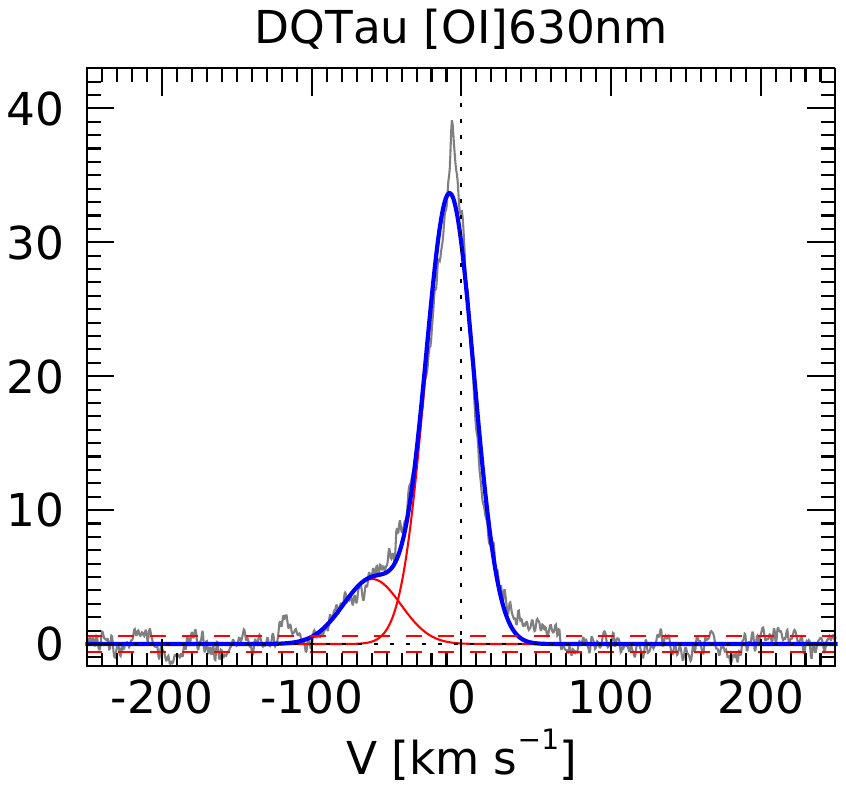}
\includegraphics[trim=80 0 80 400,width=0.2\textwidth]{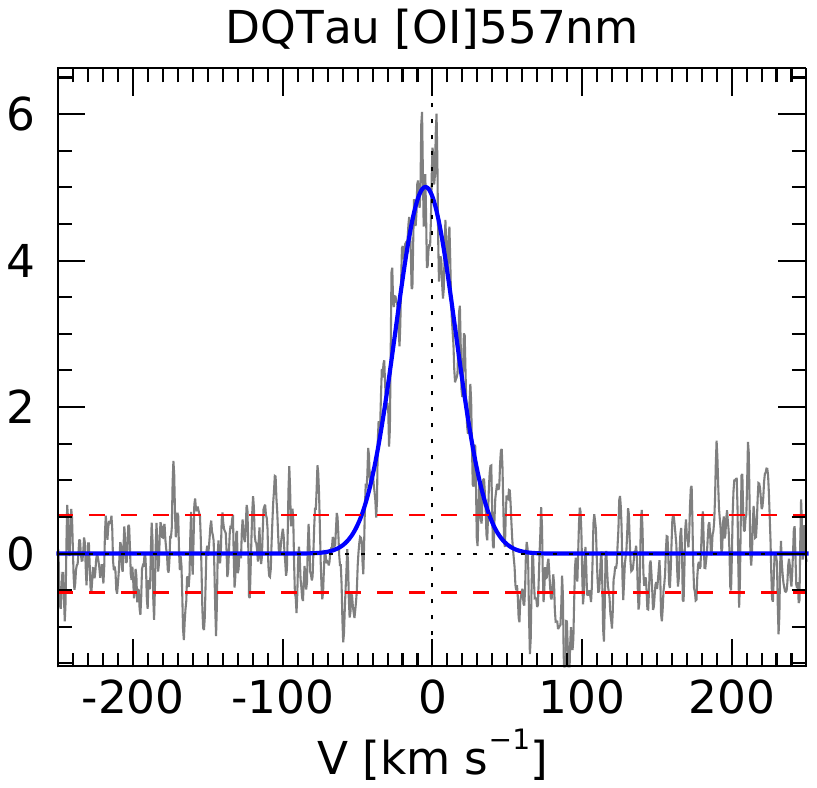}
\includegraphics[trim=80 0 80 400,width=0.2\textwidth]{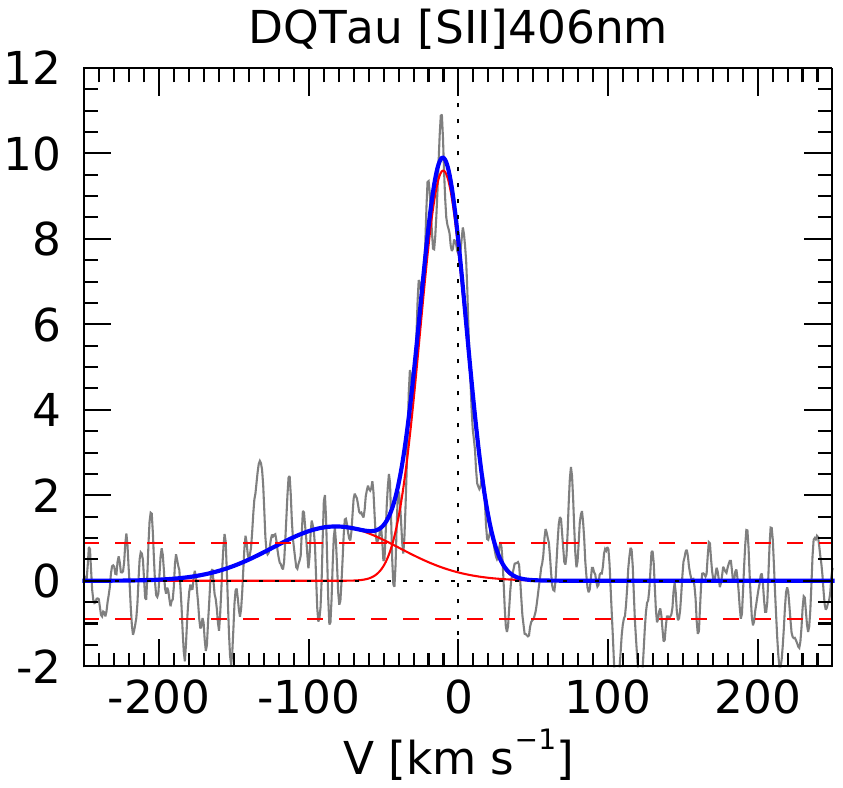}
\includegraphics[trim=80 0 80 400,width=0.2\textwidth]{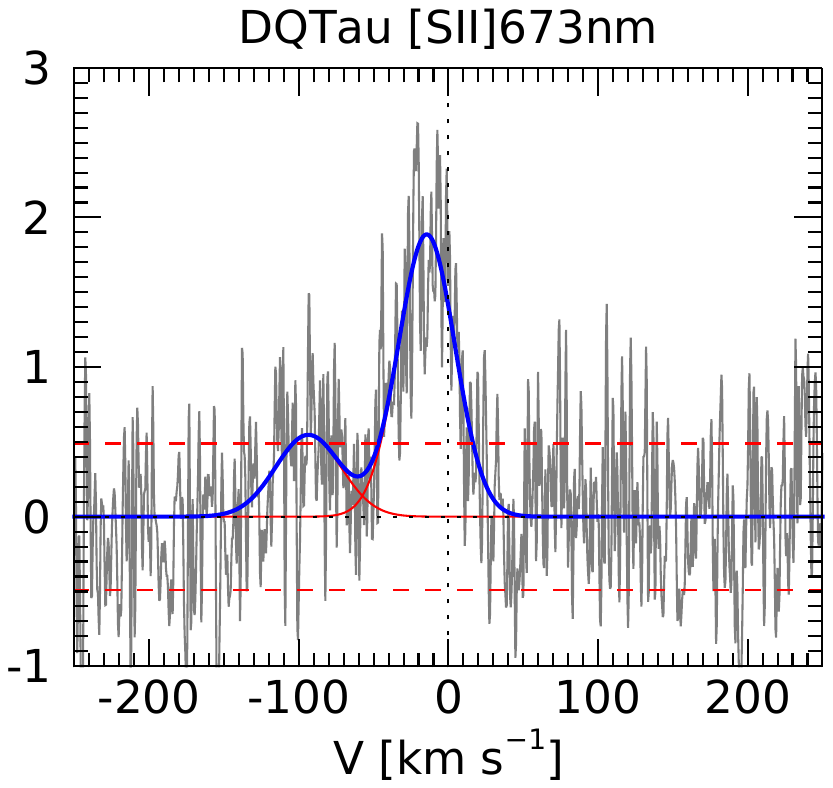}
\includegraphics[trim=80 0 80 400,width=0.2\textwidth]{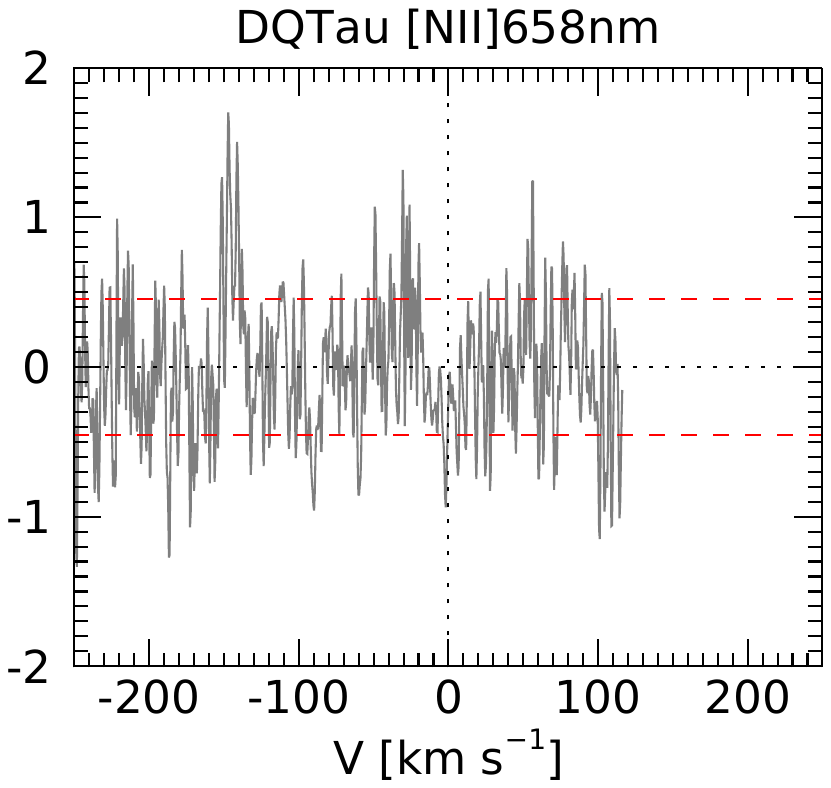}

\includegraphics[trim=80 0 80 400,width=0.2\textwidth]{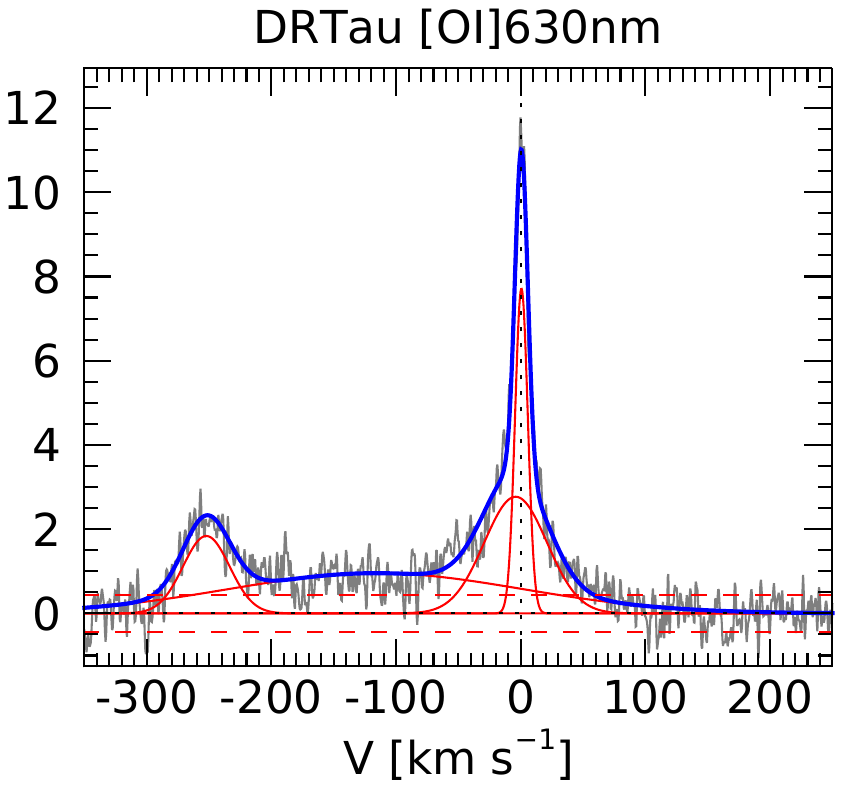}
\includegraphics[trim=80 0 80 400,width=0.2\textwidth]{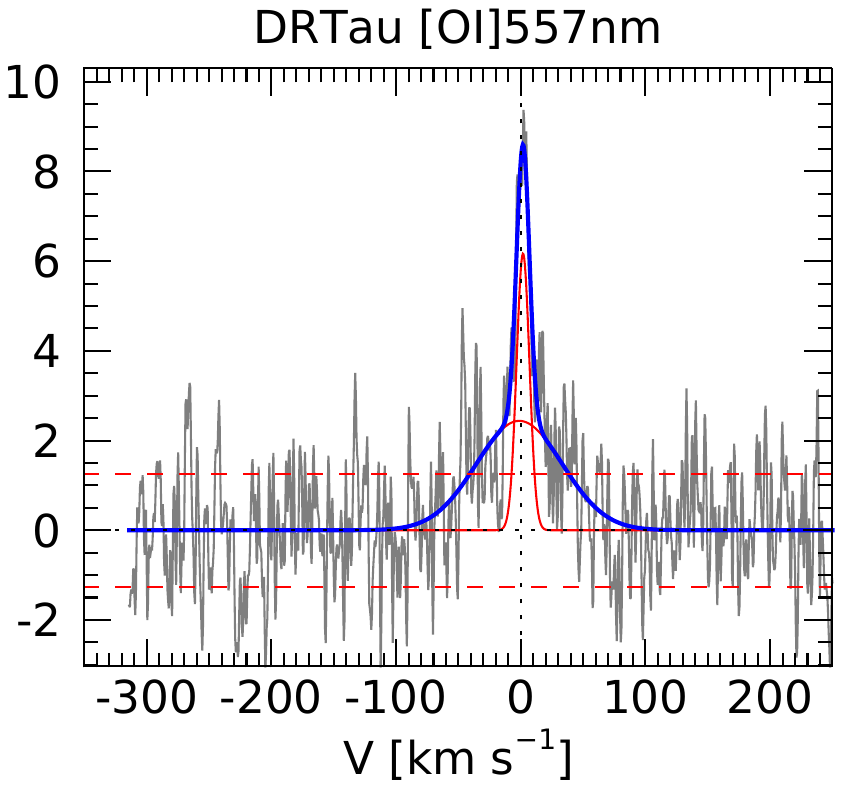}
\includegraphics[trim=80 0 80 400,width=0.2\textwidth]{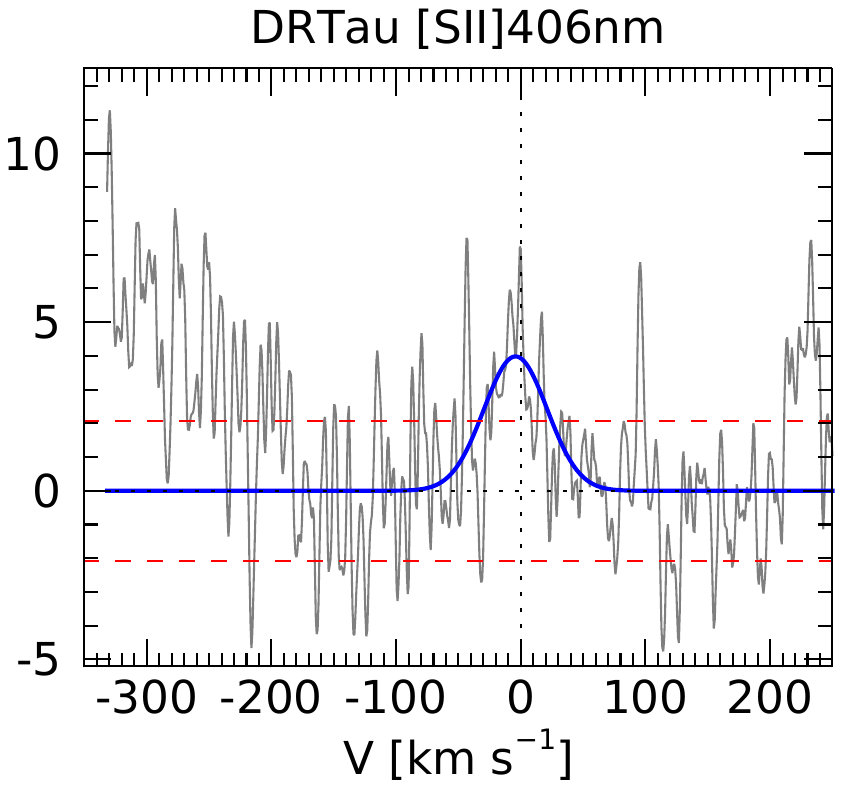}
\includegraphics[trim=80 0 80 400,width=0.2\textwidth]{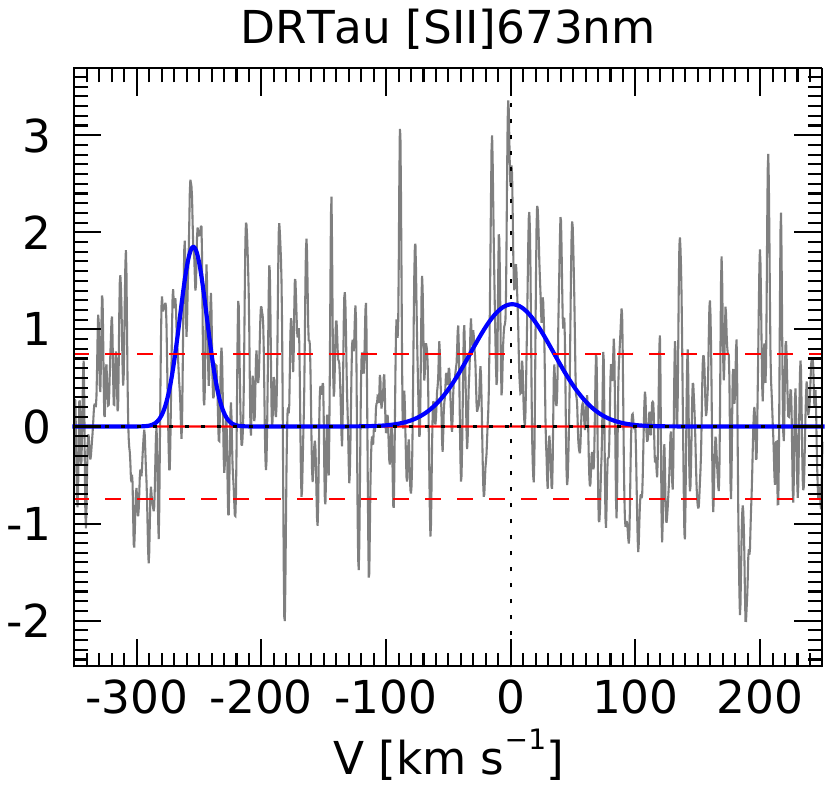}
\includegraphics[trim=80 0 80 400,width=0.2\textwidth]{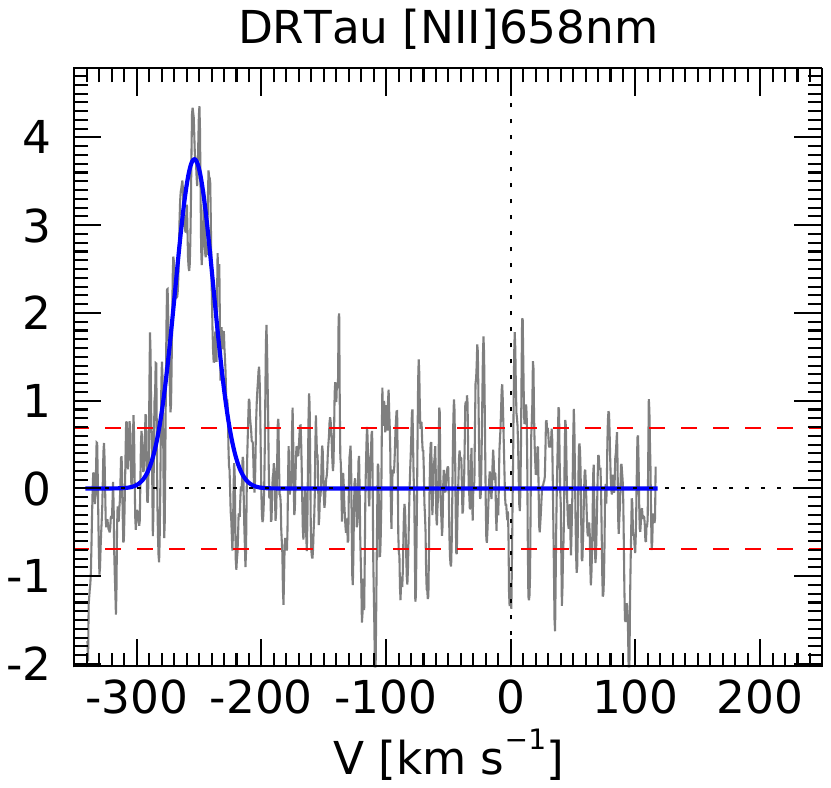}

\includegraphics[trim=80 0 80 400,width=0.2\textwidth]{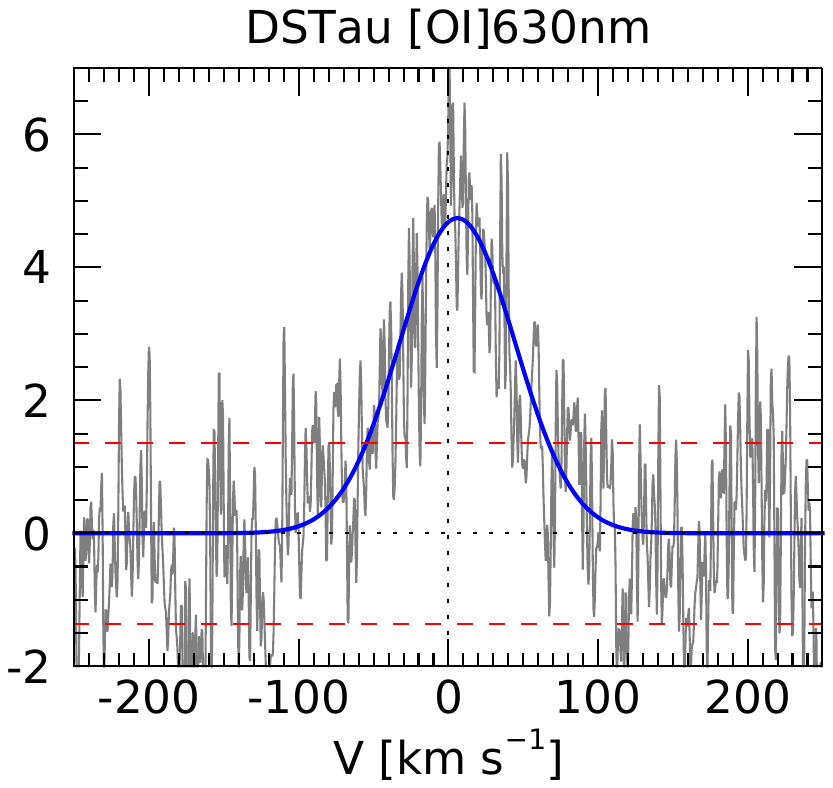}
\includegraphics[trim=80 0 80 400,width=0.2\textwidth]{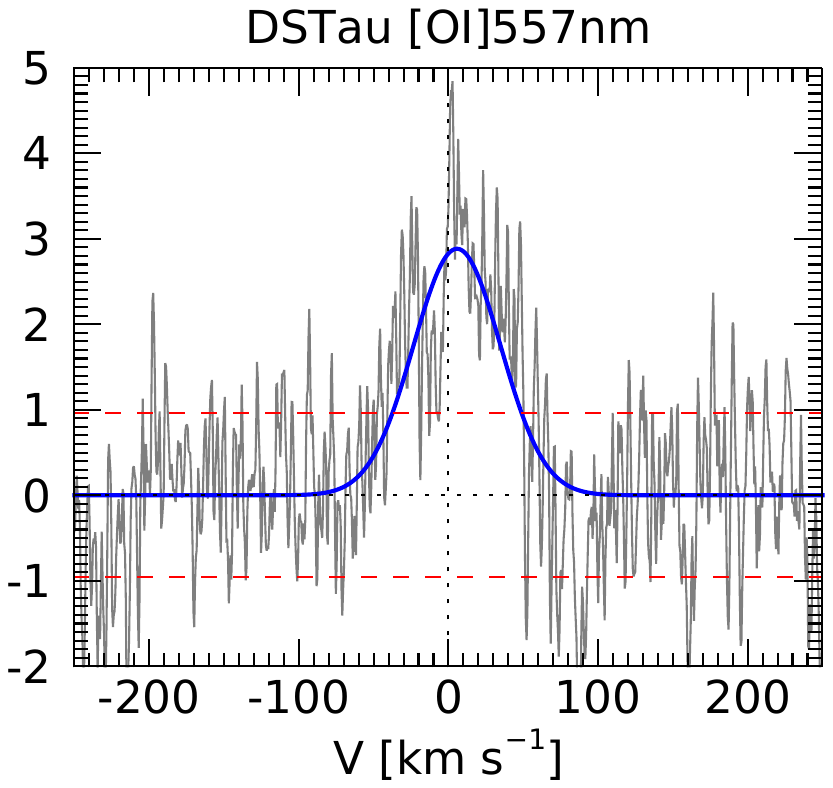}
\includegraphics[trim=80 0 80 400,width=0.2\textwidth]{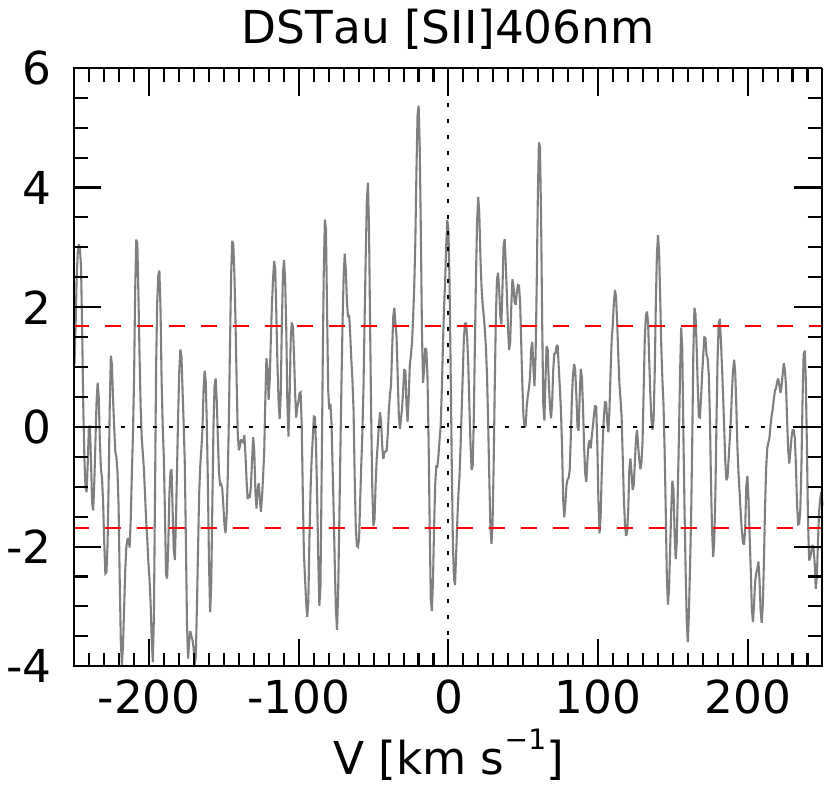}
\includegraphics[trim=80 0 80 400,width=0.2\textwidth]{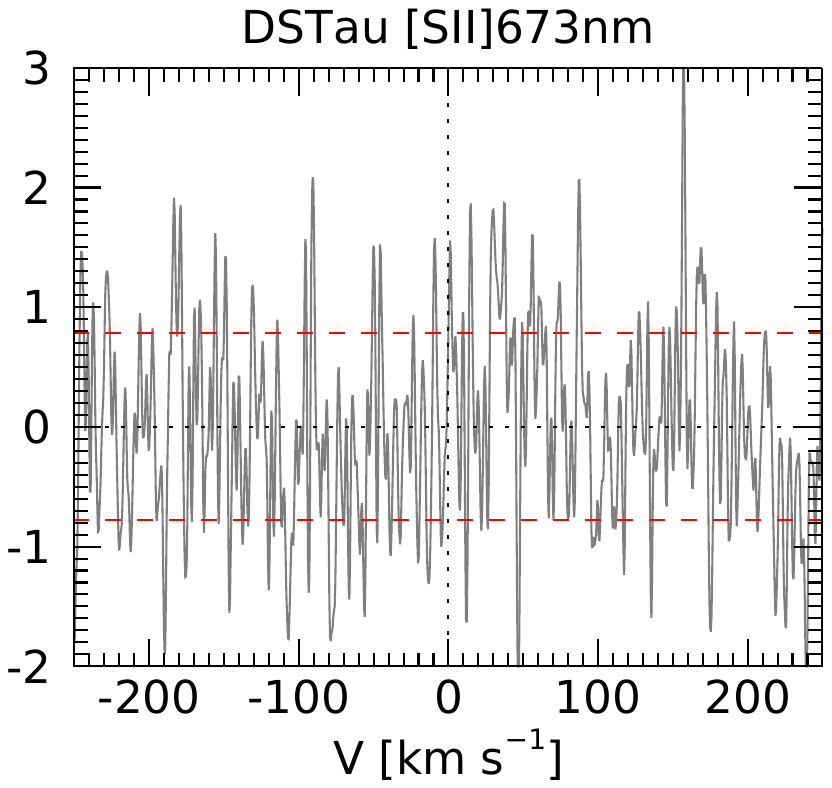}
\includegraphics[trim=80 0 80 400,width=0.2\textwidth]{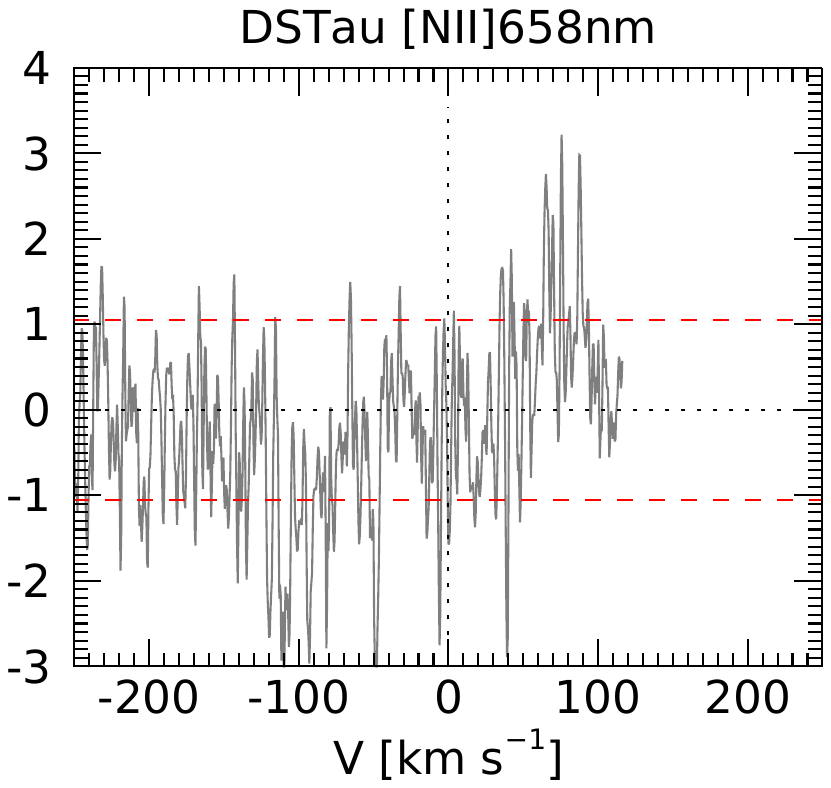}

\includegraphics[trim=80 0 80 400,width=0.2\textwidth]{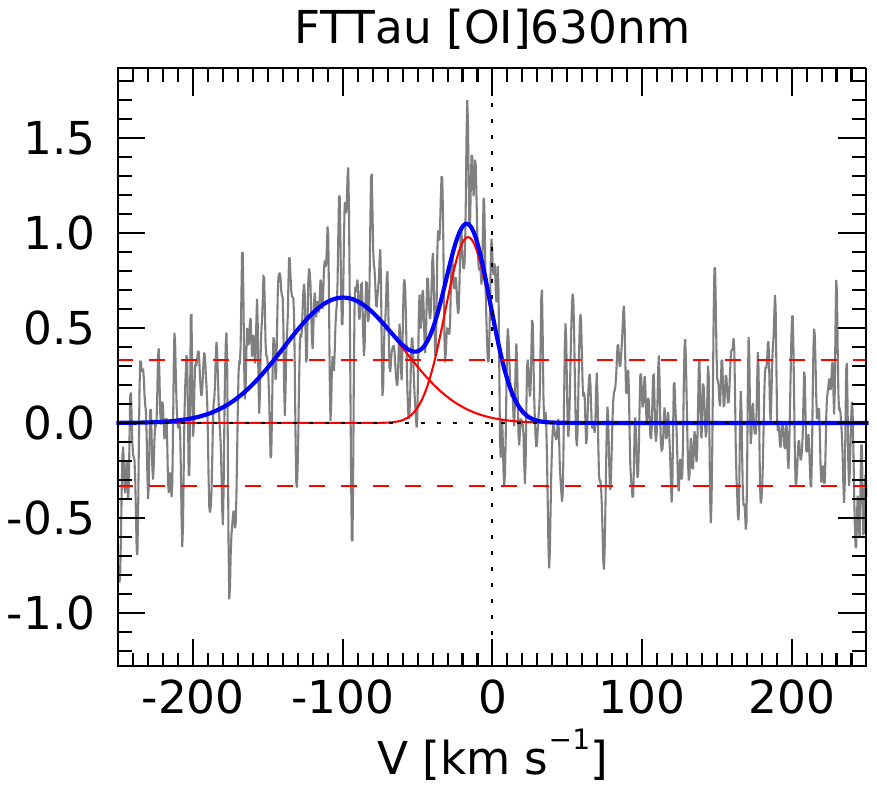}
\includegraphics[trim=80 0 80 400,width=0.2\textwidth]{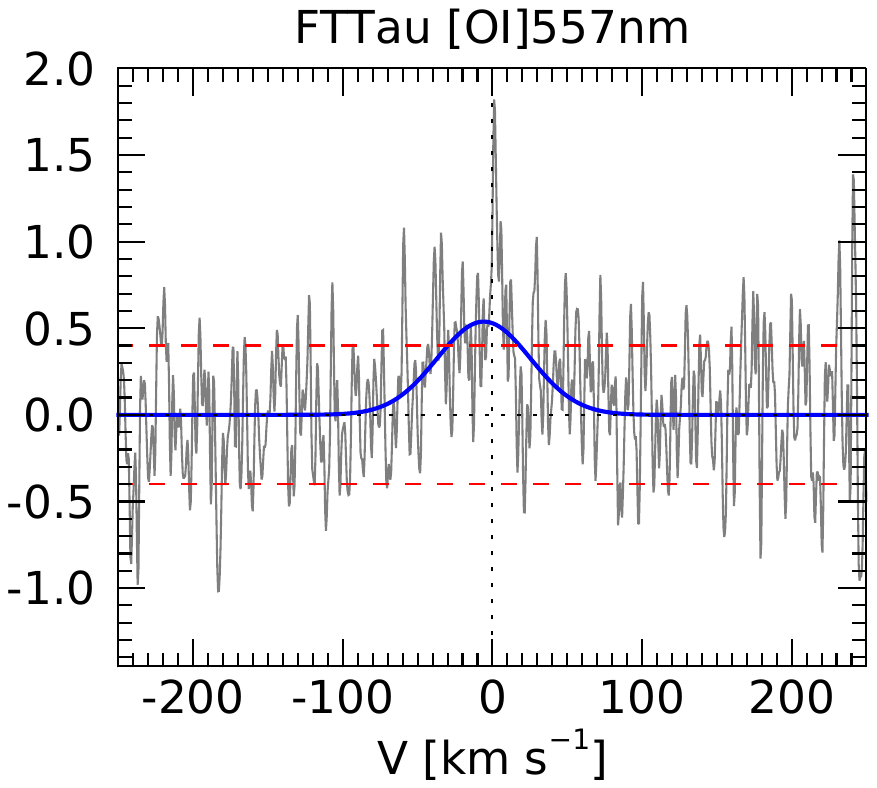}
\includegraphics[trim=80 0 80 400,width=0.2\textwidth]{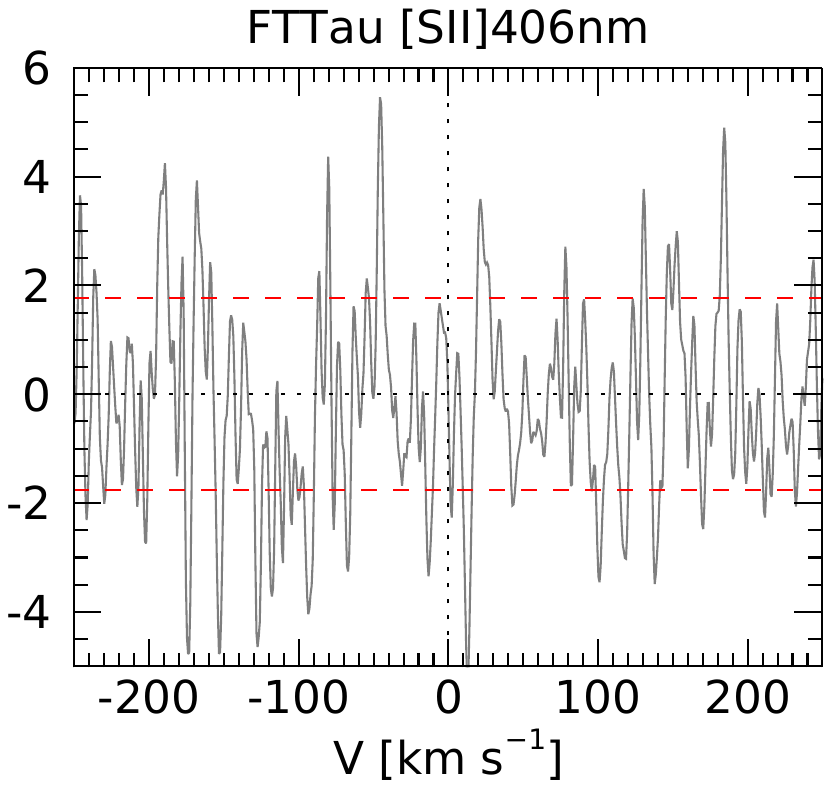}
\includegraphics[trim=80 0 80 400,width=0.2\textwidth]{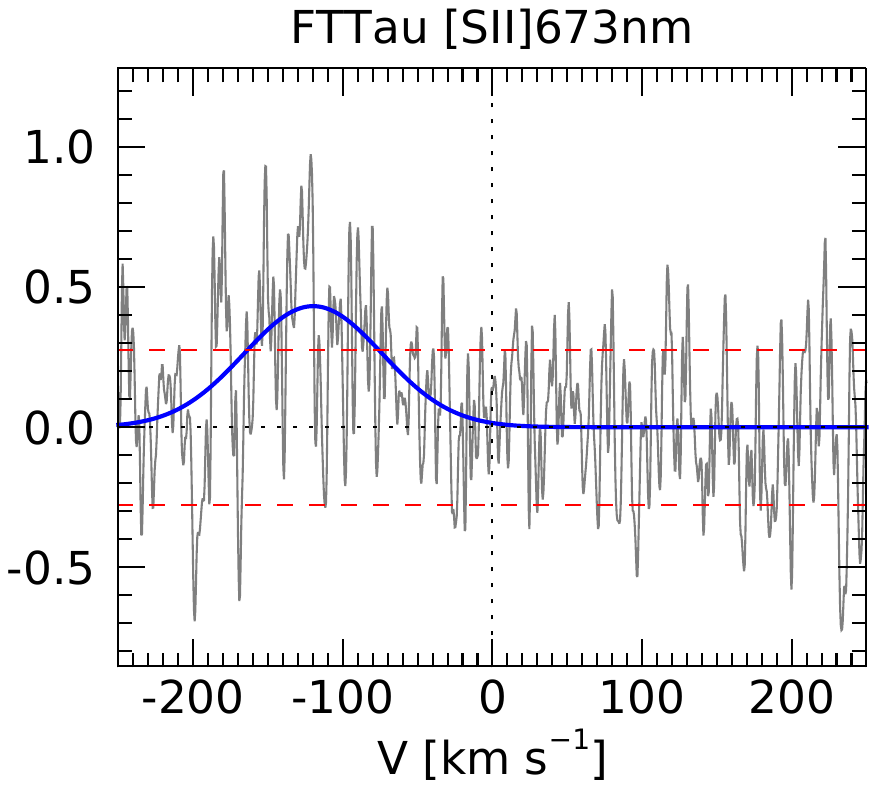}
\includegraphics[trim=80 0 80 400,width=0.2\textwidth]{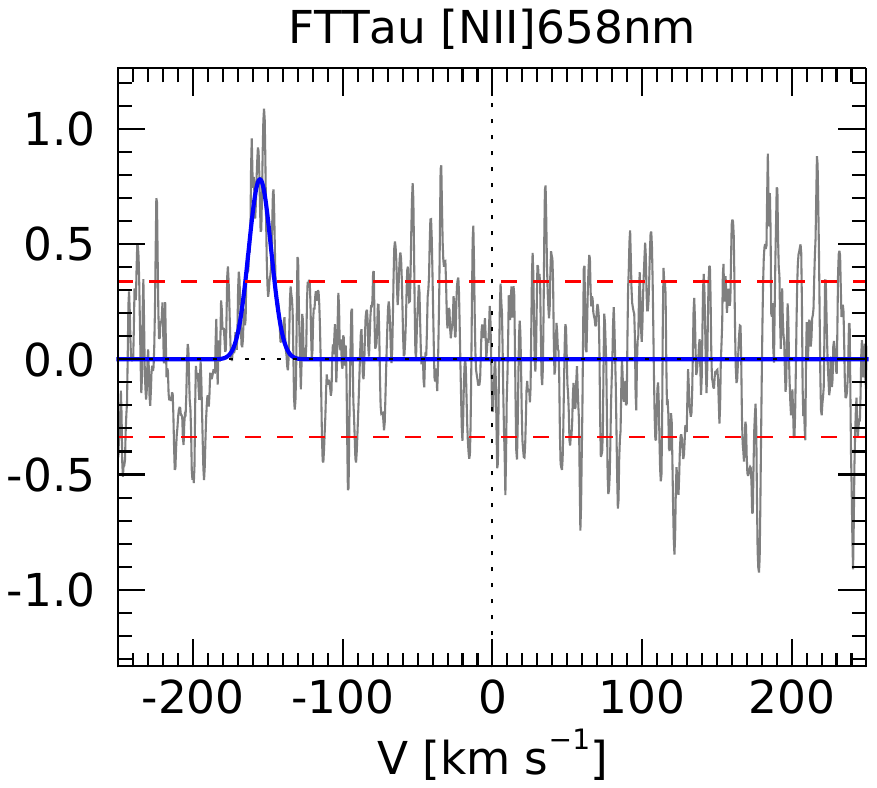}

\includegraphics[trim=80 0 80 400,width=0.2\textwidth]{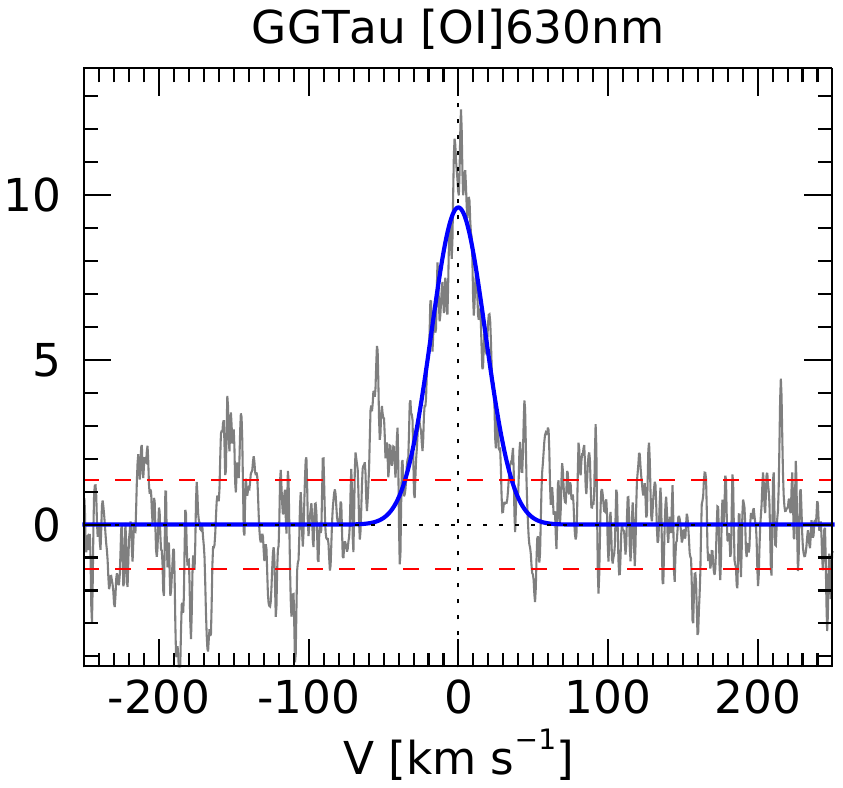}
\includegraphics[trim=80 0 80 400,width=0.2\textwidth]{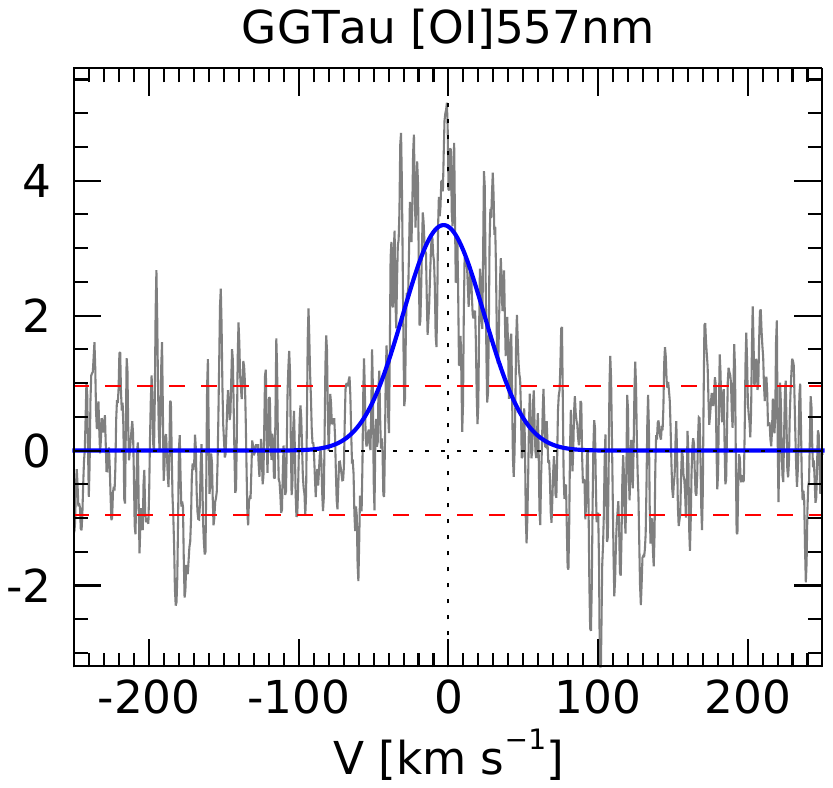}
\includegraphics[trim=80 0 80 400,width=0.2\textwidth]{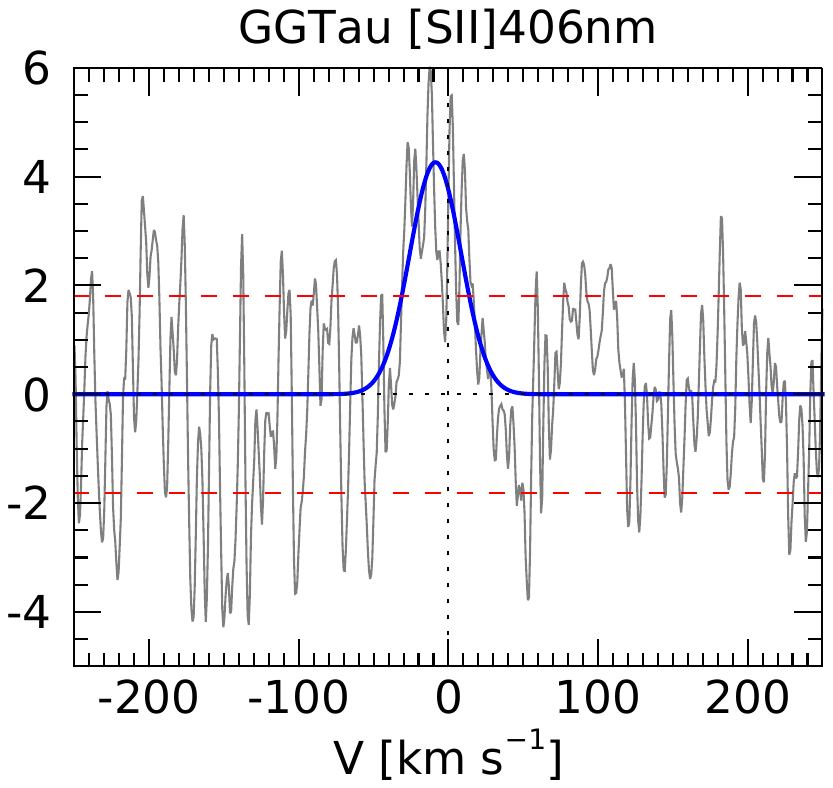}
\includegraphics[trim=80 0 80 400,width=0.2\textwidth]{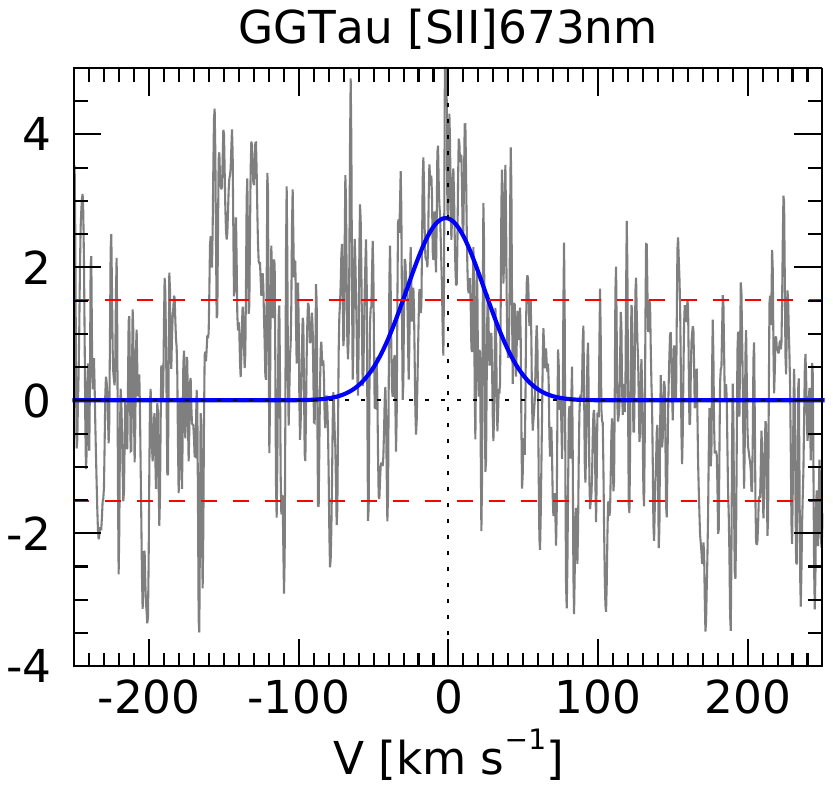}
\includegraphics[trim=80 0 80 400,width=0.2\textwidth]{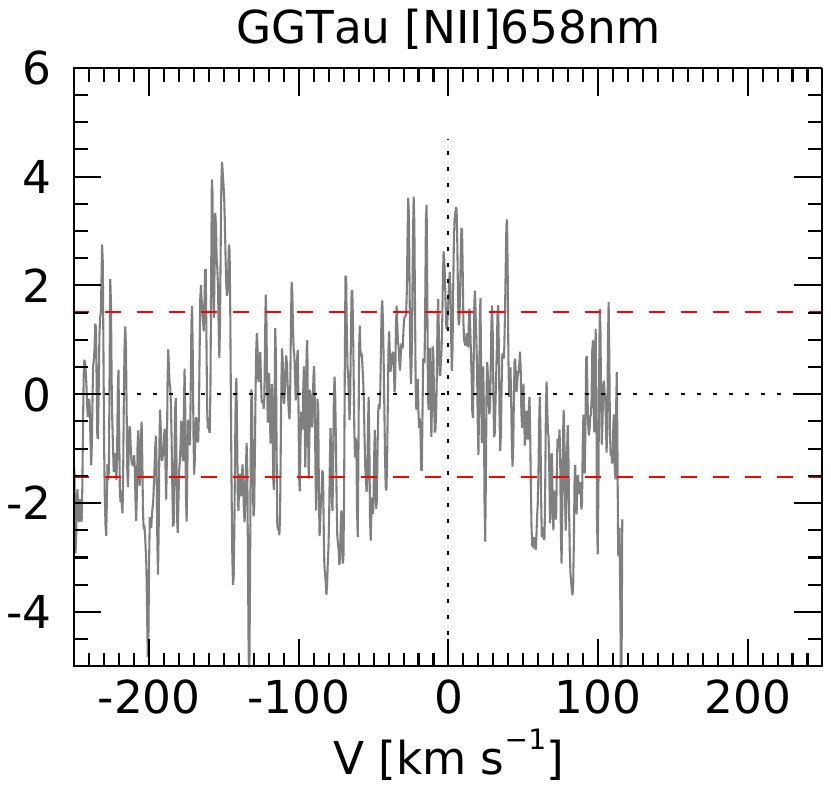}
   \caption{Continued}
   \label{fig:profiles3}
\end{figure*}

\newpage

\begin{figure*}[h]

\includegraphics[trim=80 0 80 0,width=0.2\textwidth]{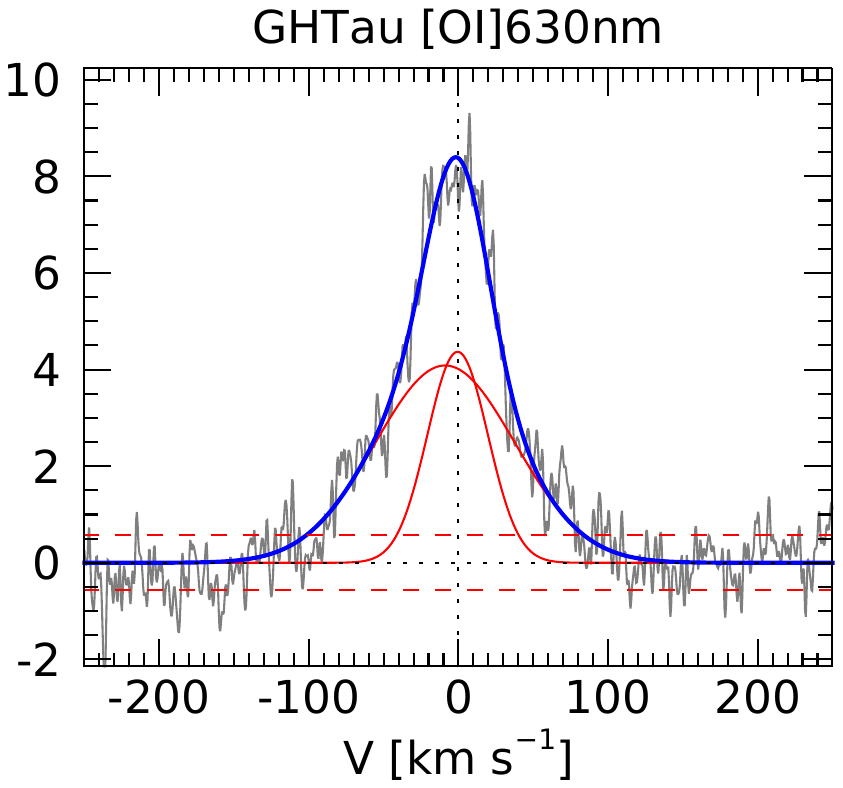}
\includegraphics[trim=80 0 80 0,width=0.2\textwidth]{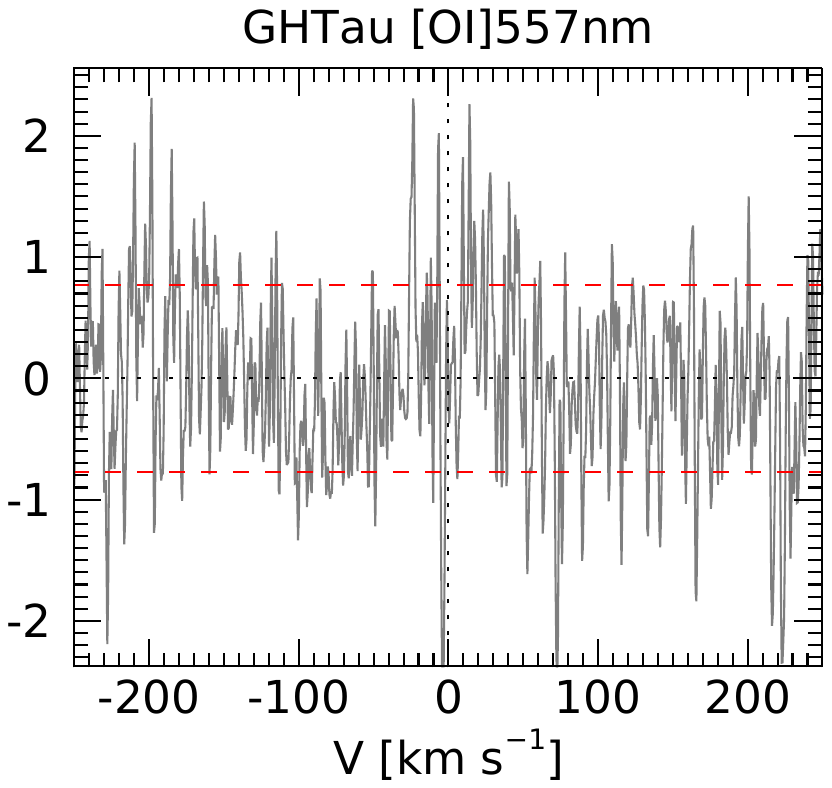}
\includegraphics[trim=80 0 80 0,width=0.2\textwidth]{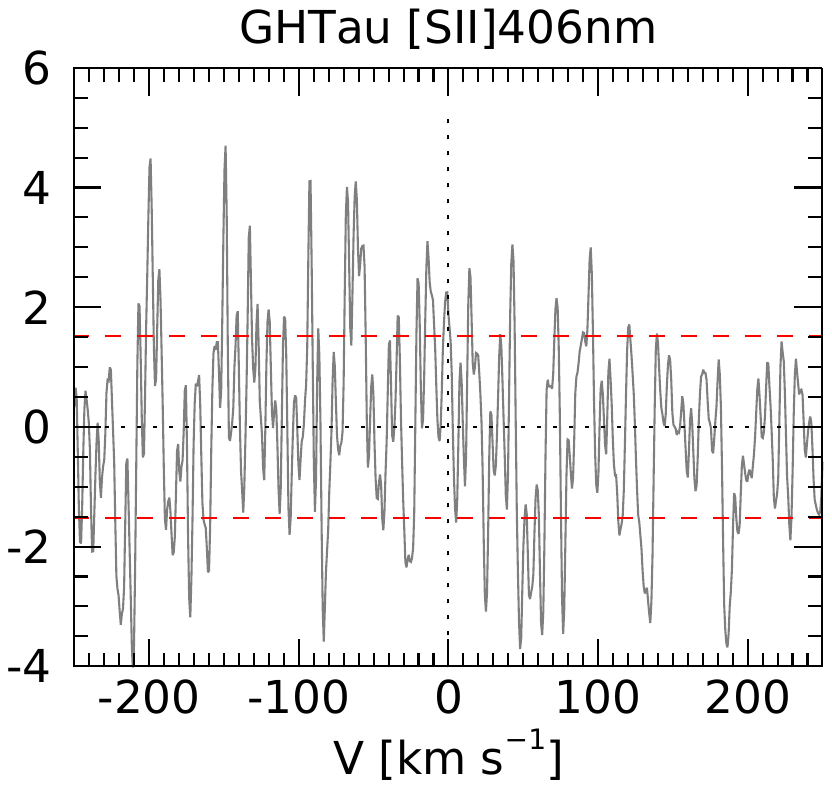}
\includegraphics[trim=80 0 80 0,width=0.2\textwidth]{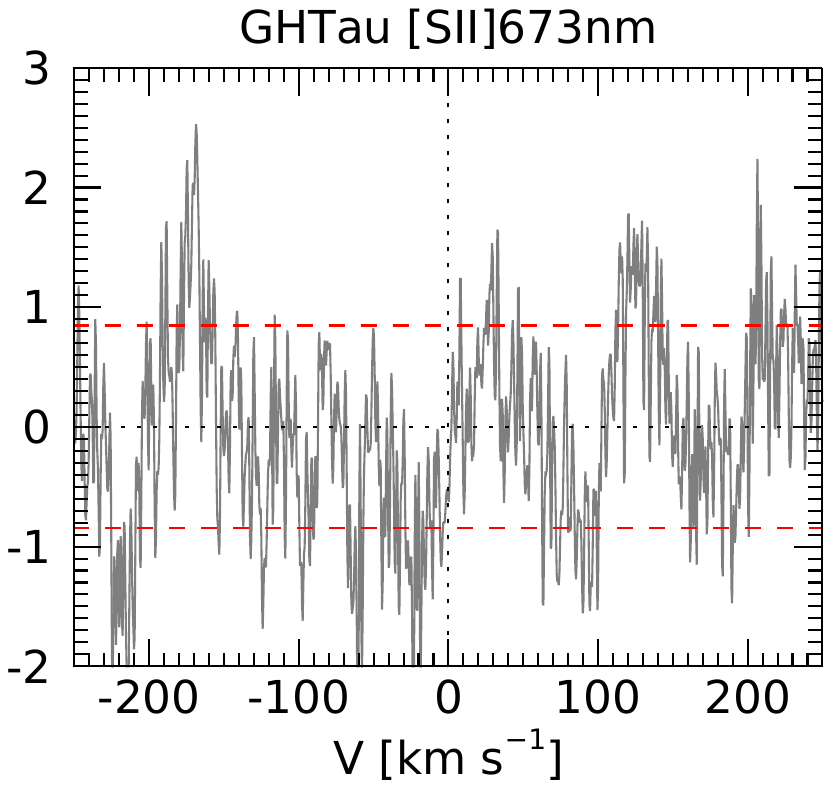}
\includegraphics[trim=80 0 80 0,width=0.2\textwidth]{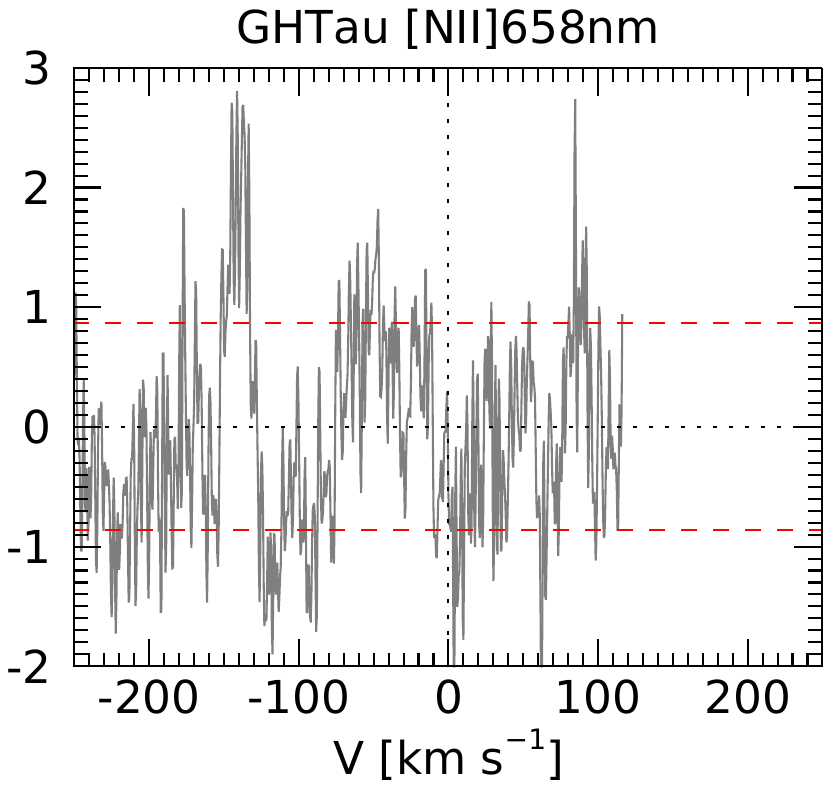}

\includegraphics[trim=80 0 80 400,width=0.2\textwidth]{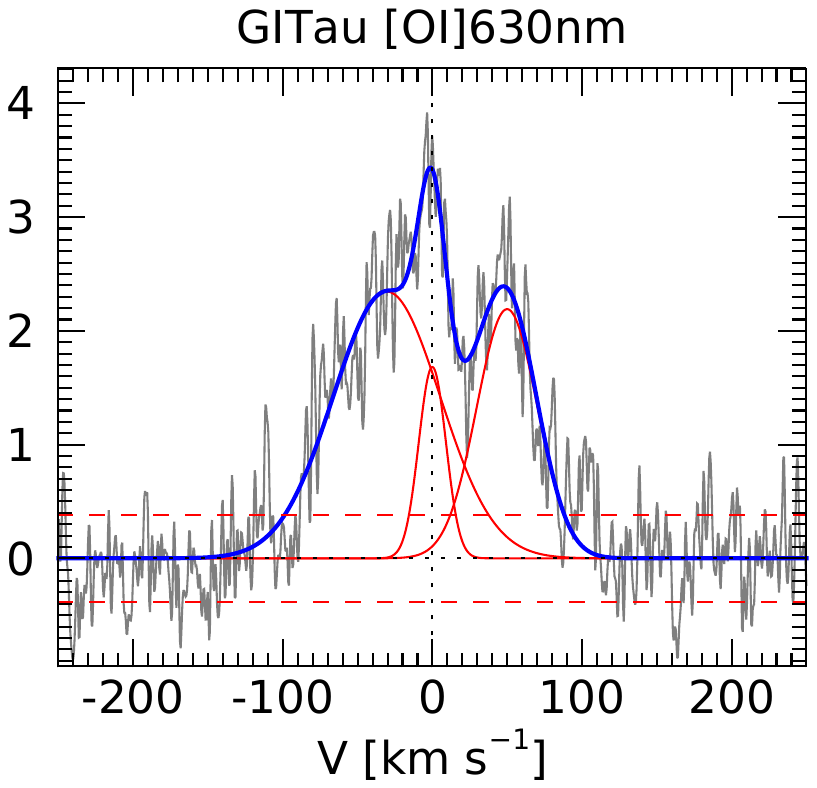}
\includegraphics[trim=80 0 80 400,width=0.2\textwidth]{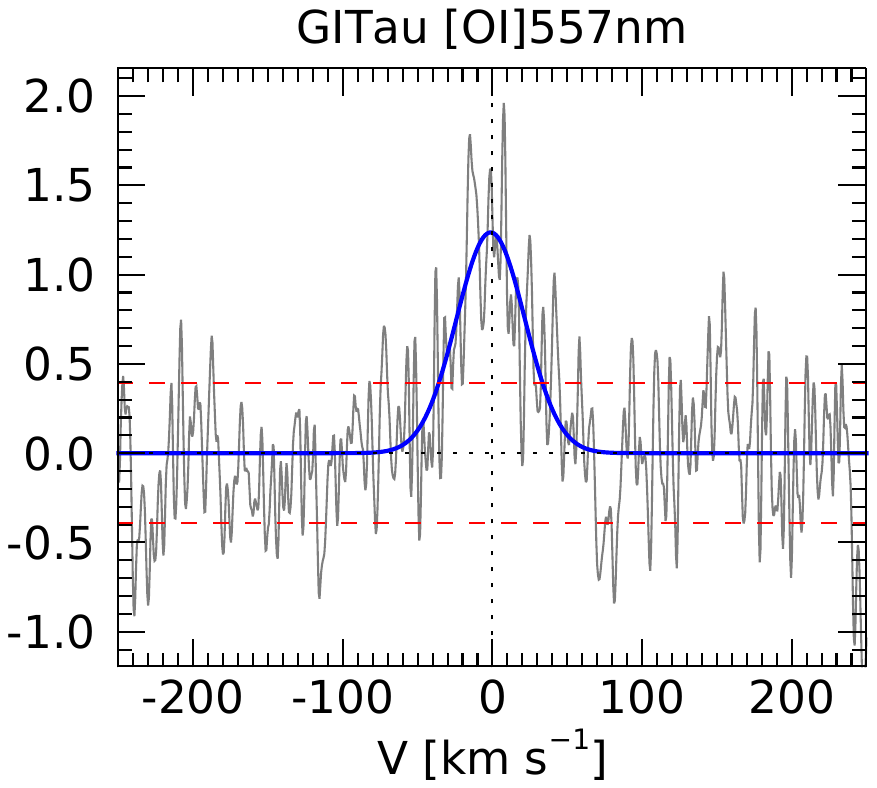}
\includegraphics[trim=80 0 80 400,width=0.2\textwidth]{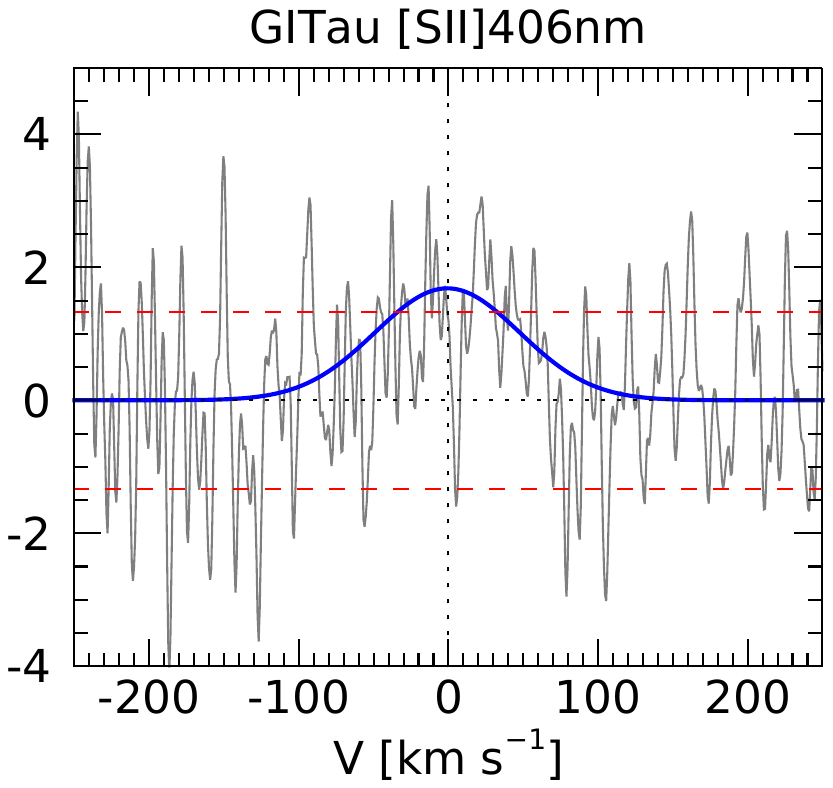}
\includegraphics[trim=80 0 80 400,width=0.2\textwidth]{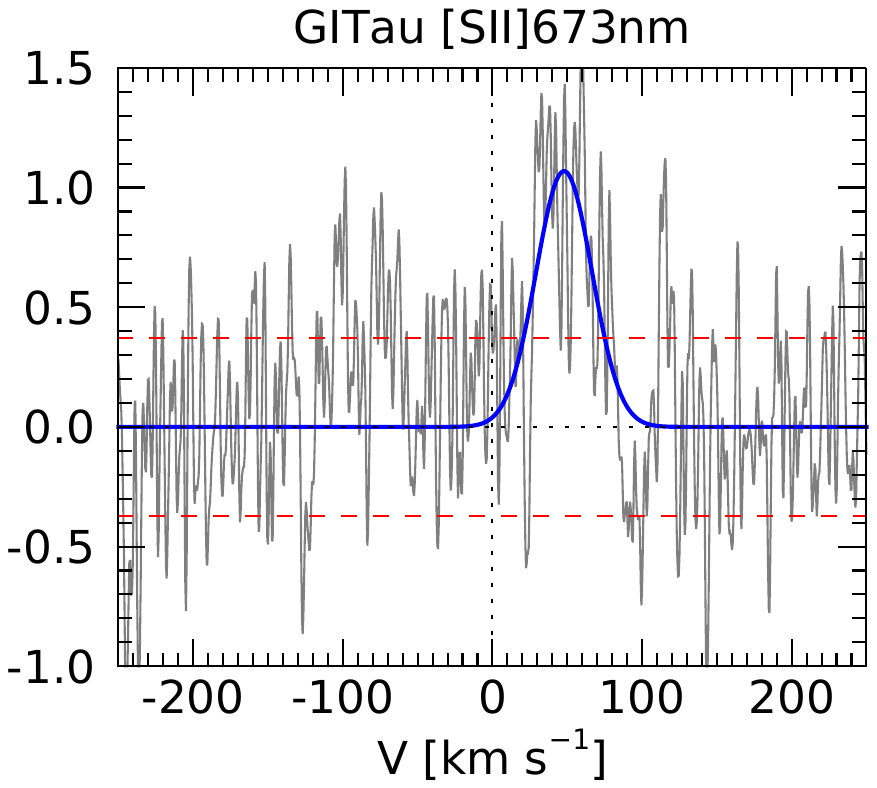}
\includegraphics[trim=80 0 80 400,width=0.2\textwidth]{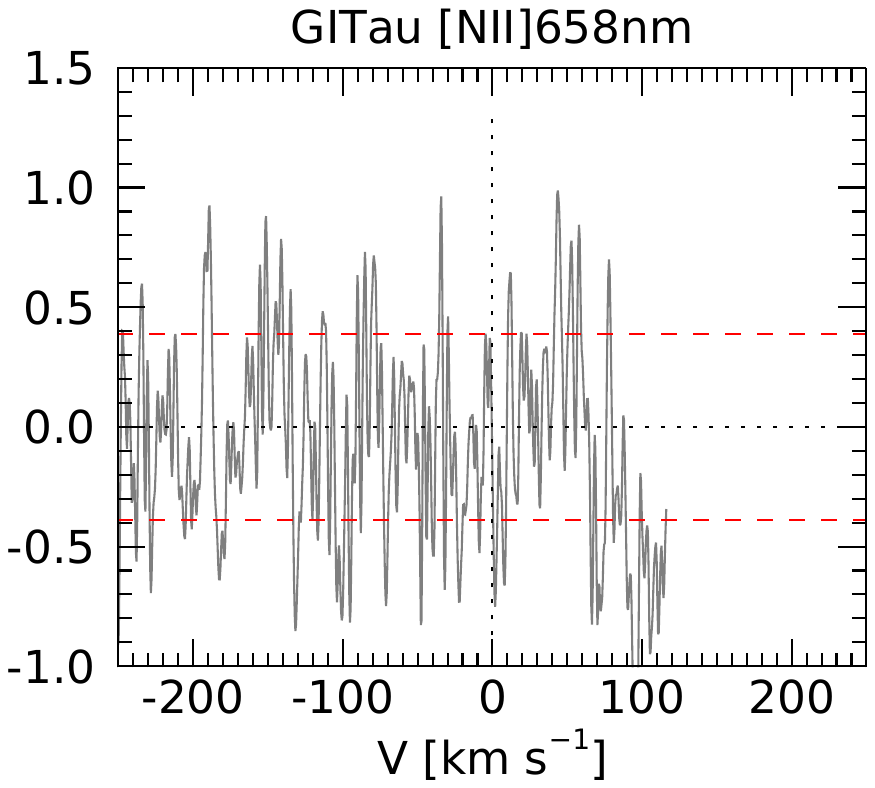}

\includegraphics[trim=80 0 80 400,width=0.2\textwidth]{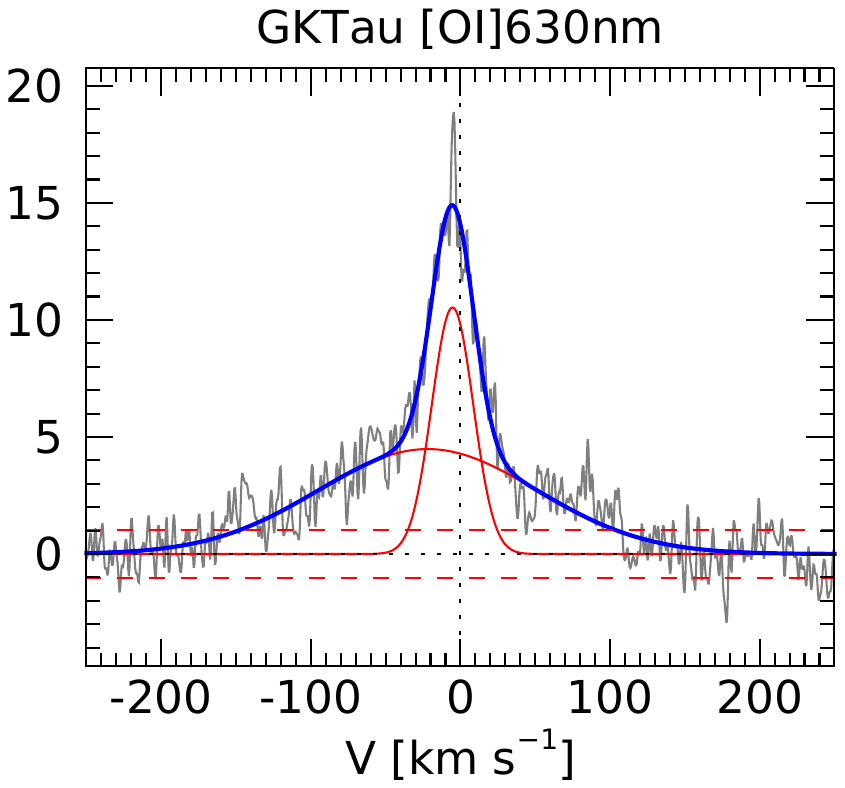}
\includegraphics[trim=80 0 80 400,width=0.2\textwidth]{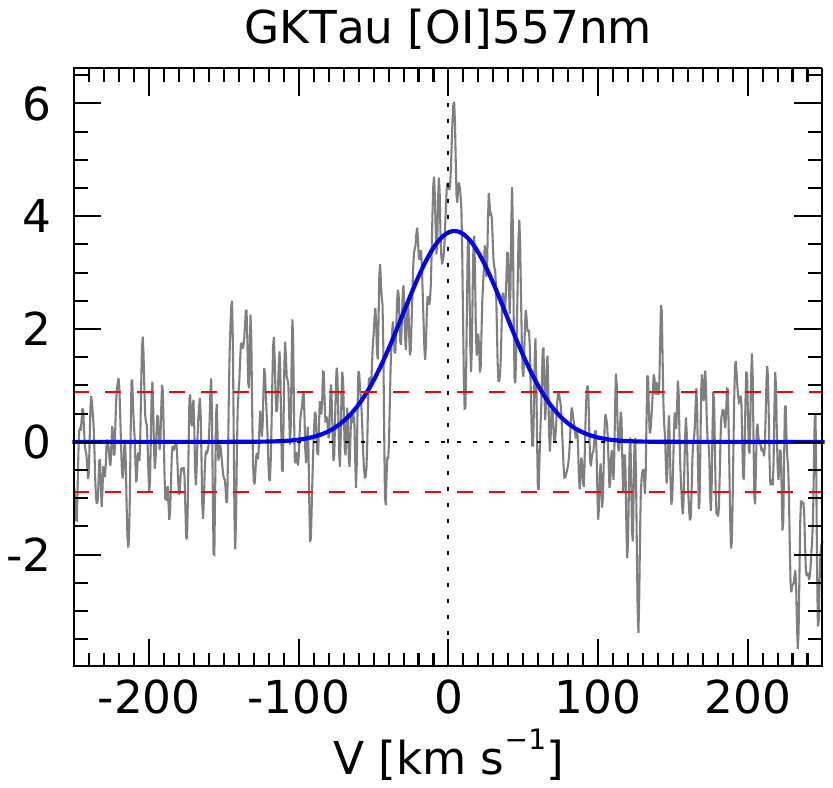}
\includegraphics[trim=80 0 80 400,width=0.2\textwidth]{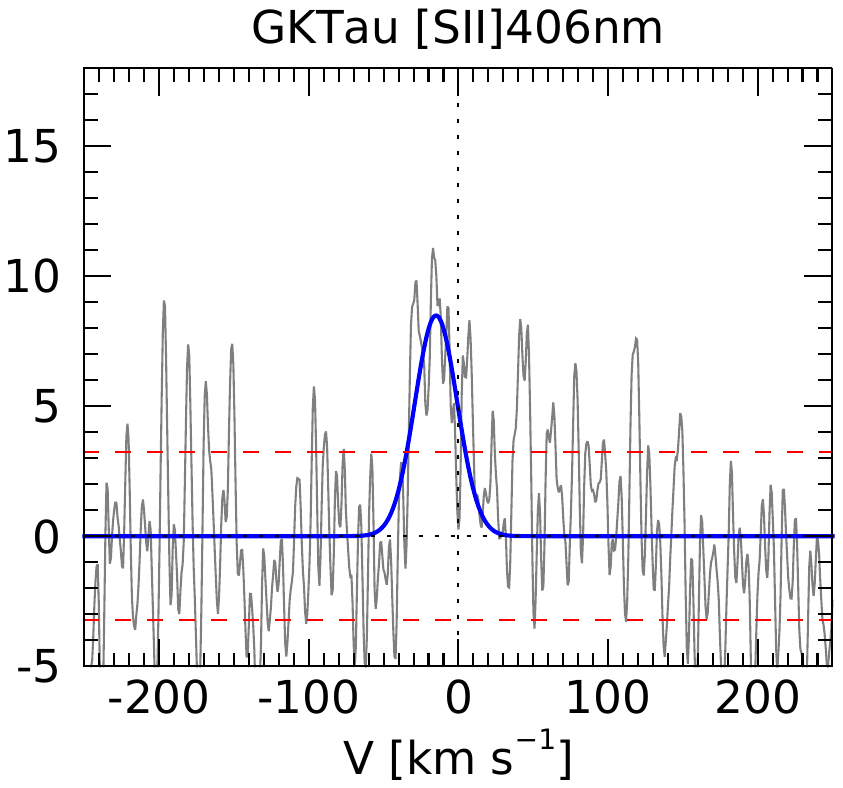}
\includegraphics[trim=80 0 80 400,width=0.2\textwidth]{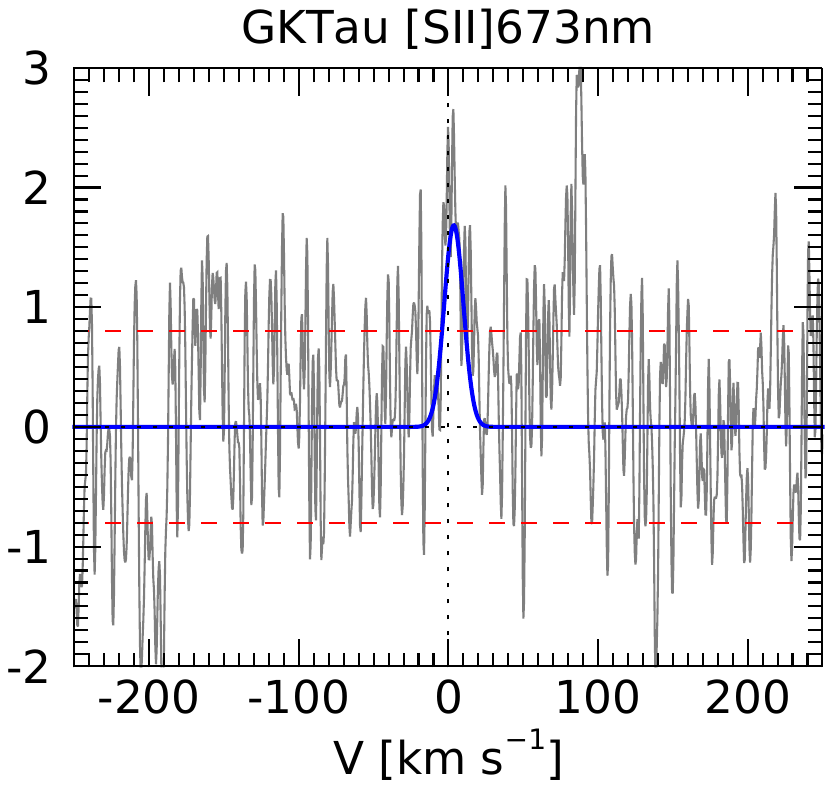}
\includegraphics[trim=80 0 80 400,width=0.2\textwidth]{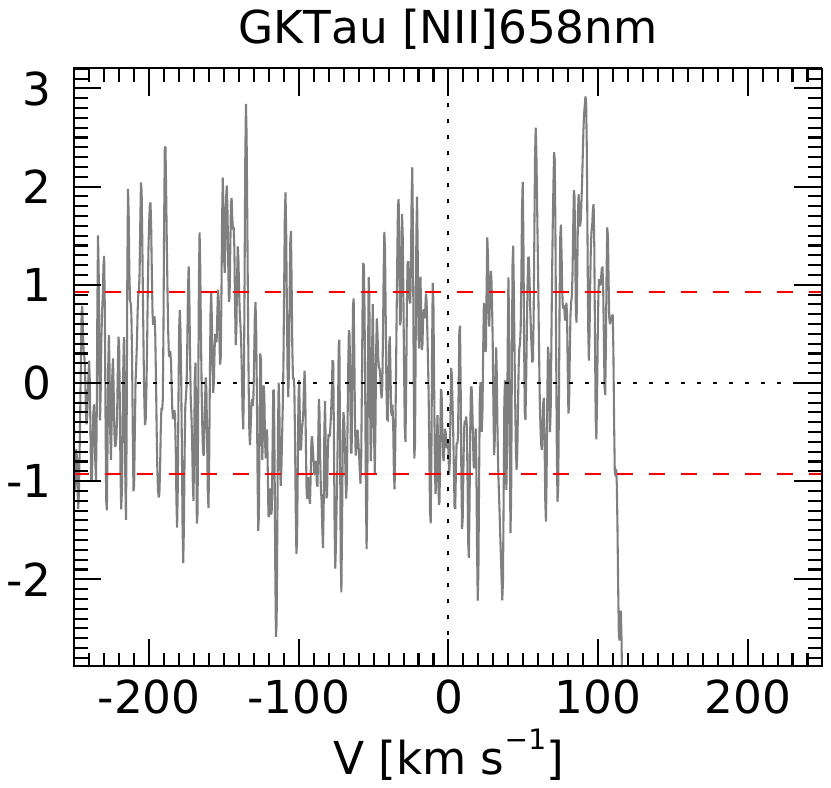}

\includegraphics[trim=80 0 80 400,width=0.2\textwidth]{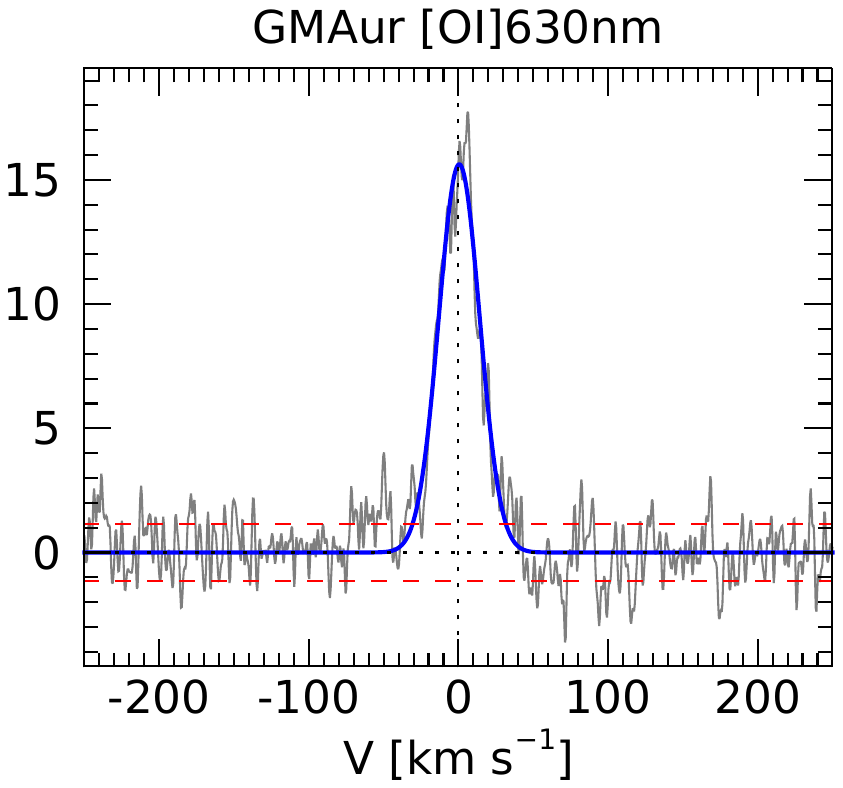}
\includegraphics[trim=80 0 80 400,width=0.2\textwidth]{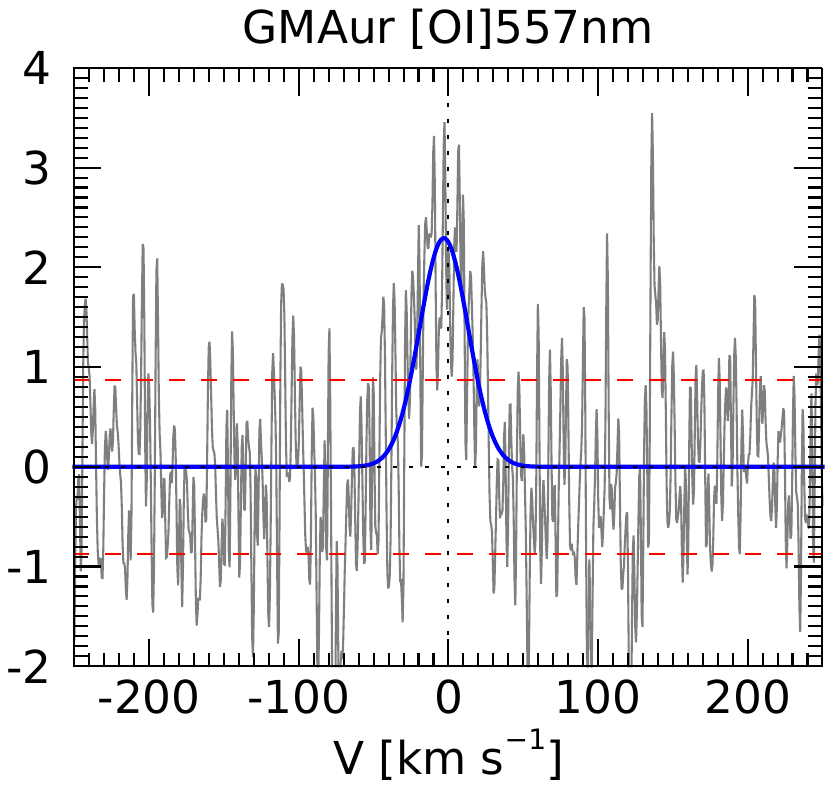}
\includegraphics[trim=80 0 80 400,width=0.2\textwidth]{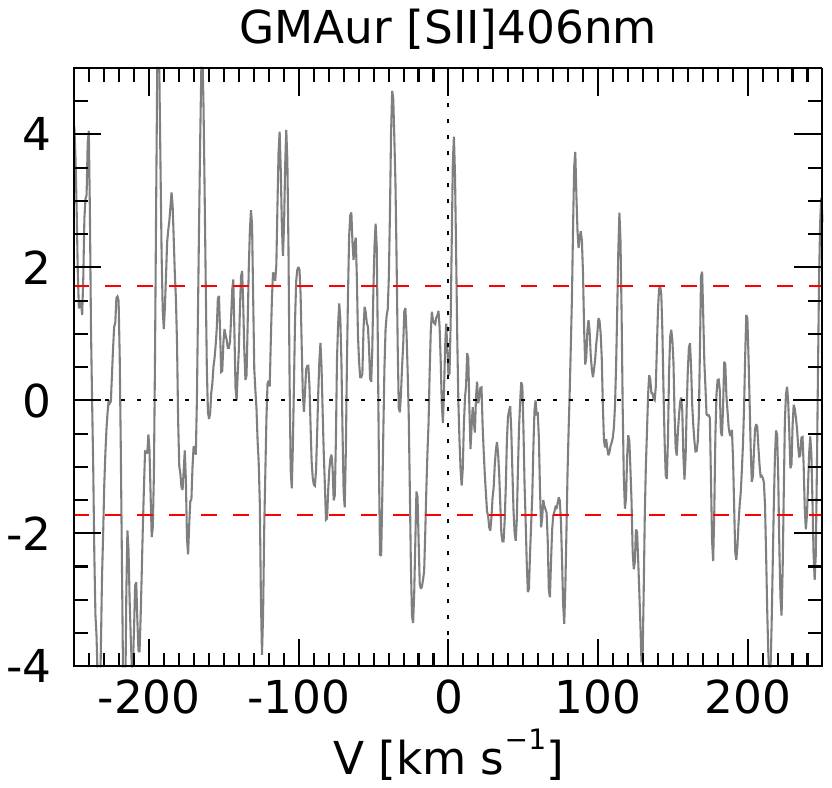}
\includegraphics[trim=80 0 80 400,width=0.2\textwidth]{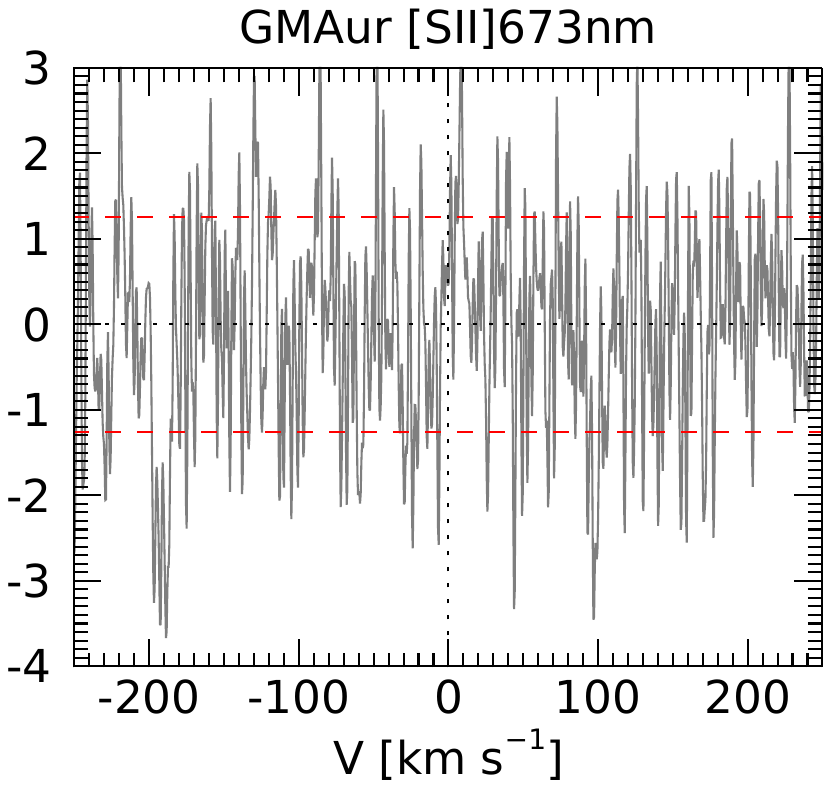}
\includegraphics[trim=80 0 80 400,width=0.2\textwidth]{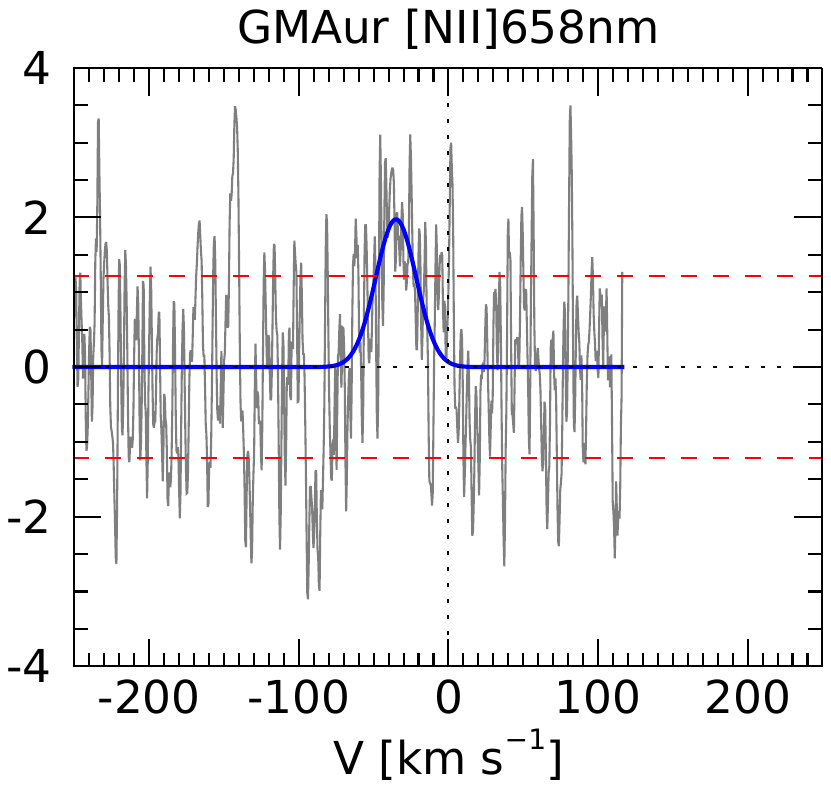}

\includegraphics[trim=80 0 80 400,width=0.2\textwidth]{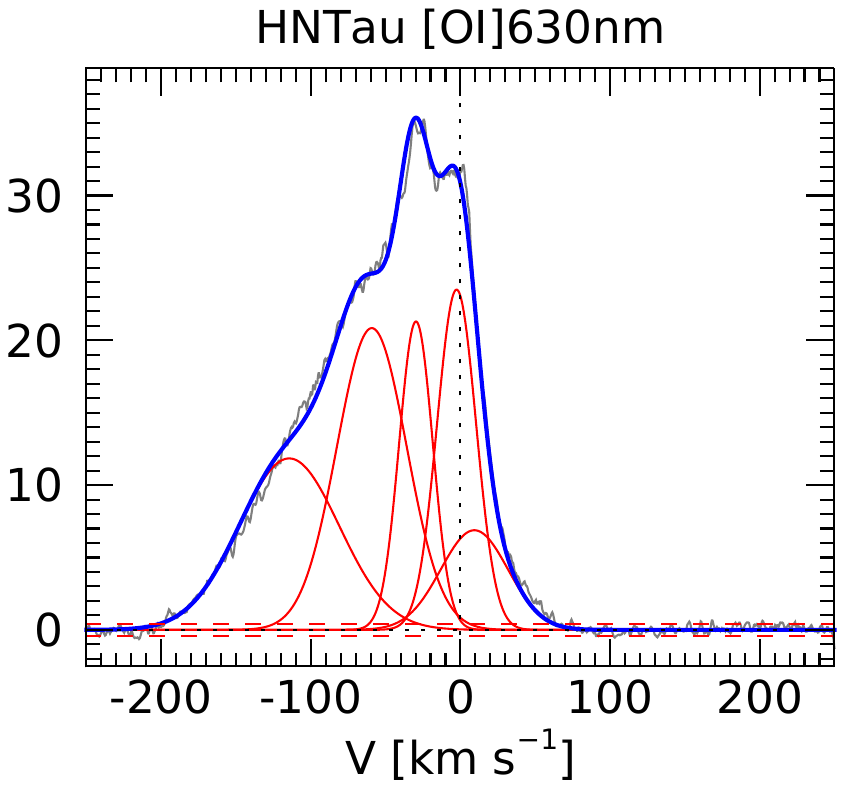}
\includegraphics[trim=80 0 80 400,width=0.2\textwidth]{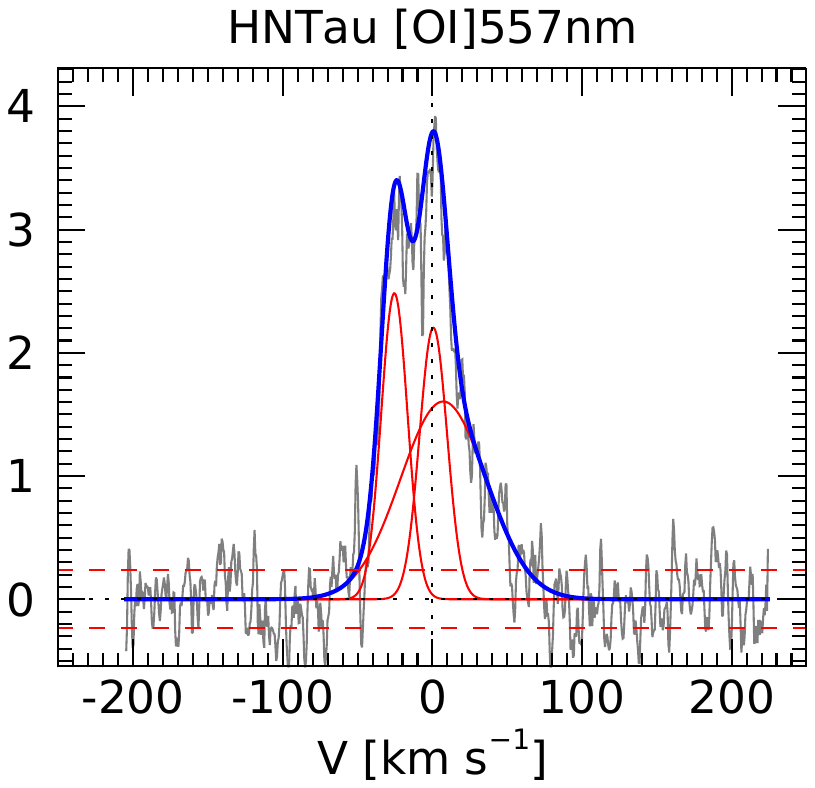}
\includegraphics[trim=80 0 80 400,width=0.2\textwidth]{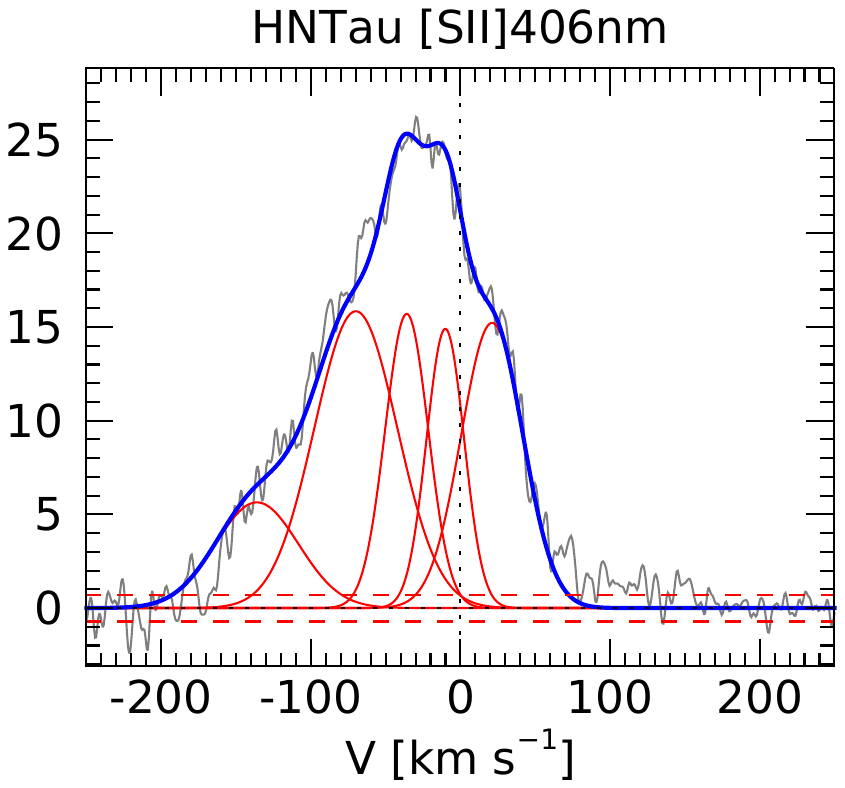}
\includegraphics[trim=80 0 80 400,width=0.2\textwidth]{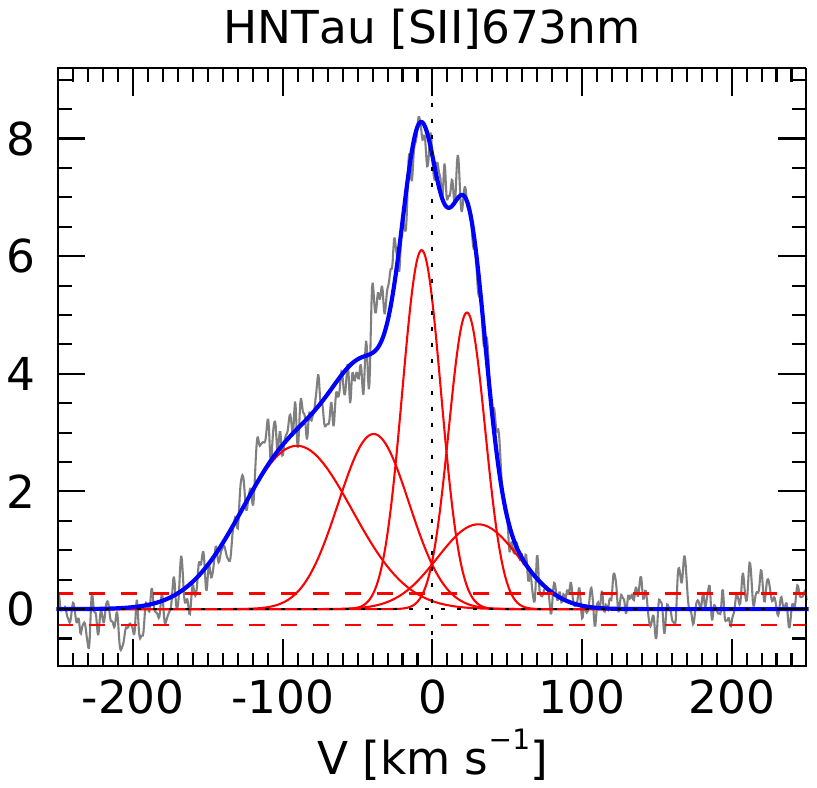}
\includegraphics[trim=80 0 80 400,width=0.2\textwidth]{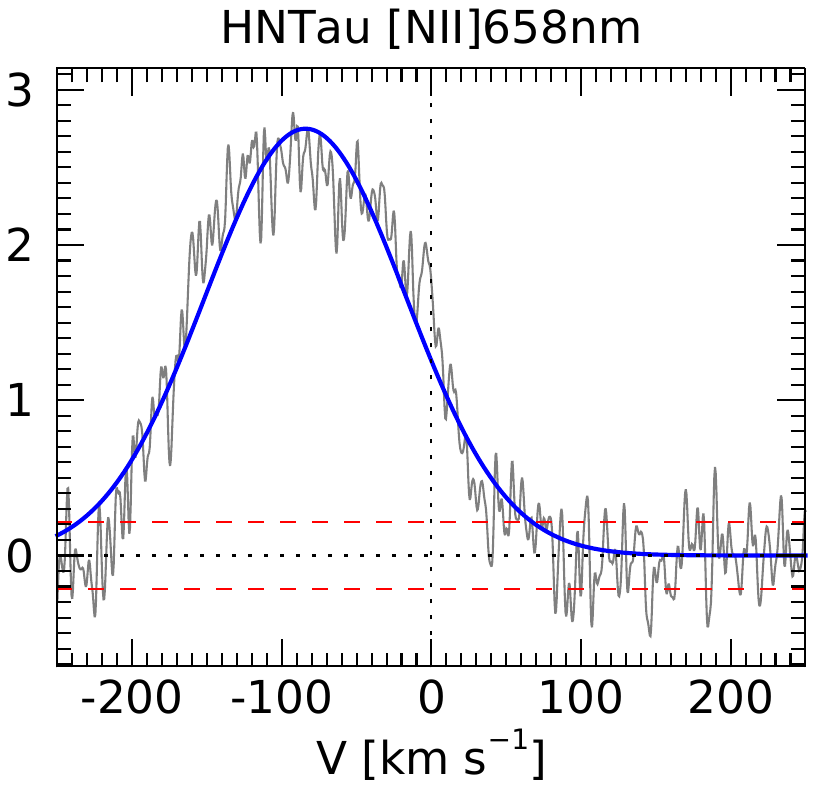}

\includegraphics[trim=80 0 80 400,width=0.2\textwidth]{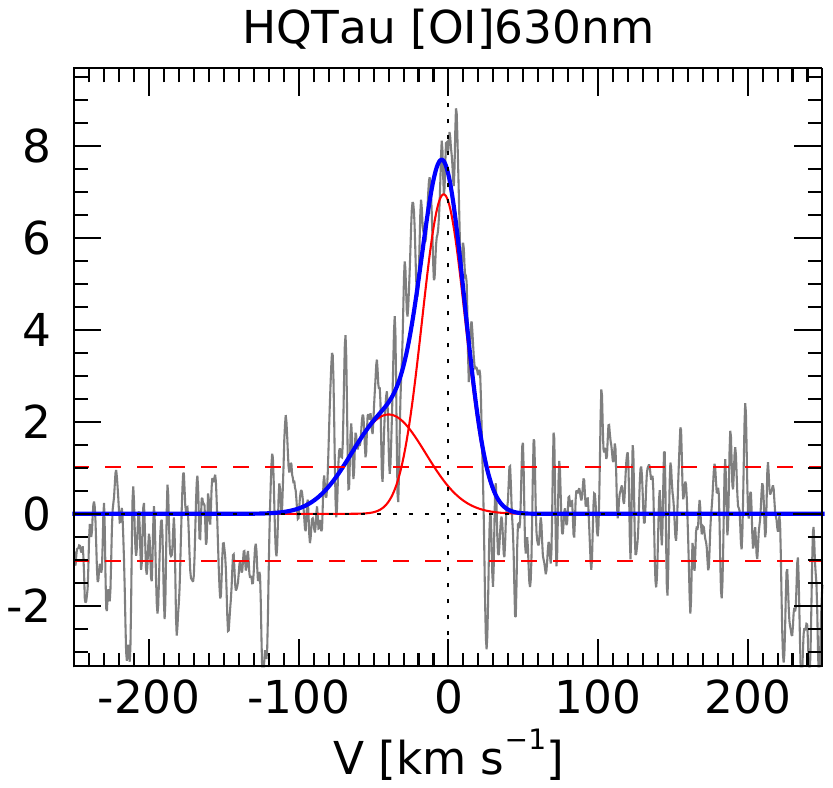}
\includegraphics[trim=80 0 80 400,width=0.2\textwidth]{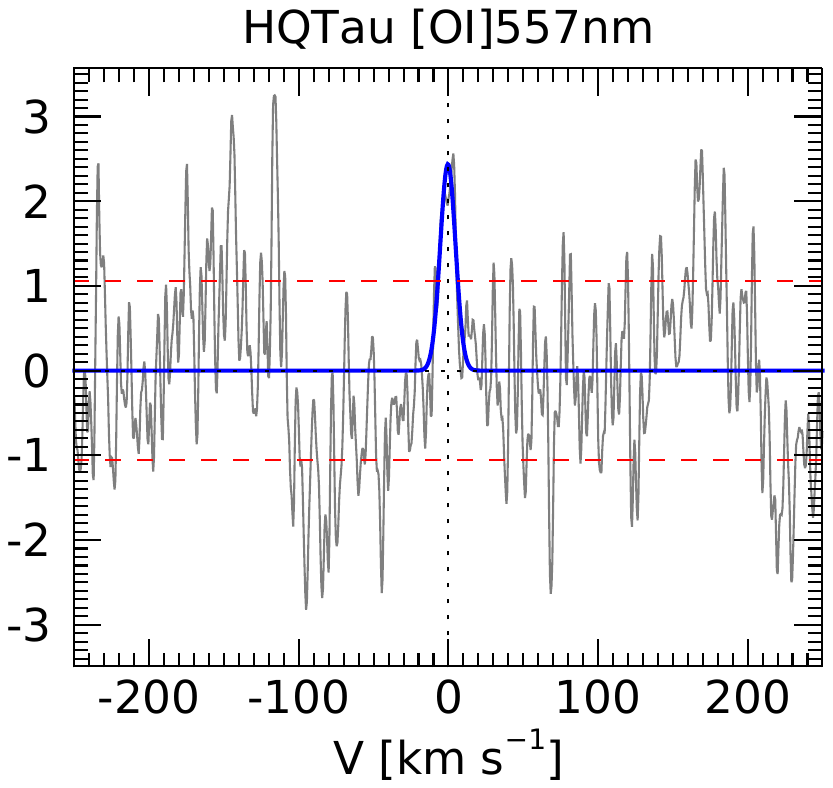}
\includegraphics[trim=80 0 80 400,width=0.2\textwidth]{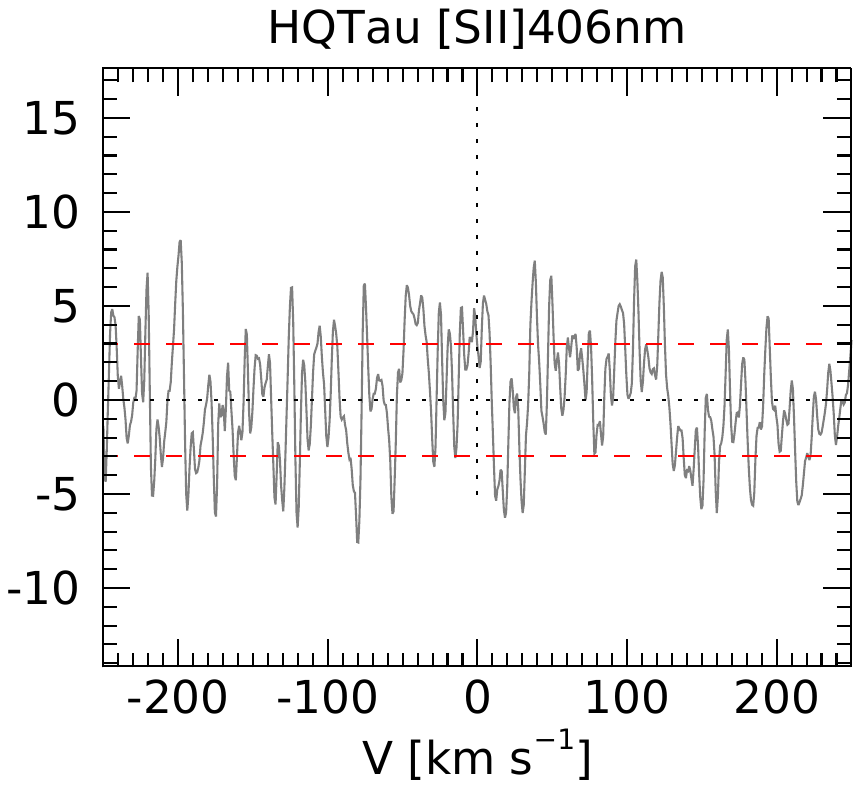}
\includegraphics[trim=80 0 80 400,width=0.2\textwidth]{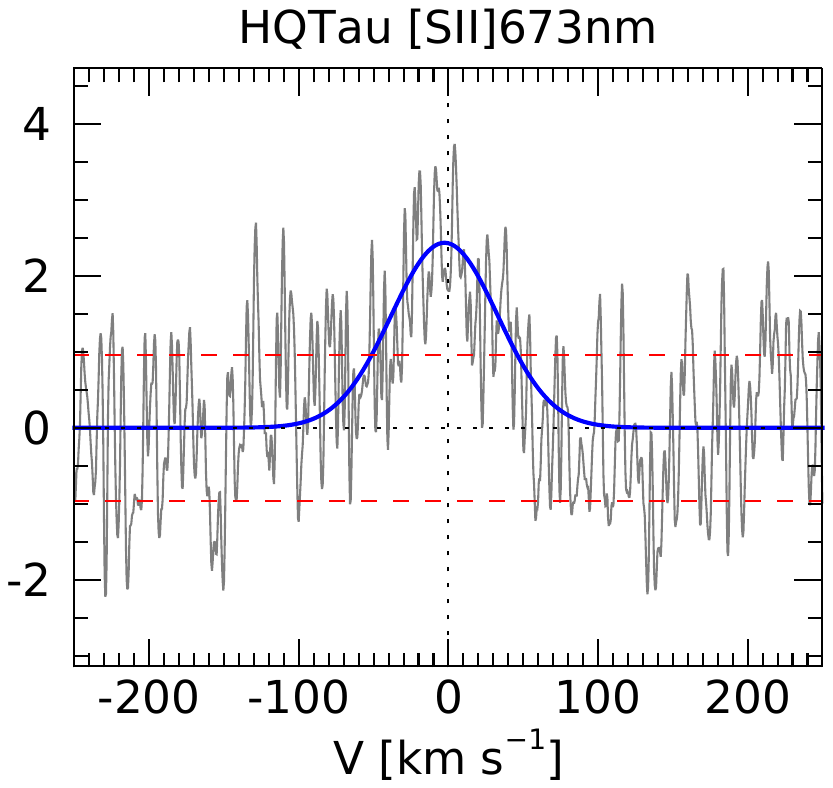}
\includegraphics[trim=80 0 80 400,width=0.2\textwidth]{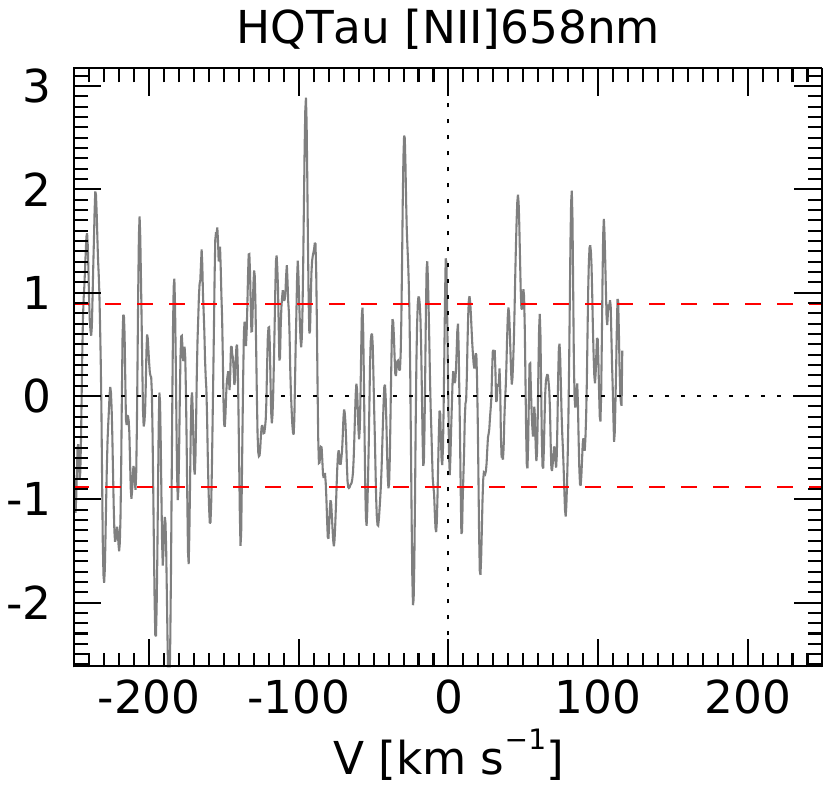}

   \caption{Continued}
   \label{fig:profiles4}
   
\end{figure*}

\newpage

\begin{figure*}[h]

\includegraphics[trim=80 0 80 0,width=0.2\textwidth]{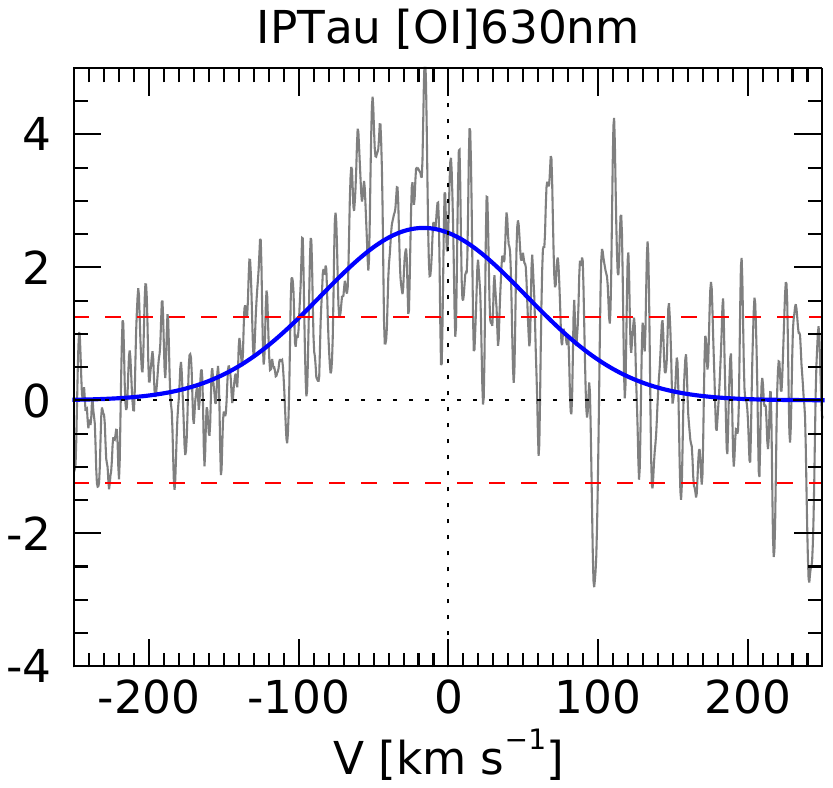}
\includegraphics[trim=80 0 80 0,width=0.2\textwidth]{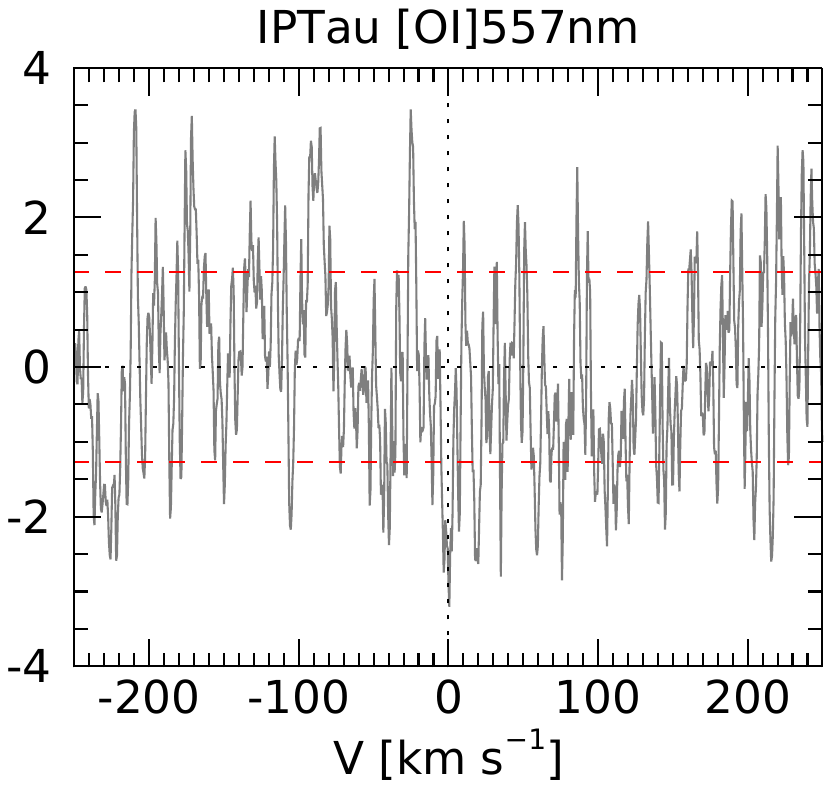}
\includegraphics[trim=80 0 80 0,width=0.2\textwidth]{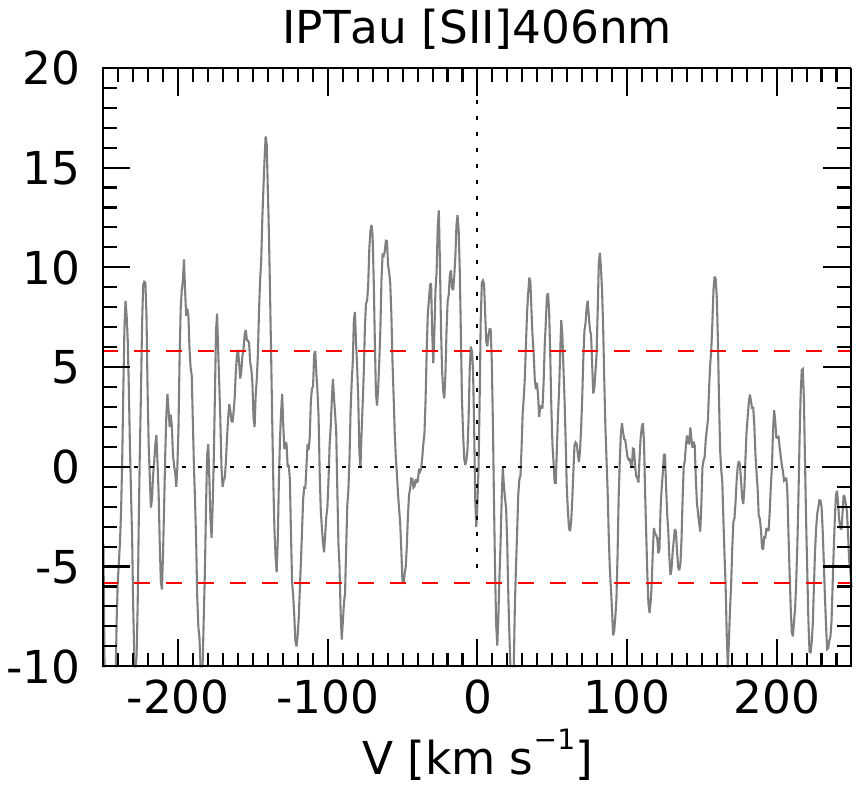}
\includegraphics[trim=80 0 80 0,width=0.2\textwidth]{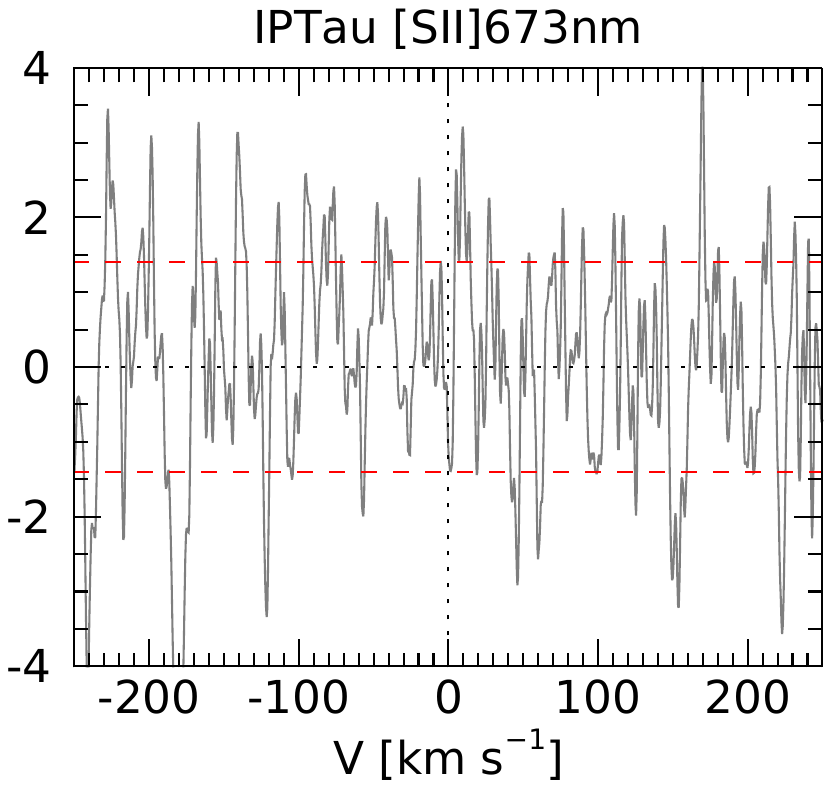}
\includegraphics[trim=80 0 80 0,width=0.2\textwidth]{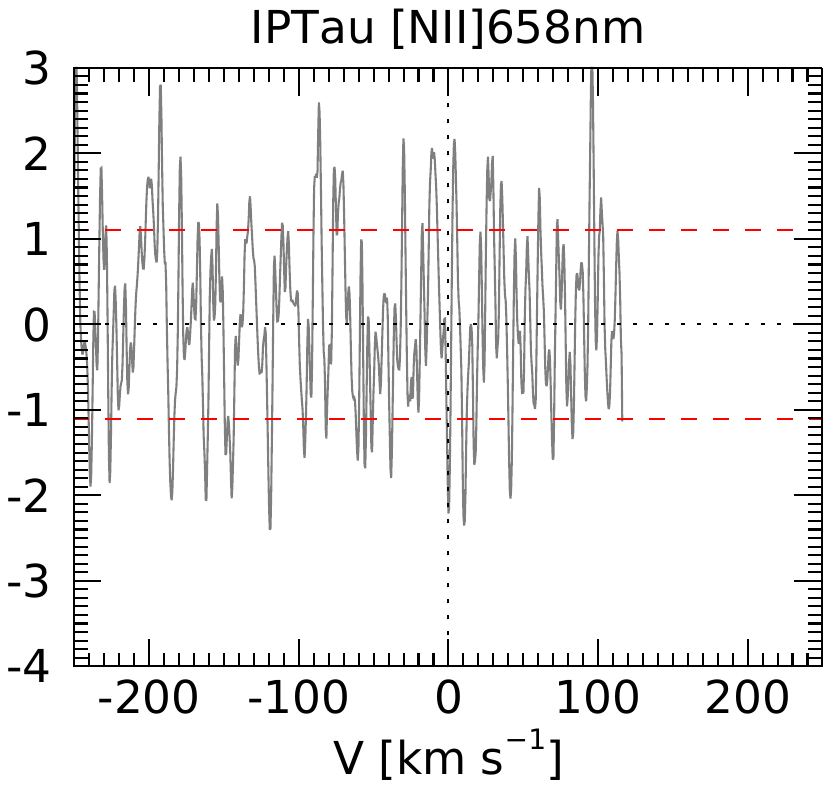}

\includegraphics[trim=80 0 80 400,width=0.2\textwidth]{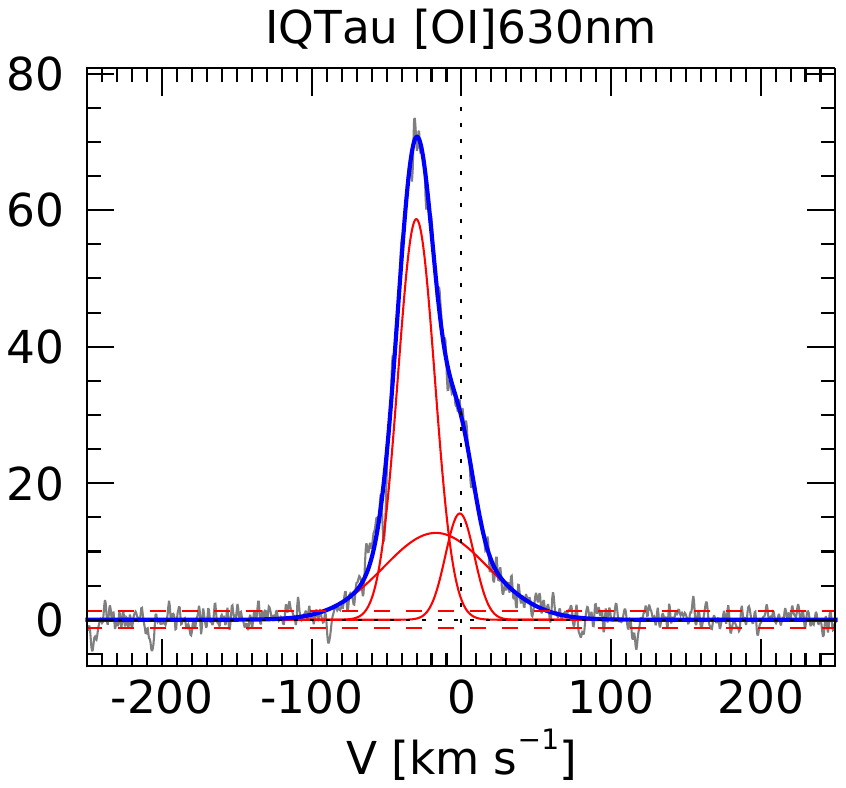}
\includegraphics[trim=80 0 80 400,width=0.2\textwidth]{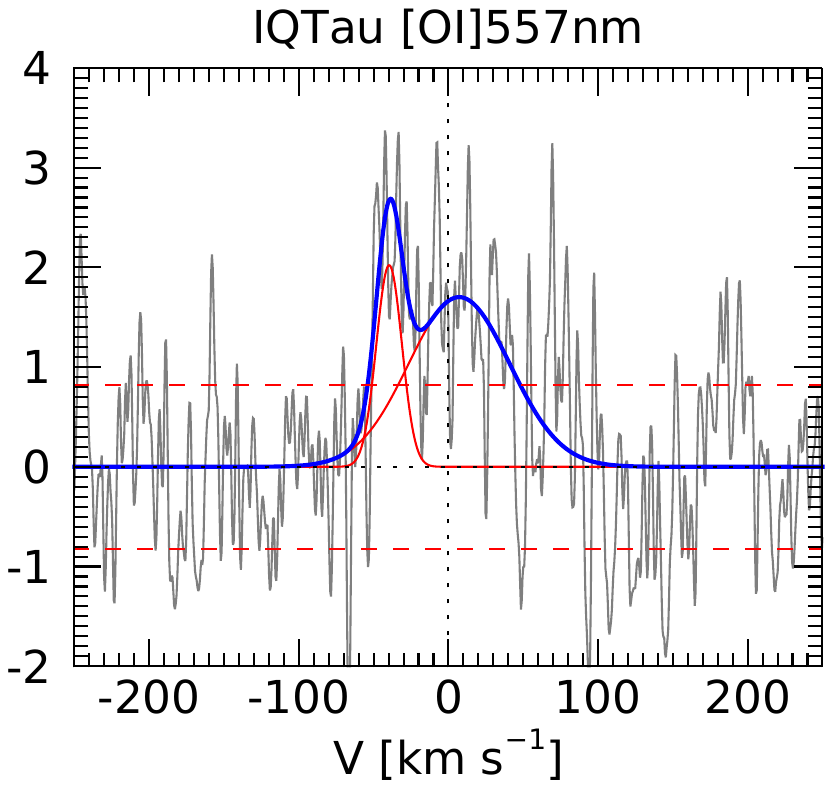}
\includegraphics[trim=80 0 80 400,width=0.2\textwidth]{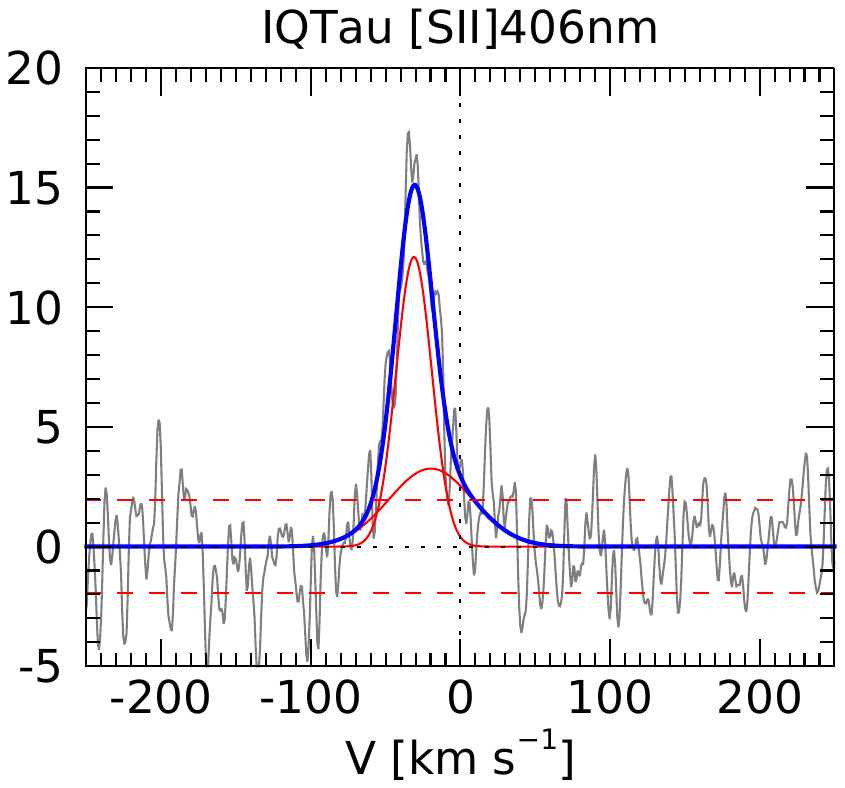}
\includegraphics[trim=80 0 80 400,width=0.2\textwidth]{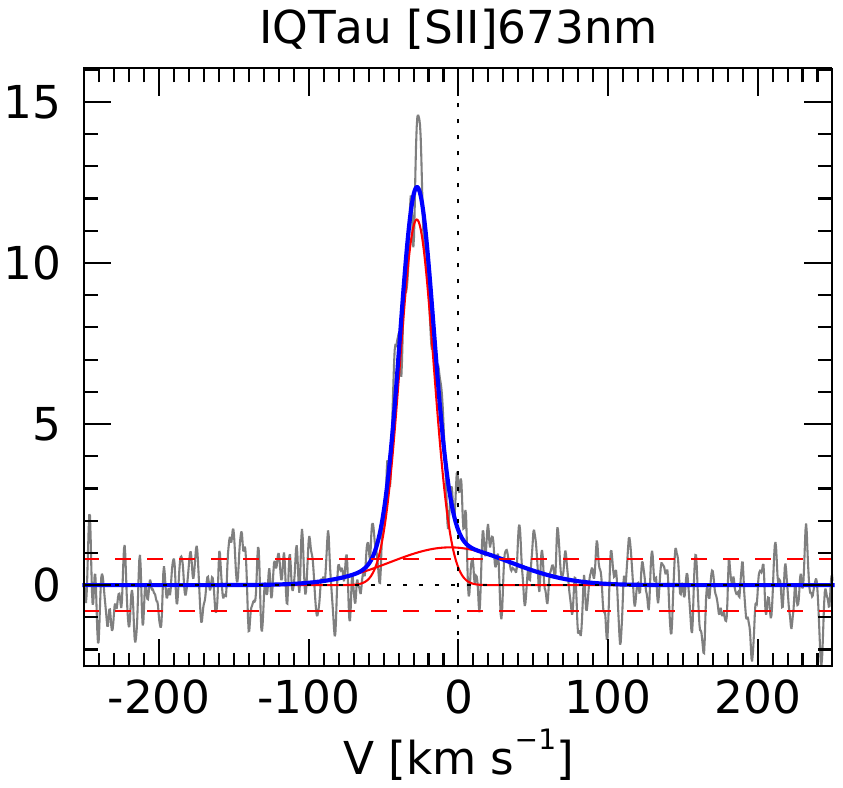}
\includegraphics[trim=80 0 80 400,width=0.2\textwidth]{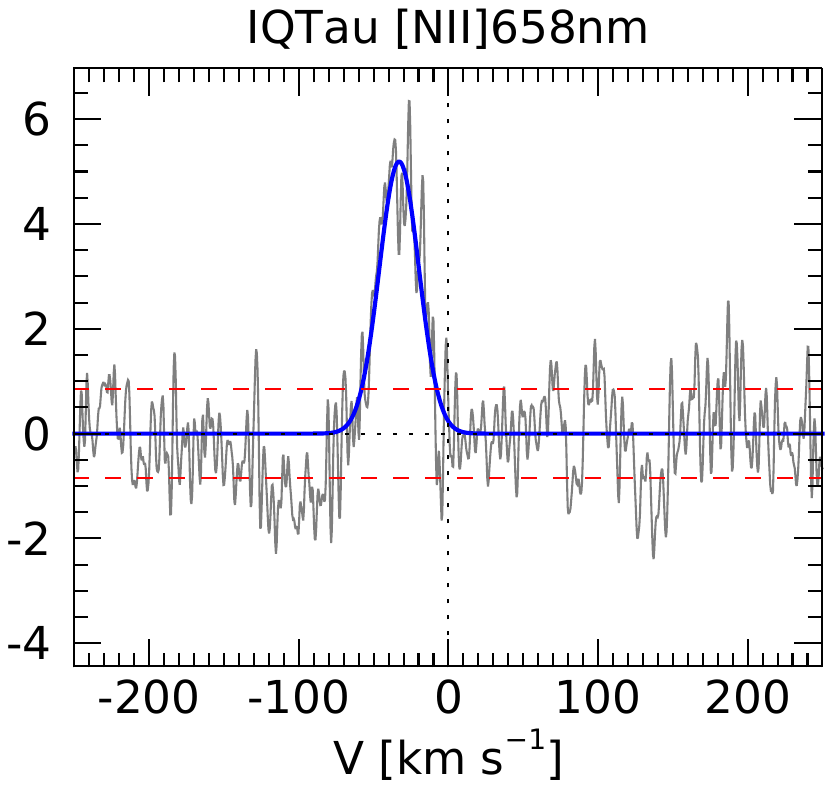}

\includegraphics[trim=80 0 80 400,width=0.2\textwidth]{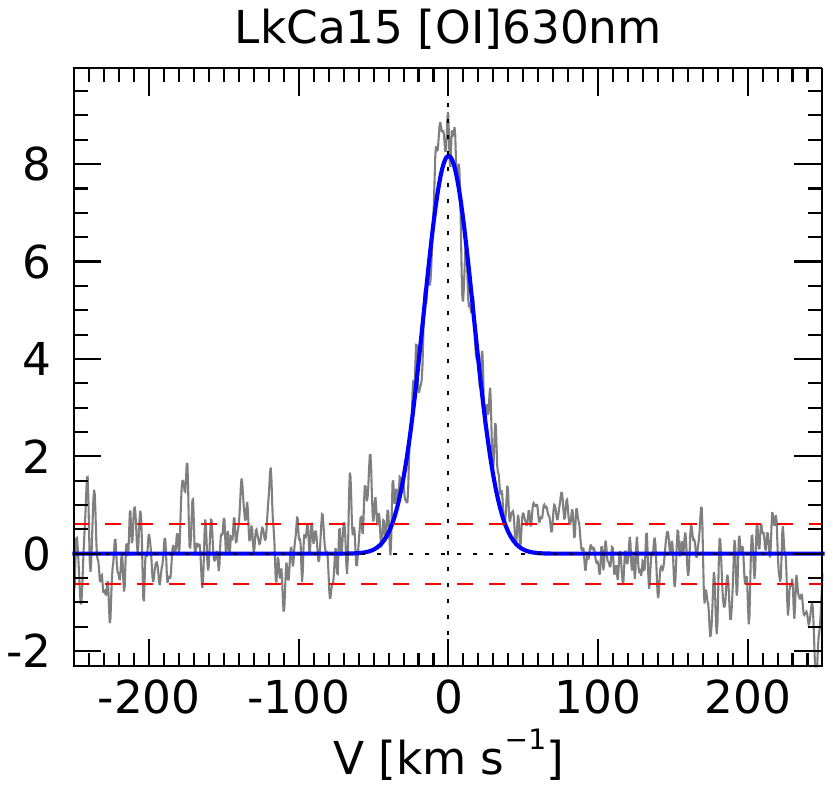}
\includegraphics[trim=80 0 80 400,width=0.2\textwidth]{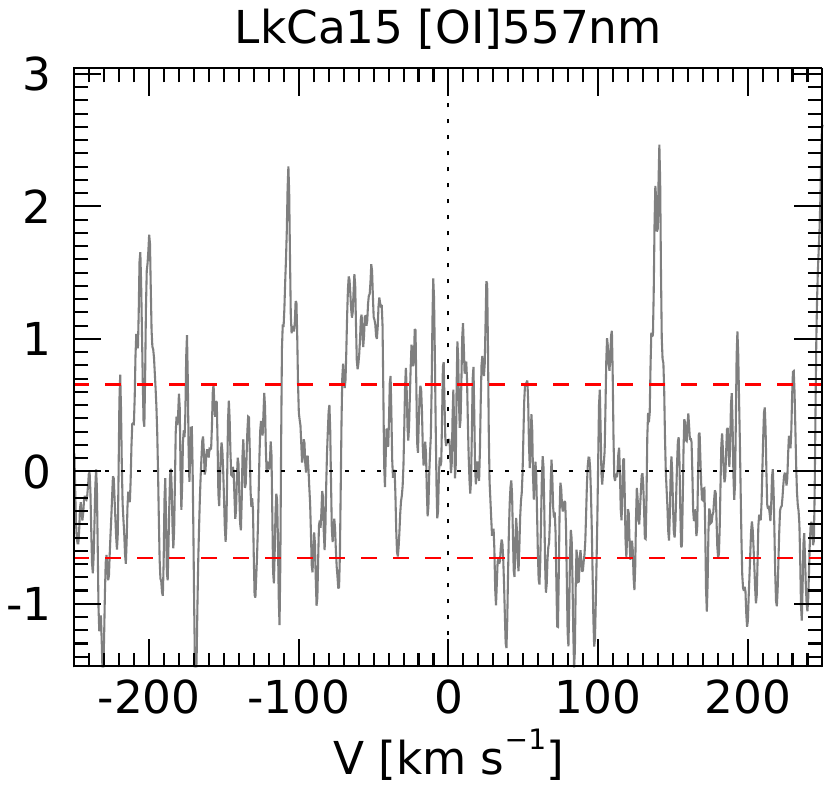}
\includegraphics[trim=80 0 80 400,width=0.2\textwidth]{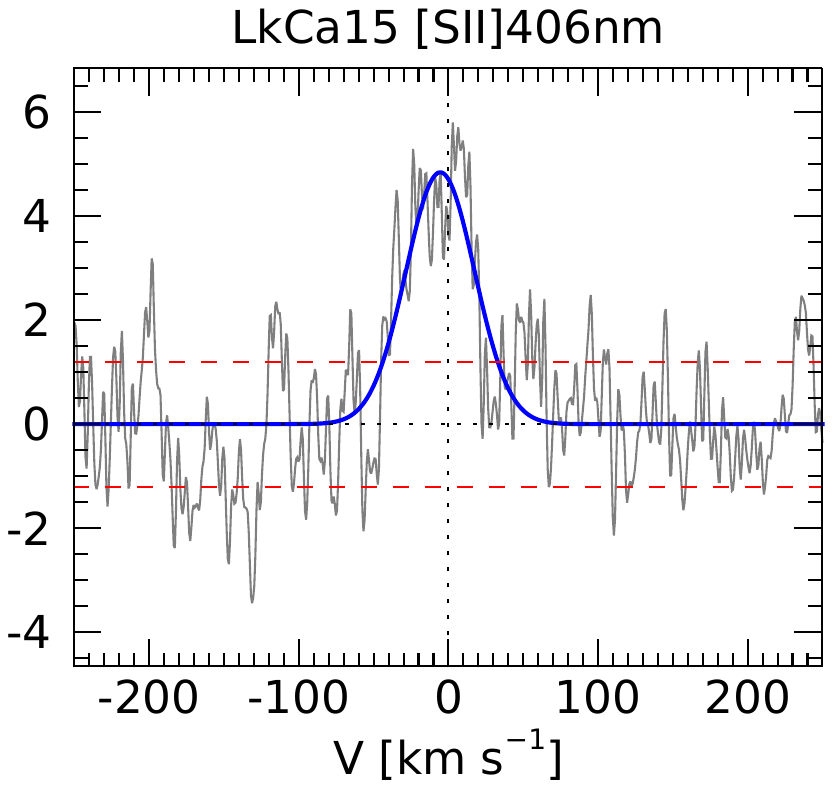}
\includegraphics[trim=80 0 80 400,width=0.2\textwidth]{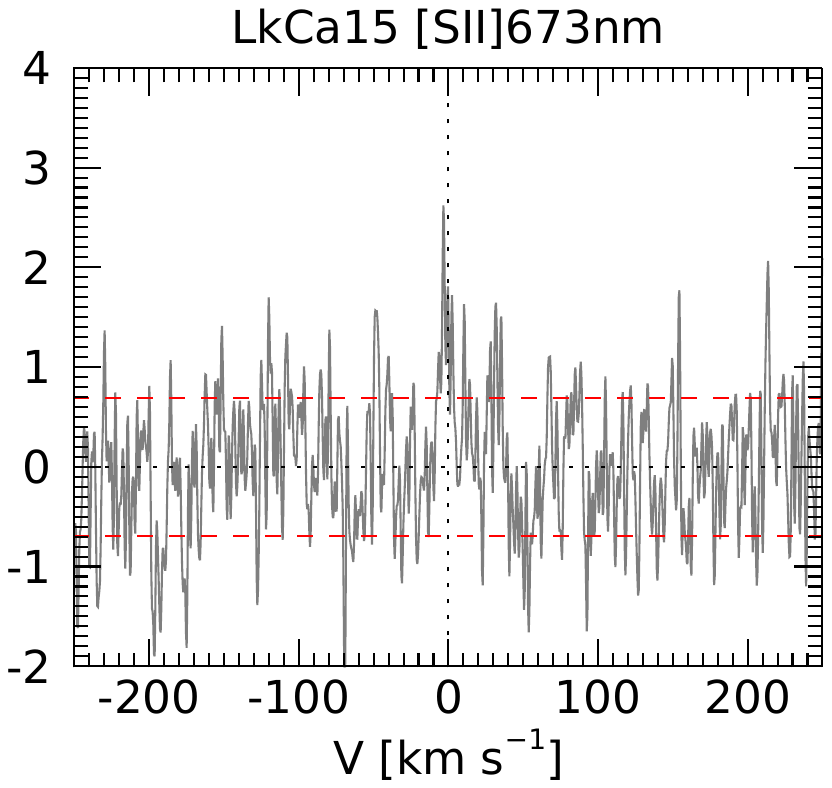}
\includegraphics[trim=80 0 80 400,width=0.2\textwidth]{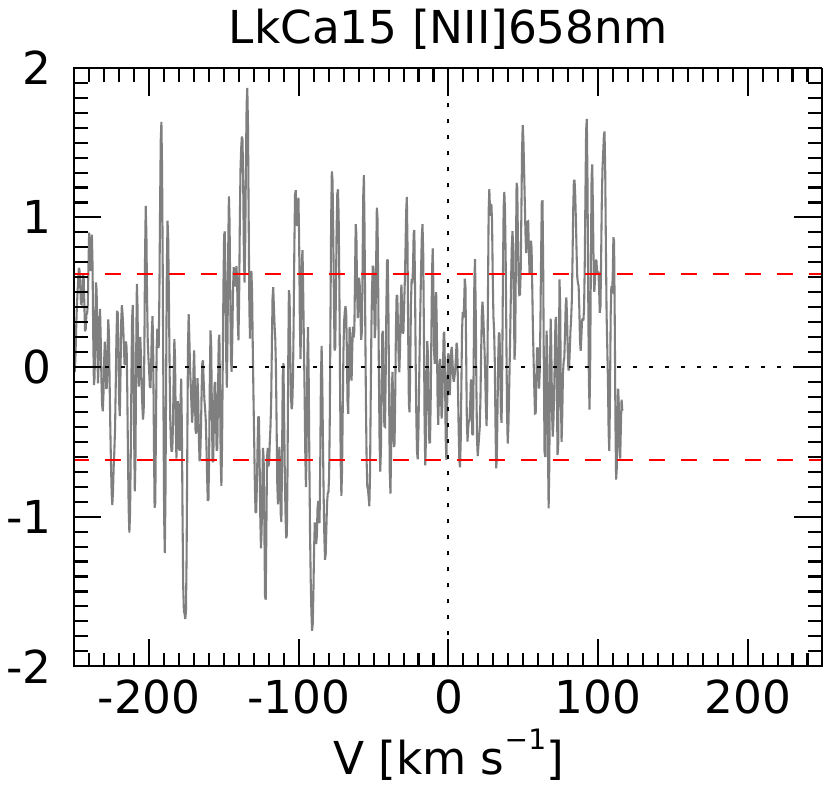}

\includegraphics[trim=80 0 80 400,width=0.2\textwidth]{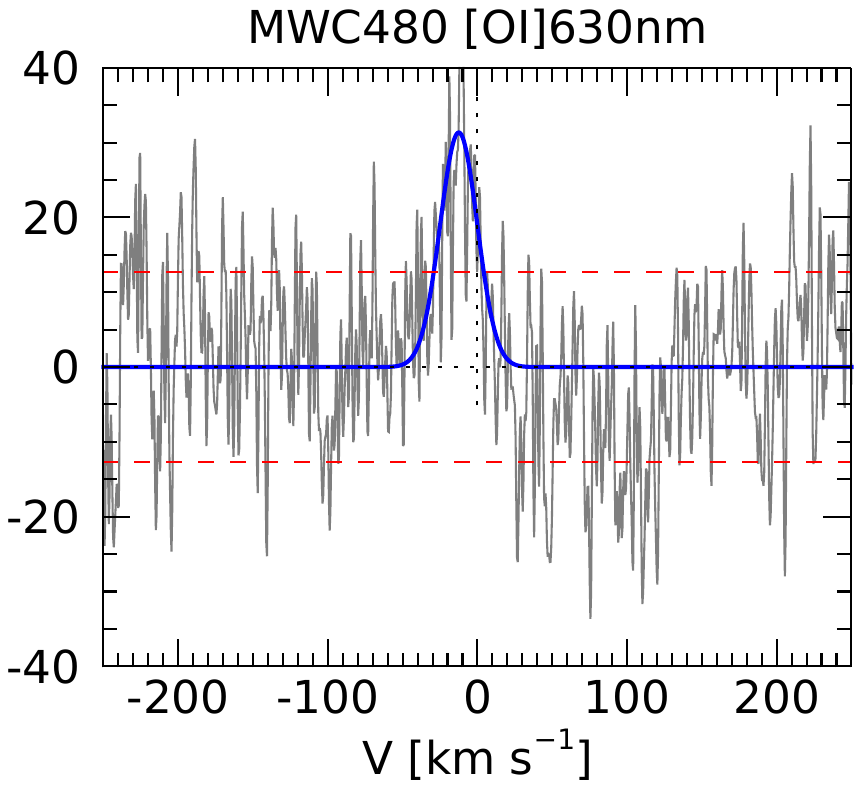}
\includegraphics[trim=80 0 80 400,width=0.2\textwidth]{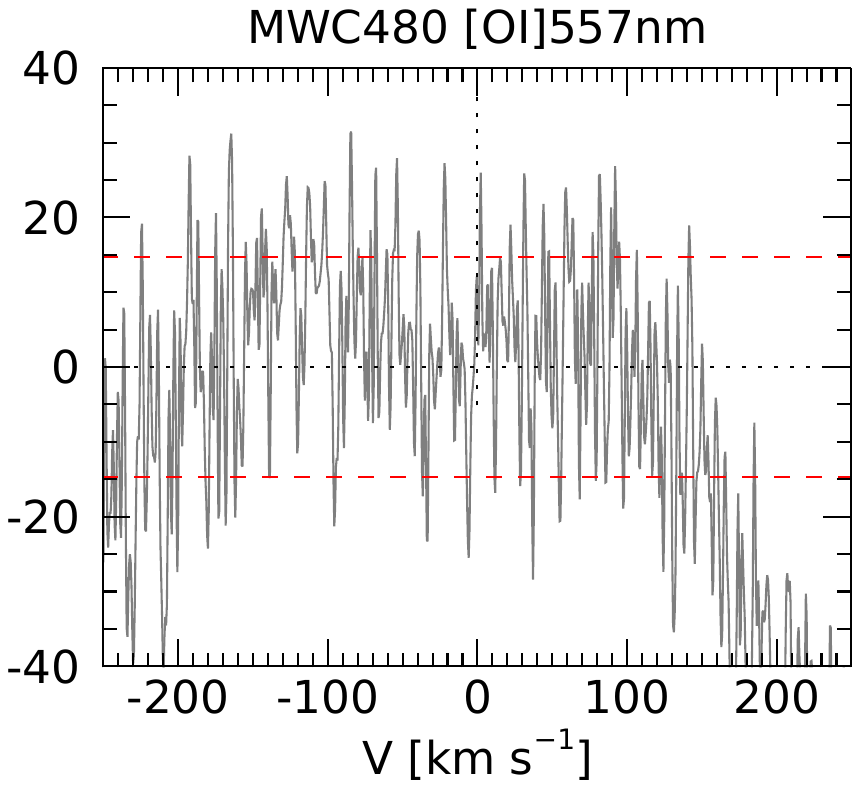}
\includegraphics[trim=80 0 80 400,width=0.2\textwidth]{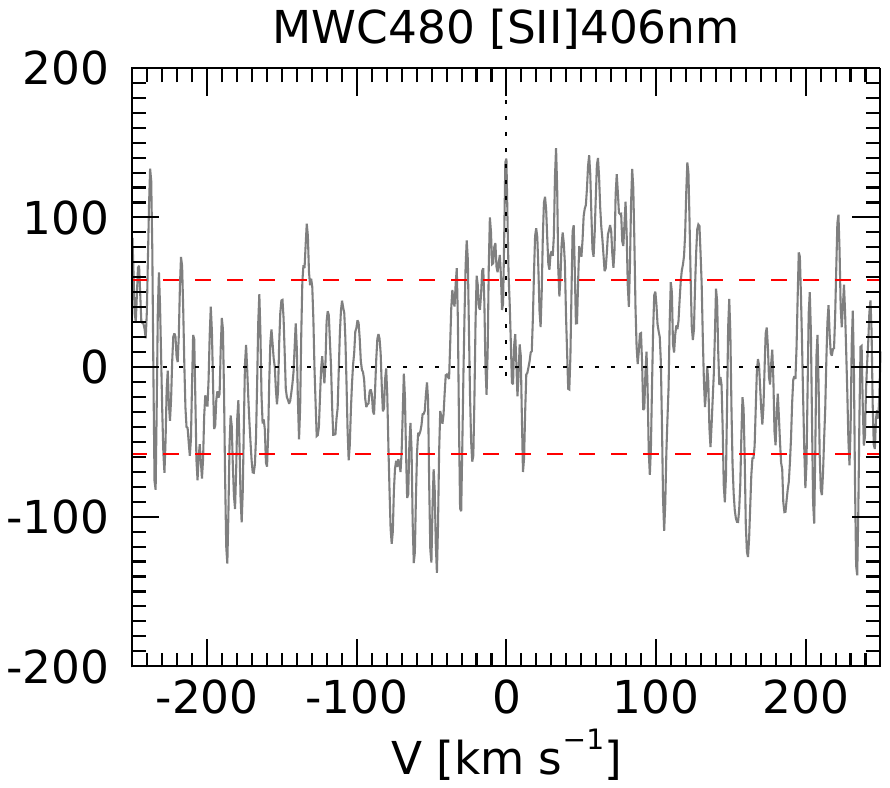}
\includegraphics[trim=80 0 80 400,width=0.2\textwidth]{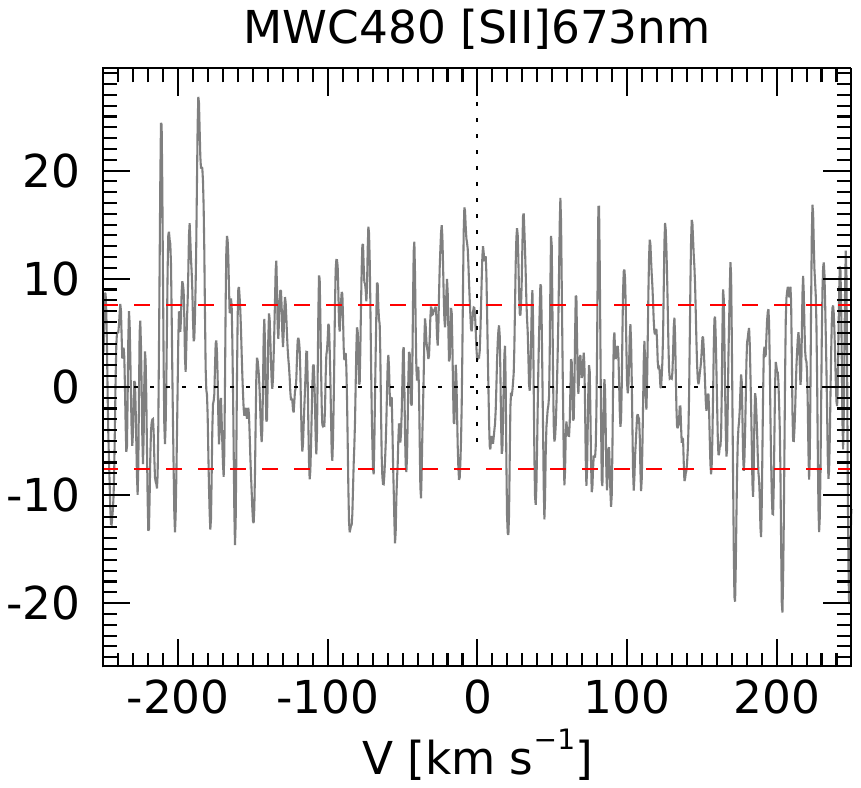}
\includegraphics[trim=80 0 80 400,width=0.2\textwidth]{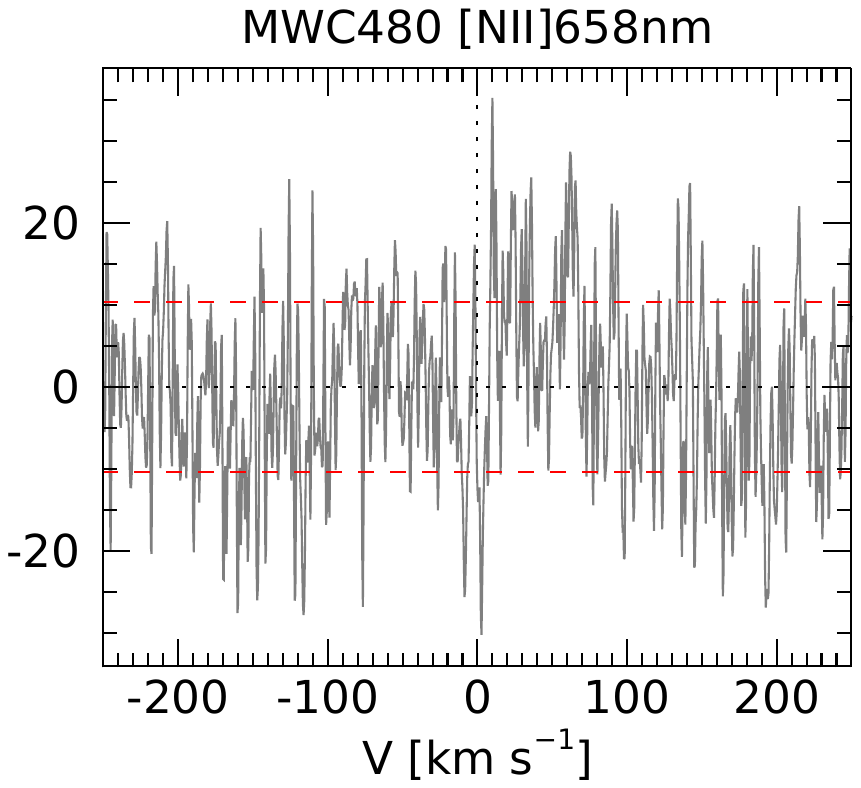}

\includegraphics[trim=80 0 80 400,width=0.2\textwidth]{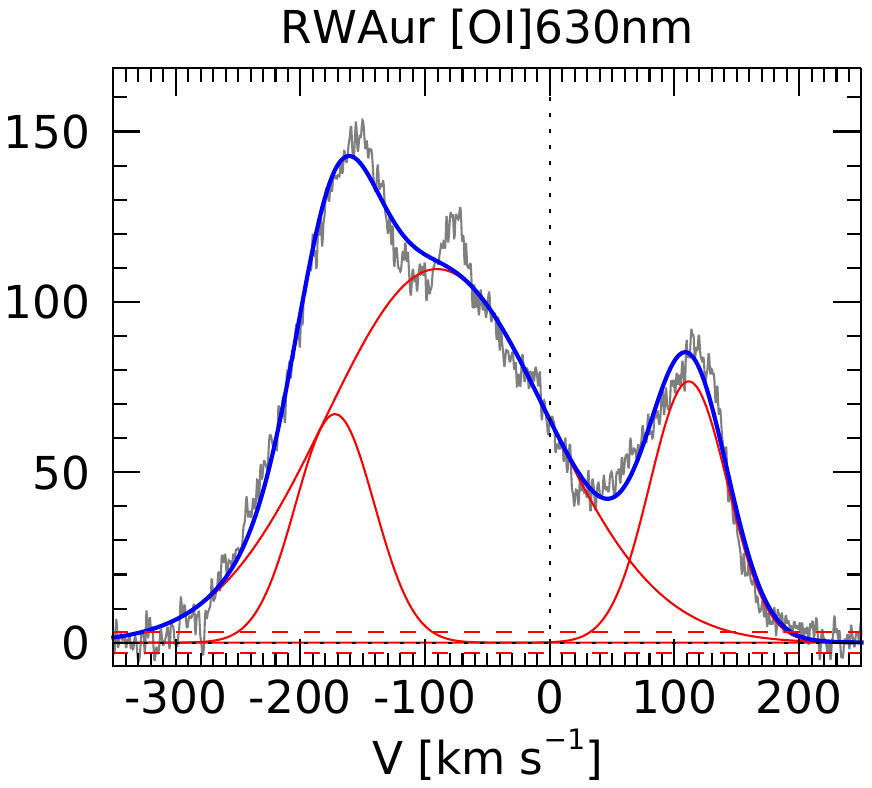}
\includegraphics[trim=80 0 80 400,width=0.2\textwidth]{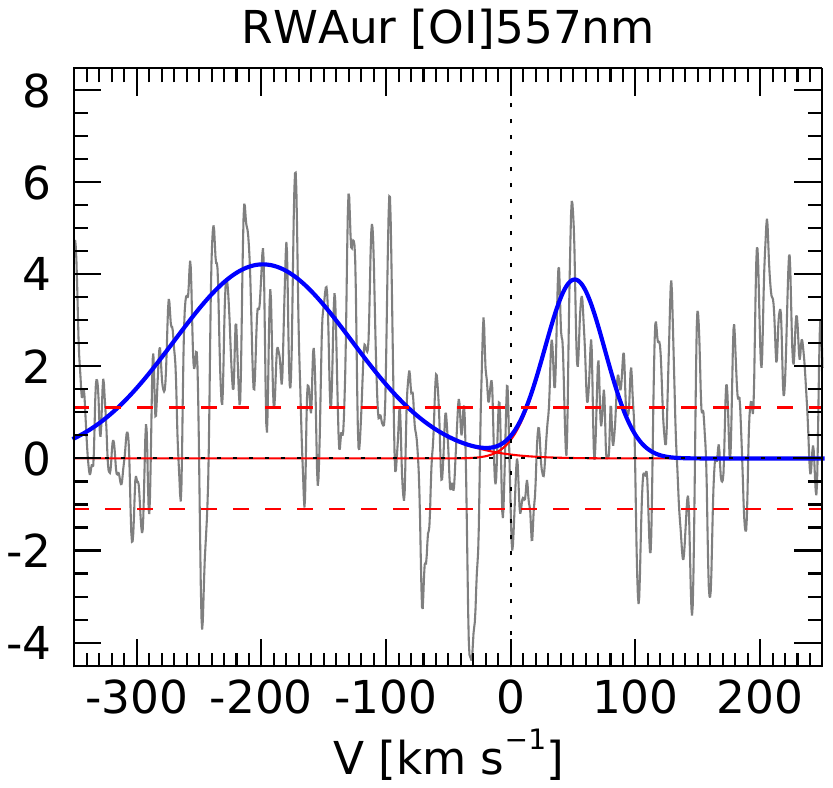}
\includegraphics[trim=80 0 80 400,width=0.2\textwidth]{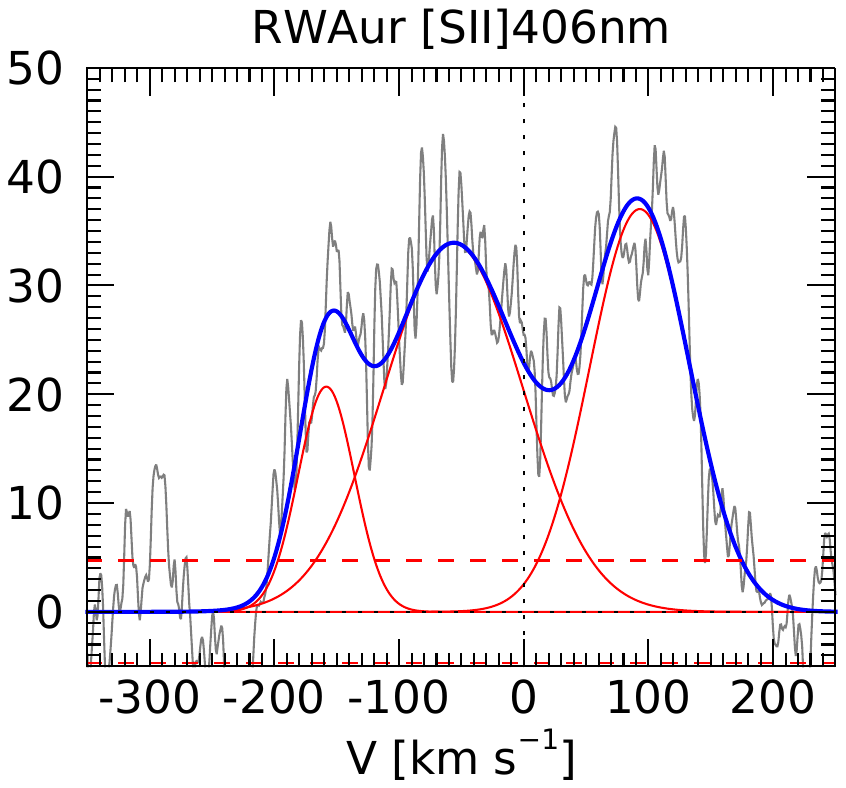}
\includegraphics[trim=80 0 80 400,width=0.2\textwidth]{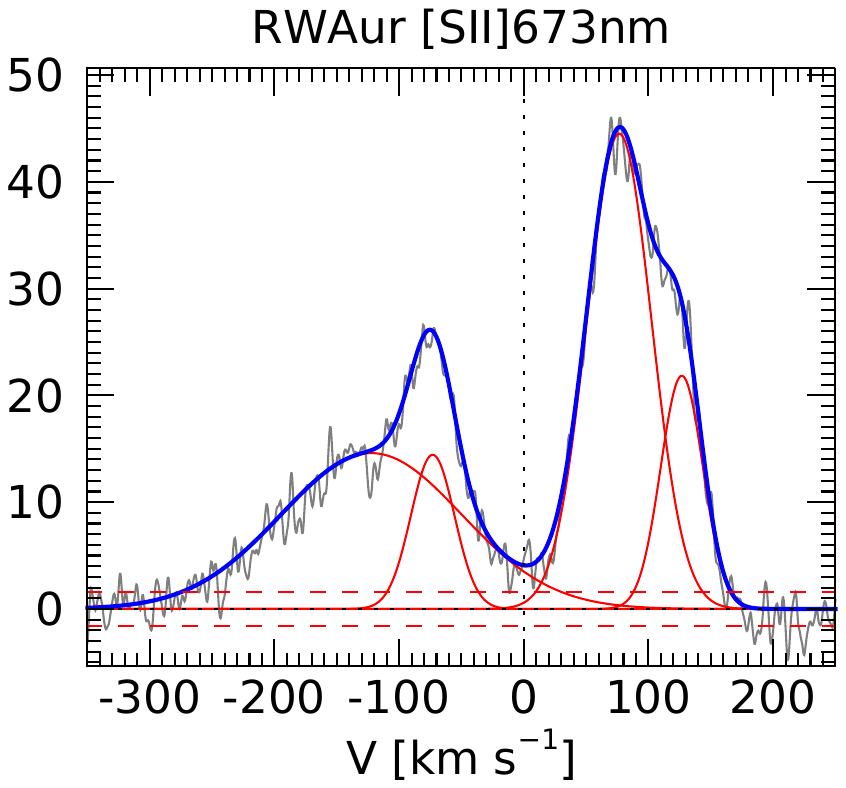}
\includegraphics[trim=80 0 80 400,width=0.2\textwidth]{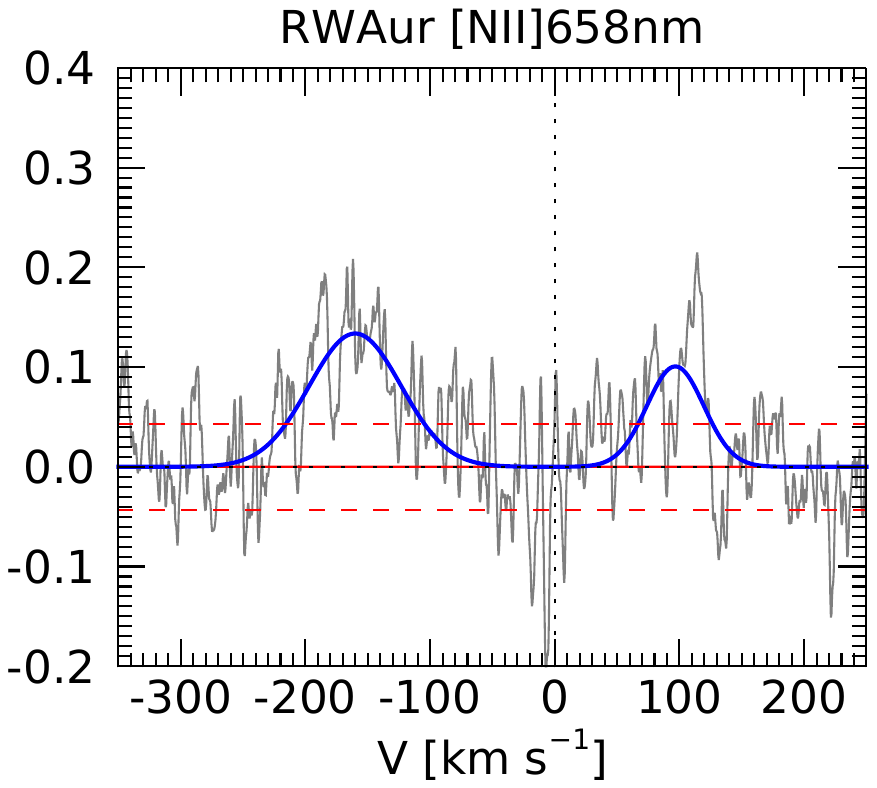}

\includegraphics[trim=80 0 80 400,width=0.2\textwidth]{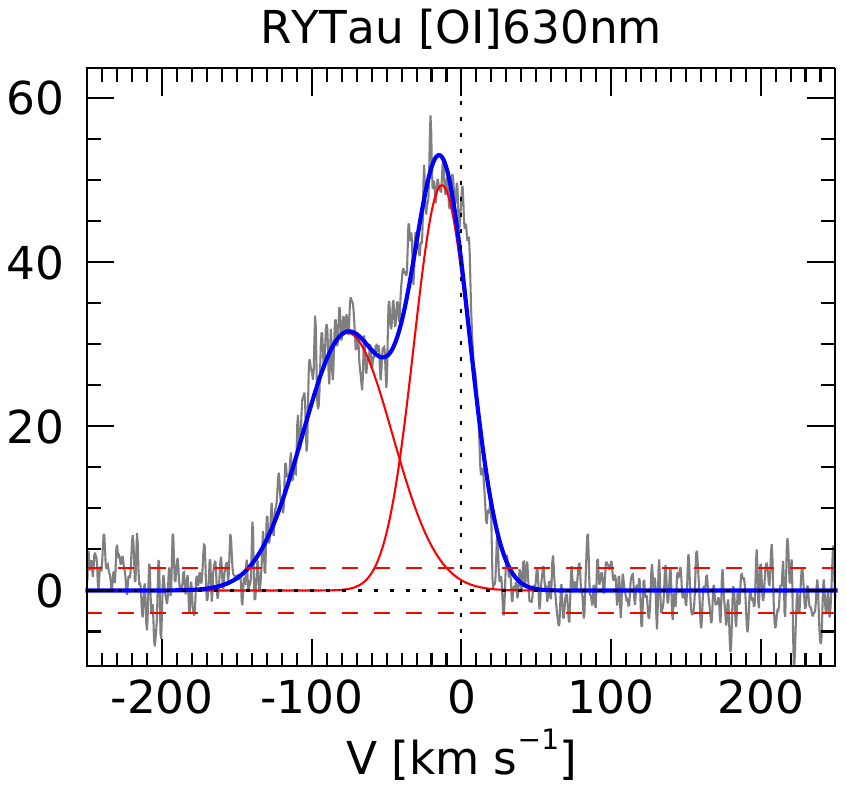}
\includegraphics[trim=80 0 80 400,width=0.2\textwidth]{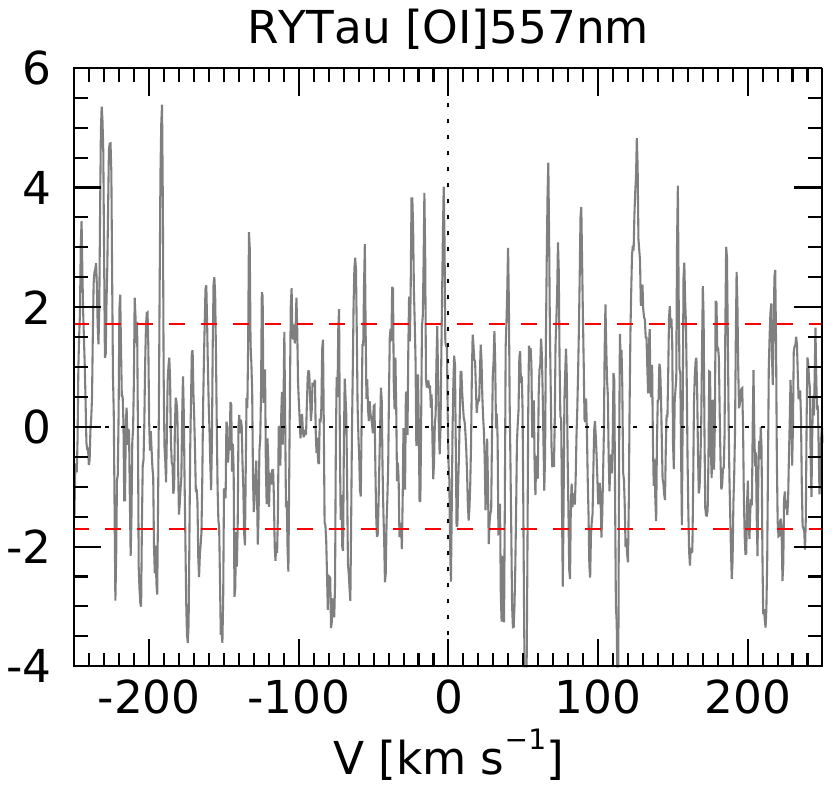}
\includegraphics[trim=80 0 80 400,width=0.2\textwidth]{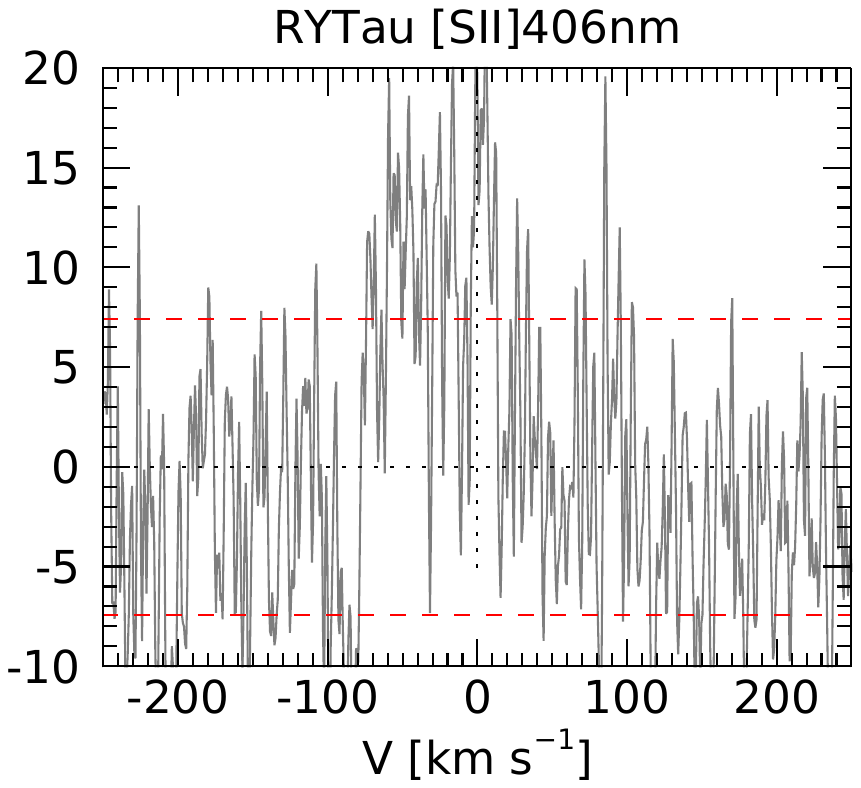}
\includegraphics[trim=80 0 80 400,width=0.2\textwidth]{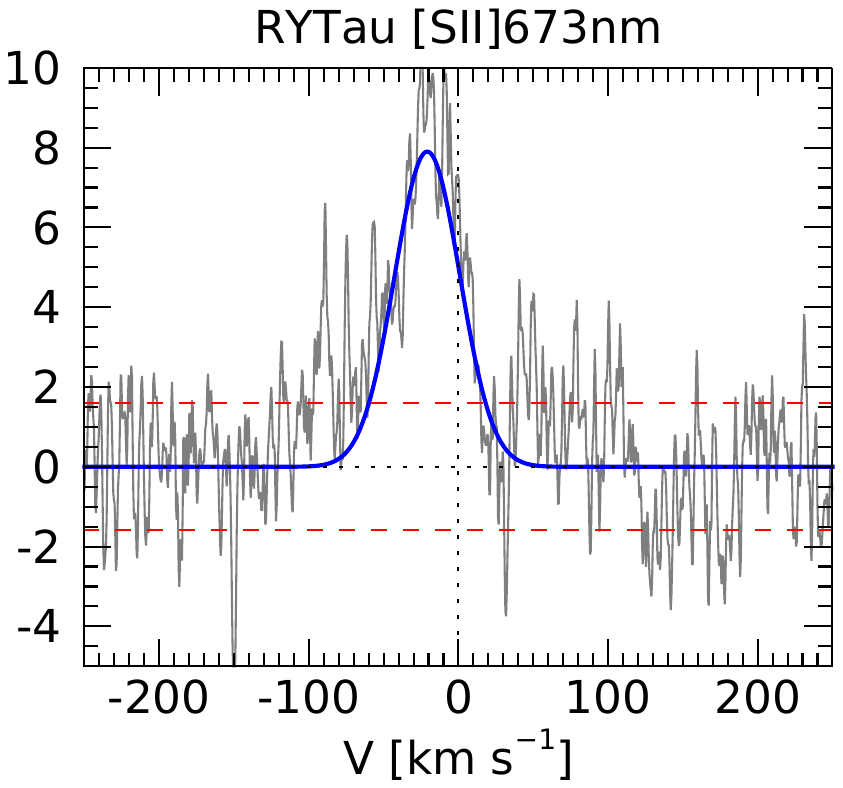}
\includegraphics[trim=80 0 80 400,width=0.2\textwidth]{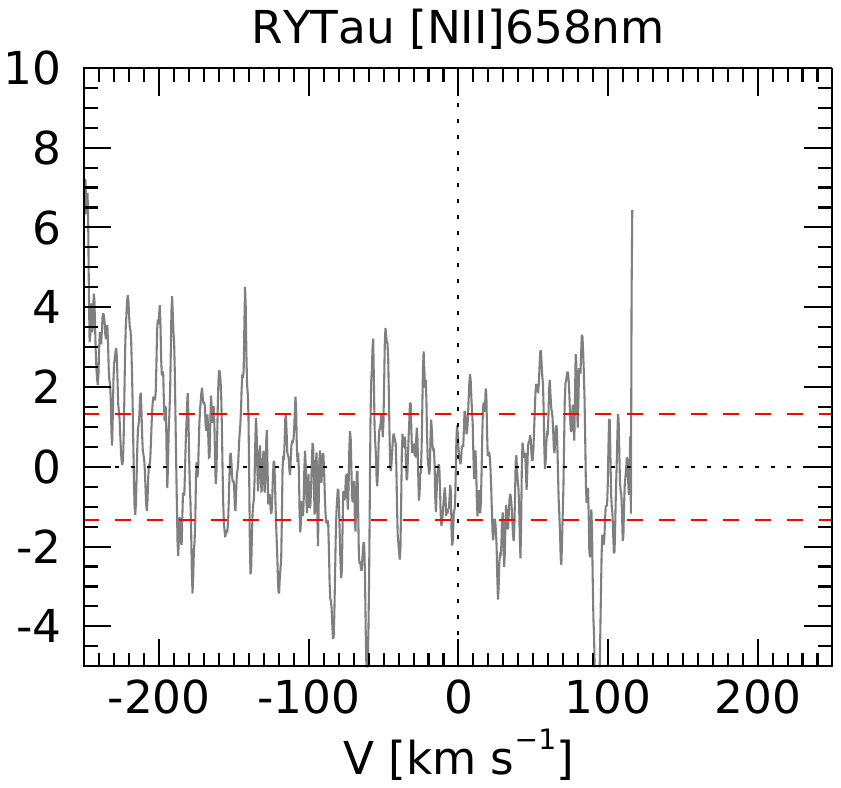}

   \caption{Continued}
   \label{fig:profiles5}
  
\end{figure*}

\newpage

\begin{figure*}[h]

\includegraphics[trim=80 0 80 0,width=0.2\textwidth]{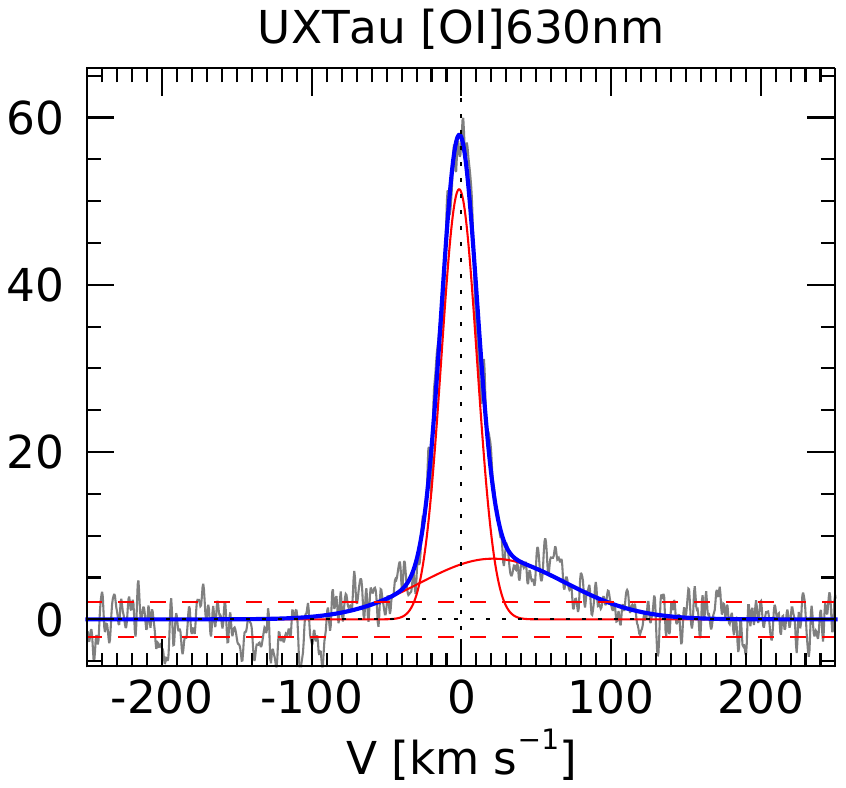}
\includegraphics[trim=80 0 80 0,width=0.2\textwidth]{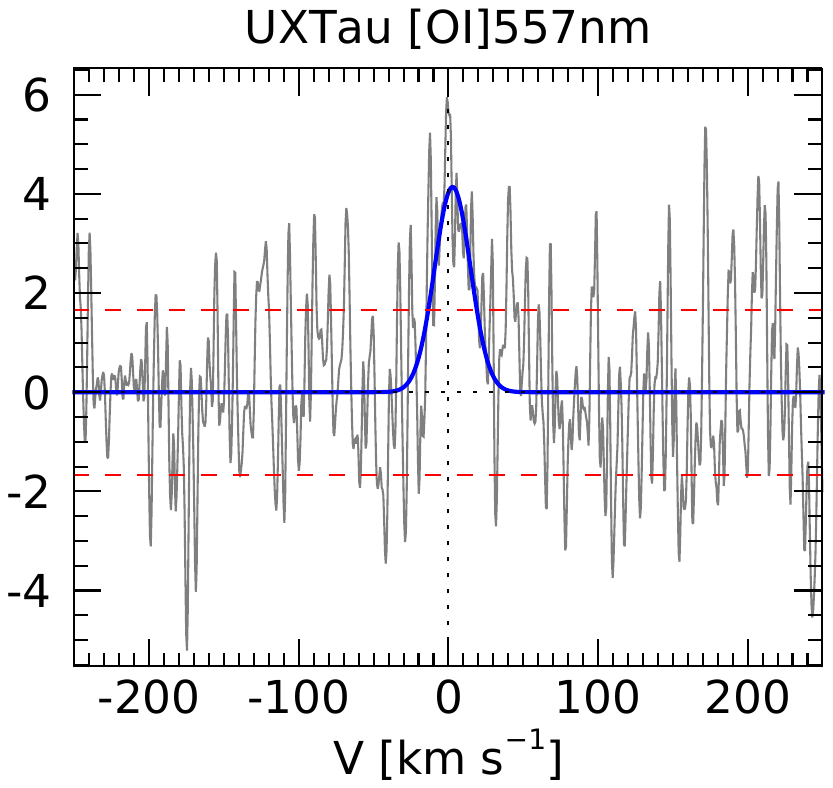}
\includegraphics[trim=80 0 80 0,width=0.2\textwidth]{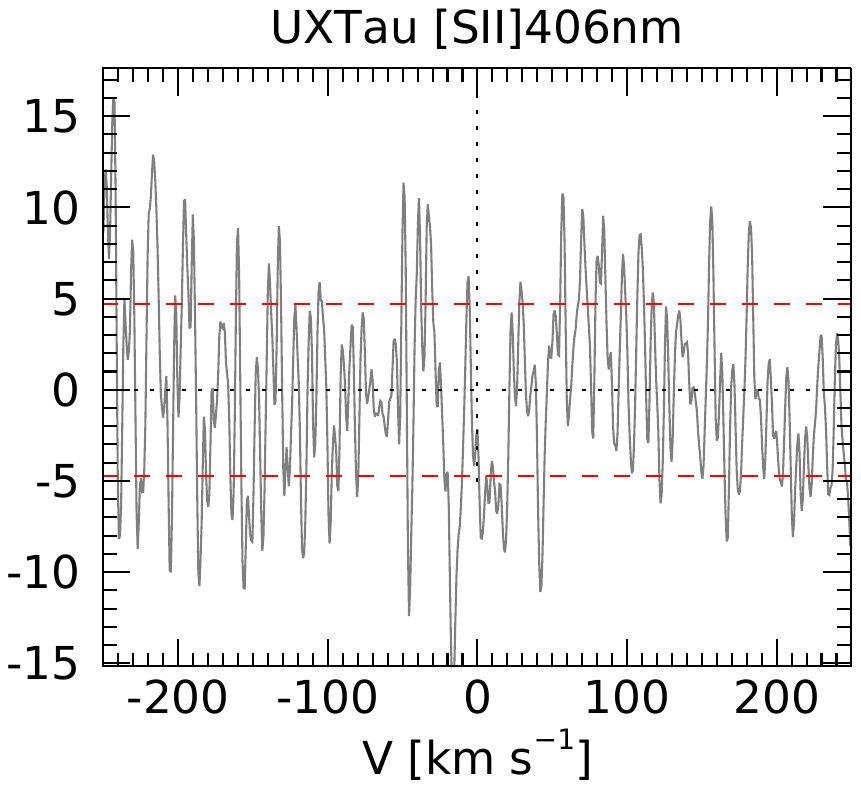}
\includegraphics[trim=80 0 80 0,width=0.2\textwidth]{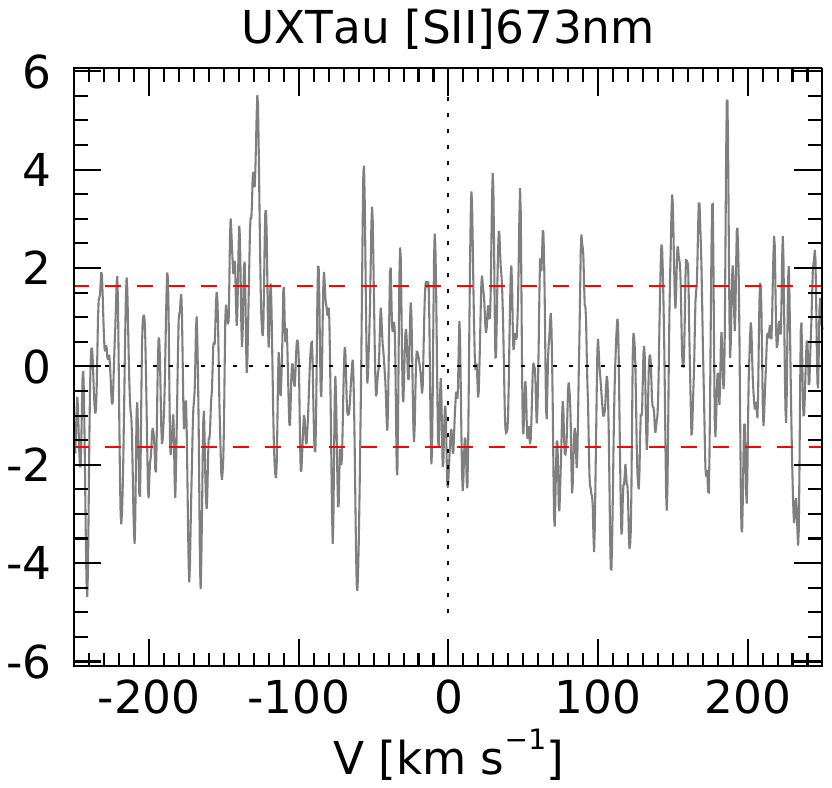}
\includegraphics[trim=80 0 80 0,width=0.2\textwidth]{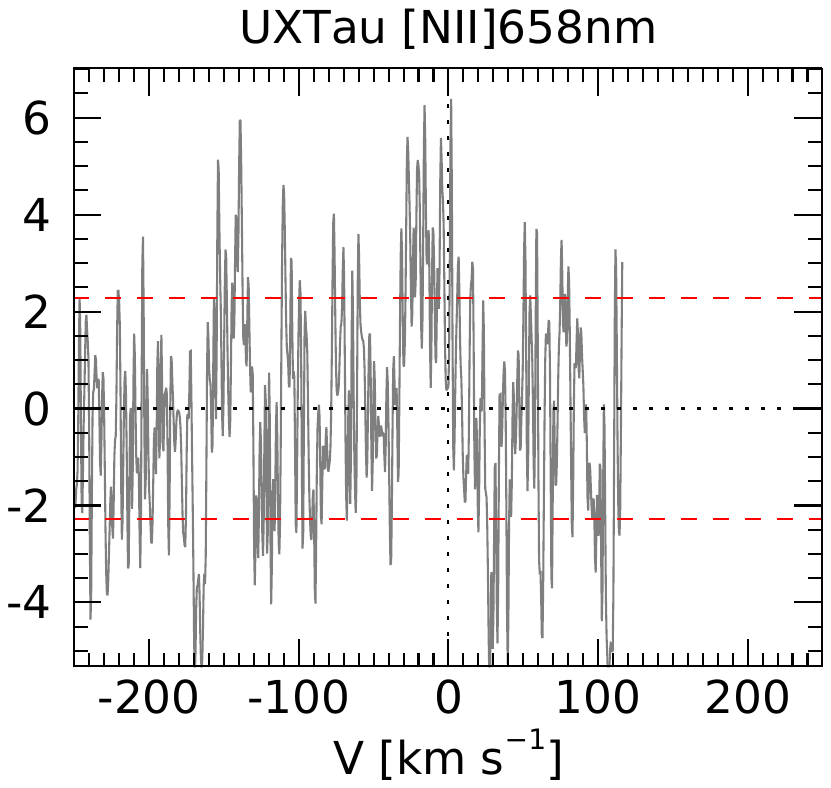}

\includegraphics[trim=80 0 80 400,width=0.2\textwidth]{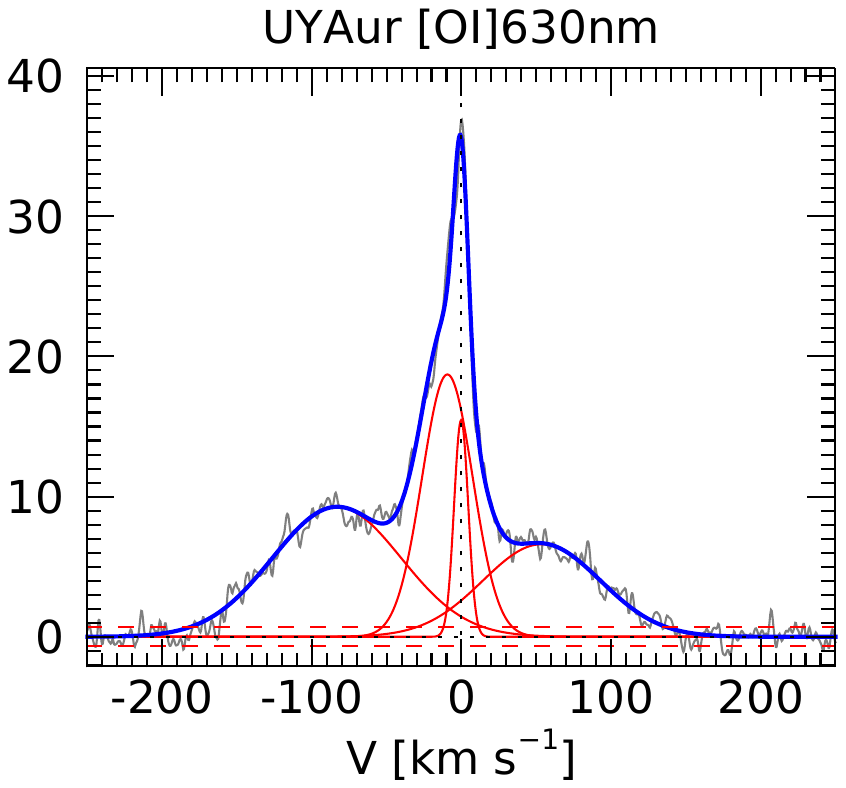}
\includegraphics[trim=80 0 80 400,width=0.2\textwidth]{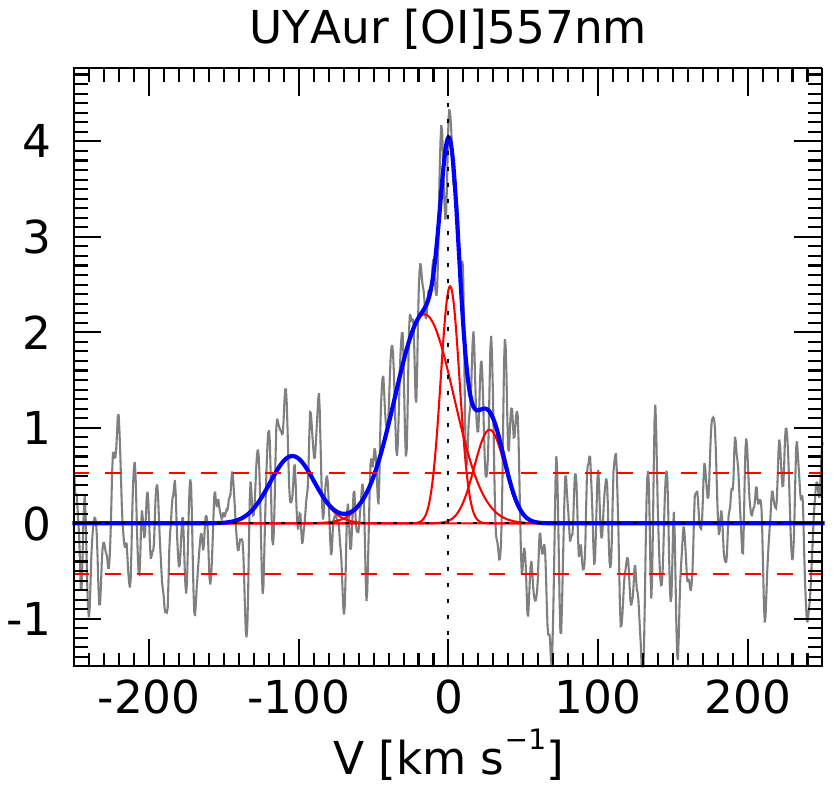}
\includegraphics[trim=80 0 80 400,width=0.2\textwidth]{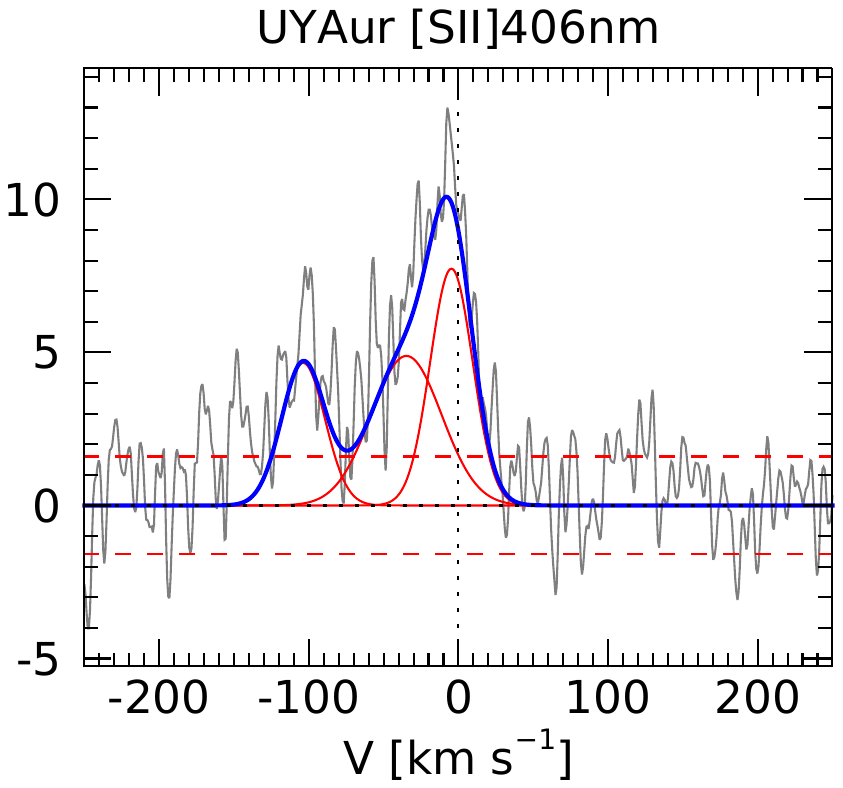}
\includegraphics[trim=80 0 80 400,width=0.2\textwidth]{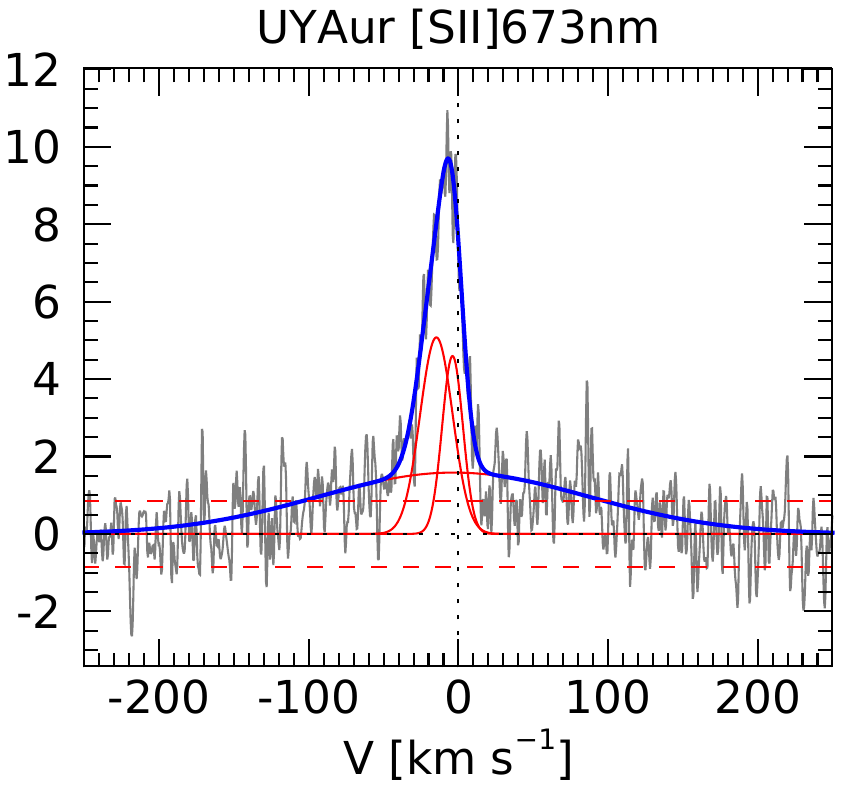}
\includegraphics[trim=80 0 80 400,width=0.2\textwidth]{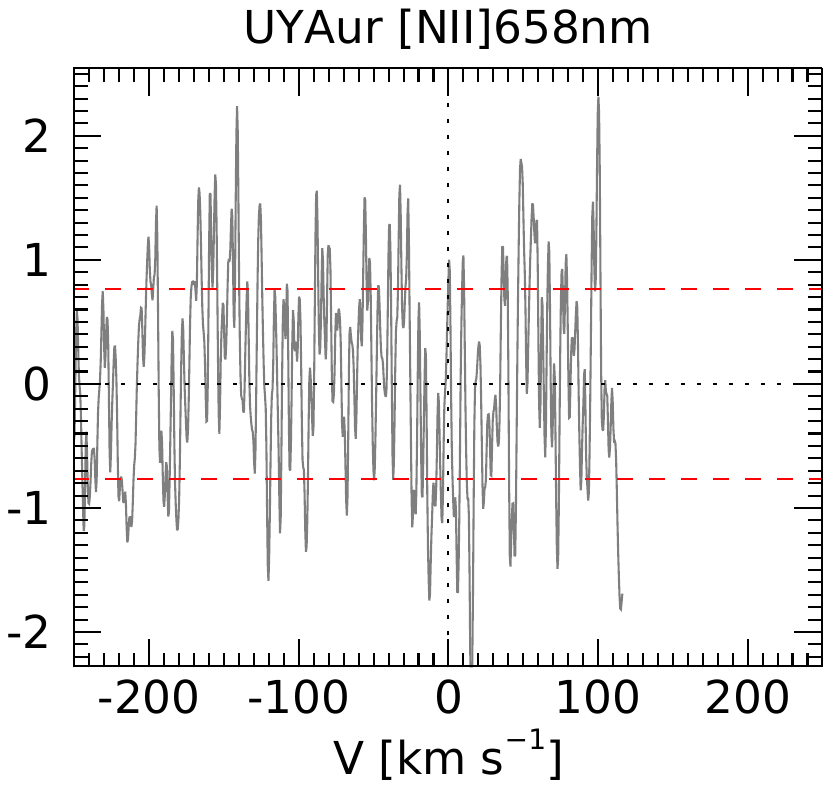}

\includegraphics[trim=80 0 80 400,width=0.2\textwidth]{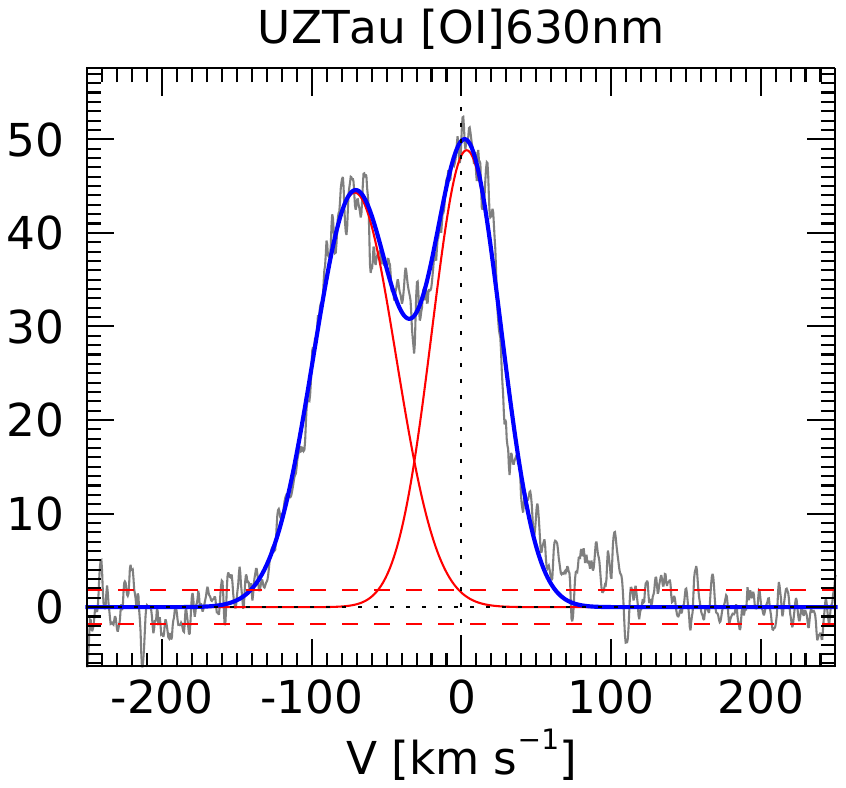}
\includegraphics[trim=80 0 80 400,width=0.2\textwidth]{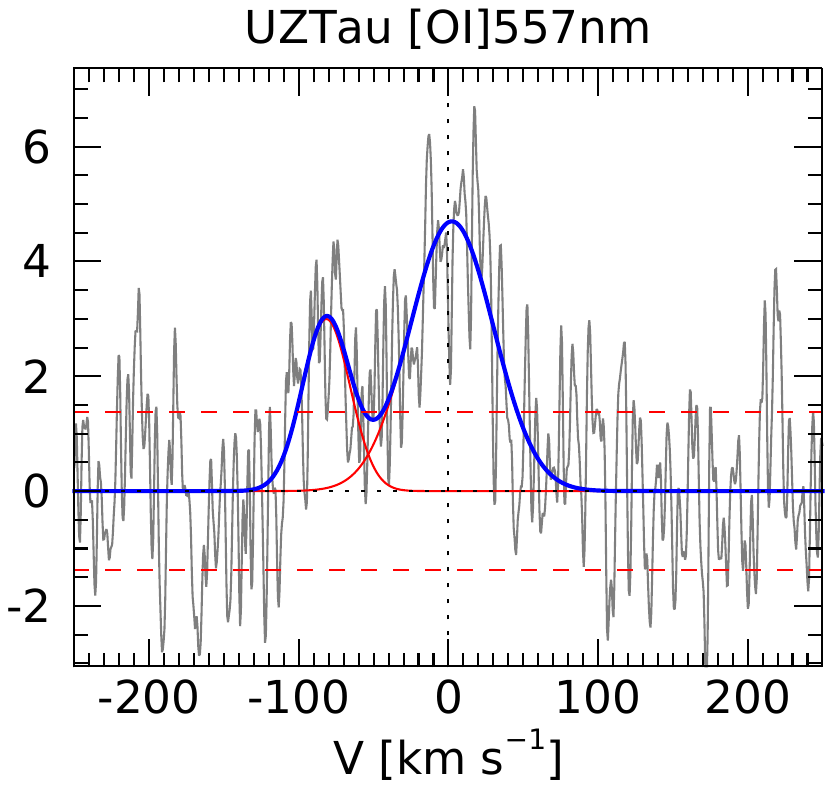}
\includegraphics[trim=80 0 80 400,width=0.2\textwidth]{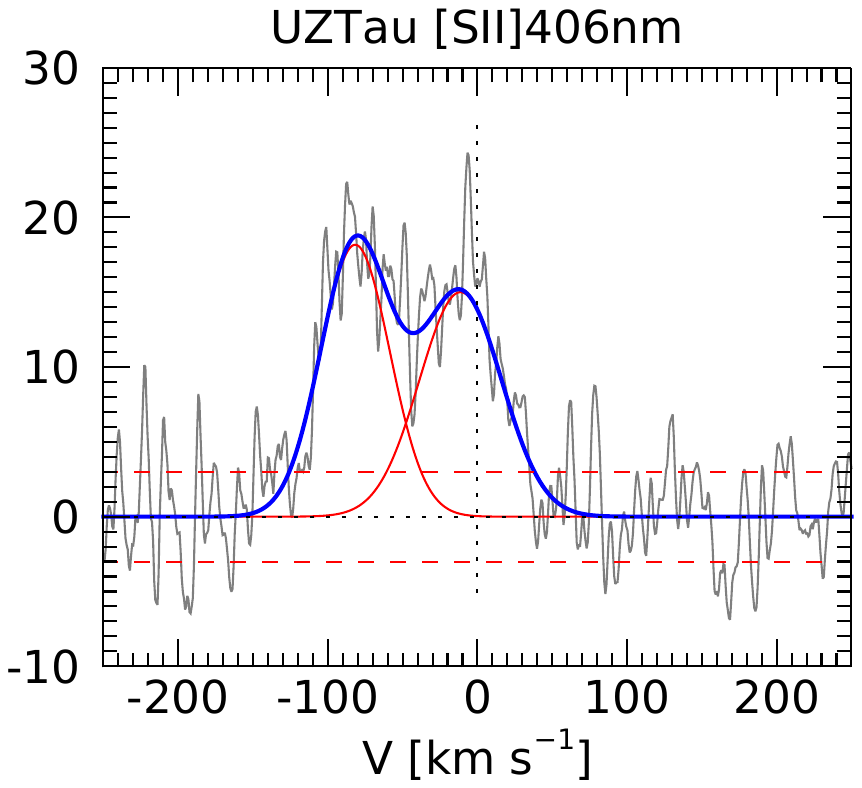}
\includegraphics[trim=80 0 80 400,width=0.2\textwidth]{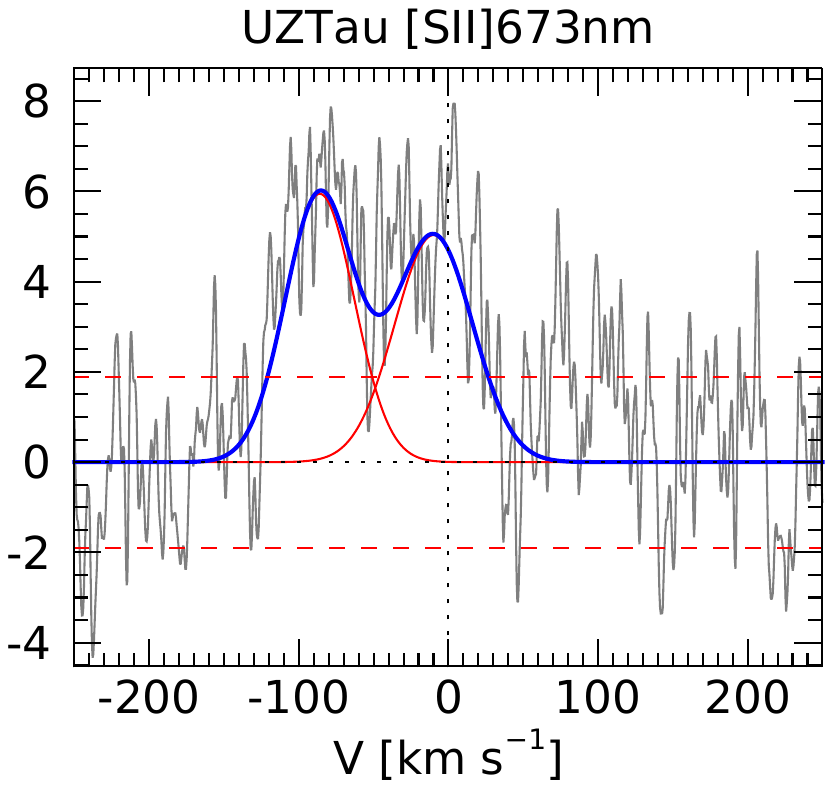}
\includegraphics[trim=80 0 80 400,width=0.2\textwidth]{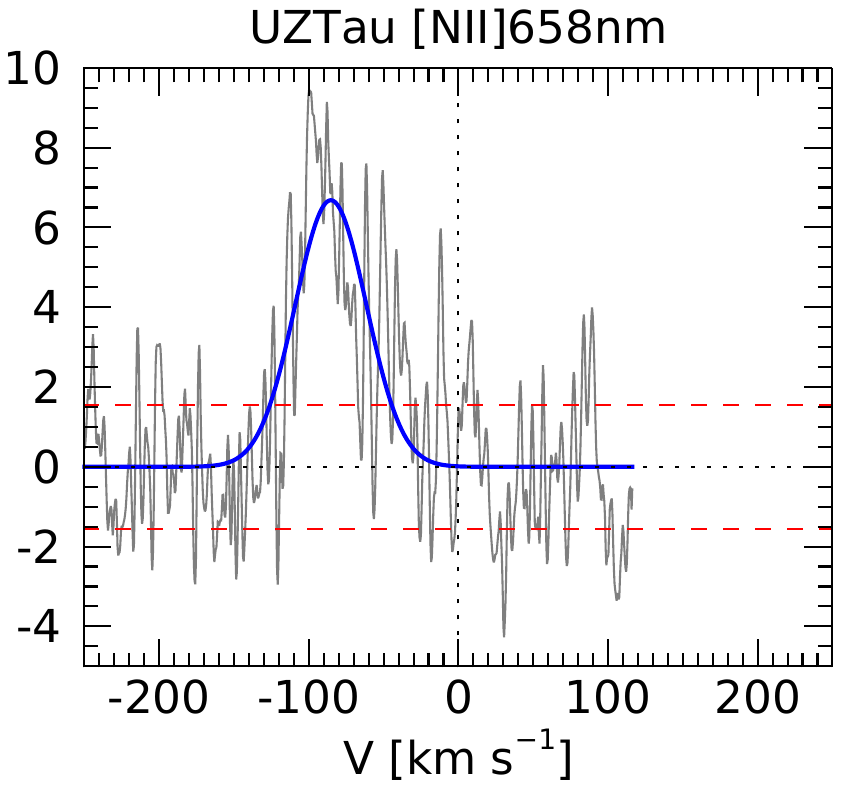}

\includegraphics[trim=80 0 80 400,width=0.2\textwidth]{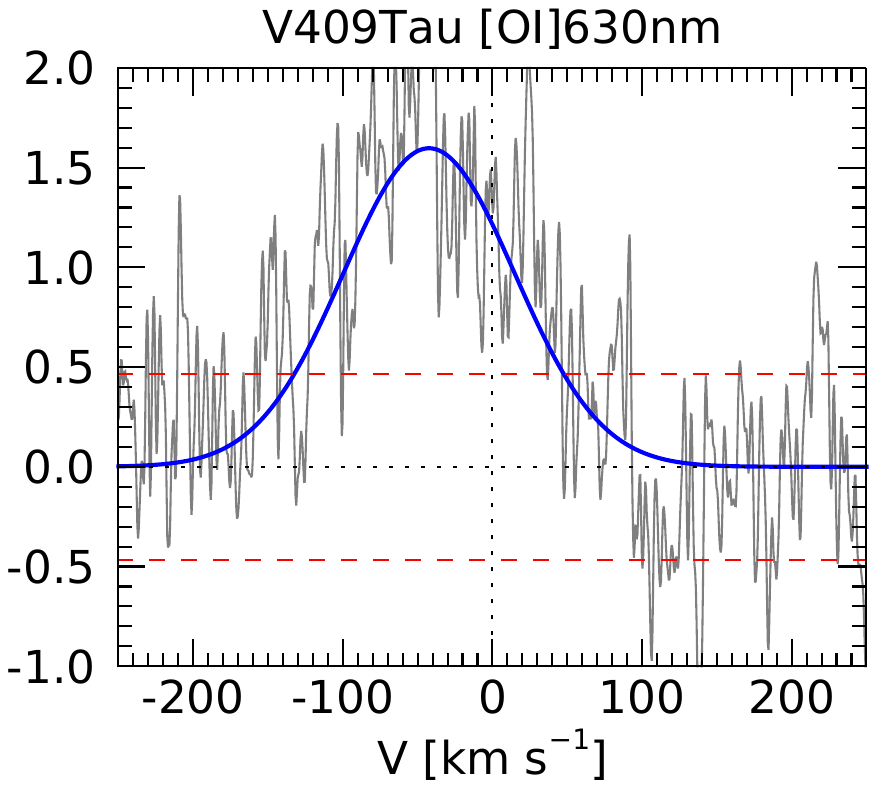}
\includegraphics[trim=80 0 80 400,width=0.2\textwidth]{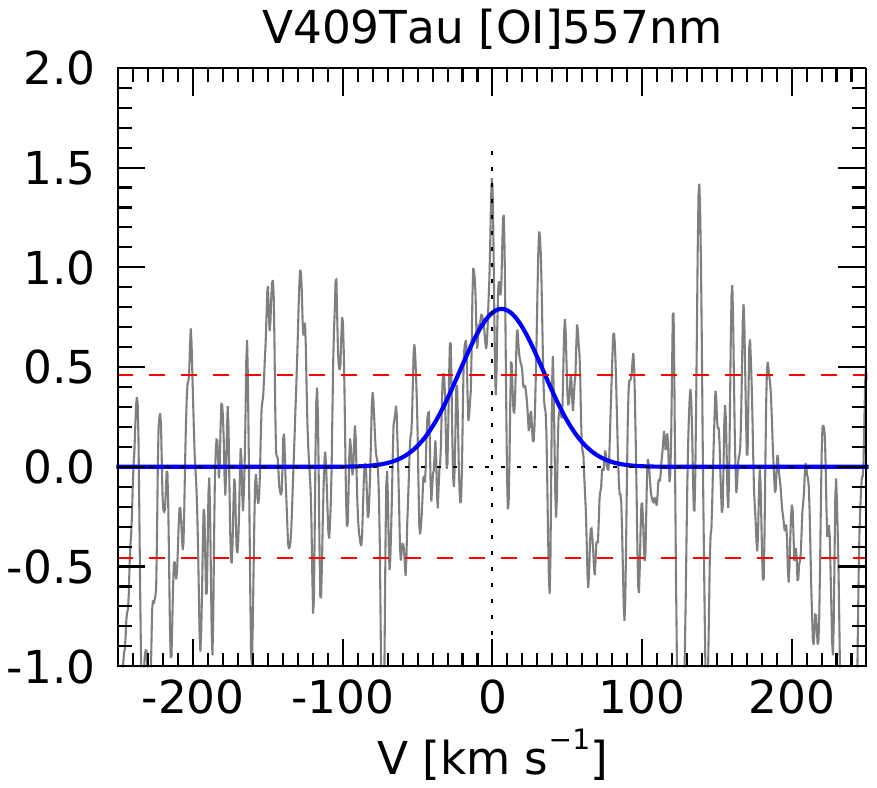}
\includegraphics[trim=80 0 80 400,width=0.2\textwidth]{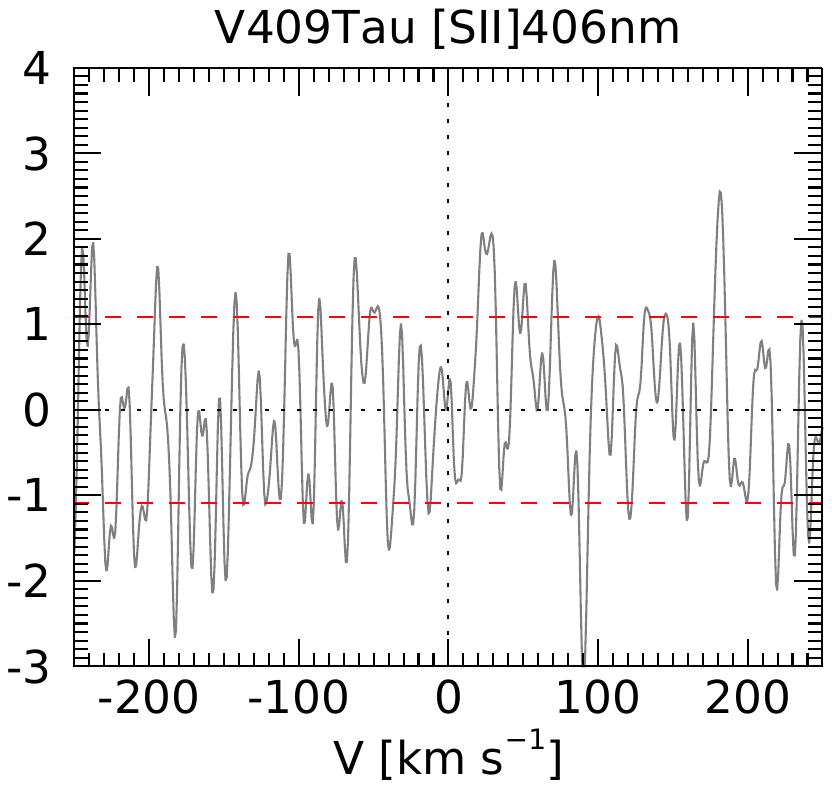}
\includegraphics[trim=80 0 80 400,width=0.2\textwidth]{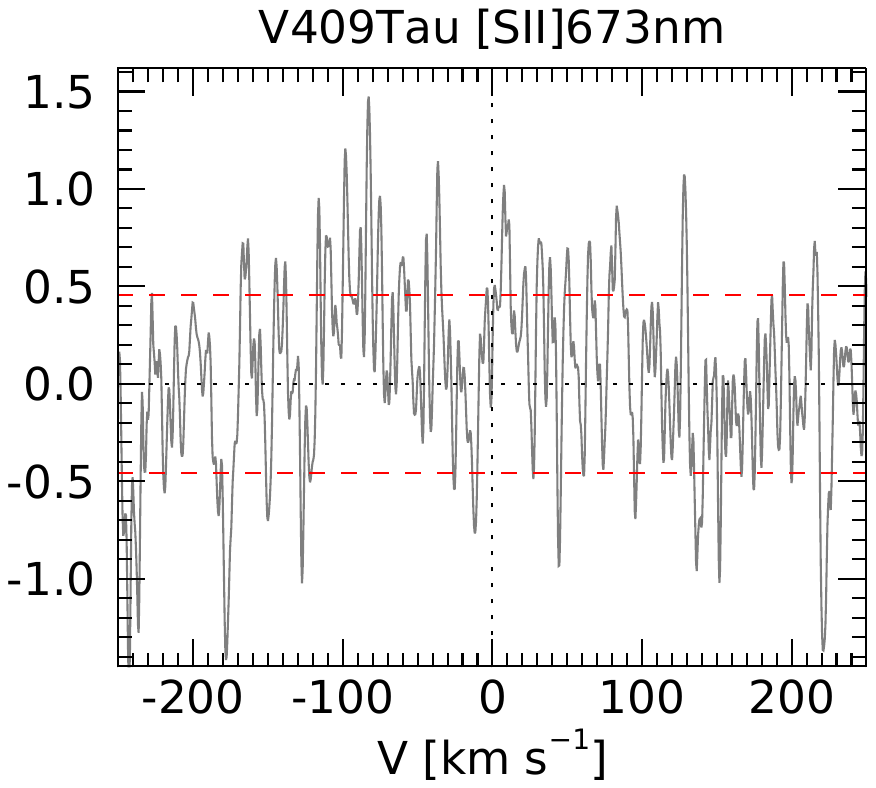}
\includegraphics[trim=80 0 80 400,width=0.2\textwidth]{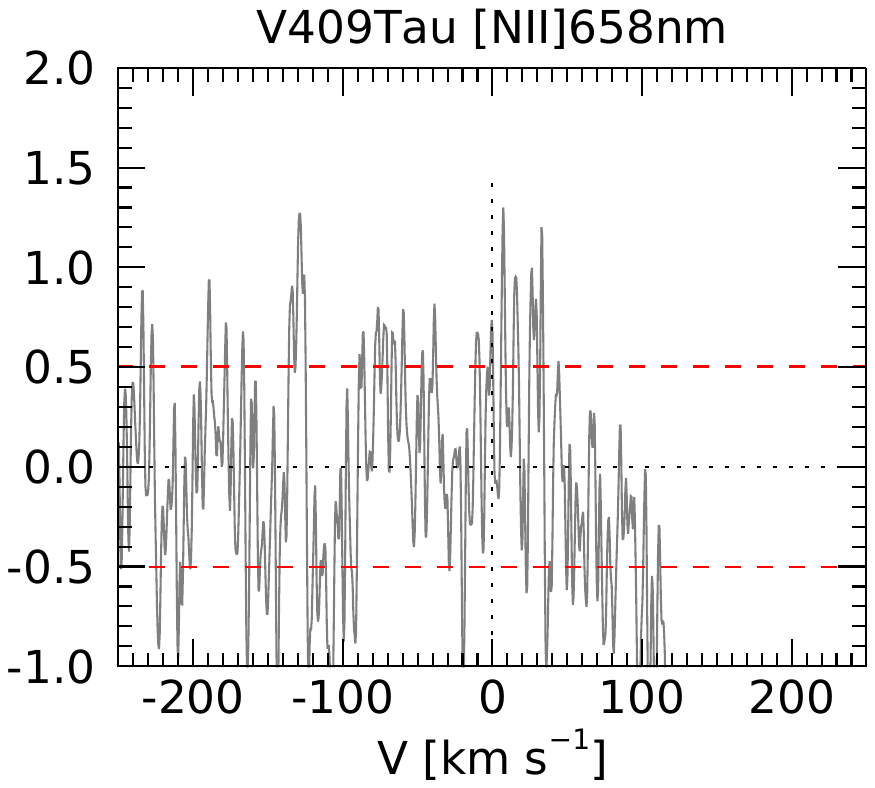}

\includegraphics[trim=80 0 80 400,width=0.2\textwidth]{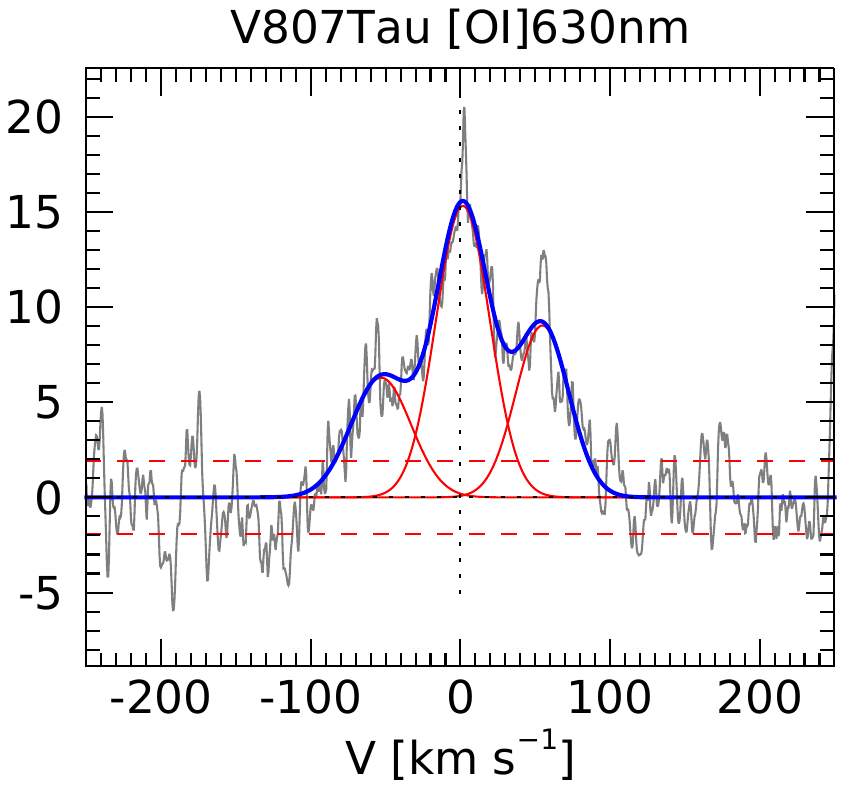}
\includegraphics[trim=80 0 80 400,width=0.2\textwidth]{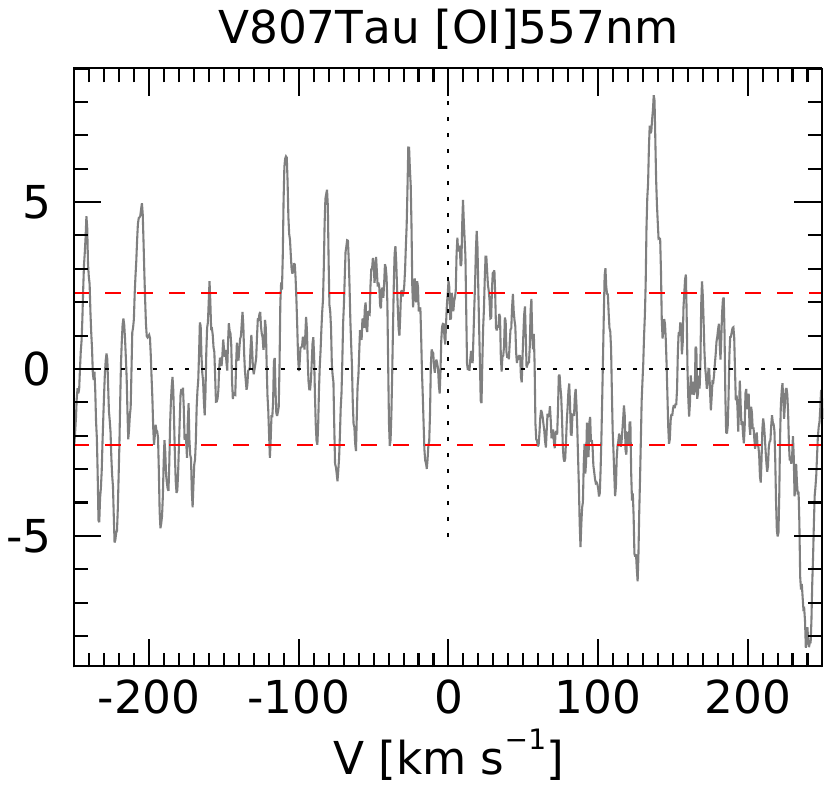}
\includegraphics[trim=80 0 80 400,width=0.2\textwidth]{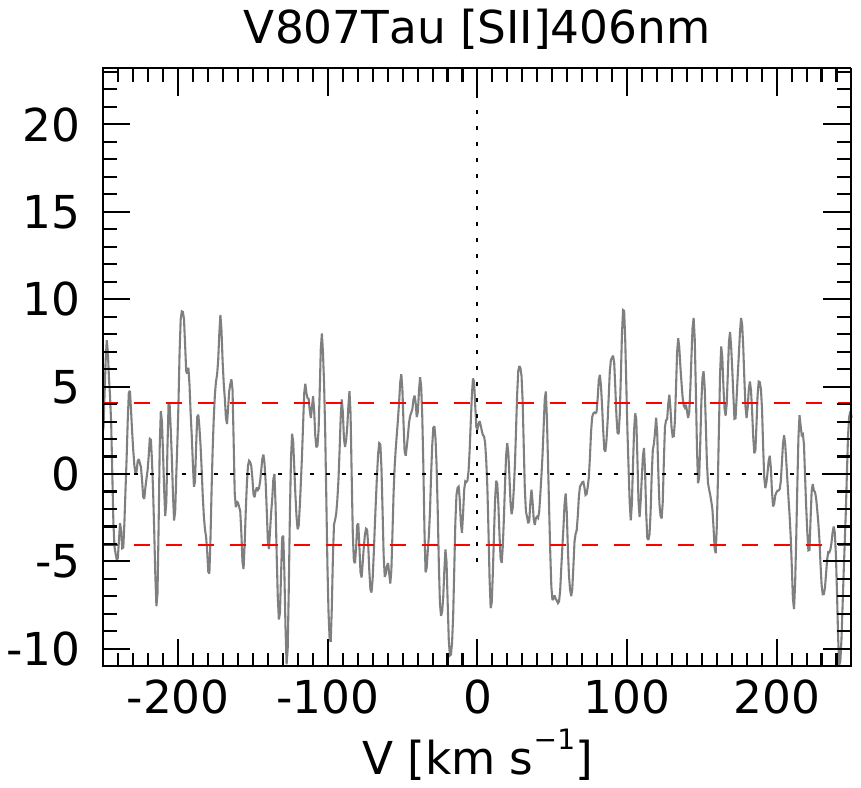}
\includegraphics[trim=80 0 80 400,width=0.2\textwidth]{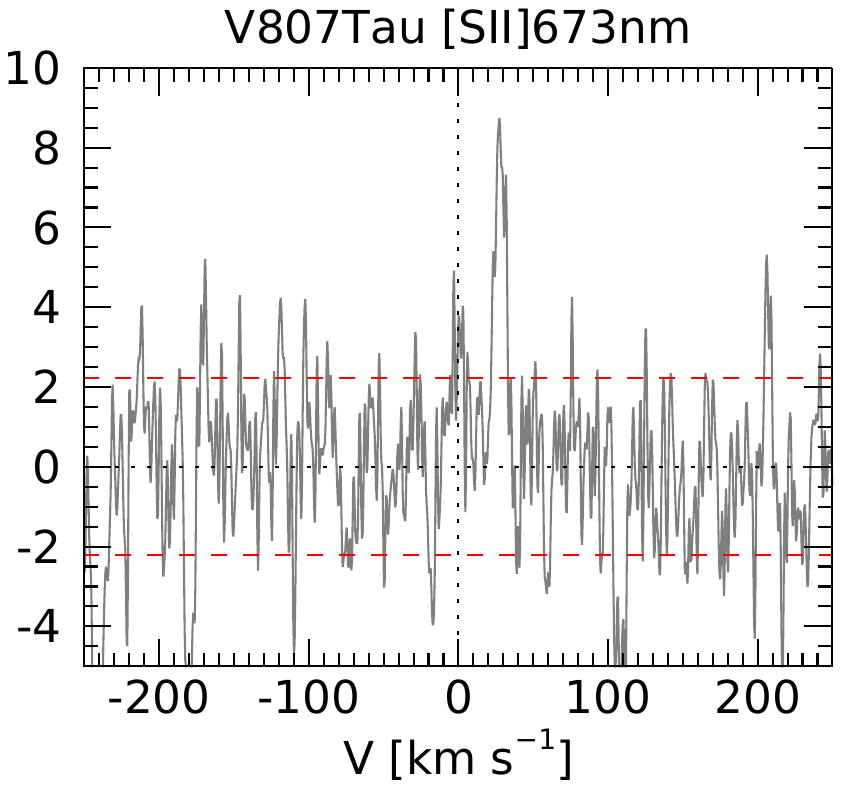}
\includegraphics[trim=80 0 80 400,width=0.2\textwidth]{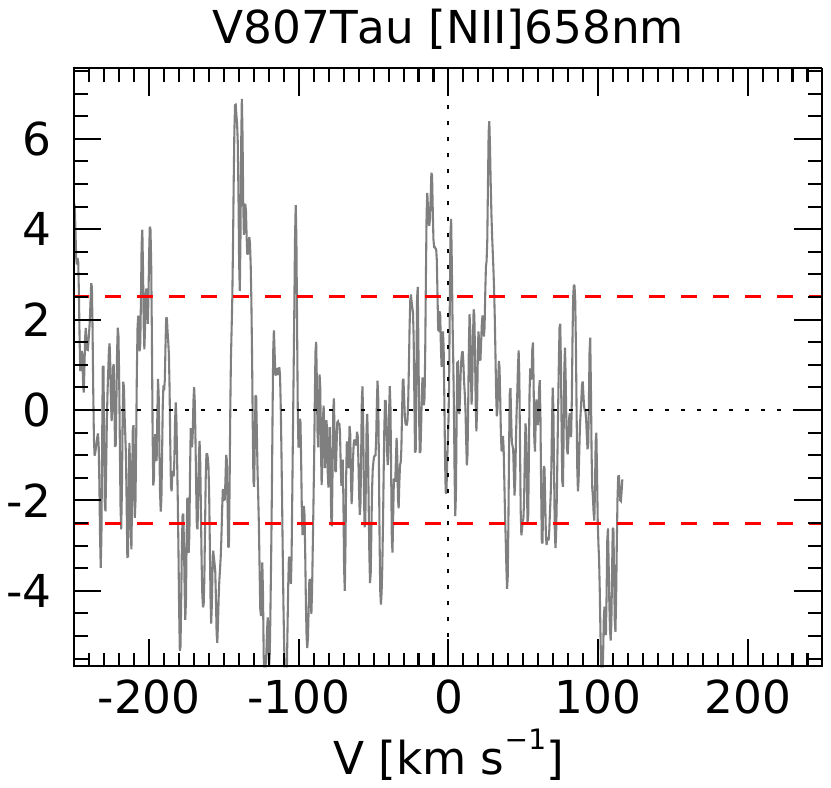}

\includegraphics[trim=80 0 80 400,width=0.2\textwidth]{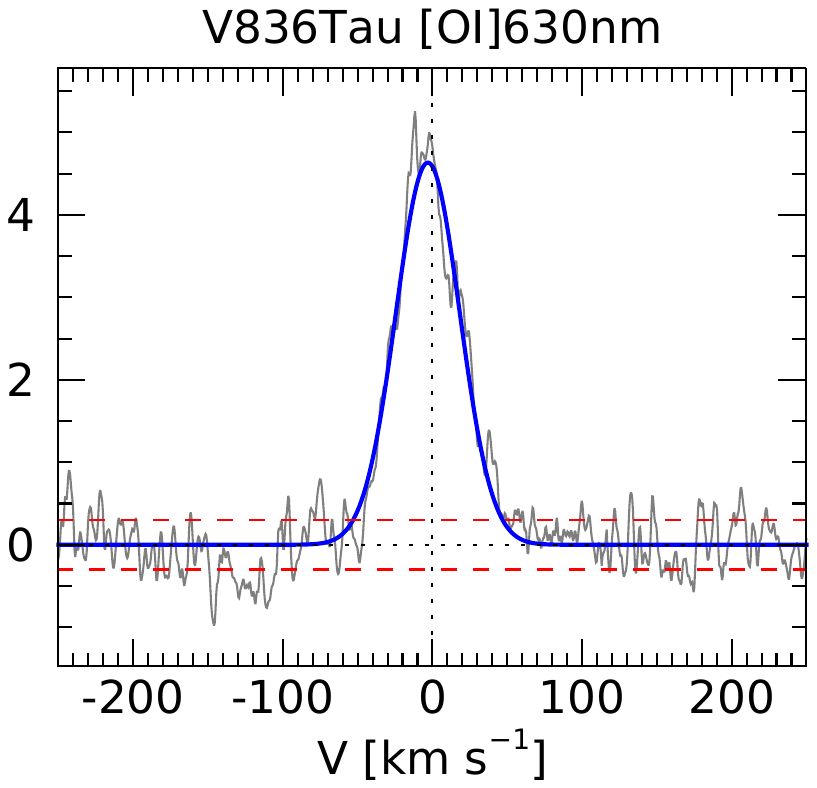}
\includegraphics[trim=80 0 80 400,width=0.2\textwidth]{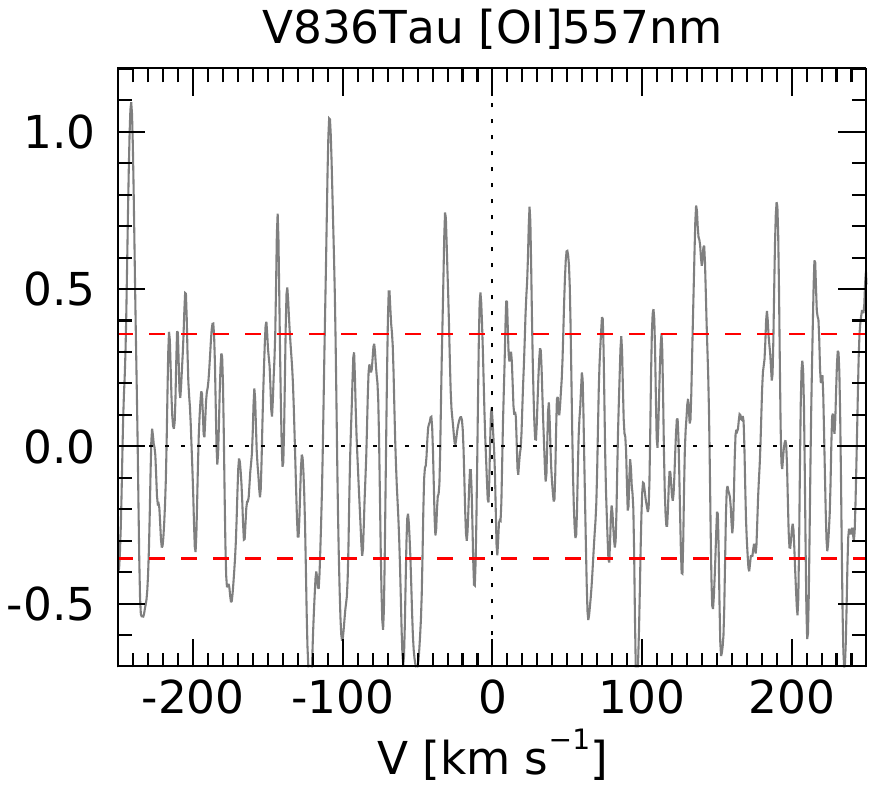}
\includegraphics[trim=80 0 80 400,width=0.2\textwidth]{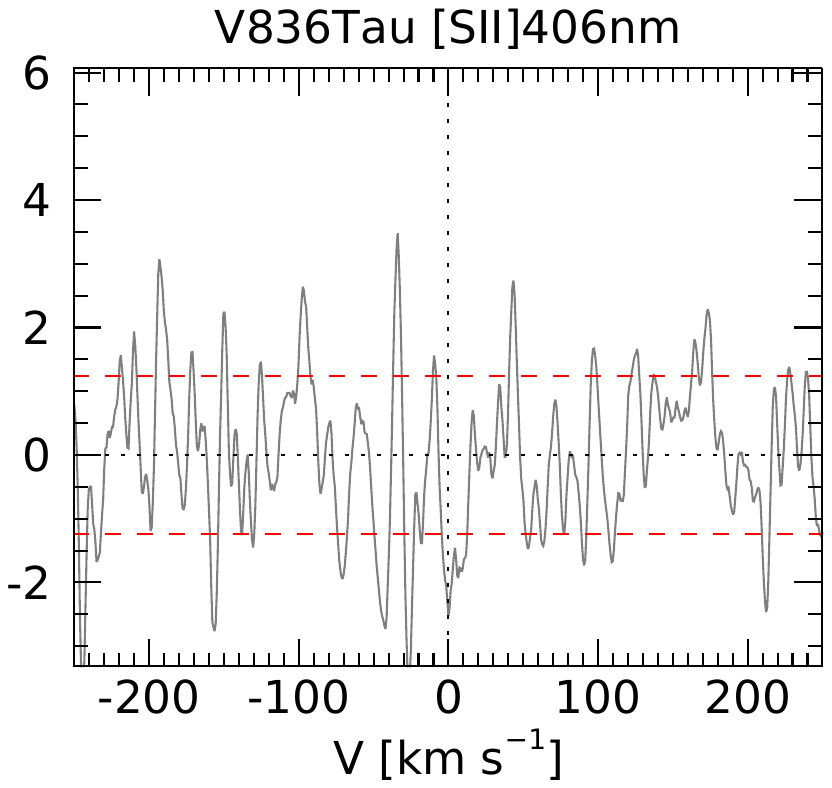}
\includegraphics[trim=80 0 80 400,width=0.2\textwidth]{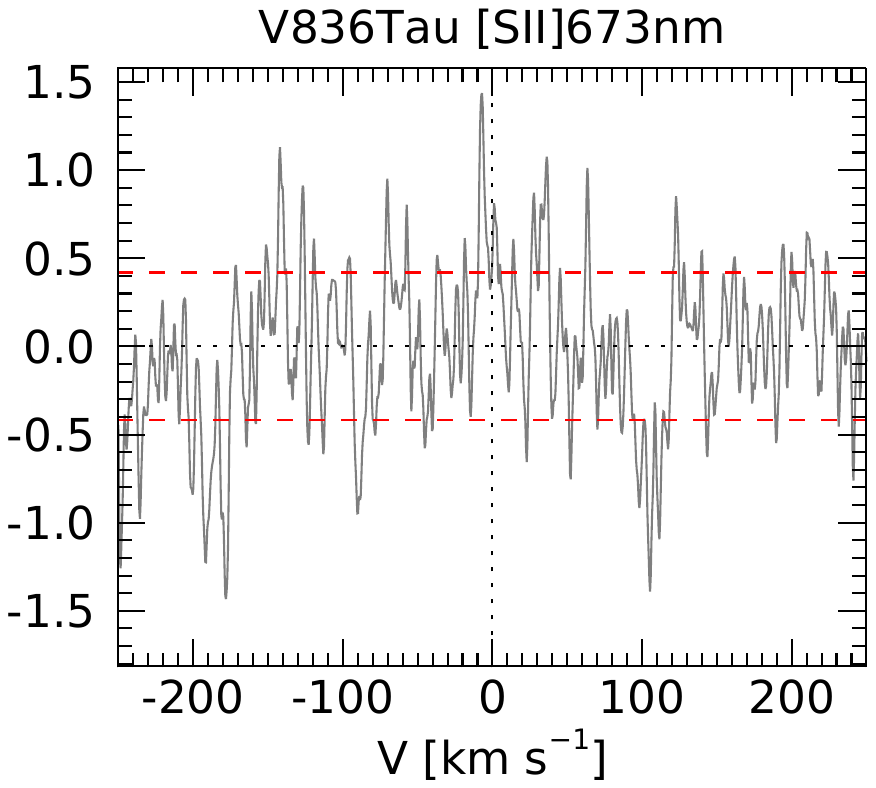}
\includegraphics[trim=80 0 80 400,width=0.2\textwidth]{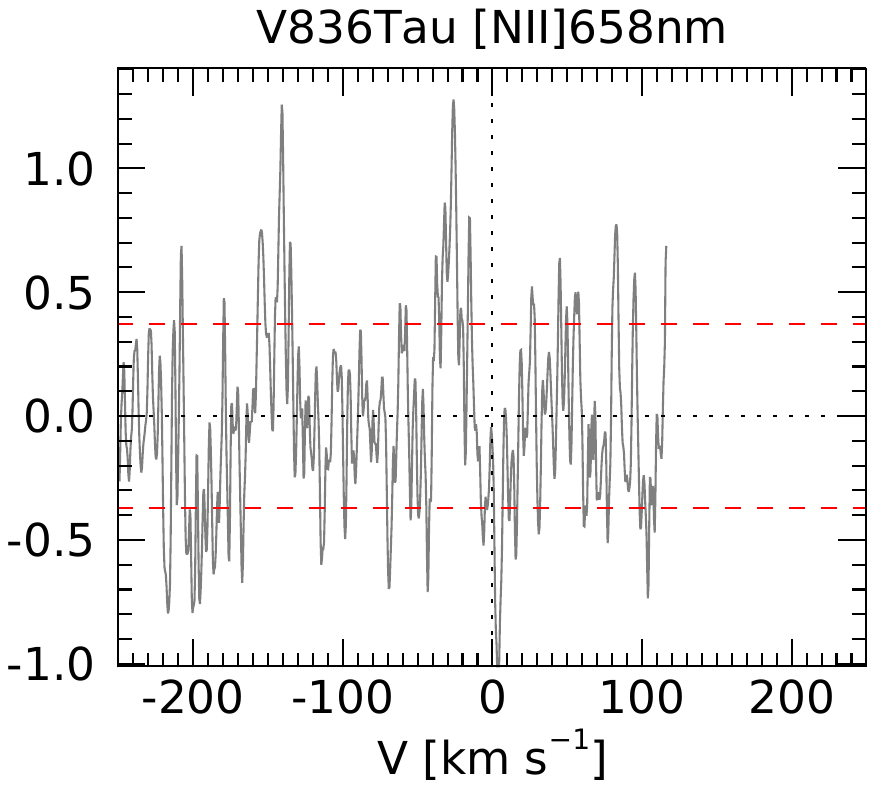}

   \caption{Continued}
   \label{fig:profiles6}
  
\end{figure*}

\newpage
\FloatBarrier

\section{Tables of line parameters}

Tables \ref{tab:line_fluxes1} and \ref{tab:line_fluxes2} report the kinematical parameters of the line profile decomposition and the calculated luminosity for each component, corrected for the extinction values given in Table \ref{tab:sources_param}. The velocity $V_p$ reported for each component has been corrected for the stellar radial velocity given in Table \ref{tab:sources_param}, while the Full Width at Half Maximum (FWHM) is corrected for the instrumental profile broadening.

\input{Tables/table8_super.tex}
\input{Tables/table9_super.tex}


\end{appendix}

\end{document}

%% file: Tables/table1.tex
\begin{table*}
\small
\center
\caption{\label{tab:sources_param} Source properties of the sample.}
\begin{tabular}{lcccccccccc}
\hline
\hline
Source    & d & $\rm A_v $ & $\rm L_\star$ & $\rm M_\star$ & V$_{rad}\rm (\pm err)$ & $\rm log $ \lacc\ $\rm (\pm err)$ & $\rm log $ \macc & $\rm i_{disk}^{a}$ &   Disk type$^{\rm b}$ &Ref. \\
          & [pc] & [mag] & [$\rm L{\odot}$] &  [$\rm M_{\odot}$]& [km\,s$^{-1}$] &[$\rm L_{\odot}$] & $\rm [M_{\odot}$ $\rm yr^{-1}]$ &  [deg]  &   &              \\
\hline
 BP Tau    &   127 &  0.2 & 0.37   &  0.48  &  16.09(0.16) & -1.34 (0.07)  & -8.25   &   38.2$\pm$ 0.5 &  Sub & 1,2 \\
 CI Tau &   160 &  2.6 & 1.04   &  1.29  &  17.06(0.35) &  0.02 (0.06)  & -7.28   &   50$\pm$ 0.3    &   Sub &  1 \\  
 CQ Tau    &   149 &  0.5 & < 15.1 & < 1.99 &  24.85(1.15) &  < -0.42      & <-7.68  &   32.2-35   &  TD &  3,9 \\
 CW Tau    &   131 &  3.6 & 0.40   &  0.90  &  16.45(1.85) & -0.07 (0.07)  & -7.47   &  57.8-59.2   &  Full & 4,11  \\
 CY Tau    &   126 &  0.0 & 0.22   &  0.36  &  16.19(0.43) & -2.00 (0.07)  & -8.85   &  28$\pm$5   &   Full & 5, 10 \\
 DE Tau    &   128 &  0.0 & 0.49   &  0.26  &  14.67(0.44) & -1.45 (0.08)  & -7.99   &   ...   &   Full & 6   \\  
 DF Tau    &   176 &  0.0 & 0.35   &  0.39  &  14.8(0.54)  & -0.75 (0.06)  & -7.55   &   52   &   Full & 7 \\
 DG Tau    &   125 &  1.5 & 0.44   &  0.70  &  14.34(1.89) & -0.25 (0.18)  & -7.35   &   37   &  	 Full & 4 \\
 DH Tau    &   133 &  0.6 & 0.20   &  0.41  &  15.57(0.33) &  -1.53 (0.07)  & -8.49   &  16.9$\pm$ 2.2  &   TD & 1 \\
 DK Tau    &   132 &  1.2 & 0.78   &  0.62  &  17.31(0.20) & -1.42 (0.07)  & -8.33   &  58$\pm$ 18    &   Full & 1,10 \\
 DL Tau    &   160 &  1.5 & 0.40   &  0.90  &  16.56(0.23) & -0.35 (0.18)  & -7.62   &  45.0$\pm$ 0.2  &   Sub & 1 \\
 DN Tau    &   129 &  0.5 & 0.49   &  0.55  &  17.44(0.21) &  -2.10 (0.09)  & -9.04   &  35.2$\pm$ 0.5  &   Sub & 1 \\
 DO Tau    &   138 &  1.4 & 0.42   &  0.50  &  18.81(0.78) &  -0.93 (0.11)  & -7.80   &  27.6$\pm$ 0.5  &   Full & 1 \\
 DQ Tau    &   195 &  1.5 & 0.98   &  0.52  &  20.86(0.52) &  -0.92 (0.07)  & -7.68   &  16.1$\pm$ 1.2  &   Full & 1 \\
 DR Tau    &   193 &  1.2 & 0.29   &  0.83  &  22.57(0.26) &  -0.56 (0.08)  & -7.93   &  5.4$\pm$ 2.6   &   Sub & 1 \\
 DS Tau    &   158 &  0.4 & 0.29   &  0.61  &  15.50(0.32) &  -1.46 (0.07)  & -8.57   & 65.2$\pm$ 0.3   &   Sub & 1 \\
 FT Tau    &   130 &  0.6 & 0.05   &  0.25  &  16.65(0.79) &  -1.93 (0.09)  & -8.92   & 35.5$\pm$ 0.3   &   Sub & 1 \\
 GG Tau    &   116 &  1.4 & 0.80   &  0.73  &  17.59(0.34) &  -0.76 (0.08)  & -7.76   &  54-61    &    TD & 3 \\
 GH Tau    &   130 &  0.8 & 0.79   &  0.42  &  16.98(0.96) &  -2.31 (0.09)  & -8.98   &  ...    &   Full & 8 \\
 GI Tau    &   129 &  1.9 & 0.27   &  0.55  &  18.64(0.47) & -1.37 (0.07)  & -8.43   & 43.8$\pm$ 1.1   &   Full & 1  \\
 GK Tau    &   129 &  1.9 & 0.87   &  1.20  &  20.96(0.32) & -1.39 (0.10)  & -8.67   & 40.2$\pm$ 6.0   &   Full & 1 \\
 GM Aur    &   158 &  1.3 & 1.16   &  1.19  &  15.98(0.37) & -1.05 (0.09)  & -8.29   & 53.21  &   TD & 3 \\
 HN Tau    &   134 &  1.2 & 2.63   &  1.58  &  27.34(2.25) & -0.12 (0.23)  & -7.31   & 69.8$\pm$ 1.4   &   Full &  1\\
 HQ Tau    &   161 &  3.7 & 5.83   &  2.00  &  17.52(1.05) & -1.24 (0.14)  & -8.43   & 53.8$\pm$ 3.2   &   Full & 1\\
 IP Tau    &   129 &  0.6 & 0.31   &  0.51  &  17.40(0.26) &  -2.15 (0.10)  & -9.14   & 45.2$\pm$ 0.9  &   TD &  1\\
 IQ Tau    &   131 &  1.6 & 0.19   &  0.55  &  17.50(0.42) & -1.40 (0.08)  & -8.54   & 62.1$\pm$ 0.5   &   Sub & 1 \\
 LkCa 15   &   157 &  0.9 & 0.60   &  1.09  &  19.43(0.17) &  -1.89 (0.09)  & -9.24   & 51.5$\pm$ 0.7  &   TD &  3,12 \\
 MWC480    &   156 &  0.0 & 10.79  &  1.71  &  12.87(1.10) &  -0.19 (0.12)  & -7.64   & 36.5$\pm$ 0.2   &   Sub & 1 \\
 RW Aur A  &   183 &  1.0 & 1.64   &  1.50  &  27.7(2.5)$^{\rm c}$        & +0.39 (0.30)  & -6.93   &  55.1$\pm$ 0.5  &   TD &  1\\
 RY Tau    &   138 &  2.2 & 8.87   &  1.80  &  21.73(1.51) &  -0.38 (0.15)  & -7.57   & 65.0$\pm$ 0.5   &   Sub &  1\\
 UX Tau    &   142 &  1.2 & 1.76   &  1.36  &  19.74(0.35) &  -1.26 (0.16)  & -8.58   & 57-88   &   TD &  9\\
 UY Aur A   &   152 &  0.5 & 0.46   &  0.51  &  14.09(0.50) &  -1.27 (0.07)  & -8.18   & 23.5$\pm$ 7.8   &   Sub & 1 \\
 UZ Tau E  &   130 &  0.7 & 0.45   &  0.41  &  $-$0.56(1.31) & -1.27 (0.07)  & -8.05   &  56.1$\pm$ 0.5  &   Sub & 1 \\
 V409 Tau  &   130 &  0.3 & 0.24   &  0.43  &  16.98(0.36) & -2.64 (0.13)  & -9.58   &  69.3$\pm$ 0.3  &   TD & 1 \\
 V807 Tau  &   184 &  1.3 & 1.69   &  1.41  &  16.75(0.15) & -0.91 (0.09)  & -8.14   & ...      &   TD & 8 \\
 V836 Tau  &   167 &  0.9 & 0.54   &  0.22  &  19.18(0.18) &  -1.93 (0.09)  & -8.37   &  43.1$\pm$ 0.8  &  Full & 1 \\
\hline\end{tabular}	
\begin{quotation}
Notes.
Stellar and accretion parameters derived from GHOsT data in \cite{Gangi22} and \cite{alcala2021}. Average uncertainties are 0.2 in $\rm A_v $, 0.2 in log $\rm L_\star$,  and 0.15 in log $\rm M_\star$.  Distances are from Gaia EDR3 with a typical uncertainty $< 10\%$. See \cite{Gangi22} and \cite{alcala2021} for details.

$\rm ^{a}$ Inclination angles refer to the outer disk estimated from millimetres observations, except for GG Tau, UX Tau and DK Tau for which an inner disk inclination different from the outer disk was measured outside observational errors (see also Table \ref{table:disk_param}). 
Errors on the inclination or a range of values is given when available.

$\rm ^{b}$ Disk type refer to the disk structure as revealed by resolved millimetre images, i.e. disks with substructures such as rings and holes (sub), full disks (full) or disks with inner sub-mm dust cavities (TD). 

$\rm ^{c}$ Radial velocity for RW Aur derived from the Li line only. 

References for disk parameters 

$\rm ^{(1)}$ \citet{long2019}, $\rm ^{(2)}$ \citet{jennings2022}, $\rm ^{(3)}$ \citet{francis2020}, $\rm ^{(4)}$ \citet{bacciotti18}, $\rm ^{(5)}$ \citet{tripathi17}, $\rm ^{(6)}$ \citet{simon2019}, $\rm ^{(7)}$ \citet{andrews2013}, $\rm ^{(8)}$ \citet{simon2016}, $\rm ^{(9)}$ \citet{bohn2022}, $\rm ^{(10)}$ \citet{nelissen2023}, $\rm ^{(10)}$ \citet{perez2015}, $\rm ^{(11)}$ \citet{ueda2022}, $\rm ^{(12)}$ \citet{pietu2007}. 

\end{quotation}
\end{table*} 

%% file: Tables/table2.tex
%

\begin{table*}
\caption{\label{tab:lines}Observed lines and rate of detection of different components}             
\label{table:lines}      
\centering          
\begin{tabular}{l c c c c c c }     
\hline\hline       
Transition & $n_{cr}$[cm$^{-3}$] & Total & NLVC & BLVC & HVC-B & HVC-R \\
\\
\hline                    
\osix & 1.6\,10$^{6}$ & 36(100\%) & 31(86\%) & 12(33\%)  & 20(55\%)   & 7(19\%)  \\
\ofive& 9.2\,10$^{7}$ &27 (75\%) & 19(53\%)  & 14(39\%)  & 6(17\%)  & 3(8\%)  \\
\sfour& 1.9\,10$^{6}$ &19 (53\%) & 9 (25\%) & 6(17\%)  & 8(22\%)  & 2(5.5\%)  \\
\ssix & 1.6\,10$^{4}$ &14(39\%)  & 5(14\%)  & 6(17\%)  & 12(33\%)  & 3(8\%)   \\
\nsix & 8.0\,10$^{4}$ &13(36\%)  & 0(0\%)  & 0(0\%)  & 12(33\%)  & 2(5.5\%)   \\
\hline                  
\end{tabular}

\begin{quotation}
$^{a}$ Critical densities are from \cite{giannini2019}
\\
The velocity separation among the different components has been defined as follow (see text): Narrow Low Velocity Component (NLVC): $\lvert{V_p}\rvert <$ 30 km\,s$^{-1}$, FWHM$<$55 km\,s$^{-1}$, Broad Low Velocity Component (BLVC): $\lvert{V_p}\rvert <$ 30 km\,s$^{-1}$, FWHM$>$55 km\,s$^{-1}$, High Velocity Component-Blue (HVC-B): ${V_p} < -30$ km\,s$^{-1}$, High Velocity Component-Red (HVC-R): $V_p > +30$ km\,s$^{-1}$
\end{quotation}
\end{table*}

%% file: Tables/table3.tex
\begin{table*}
\small
\caption{\label{tab:disk_param} Disk properties of TD of the sample.}
\begin{tabular}{lccccc}
\hline
\hline
Source    &  $\rm i_{disk,out}^{a}$ & $\rm i_{disk,in}^{b}$ & P.A.(out)$^{a}$ & P.A.(in)$^{b}$ &  $\rm R_{cav}$  \\
          & [deg] & [deg] & [deg]  & [deg] & [au]     \\
\hline      
 CQ Tau    &   35(3) -32.2(4)       &  29.25(4)    &233.9(4)&140.4(4) &  50(3)-53(6)	 \\
 GG Tau    &   36(3) &  54-61(3) & 36(3)  & 98(3)  & 224(3)-192(7) \\
GM Aur    &   53.2 (1) &  68.04(4) & 57.2(4)& 36.9(4) & 40(3)-35(8)    \\
 IP Tau    &   45.2(1)   &  61.7(4)     & 173(1) & 163.13(4)  & 25(3)-42.15(1)\\
 LkCa 15   &   51.5(5)-55(3) &   61.02(4) & 60(3) &	101.3(4)  & 76 (3) \\
UX Tau    &  38.0(4)   &  57-88(4) & 167(3) &	115.67(4)  &  31(3) \\
 UZ Tau E &   56.15 (1)  &   & 23.06(1)   &      & 9.46   \\
 RY Tau     &  65(1)     &   &23(1)      &       &   13.8(1)-27(3) \\
\hline
\end{tabular}	
\\
\\
\textbf{Notes:} 
\\
$^{a}$ Inclination and position angle measured from sub-mm resolved images of the outer disk.\\
$^{b}$ Inclination and position angle measured from IR and sub-mm observations of the inner disk. A range of values is given only for sources where $\rm i_{disk,in}$ differs from $\rm i_{disk,out}$ outside the uncertainties.\\
\textbf{References.} (1)\cite{long2019}; (3)\cite{francis2020}, (4)\cite{bohn2022}, (5)\cite{pietu2007}, (6)\cite{uberia2019}, (7)\cite{phuong2020}, (8)\cite{marcias2018}
\label{table:disk_param}
\end{table*} 

%% file: Tables/table4.tex
%

\begin{table*}
\caption{Results of the fit with the disk model}             
\label{table:disk_fit}      
\centering          
\begin{tabular}{l c c c c c c}     
\hline\hline       
Source & $R_i$ & $R_{o}$ & $\alpha$ & $R_{kep}$ & $F(R_{kep})/F_{tot}$ & Notes\\
\\
\hline                    
BP Tau & 0.02$\pm$0.01 & 8.0$\pm$0.9 & 2.34$\pm$0.02 & 0.77 & 0.82 & NLV\&BLV\\
CQ Tau & 0.78$\pm$0.19 & 42$\pm$21 & 2.22$\pm$0.21 & 3.2 & 0.46 &\\
CW Tau & 0.25$\pm$0.01  & 49$\pm$3 & 2.38$\pm$0.01 & 2.91 & 0.70 & NLV\&BLV\\
CY Tau & 0.03$\pm$0.01 & 7$\pm$2 & 2.40$\pm$0.03 & 0.14 & 0.53 &\\
DH Tau & 0.01$\pm$0.01 & 6.4$\pm$1.5 & 2.39$\pm$0.02 & 0.32 & 0.81 & NLV\&BLV\\
DN Tau & 0.01$\pm$0.01 & 1.6$\pm$0.1   & 1.91$\pm$0.04 &  0.24 & 0.58 & \\
DR Tau & 0.01$\pm$0.01 & 10$\pm$3  & 2.24$\pm$0.02 &  0.19 & 0.64  & NLV\&BLV\\
GG Tau & 0.69$\pm$0.09 & 70$\pm$26 & 2.50$\pm$0.05 & 2.4 & 0.52 & \\
GI Tau & 0.41$\pm$0.47 & 9.0$\pm$4.4 & 1.55$\pm$0.75 & 2.06  & 0.35 \\
GM Aur & 0.36$\pm$0.10 & 22$\pm$23  & 2.0$\pm$0.1 & 2.62 & 0.48 & \\
LkCa 15 & 0.25$\pm$0.02 & 26$\pm$3 & 2.24$\pm$0.04  & 1.49 & 0.52 & \\
UX Tau & 1.57$\pm$0.09  & 52$\pm$6 & 2.25$\pm$0.06 & 5.66 & 0.47 &\\
\hline                  
\end{tabular}
\end{table*}

%% file: Tables/table5.tex
%
\begin{table}
\caption{Average values of physical parameters within the emission regions of the \osix\, and \ofive\, lines for the PE wind models of Weber et al. 2020}             
\label{table:model}        
\begin{tabular}{lcccc}     
\hline\hline
& \multicolumn{2}{c}{$L_{acc} = 0.026 L_*$} &  \multicolumn{2}{c}{$L_{acc} = L_*$} \\
 & 630nm  & 557nm   & 630nm  & 557nm\\
\hline
$n_e$ [cm$^{-3}$] & 5.9\,10$^{5}$  & 1.9\,10$^{7}$ & 5.9\,10$^{5}$ & 1.9\,10$^{7}$\\
$n_{H}$   & 9.45\,10$^{6}$   & 3.15\,10$^{8}$ & 3.3\,10$^{6}$ & 5.6\,10$^{7}$  \\
$x_e$     & 0.09            & 0.13   & 0.2  & 0.5        \\              
$T_e$ [K]  & 6700          & 7500   & 6300   & 8800         \\
\hline                  
\end{tabular}
\\
\end{table}
\begin{table}
\caption{Synthetic parameters derived from the PE wind models that assume a total $\dot{M}_{loss}$ = 10$^{-8}$ M$_\odot$\,yr$^{-1}$ and an inclination angle of 60$^{\circ}$  }             
\label{table:model}        
\begin{tabular}{lcccc}     
             
\hline\hline
& \multicolumn{2}{c}{$L_{acc} = 0.026 L_*$} &  \multicolumn{2}{c}{$L_{acc} = L_*$} \\      
Parameter & Empirical$^a$  & Model$^b$ & Empirical$^a$ & Model$^b$  \\
\hline
$n_e$ [cm$^{-3}$] & 3\,10$^{6}$        & 5.9\,10$^{5}$    & 3\,10$^{6}$  & 5.4\,10$^{5}$\\
$n_{H}$   & 3\,10$^{7}$   & 9.5\,10$^{6}$    & 3\,10$^{7}$   & 3.3\,10$^{6}$ \\
$R_w$ [au]  & 1.02          & 1.5             & 5.47       & 4 \\
$\dot{M}_{wind}$(OI)$^c$  & 5.8\,10$^{-10}$  & 4.\,10$^{-10}$   & 1\,10$^{-8}$      &6.4\,10$^{-10}$ \\
$V$ [au$^3$]  & 0.68   &  11.2  & 6.4  & 536 \\
$l_W$ [au]    & 0.21   &  3     & 0.13  & 3.5 \\
\hline                  
\end{tabular}
\\

$^a$"Empirical" refers to the values of parameters estimated applying equations (1) and (2) to the synthetic observations produced by the model.\\
$^b$"Model" refers to the values of parameters estimated directly from the structure of the \osix\, emission region predicted by the model. \\
$^c$ Mass loss rate from the \osix\, line emission in units of M$_\odot$\,yr$^{-1}$
\end{table}

%% file: Tables/table6.tex
%
\begin{table}
\caption{Mass loss rate $\rm \dot{M}_{jet}$ in the HVC}             
\label{table:MHVC}        
\begin{tabular}{l cccc }     
\hline\hline       
             & \multicolumn{2}{c}{High Velocity} &  \multicolumn{2}{c}{Medium Velocity} \\
             
Source &  $V_{tan}$$^a$ & log($\rm \dot{M}_{jet}$)$^b$ & $V_{tan}$$^a$ & log($\rm \dot{M}_{jet}$)$^b$  \\
       & \kms & $\rm M_{\odot}$ $\rm yr^{-1}$ &  \kms & $\rm M_{\odot}$ $\rm yr^{-1}$\\
\hline
BP Tau & 45(5) & $-$10.4 & ... & ... \\
CI Tau & $-$96(7)& $-$9.0 &  ... & ... \\
CW Tau & $-$190(2) & $-$7.9 &  ... & ... \\
DE Tau  & $-$132(1)$^c$ & $-$10.1 &  ... & ... \\
DF Tau & $-$138.4(0.8) & $-$9.2& $-$71(3) & $-$9.3  \\
DG Tau & $-$97.1(0.2) & $-$7.7 & $-$46.4(0.1) & $-$8.1 \\
DK Tau & $-$142(16) & $-$9.2  & ... & ... \\
DL Tau &  $-$165(5) & $-$9.8 &$-$62(3) &$-$8.1 \\
DO Tau & $-$53.6(0.2)& $-$9.0 &$-$48.1(0.1) & $-$8.7 \\
DR Tau & $-$11(3) & $-$10.1 &$-$23.8(0.1) & $-$10.3\\
FT Tau & $-$71(2)& $-$10.5 & ... & ... \\
GI Tau & 48(1)& $-$9.9 & ... & ... \\
HN Tau & $-$188(1)& $-$8.2 &  ... & ... \\
HQ Tau & $-$55(10)& $-$8.9 & ... & ... \\
IQ Tau & $-$56.9(0.2)& $-$8.8 & ... & ... \\
RW Aur & $-$246.9(0.4) & $-$7.6 &$-$129.5(0.7) & $-$7.2\\
RY Tau & $-$163(2) & $-$8.3 & ... & ... \\  
UY Tau & $-$36.2(0.5)  & $-$9.5 & ... & ... \\
UZ Tau & $-$106(2)& $-$8.6 & ... & ... \\
V409 Tau & $-$112(4)& $-$9.9 & ... & ... \\
V807 Tau & $-$53(3)$^c$& $-$9.4 & ... & ... \\
\hline                  
\end{tabular}
\\
$^a$ Jet tangential velocity estimated from the peak radial velocity of the HVC gaussian fit, corrected for jet inclination assumed perpendicular to the inclination of the outer disk. Errors are given in parenthesis.\\
$^b$ $\rm \dot{M}_{jet}$ of the blue-shifted jet component, computed from the luminosity of the \osix\, line assuming $T_e = 10^4$ K and $n_e = 5\,10^4 $\cmt . A typical uncertainty of an order of magnitude due to the adopted assumptions is estimated (see Section 9.2).\\
$^c$ Value computed assuming $i_d$ = 45$^o$ \\
\end{table}

%% file: Tables/table7.tex
\begin{table}
\caption{Comparison with literature values of $\rm \dot{M}_{jet}$ from spatially resolved observations}             
\label{table:mlosscomp}           
\begin{tabular}{l c c c }     
\hline\hline       
Source & log($\rm \dot{M}_{jet}$) & Ref &  Notes \\
       & $\rm M_{\odot}$ $\rm yr^{-1}$  & &  \\
\hline          
\hline          
DG Tau HV& $-$7.7 & &\\
       & $-$8.1 & (1) & Total\\
       & $-$7.8 & (2) & \\
DG Tau MV& $-$8.1 &  &\\
       & $-$7.8 & (2) & \\
\hline
DO Tau & $-$8.5 &  (HV+MV)&\\
       & $-$8.3 & (3) & \\
\hline 
RW Aur A & $-$7.1 &  (HV+MV)& \\
         & $-$8.7 & (4) & BE \\    
\hline
CW Tau   & $-$7.9 &   &      \\ 
         & $-$8.4 & (5) & BE   \\
\hline
HN Tau   & $-$8.2 &   &      \\   
         & $-$9.3/$-$8.34 & (6) & (a) \\
\hline
RY Tau   & $-$8.3  &      &  \\
         & $-$8.6   & (7)  & CIV lines\\
         & $-$8.8   & (8) & \\   
\hline
DF Tau  HV & $-$9.2  &      &  \\   
        HV   & $-$8.85  & (9)  & (b)  \\
DF Tau  MV & $-$9.3  &      &  \\ 
        MV & $-$8.85  & (9)  & (b)  \\   
\hline
UY Aur A   & $-$9.5  &     &  \\   
           & $-$9.3   &  (9)  & (b)  \\         
\hline
\hline              
\end{tabular}
\\
Values in boldface are those derived in this work. HV and MV refer to the components at high and medium velocity. BE indicates derivations with the BE technique \citep{bacciotti99}\\

References. $(1)$ \citet{maurri14}, $(2)$ \cite{agra-amboage11}, $(3)$ \cite{erkal2021}, $(4)$ \cite{melnikov09}, $(5)$ \cite{coffey2008}, $(6)$ \cite{hartigan04}, $(7)$ \cite{skinner2018},$(8)$ \cite{agra-amboage2009}, $(9)$ \cite{uvarova2020}.

$(a)$ The two values refer to line luminosity and shock method, respectively.\\
$(b)$ Assuming $n_e$ = 10$^{3}$\cmt . 
\end{table}

%% file: Tables/table8_super.tex
\onecolumn
\footnotesize

\tablecaption[ ]{\label{tab:line_fluxes1} Kinematical parameters and luminosities of Gaussian components for the \osix , \ofive\, and \sfour\, lines}

\tablefirsthead{%
\\
   &  \multicolumn{3}{c}{[\ion{O}{i}]\,630nm}   && \multicolumn{3}{c}{[\ion{O}{i}]\,557nm}  &  & \multicolumn{3}{c}{[\ion{S}{ii}]\,406nm}  &ID\\
   \cline{2-4} \cline{6-8} \cline{10-12} \\ 
Name     &$V_{\rm p}$ & FWHM  & $L_{\rm 630}$ & &$V_{\rm p}$ & FWHM & $L_{\rm 557}$ & & $V_{\rm p}$ & FWHM & $L_{\rm 406}$ &  \\
&[\kms]&[\kms]  &[10$^{-5} L_{\odot}$] & &[\kms] &[\kms] &[10$^{-5} L_{\odot}$]& & [\kms] &[\kms] & [10$^{-5} L_{\odot}$] & \\
\\
   \cline{2-4} \cline{6-8} \cline{10-12} \\ 
}

\tablehead{%
\multicolumn{15}{c}%
{{\bfseries \tablename\ \thetable{} -- continued from previous page}} \\

\\
   &  \multicolumn{3}{c}{[\ion{O}{i}]\,630nm}   & & \multicolumn{3}{c}{[\ion{O}{i}]\,557nm}  &  & \multicolumn{3}{c}{[\ion{S}{ii}]\,406nm}  &ID\\
   \cline{2-4} \cline{6-8} \cline{10-12} \\ 
Name     &$V_{\rm p}$ & FWHM  & $L_{\rm 630}$ & &$V_{\rm p}$ & FWHM & $L_{\rm 557}$ & & $V_{\rm p}$ & FWHM & $L_{\rm 406}$ &  \\
&[\kms]&[\kms]  &[10$^{-5} L_{\odot}$] & &[\kms] &[\kms] &[10$^{-5} L_{\odot}$]& & [\kms] &[\kms] & [10$^{-5} L_{\odot}$] & \\
\\
   \cline{2-4} \cline{6-8} \cline{10-12} \\ 
}

\tabletail{%
\hline
\multicolumn{15}{c}{{Continued on next page}} \\
}
\tablelasttail{\hline}

\begin{supertabular}{lcccccccccccccc}
 BP Tau	  &    0.2(0.3)  &    29.0(0.5)  &   0.40(0.02)    &&   -0.4(0.4) &  62.1(0.8)  &    0.58(0.01)  &&    -0.3 (0.4) &  57.8(0.7)  &1.16(0.03)   &  NLV\\
          &    7.9 (0.5) &     88.3(1.0) &   1.20 (0.04)   &&      ...	  &   	...	&       ...      &&    ...       &   ...       &  ...        &  BLV\\
          &  57.3(0.7)   &   35.3(1.5  ) &  0.25(0.01)    &&   45.2(1.0) &  22(2 ) & 0.06(0.01)     &&    63.6(0.9) &  71.9(2.5)  &0.39(0.04)   & HV-R \\
 CI Tau   &   -6.8(0.8)  &    35.0(3.2)  &   1.47(0.01)   &&     ...      &   ...	&   $<$0.54      &&    ...       &   ...       &   $<$3.2    &  NLV\\
	      &   -81.3(5.5) &   140(8)  &    3.14(0.04)  &&      ... &  	...	&     ...        &&   ...        &  ...        &  ...        & HV-B \\
 CQ Tau   &    3.3(1.2)  &    29.3(0.8)  &   1.20 (0.02)   &&     ...	  &  	...	&   $<$0.94      &&   ...        &  ...        &  $<$3.2     & NLV\\
	  &  -25.5(1.2)	 &   14.3(0.8)	 & 0.30(0.02)	  &&	 ...	  &	...	&	...	 &&	   ...	 &	...    &	...  & NLV\\
 CW Tau	  &  -0.65(0.2)  &    28(2)  &   15.88(0.01)  &&     -4.7(2.1)&   24.5(4.1) &    3.01(0.01)  &&     -0.4(2.0) &   22(4) &  2.44(0.01) &  NLV  \\
          &   -3.3(0.2)  &    76.6(3.5)  &   41.93(0.03)  &&      6.1(1.9)&   66.7(0.7) & 8.38(0.01)	 &&    -1.1(2.6)  &   74(5) &  7.78(0.03) & BLV\\
          & -117(2)  &    88(3)  &   20.39(0.03)  &&      ...	  &  	...     &	...	 && -110(3)  &   77(5) &  4.74(0.03) & HV-B\\
 CY Tau   &    1.3(0.5)  &    44(1)  &    0.32(0.01)  &&     1.3(0.5) &   53(1) &    0.10(0.01)  &&	...	 &    ...      &   $<$0.04   & NLV\\
 DE Tau   &   -7.0(0.5)  &    19.7(0.5)  &    0.14(0.01)  &&    -0.04(1.6) &   62.0(2.3) &    0.15(0.02)  &&   -14.2(1.5) &  31(2)  & 0.09(0.01)  & NLV\\
          &  -28.9(2.5)  &   123(4)  &    0.30(0.05)   &&     ...	  &  	...     &	...	 &&  ...	 & ...         &  ...	     & BLV\\
          & -131.9(1.5)  &    51(2)  &   0.16(0.02)   &&     ...	  &  	...     &	...	 &&  -112(2) &   20(9) &  0.05(0.01) &  HV-B \\
 DF Tau   &   2.2(0.6)  &    13.5(0.4)  &    0.90(0.02)  &&    -2.1(0.7) &   49(2) &    1.13(0.04)  &&	...	 &    ...      &   $<$0.2    & NLV\\
          &  -13.2(0.9)  &    26(1)  &    1.20(0.04)  &&        ...	  &  	...     &	...	 &&   -26.9(4.9) &  47.6(10.5) &  0.30(0.07) & NLV\\
          &  -53.1(2.2)  &    73.0(9.3)  &    5.65(0.10)  &&   -80.1(0.8) &  89(1) &   0.32(0.08)   &&    ...	 &   ...       &    ...      & HV-B\\
          & -105.6(0.7)  &    42(2)  &    3.42(0.06)  &&        ...	  &  	...     &	...	 &&  -103(1) &   38(3) &  0.81(0.05) & HV-B\\
          &   26(4)  &    59(7) &    2.00(0.09)  &&         ...  &   	...	& 	...      &&  ...	 &    ...      &     ...     & HV-R\\
          &  129.7(0.5)  &    48(1)  &    1.67(0.09)  &&         ...  &   	...	& 	...      &&  ...	 &    ...      &     ...     & HV-R\\
 DG Tau   &   -3.9(1.9)  &    12.3(0.1)  &    9.20(0.01)  &&    -5.2(1.9) &   13.5(0.3) &    1.03(0.01)  &&     ...      &    ...      &   $<$0.15   & NLV\\
          &  -16.9(1.9)  &    24.0(0.1)  &   13.77(0.03)  &&   -14.1(1.9) &   36.6(1.2) &    1.25(0.01)  &&   -29.6(2.4) &   48(5) &  5.55(0.05) & NLV\\
          &  -60.0(1.9)  &    78.0(0.6)  &  132.65(0.06)  &&       ...	  &  	...	& 	...      &&-82.6(3.6)    &   69(3) & 11.13(0.07) & HV-B\\ 
          & -127.2(1.9)  &    51.5(0.2)  &   55.18(0.08)  &&  -123(2) &   59(1) &    1.35(0.02)  &&-130(2)   &   43.6(0.9) &  11.23(0.04)& HV-B\\
 DH Tau   &   -0.05(0.5)  &    20(2)  &    0.19(0.01)  &&    -3.0(0.7) &   54(1)&    0.32(0.01)  &&   ...        &  ...        &   ...       & NLV\\
          &    3.5(0.8)  &    69(4)  &   0.67(0.01)   &&    38.0(0.9) &   26(1) &    0.05(0.01)  &&  30(4)   &   52(8) &   0.22(0.02)& BLV\\   
 DK Tau   &   -7.0(0.5)  &    56(12) &     3.22(0.04) &&     -3.7(0.7)&    56(1)&    0.92(0.02)  && -20.4(3.0)   &  133(8) &   1.15(0.06)& BLV\\  
          &  -89.6(10.1) &    92(12) &    3.01(0.06)  &&   -79.4(0.2) &   17.4(0.3) &    0.08(0.01)  &&   ...        &  ...        &   ...       & HV-B\\
          &   62(29) &    90(26) &    1.90(0.06)  &&    39.5(0.9) &   21.9(0.7) &    0.14(0.01)  &&   ...        &  ...        &   ...       & HV-R\\ 
 DL Tau   &   -3.0(0.55)  &    38(2)  &    1.44(0.01)  &&     4.8(2.5)&   50(50)&    0.48(0.01)  &&   ...        &  ...        & $<$0.94     & NLV\\
          &  -63.6(0.5)  &   106(12) &    1.94(0.03)  &&        ...	  & 	...	&   	...      &&      ...     &     ...     &      ...    & HV-B\\
          &  -165.7(0.5) &    90.5(4.2)  &   2.51(0.03)   &&        ...   & 	 ...	&   	...      &&      ...     &     ...     &      ...    & HV-B\\
 DN Tau   &   -0.30(0.45)  &    51.4(0.9)  &   0.99(0.03)   &&    8(2)  &   53.7(3.5) &    0.20(0.03)  &&   ...        &    ...      &   $<$0.40   & NLV\\
 DO Tau   &  -19.2(0.8)  &    40(1)  &    4.34(0.03)  &&   -11.7(1.0) &   34.3(1.5) &   0.53(0.01)   &&  ...         &    ...      &   $<$0.18   & NLV\\
          &  -92.9(0.8)  &    53.3(0.8)  &   12.80(0.05)  &&         ...  &  	...	&   	...      &&    -69.2(0.8)&   42(13)&   1.46(0.06)& HV-B\\	
          & -103.4(0.8)  &    21.3(0.5)  &    5.16(0.02)  &&         ...  &  	...	&   	...      &&   -104(1)&    33.3(1.5)&   6.17(0.05)& HV-B\\	
 DQ Tau   &   -7.15(0.50)  &    38(2)  &   13.40(0.02)   &&    -4.0(0.6) &   47.3(0.6) &    2.21(0.02)  && -9.5(0.6)   &   36(1) &   2.38(0.02)& NLV\\
          &  -59.1(0.5) &   45(8)   &   2.30(0.06)   &&        ...   & 	...	&   	...      &&   -81.4(1.3)&   100(20)&   0.86(0.06)& HV-B\\	 
 DR Tau   &    0.2(0.3)  &    12(1)  &    0.71(0.03)  &&     1.5(0.4) &   12(1) &    0.51(0.01)  &&   ...        &     ...     &      ...    & NLV\\
          &   -4.5(2.1)  &    59(17) &    1.29(0.02)  &&    -1.5(2.4) &   81(4) &   1.38(0.08)   && -4.5(4.2)    &   60(21)&   1.2(0.1)& BLV\\  
          & -117(33) &   277(37) &    2.1(0.1)  &&        ...	  & 	..	&   	...      &&      ...     &     ...     &      ...    & HV-B\\
          & -252(1)  &    44(4) &    0.63(0.02)  &&        ...	  & 	..	&   	...      &&      ...     &     ...     &      ...    & HV-B\\
 DS Tau   &    5.0(1.0)  &    91(3)  &    1.08(0.07)  &&      4.8(1.2)&   69(2) &     0.44(0.03) &&    ...       &     ...     &    $<$0.28  & BLV\\
 FT Tau   &  -15.7(1.4)  &    36(3)  &    0.07(0.04) &&     -5.8(2.7)&   73(9) &    0.07(0.01)  &&   ...        &     ...     &    $<$0.25  & NLV\\
          &  -99.3(3.1)  &    93(5)  &    0.12(0.01)  &&        ...	  & 	..	&   	...      &&      ...     &     ...     &      ...    & HV-B\\
 GG Tau   &    -0.01(0.7)  &    29(2)  &    0.76(0.01)  &&    -3.2(1.3) &  63.6(1.5)  &    0.64(0.01)  &&  -8.6(1.4)   &   40(2) &   0.38(0.02)& NLV\\  
          &  -10.1(4.9)  &  102(12)  &    0.93(0.05)  &&         ...  &     ...	&  	...      &&  ...         & ...         &  ...        & BLV\\	 
 GH Tau   &   -0.52(1.04)  &    48(1)  &    0.51(0.01)  &&         ...  &     ...	&    $ <$0.07    &&   ...        &  ...        & $<$0.11     &    NLV\\  
          &   -8.9(1.3)  &   110(3)  &   1.10(0.03)   &&        ...   &    ...	&  	...      &&  ...         & ...         &  ...        & BLV\\	 
 GI Tau   &   -0.72(1.0)  &    21(3)  &   0.24(0.03)  &&   -1.9(1.3)  &   54(3) &     0.39(0.01) &&    ...       &   ...       &    ...      & NLV\\   
          &  -31.2(4.1)  &    84(6)  &    1.32(0.01)  &&         ...  &   	...	& 	...      && -1.4(4.6)     & 113(14) &  0.83(0.03) & HV-B\\	 
          &   49.2(1.1)  &    48(4)  &    0.70(0.01)  &&         ...  &   	...	& 	...      &&  ...         & ...         &  ...        & HV-R\\	 
 GK Tau   &   -6.5(2.5)  &    32.5(0.8)  &    2.29(0.02)  &&     2.8(0.8) &   81(2) &     1.79(0.02) &&  -16.2(1.1)  & 33(2)   & 1.22(0.03)  & NLV\\	 
          &  -21.9(1.7)  &   172(4)  &    5.2(0.1)  &&         ...  &   	...	& 	...      &&  ...         & ...         &  ...        & BLV\\
 GM Aur   &   -1.3(0.5)  &    32.1(0.6)  &    2.89(0.02)  &&    -4.9(0.4) &   38.2(0.2) &      0.45(0.02)&&     ...      &    ...      &     $<$2.35 & NLV\\  
 HN Tau  & -2.6(2.3)  &  31(1)   &  2.8(0.07)       &&   -1.0(2.3)   &  21.5(0.5)  &   0.17(0.05)  &&  -10.1(0.5)  &  29.4(0.6)  &  1.12(0.01) & NLV\\
         & -29.7(2.3)  & 26(1)   &  2.22(0.06)      && -27.1(0.3)    &  20.6(0.4)  &   0.18(0.01)   &&  -31(2)  & 33.9(0.7)   &  1.36(0.01) &  NLV  \\
         & 9.3(1.5)    & 53(7)    &  1.44(0.01)      &&  5.6(2.3)    &  67.5(1.2)  &   0.38(0.01)   &&  21(2)   &  48.5(0.6)  &  1.89(0.02) & BLV\\
         & -59.2(2.3)  & 54.9(2.3)    &  4.54(0.01)      &&       ...     &	...	&	...      &&   -69.9(2.5) &  65.3(0.9)  &  2.65(0.02) & HV-B\\ 
         & -114.5(1.0)  & 79(3)    & 3.70(0.02)     &&       ...     &	...	&	...      &&  -136(3)& 65(2)   & 0.94(0.02)  & HV-B\\  
 HQ Tau   &   -4.4(2.5)  &    34(4)  &  12.85(0.02)   &&    -0.3(0.6) &  11.8(0.8)  &    1.39(0.01)  &&   ...        &  ...        &  $<$39.9    & NLV\\ 
          &  -41.2(8.2)  &   59(6)   &  6.91(0.04)    &&       ...    & 	...	& 	...      &&  ...	 & ...	       &  ...        & BLV\\ 
 IP Tau   &  -18.5(2.7)  &   161(6)  &    0.84(0.09)  &&         ...  &   	...	& 	$<$0.17  &&  ...	 & ...	       &  $<$0.72    & BLV\\ 
 IQ Tau  &  -3.5(0.5)  &  22.7(0.5)   &  1.83(0.01)      & &   4.7(2.5)   &  79.6(0.6)  &   0.63(0.02)   &&    ...       &   ...       &   $<$0.12   & NLV\\ 
         & -19.5(0.4)  & 82(1)   & 5.46(0.05)        & &       ...    &	...	&	...      &&    ...       &   ...       &   ...       & BLV\\ 
         & -32.5(0.4)  &  29(1)  &  8.87(0.05)       & & -42.0(1.1)   &  20.0(2.5)  &   0.19(0.01)   && -33.5(1.1)   &  28(1)  &  1.14(0.02) & HV-B\\ 	   
 LkCa 15  &    -1.1(0.3)  &    38.9(0.5)  &    1.25(0.02)  &&         ...  &   	...	& 	$<$0.19  &&-6.6(1.2)     &  54.0(0.3)  &  0.67(0.03) & NLV\\
 MWC480   &  -5.8(2.1)  &    29(6)  &   1.6(0.2)   &&    ...       &	    ...	& 	...      &&  ...         & ...         &  $<$3.94    & NLV\\ 
 RW AurA  & -172.3(0.3)  &    74.3(0.7)  &   29.3(0.2)  &&  -199(29)& 167(61) &    3.7(0.2)  &&-158.4(0.9)   &   53(1) &  4.2(0.1)  & HV-B\\  
          &  -90.4(0.5)  &   209(1)  &  135(0.6)  &&        ...	  &  ...	&         ...    &&  -56.4(1.2)  &131(3)   & 16.9(0.3)  & HV-B\\
          &  111.4(0.2)  &    73.6(0.5)  &   76.7(0.2)   &&    51(4) &  57(10) &     1.15(0.06) &&   93.1(0.9)  &  95(2)  & 13.3(0.2)  & HV-R\\  
 RY Tau   &  -14.9(1.5)  &    42.7(0.7)  &   19.51(0.09)  &&        ...	  &  ...	&        $<$1.36 &&       ...    &      ...    &  $<$10.8    & NLV\\ 
          &  -78.1(1.9)  &    59(2)  &   19.17(0.09)  &&         ...  &   	...	&       ...      &&  ...         & ...         &  ...        & HV-B\\ 
 UX Tau   &   -2.1(0.4)  &    27.9(0.2)  &    6.10(0.03)  &&     2.2(1.6) &  28(5)  &    0.44(0.02)  &&   ...        &  ...        &  $<$1.87    & NLV\\	
          &   20.5(1.6)  &   114(2)  &    3.5(0.1)  &&        ...	  &  ...	&         ...    &&       ...    &      ...    &  ...        & BLV\\	 
 UY Aur   &   0.3(0.6)  &    10.9(0.6)  &    0.43(0.05) &&     2.2(0.9) &   15(4) &      0.08(0.01)&&    -0.5(0.9) &   19(2) &  0.13(0.01) & NLV\\
          &   -9.4(0.4)  &    40(1)  &    1.91(0.02)  &&   -15(5) &   46(6) &     0.23(0.01) &&    ...       &   ...       &  ...        & NLV\\
          &  -83.0(1.1)  &   101(2) &   2.40(0.05)   &&  -103(2) &  35(3)  &     0.06(0.01) && -102(2)  &  70(3)  &  0.45(0.04) & HV-B\\
          &   54.0(1.5)  &    93(3)  &   1.56(0.05)   &&    29(3) &   24(3) &     0.05(0.01) &&    ...       &   ...       &  ...        & HV-R\\
 UZ Tau E &    7.0(1.3)  &    54.4(0.5)  &   48.83(0.05)  &&     6(2) &   66(4) &     0.62(0.03) &&  -7.7(3.4)  &  65(6)  &  1.41(0.06) & NLV\\ 
          &  -67.7(1.8)  &    64.8(0.6)  &   44.30(0.06)  &&   -78.0(2.2) &   38(5) &     0.22(0.02) &&  -78.1(2.5)  &  57(5)  &  1.51(0.05) & HV-B\\	 
 V409 Tau &  -43.1(1.4)  &   135(3)  &    0.33(0.02)  &&  5.3(2.5) &   66(7) &     0.07(0.01) &&    ...       &   ...       &  $<$0.09    & HV-B\\
 V807 Tau &    0.83(1.1)  &    42(3)  &    5.09(0.09)  &&        ...	  &  ...	&        $ <$1.39&&       ...    &      ...    &  $<$3.37    & NLV\\
          &  -53.9(3.5)  &    47(5)  &    2.3(0.1)  &&        ...	  &  ...	&         ...    &&       ...    &      ...    &  ...        & HV-B\\
          &   53.9(3.5)  &    42(4)  &    2.95(0.09)  &&        ...	  &  ...	&         ...    &&        ...   &       ...   &  ...        & HV-R\\
 V836 Tau &   -3.7(0.4)  &    50.2(0.7)  &    1.04(0.01)  &&        ...	  &  ...	&        $ <$0.13&&       ...    &      ...    &  $<$0.61    & NLV \\
\hline                                                    
\end{supertabular}
\\
\\
 \footnotesize{Notes:
 ID identify the type of component relative to the \osix\, line. NLVC: $\lvert{V_p}\rvert <$ 30 km\,s$^{-1}$, FWHM$<$55 km\,s$^{-1}$. BLVC: $\lvert{V_p}\rvert <$ 30 km\,s$^{-1}$, FWHM$>$55 km\,s$^{-1}$. HVC-B: ${V_p} < -30$ km\,s$^{-1}$. HVC-R: $V_p > +30$ km\,s$^{-1}$. 
 }
\normalsize
 

%% file: Tables/table9_super.tex
\onecolumn
\footnotesize

\tablecaption{\label{tab:line_fluxes2}Kinematical parameters and luminosities of Gaussian components for the \ssix\, and \nsix\, lines} 

\tablefirsthead{%
\\
   &   \multicolumn{3}{c}{[\ion{S}{ii}]\,673nm} &~~~~ & \multicolumn{3}{c}{[\ion{N}{ii}]\,658nm}\\
   \cline{2-4} \cline{6-8} \\ 
Name     &$V_{\rm p}$ & FWHM & $L_{\rm 673}$ & &$V_{\rm p}$ & FWHM & $L_{\rm 658}$ \\
&[\kms]&[\kms]  &[10$^{-5} L_{\odot}$] & &[\kms] &[\kms] &[10$^{-5} L_{\odot}$] \\
\\
   \cline{2-4} \cline{6-8} \\ }
\tablehead{%
\multicolumn{8}{c}%
{{\bfseries \tablename\ \thetable{} -- continued from previous page}} \\
\\
   &   \multicolumn{3}{c}{[\ion{S}{ii}]\,673nm} &~~~~ & \multicolumn{3}{c}{[\ion{N}{ii}]\,658nm}\\
   \cline{2-4} \cline{6-8} \\ 
Name     &$V_{\rm p}$ & FWHM & $L_{\rm 673}$ & &$V_{\rm p}$ & FWHM & $L_{\rm 658}$ \\
&[\kms]&[\kms]  &[10$^{-5} L_{\odot}$] & &[\kms] &[\kms] &[10$^{-5} L_{\odot}$] \\
\\
   \cline{2-4} \cline{6-8} \\}

\tabletail{%
\hline
\multicolumn{8}{c}{{Continued on next page}} \\
}
\tablelasttail{\hline}

\begin{supertabular}{lccccccc}
 BP Tau	 &     ...	 &     ...	 &     $<$0.07   &&	...	&     ...	&    $<$0.07  \\
         &	...	 &     ...	 &     ...	 &&	...	&     ...	&    ...      \\
 CI Tau  &	...	 &     ...	 &     $<$0.59   &&	...	&     ...	&    $<$0.18  \\
	 &      ...      &    ...	 &     ...	 &&	...	&     ...	&    ...      \\
 CQ Tau  &	...	 &     ...	 &     $<$0.25   &&	...	&     ...	&   $<$0.17  \\
 CW Tau	 &    -0.38(1.9)  &    24(3)  &    2.04(0.01) &&	...	&     ...	&   $<$0.03       \\	     \\
         &   -78.1(1.9)  &    130(13)&    5.45(0.03) &&	 ...	&      ...	&     ...    \\	     
         &  -116.7(1.9)  &    24(3)  &    2.04(0.01) &&  -111.9(1.5)&   43(4) & 2.32(0.01) \\
 CY Tau  &	...	 &     ...	 &     $<$0.03   &&	...	&     ...	&  $<$0.03  \\
 DE Tau  &	...	 &     ...	 &     $<$0.13   &&	...	&     ...	&  $<$0.02   \\
         &	...	 &     ...	 &     ...	 &&	...	&     ...	&    ...     \\
         &	...	 &     ...	 &     ...	 &&  -138.7(0.8)&   17(4)	& 0.06(0.08) \\
 DF Tau  &	...	 &     ...	 &    $<$0.09	 &&	...	&     ...	&   $<$0.1\\
         &	...	 &     ...	 &     ...	 &&	...	&     ...	&    ...      \\
         &	...	 &     ...	 &     ...	 &&  -52.0(1.3) &   61(5)	& 0.54(0.06)  \\
         &  -105.1(1.1)  &    32(2)  &   0.26(0.03)  && -114.1(1.2) &   28(1)	& 0.22(0.02)  \\
         &	...	 &     ...	 &     ...	 &&   18(3) &   43(5)	& 0.22(0.04)  \\
         &	...	 &     ...	 &     ...	 &&	...	&     ...	&    ...     \\
 DG Tau  &	...	 &     ...	 &     $<$0.09   &&	...	&     ...	&   $<$0.1    \\
         &   -17.1(1.9)  &    28.3(0.2)  &    5.13(0.01) &&	...	&     ...	&    ...    \\
         &   -50.4(1.9)  &    69.6(0.5)  &   10.85(0.03) &&  -99(1) &   137(2)  & 7.06(0.07)  \\ 
         &  -126.4(1.9)  &    47.3(0.3)  &    8.82(0.02) &&  -136.7(0.1)&    43.9(0.6)  &  5.98(0.02) \\
 DH Tau  &	...	 &     ...	 &    $<$0.11	 &&	...	&     ...	&  $<$0.11   \\
         &	...	 &     ...	 &     ...	 &&	...	&     ...	&    ...     \\	            	     
 DK Tau  &	...	 &     ...	 &     $<$0.09   &&	...	&     ...	&   $<$0.03   \\  
         &   -69.0(3.8)  &   56.2(10.2)  &   0.51(0.03)  &&    ...	&    ...	&   ... \\
         &  -131.8(3.0)  &   44(4)   &   0.52(0.02)  && -141.1(3.9) &   33(7)	& 0.24(0.01)  \\
 DL Tau  &	...	 &     ...	 &    $<$0.94	 &&	...	&     ...	&   $<$0.94   \\
         &  -113.7(1.2)  &  129(8)   &  1.00(0.05)   &&	...	&     ...	&    ...  \\
         &  -156.0(1.0)  &  34(5)	 &  0.54(0.01)   &&-165.5(0.7)  &   20(1)	& 0.18(0.06) \\
 DN Tau  &	...	 &     ...	 &    $<$0.12	 &&	...	&     ...	&  $<$0.10    \\
 DO Tau  &	...	 &     ...	 &    $<$0.22	 &&	...	&     ...	&   $<$0.22   \\
         &   -86.2(1.1)  &   58(2)   &    1.92(0.02) &&	...	&     ...	&    ...           \\
         &  -102.6(0.8)  &   17.4(0.4)   &    1.32(0.01) &&	...	&     ...	&    ...    \\   
 DQ Tau  &   -9.5(0.9)  &   44.3(1.5)   &    0.94(0.02) &&	...	&     ...	&  $<$0.27  \\   
         &   -93.0(2.7)  &   51(6)   &    0.32(0.02) &&	...	&     ...	&    ...    \\
 DR Tau  &     1.1(1.9)  &   78(10)   &   0.82(0.06)  &&    ...	&    ...	& $<$0.06  \\	 
         &	...	 &     ...	 &     ...	 &&	...	&     ...	&    ...    \\
         &	...	 &     ...	 &     ...	 &&	...	&     ...	&    ...    \\  
         &  -254.9(0.8)  &   25(2)   &   0.39(0.02)  && -253.8(0.5) &    36(1)  &  1.11(0.02)\\
 DS Tau  &     ...	 &     ...	 &    $<$0.14	 &&	...	&     ...	&  $<$0.16  \\
 FT Tau  &	...	 &     ...	 &    $<$0.1	 &&	...	&     ...	&   $<$0.1  \\		
         &  -119.2(3.2)  &  108.9(8.5)   &   0.10(0.01)  && -155.4(0.7) &   18(2)	& 0.030(0.003)    \\
 GG Tau  &    -8.6(5.3)  &  61(17)   &   0.61(0.03)  &&    ...	&    ...	& $<$0.30 \\
         &	...	 &     ...	 &     ...	 &&	...	&     ...	&    ...   \\  		      
 GH Tau  &	...	 &     ...	 &    $<$0.11	 &&	...	&     ...	&  $<$0.11  \\         	  
         &	...	 &     ...	 &     ...	 &&	...	&     ...	&    ...   \\	   
 GI Tau  &	...	 &     ...	 &    $<$0.03	 &&	...	&     ...	&  $<$0.14  \\	    
         &	...	 &     ...	 &     ...	 &&	...	&     ...	&    ...       \\		       
         &    47.2(4.8)  &   44(32)  &    0.34(0.01) &&	...	&     ...	&    ...    \\      
 GK Tau  &    3.6(3.8)   &   15(17)  &    0.18(0.01) &&	...	&     ...	&  $<$0.32 \\     
         &	...	 &     ...	 &     ...	 &&	...	&     ...	&    ...    \\     
 GM Aur  &	...	 &     ...	 &    $<$0.34	 &&	...	&     ...	&  $<$0.42  \\		     
 HN Tau  &  -7.2(0.9)    & 30(4)	 &  0.780(0.004)  &&	...	&     ...	& $<$0.06  \\ 
         &  23.1(1.1)    & 28.5(5.0)	 &  0.600(0.003)   &&	...	&     ...	&  ...    \\	 
         &  30.7(5.3)    & 64(13)	 &  0.390(0.008)  &&	...	&     ...	&  ...  \\	  
         &  -39(2.6)	 & 56(4)	 &  0.700(0.007)   &&	...	&     ...	&  ...   \\	  
         &  -90.5(6.8)   & 87(6)	 &  1.02(0.01)   &&  -84.1(6.9) &   158(13) &   1.80(0.02)\\	   	     
 HQ Tau  &   -3.8(1.8)   &   84.0(3)   &   11.92(0.06) &&	...	&     ...	&   $<$1.82  \\ 
         &	...	 &     ...	 &     ...	 &&	...	&     ...	&    ...    \\ 
 IP Tau  &	...	 &     ...	 &   $<$0.15	 &&	...	&     ...	&   $<$0.13 \\ 
 IQ Tau  &   -8.3(0.8)   &   94(15)  &   0.60(0.04)   &&    ...	&    ...	&  $<$0.03 \\ 
         &   -30.3(0.5)  &   26(1)   &    1.70(0.01)  &&   -32.9(0.4)&   31.1(0.7)	& 0.88(0.01)  \\ 
 LkCa 15 &	...	 &     ...	 &    $<$0.16	 &&	...	&     ...	&  $<$0.11  \\ 
 MWC480  &	...	 &     ...	 &    $<$0.98	 &&	...	&     ...	&   $<$1.41 \\
 RW AurA &   -122.21(1.6)&   170(3)  &  15.6(0.3)  &&  -160.1(1.6)&   86.8(2.1)	& 0.007(0.004)    \\ 
         &    -73.1(0.4) &   41(1)   &    3.72(0.07) &&	...	&     ...	&    ...   \\  
         &     76.5(0.5) &   61.1(0.9)   &   17.1(0.1) &&  96.7(0.5)  &   53.9(0.8)	& 0.003(0.002) \\
 RY Tau  &	-20.82(0.9)	 &  51.3(2.5) &  4.60(0.04)	 &&	...	&     ...	&   $<$1.3   \\  
         &	...	 &     ...	 &     ...	 &&	...	&     ...	&    ...       \\ 
 UX Tau  &	...	 &     ...	 &    $<$0.51	 &&	...	&     ...	&  $<$0.37  \\ 		     
         &	...	 &     ...	 &     ...	 &&	...	&     ...	&    ...     \\    		      
 UY Aur  &     -0.5(0.6) &   17(2)   &    0.20(0.01) &&	...	&     ...	&  $<$0.13   \\     
         &    -13.7(0.6) &   25.5(1.7)   &    0.35(0.01) &&	...	&     ...	&    ...   \\
         &	...	 &     ...	 &     ...	 &&	...	&     ...	&    ...    \\
         &	...	 &     ...	 &     ...	 &&	...	&     ...	&    ...    \\
 UZ Tau E&     -7.7(3.2) &    61.5(4.5)  &   0.74(0.05)  &&    ...	&    ...	&  $<$0.12 \\		     
         &    -82.4(2.2) &    54(5)  &   0.78(0.04)  && -85(2)  &  56.0(4.4)	& 0.88(0.04) \\ 		     
 V409 Tau&	...	 &     ...	 &    $<$0.05	 &&	...	&     ...	&   $<$0.05  \\    
 V807 Tau&	...	 &     ...	 &    $<$0.89	 &&	...	&     ...	&   $<$0.88  \\
         &	...	 &     ...	 &     ...	 &&	...	&     ...	&    ...       \\
         &	...	 &     ...	 &     ...	 &&	...	&     ...	&    ...      \\
 V836 Tau&	...	 &     ...	 &    $<$0.15	 &&	...	&     ...	&   $<$0.11 \\	     
\hline                                                    
\end{supertabular}
\normalsize